\documentclass[aps,pre,subeqn,twocolumn,epsfig,dvips,groupedaddress,amsmath,amssymb,eqsecnum]{revtex4}
\usepackage{latexsym}           
\usepackage[dvips]{graphicx}            
\usepackage{rotating}           
\usepackage{amsmath}
\usepackage{mathrsfs}
\usepackage{endnotes}
\newcommand{\rv}{\vec{r}}
\newcommand{\xv}{\vec{x}}
\newcommand{\Xv}{\vec{X}} 

\newcommand{\eh}{\hat{e}}
\newcommand{\uv}{\vec{u}}

\newcommand{\qv}{\vec{q}}

\newcommand{\nh}{\hat{n}}
\newcommand{\mm}[1]{{\bf #1}}

\newcommand{\Qm}{{\mm{Q}}}
\newcommand{\Pn}{{\mm{P}_{n}}}
\newcommand{\Pm}{{\mm{P}^{\perp}_n}}
\newcommand{\Lm}{{\boldsymbol{\Lambda} }}
\newcommand{\lm}{{\boldsymbol{\lambda} }}
\newcommand{\pv}{\vec{p}}
\newcommand{\Tr}{{\rm Tr}}

\newcommand{\pz}{{\partial_z}}
\newcommand{\ppi}{{\partial_i}}
\newcommand{\ppj}{{\partial_j}}
\newcommand{\ppz}{{\partial_z}}
\newcommand{\ppk}{{\partial_k}}
\newcommand{\ppl}{{\partial_l}}
\newcommand{\ppa}{{\partial_a}}

\newcommand{\calH}{{\mathcal H}}

\newcommand{\cc}{{\mathscr C}}

\newcommand{\ct}{{\mathscr T}}
\newcommand{\zp}{\zeta_{\parallel}}
\newcommand{\zt}{\zeta_{\perp}}

\def\rf#1{(\ref{#1})}
\def\rfs#1{Eq.~\rf{#1}}

\begin{document}
\title{Nonlinear Elasticity, Fluctuations and Heterogeneity of 
Nematic Elastomers}
\author{Xiangjun Xing}
\affiliation{Department of Physics, Syracuse University, Syracuse, NY 13244} 
\author{Leo Radzihovsky}
\affiliation{Department of Physics, University of Colorado, Boulder,
  CO 80309}%

\date{\today}

\begin{abstract}
  Liquid crystal elastomers realize a fascinating new form of soft 
  matter that is a composite of a conventional crosslinked polymer gel
  (rubber) and a liquid crystal.  These {\em solid}\; liquid crystal
  amalgams, quite similarly to their (conventional, fluid) liquid
  crystal counterparts, can spontaneously partially break
  translational and/or orientational symmetries, accompanied by novel
  soft Goldstone modes.  As a consequence, these materials can exhibit
  unconventional elasticity characterized by symmetry-enforced
  vanishing of some elastic moduli.  Thus, a proper description of
  such solids requires an essential modification of the classical
  elasticity theory.  In this work, we develop a {\em rotationally
    invariant}, {\em nonlinear} theory of elasticity for the nematic
  phase of ideal liquid crystal elastomers.  We show that it is
  characterized by soft modes, corresponding to a combination of long
  wavelength shear deformations of the solid network and rotations of
  the nematic director field. We study thermal fluctuations of these
  soft modes in the presence of network heterogeneities and show that
  they lead to a large variety of anomalous elastic properties, such
  as singular length-scale dependent shear elastic moduli, a divergent
  elastic constant for splay distortion of the nematic director,
  long-scale incompressibility, universal Poisson ratios and a
  nonlinear stress-strain relation for arbitrary small strains.  These
  long-scale elastic properties are {\em universal}, controlled by a
  nontrivial zero-temperature fixed point and constitute a qualitative
  breakdown of the classical elasticity theory in nematic elastomers.
  Thus, nematic elastomers realize a stable ``critical phase'',
  characterized by universal power-law correlations, akin to a
  critical point of a continuous phase transition, but extending over
  an entire phase.
\end{abstract}


\maketitle



%



\section{Introduction}
\label{Sec:Intro}

\subsection{Classical and generalized elasticity}

The classical theory of elasticity \cite{EL:Love,EL:Landau} is a
well-established discipline developed in the mid-nineteen century by
Hooke, Cauchy, Poisson, Green and many others.  It models conventional
solid materials as continuous media with an energetically preferred,
stress-free, reference configuration.  In this theory, the elastic
energy of a solid is a functional of the deformation tensor, defined
relative to the aforementioned stress-free reference state.  As shown
by Cauchy, the conservation of angular momentum requires the stress
tensor in conventional solids to be symmetric, encoding the underlying
physics that torque cannot be transmitted through the media without
the transmission of force.

The development of solid-state physics provided a microscopic
understanding of inter-atomic cohesive forces inside crystalline
solids, and allowed a first-principle derivation of their elasticity,
as well as the computation of the associated elastic constants.

In another major development, complementing this microscopic
description, it was appreciated that the rigidity of crystalline
solids, which distinguishes them from liquids, arises as a consequence
of spontaneous breaking of translational symmetry. Transcending
microscopics, the functional form of the associated elastic energy is
completely determined by the symmetry, i.e., the space symmetry group,
of the resulting ordered crystalline state. From this perspective, the
spontaneous nature of crystalline ordering requires the existence of
the low-energy, long-wavelength excitations, i.e., phonons, that are
the Goldstone modes \cite{CMP:Anderson} associated with the
spontaneous breaking of translational symmetry \cite{comment:Halperin}.

This intimate relation between symmetry breaking, generalized rigidity
and elasticity is one of the most important cornerstones of modern
condensed matter physics \cite{CMP:Anderson}.  It extends to a wide
class of system with translational, orientational, and other more
exotic orders, such as magnets, superfluids, and liquid 
crystals \cite{PhysicsTodayPalffy}. Each system is characterized by its own set
of Goldstone modes and generalized rigidity, corresponding to the
symmetry spontaneously broken by the particular ordered state. For
example, a nematic liquid crystal breaks the rotational, but not
translational, symmetry, and, as a result, is able to transmit a
torque, but not a force.  It thereby admits an asymmetric stress
tensor, i.e., contains both a symmetric and an antisymmetric part,
in strong contrast to an ordinary isotropic liquid.  A smectic liquid
crystal, on the other hand, breaks translational symmetry in one
direction, along which it is able to transmit force.  The term
(generalized) elasticity, therefore gains much wider meaning in the
context of modern condensed matter physics, applying to {\em all}
thermodynamic phases that exhibit a spontaneous long-range order.

\subsection{Classical rubber elasticity}

Despite the tremendous success of this symmetry breaking description
there are classes of solids whose elasticity do not fit into this
paradigm. One example is a structural glass, which, despite of not
breaking any spatial symmetries, exhibits a macroscopic isotropic
solid-like elasticity on experimental time scales \cite{comment:glass}. 

Elastomers (rubber) constitute another class of materials that break
no spatial symmetries but nevertheless, macroscopically are true
three-dimensional solids, in the sense that they do not flow and have a
finite shear modulus.  Microscopically, elastomers are polymer
networks formed by crosslinking polymer melts.  Strain deformations
decrease the entropy of constituent fluctuating polymer chains, and
therefore cost elastic free energy.  The shear modulus of elastomers
is proportional to temperature \cite{EL:Treloar}, and is typically $4-5$
orders of magnitude smaller than that of crystalline solids.  Due to
the flexibility of polymer chains, elastomers can support extremely
large (typically several hundred percent) elastic shear
deformations.

The elasticity of isotropic rubber is usually understood within the
framework of {\em classical theory} of rubber elasticity
\cite{EL:Treloar}, developed long time ago by Kuhn, Wall, Flory,
Treloar and others.  This theory treats, in an obviously erroneous
way, the crosslinks of polymer networks as stationary objects that
nevertheless deform affinely with the network when it is subjected to
an external strain deformation.  The total entropy of the network is
then simply the sum of those of individual chains.  This theory
explains the stress-strain relation of rubber for small strain
reasonably well but fails at larger deformations.  A number of
extensions of the classical rubber theory \cite{EL:Treloar} have been
developed subsequently, which attempt to include non-Gaussian chain
statistics, irreversibility, short-scale nematic and crystalline
ordering, and in particular polymer entanglements 
\cite{Edwards-tube,Rubinstein-tube,Terentjev-tube}.  The success of
these efforts however are rather difficult to assess, largely due to
the fact that they do not make prediction other than stress-strain
relations.  Thus, despite their technological importance and abundant
existence, our understanding of elastomers remains rather incomplete,
and in part is due to the lack of proper description of their random
structures \cite{comment:vulcanization}.  

A recent exciting development is the observation and the subsequent
resolution \cite{XGR-rubber} of internal inconsistency of the
classical rubber theory.  Namely, that the entropy of fluctuating
crosslinks, ignored by the classical theory, is comparable to that of
the polymer chains.  Taking into account phonon fluctuations on scales
longer than the network mesh size, while incorporating rubber's near
incompressibility, one finds quantitative agreement with the
stress-strain relation of rubber in the large deformation regime
\cite{XGR-rubber}, without appealing to any other more complicated and
poorly characterized mechanism.

\subsection{Nematic liquid-crystalline elastomers}
Another fascinating, relatively new class of elastic materials that
falls outside the scope of classical crystal elasticity is that of
liquid crystalline elastomers \cite{LCE:WT,WarTer96,Terentjev99,PhysicsTodayPalffy}.
\begin{figure}[!htp]
\begin{center}
\includegraphics[width=8cm]{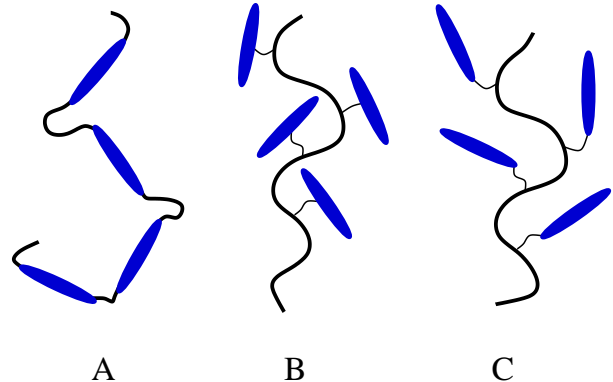}
\caption{An illustration of (A) main-chain liquid crystalline polymer and (B), (C) side-chain liquid crystalline polymers.}
\label{LC-polymer}
\end{center}
\end{figure}
These are crosslinked networks of liquid crystalline polymers that incorporate mesogenic groups, i.e., chemical groups with a sufficient anisotropy to exhibit liquid crystalline orders.  These mesogens may either be parts of the polymer backbone, separated by flexible chemical spacer groups, as in the case of main-chain liquid crystalline polymers, or be attached to the main polymer backbone, as in the case of side-chain liquid crystalline polymers, illustrated in Fig.~\ref{LC-polymer}.  Quite similarly to their liquid crystal counterparts, liquid crystalline elastomers may also spontaneously partially break translational and/or rotational symmetry, when the anisotropic interaction between mesogenic groups is sufficiently strong.  The resulting phases have long-range liquid crystalline order in the background of a random, statistically isotropic polymer network.  The coupling between network elasticity and liquid crystalline order leads to a rich phenomenology \cite{LCE:WT,PhysicsTodayPalffy} and an exciting possibility of controlling one using the other.  

Nematic elastomers \cite{WarTer96,LCE:WT} are nematically-ordered rubbery materials, prepared by crosslinking nematic polymer melts or solutions under appropriate conditions.  In an idealized approach, one would crosslink a homogeneous polymer melt in the isotropic phase, and subsequently lower the temperature so that the nematic order appears spontaneously.  We shall refer to such a fictitious nematic elastomer as {\em ideal}.  In reality, however, network heterogeneity interferes with the nematic ordering and a system prepared this way only exhibits a finite-range nematic order at low temperature.  The precise mechanism leading to suppression of long-range nematic order is, however, not yet completely understood \cite{comment:possibilities}.  
To obtain a single-domain nematic elastomer, the crosslinking is instead done slightly in the nematic phase or under a weak strain \cite{KupferFinkelmann91}.  Hence  a weak nematic order is permanently imprinted into the system, leading to a small free energy cost for the Goldstone modes of the ideal systems, a property referred to as semi-softness \cite{SemiSoftLubensky}.   In this work, however, we shall only consider ideal homogeneous nematic elastomers crosslinked under isotropic conditions.  We shall consider the effects of network heterogeneity toward the end of this paper.

\begin{figure}[!htp]
\label{IN-transition}
\begin{center}
\includegraphics[width=8cm]{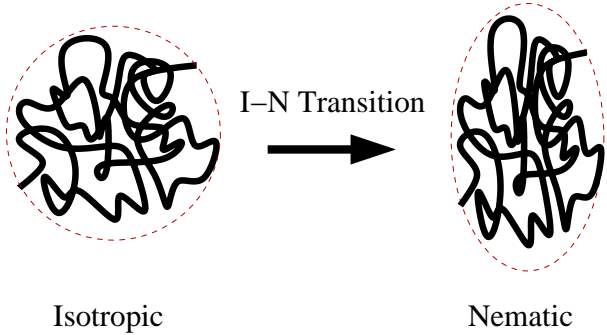} 
\caption{A cartoon of a liquid crystalline polymer coil undergoing the
  isotropic-nematic transition, driven by the nematic ordering of the
  mesogenic groups (not shown in the figure).  In the isotropic phase
  the mesogens are randomly oriented in the space, while in the
  nematic phase they are align on average.  In an elastomer, this is
  accompanied by a uniaxial distortion of polymer coils that in turn
  leads to a spontaneous uniaxial deformation of the underlying
  polymer network, i.e., the solid itself. }
\end{center}
\end{figure}

The interplay between liquid crystalline order and network elasticity 
becomes most interesting when the liquid crystalline order appears
spontaneously, and the Goldstone theorem dictates that there must
exist corresponding soft modes whose elastic energy vanishes in the
long wavelength limit.  
This fascinating possibility has been explored in most detail in
nematic elastomers.  Due to the coupling between the mesogenic groups
and the polymer backbones, the isotropic-nematic (I-N) transition in a
liquid crystalline polymer melt is accompanied by a spontaneous
uniaxial deformation of all polymer coils, illustrated in
Fig.~\ref{IN-transition}
In a crosslinked network of polymers, such a shape change of all
polymer coils implies the same shape change of the whole network,
i.e. a spontaneous uniaxial stretch of the elastomer
\cite{KupferFinkelmann91,Selinger02}.  Since the rotational symmetry
is broken {\em spontaneously}, there exists a continuous manifold of
degenerate ground states inside the nematic phase, corresponding to
distinct choices of the uniaxial axis in the statistically isotropic
solid.  It is important to note that these degenerate states are {\em
  not} simply related to each other by global rotations of the sample.
Only a special combination of shear deformation of the solid and
rotation of nematic director can take the system from one such ground
state to another.  These soft modes, first predicted by Golubovi\'{c}
and Lubensky \cite{GolLub89} based purely on symmetry consideration,
are responsible for most of the unusual properties of nematic
elastomers.  As noted by these authors, the effective elastic free
energy of the nematic elastomer resembles that of a uniaxial solid,
but with a strict vanishing of one of the five elastic constants,
enforced by the spontaneous breaking of rotational symmetry and
encoding the presence of the corresponding Goldstone mode.  This then
leads to a striking elastic response, where under a shear transverse
to the nematic axis, the stress-strain relation of an ideal nematic
elastomer exhibits an extended plateau with a vanishing stress
\cite{KupferFinkelmann94,Clarke01,Warner99}.

Complementary to this pioneering work, a very successful molecular
level elasticity theory of ideal nematic elastomers, referred to as
the {\em neo-classical} theory, was developed by Warner and Terentjev
\cite{WarTer96,LCE:WT}.  Essentially a spontaneously anisotropic
generalization of the classical theory of rubber elasticity
\cite{EL:Treloar}, it successfully captures the soft modes of ideal
nematic elastomers.  Characterized by only two parameters, the shear
modulus $\mu$ and the nemato-elastic coupling constant, which
determines the spontaneous distortion ratio, the neo-classical theory
provides a minimal model remarkably useful for practical calculations
\cite{LCE:WT}.  It is important to note that this elastic energy
cannot be expressed in terms of any {\em symmetric} strain tensor,
linear or nonlinear.  This reflects the facts that nematic elastomers
are not ordinary solids as understood by Cauchy and his
contemporaries, and that a correct elastic description of these
materials necessarily leads to physics beyond the framework of
classical elasticity theory.

It is worthy to note that theoretical attempts to go beyond the
classical elasticity theory date back to early nineteenth century, by
Voigt and Cosserat brothers \cite{comment:history}.  These authors
considered fictitious elastic media, i.e., the so-called Cosserat
medium, that consists of three-dimensional rigid constituents with
both translational and rotational degree of freedom, quite analogous
to the modern realization in nematic elastomers.  It was understood
quite early that these media can transmit torque independent of force,
and therefore admit asymmetric stress tensors.  Receiving little
attention for a long time, research in Cosserat elasticity was revived
about half a century later in the engineering community under the new
name of {\em micropolar media}.  However, almost all of theses studies
focus on isotropic micropolar elastic media, lacking the
strain-rotation coupling.  As discussed above, this coupling is the
most interesting ingredient of the elasticity of micro-polar media and
underlies all of their unusual properties.

\subsection{Thermal fluctuations, heterogeneity and anomalous
  elasticity: critical phases}

\begin{figure}[!htp]
\begin{center}
\includegraphics[width=7cm]{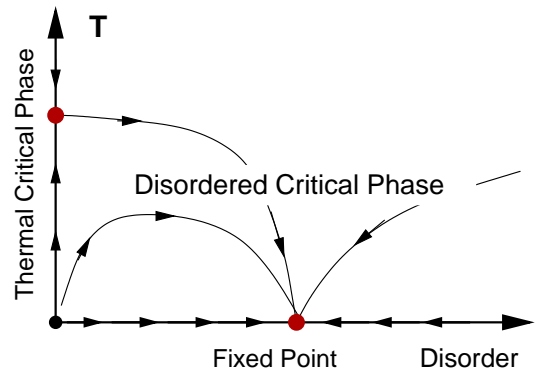}
\caption{A schematic renormalization group flows for an anomalous
  solid, such as a nematic elastomer.  The elasticity of an ideal
  homogeneous system is controlled by a finite temperature
  non-Gaussian fixed point on the vertical temperature axis.  This
  fixed point is unstable with respect to infinitesimal heterogeneity.
  The elasticity of a heterogeneous system is controlled by a
  zero-temperature, finite disorder non-Gaussian fixed point, that
  endows the nematic gel with its anomalous universal elastic
  properties. }
\label{flow_figure}
\end{center}
\end{figure}

In spite of its remarkable success \cite{LCE:WT}, the principle shortcomings of the neo-classical theory are evident.  As
already discussed above, being built on the classical theory of rubber
elasticity it misses a substantial entropic contribution associated
with fluctuations of the network crosslinks \cite{XGR-rubber}.
Neither does it address network heterogeneity, i.e., elastomers
quenched random structure.  As have been recently demonstrated in a
series of publications \cite{Xing-Radz1,Xing-Radz2,StenLub,StenLub-2},
thermal fluctuations and network random heterogeneity have even more
drastic effects on the macroscopic elasticity of {\em nematic}
elastomers, as compared with the conventional isotropic rubber.  This
is mainly due to the aforementioned existence of soft Goldstone modes,
together with the resulting strict vanishing of one of the shear
moduli \cite{GolLub89,LMRX}.  As we will show in detail below, the
presence of local fluctuations of these soft modes, relevant elastic
nonlinearities lead to a {\em qualitative} modification of the
macroscopic elasticity of nematic elastomers.  As a result, at long
scales these materials exhibit a variety of anomalous elastic
properties.  Namely, all shear moduli, as well as the splay constant
for nematic distortion, become singular functions of the probing
length-scale and macroscopic strain deformation, and either diverge or
vanish in the thermodynamic limit.  Related to these, the anomalous
solid exhibits a long-scale incompressibility, universal Poisson
ratios, as well as nonlinear stress-strain relation for arbitrary
small strains.  These properties are {\em universal}, controlled by a
non-Gaussian, finite-temperature (for idealized homogeneous
elastomers) fixed point, as illustrated in the vertical temperature
axis in Fig.~\ref{flow_figure}, and constitute a qualitative breakdown
of classical elasticity theory.  Similar phenomena have been
previously shown to appear in other ``soft'' solids, such as smectic
and columnar liquid crystals \cite{GP1,RTaerogel1,RTaerogel6},
polymerized membranes \cite{NelsonPeliti,membrane_AGL} and putative
spontaneous vortex lattices in ferromagnetic superconductors
\cite{RT:MSC,RT:MSC-2}.

These effects are similar but qualitatively stronger in (more
realistic) elastomers with random heterogeneity, controlled by a
zero-temperature, finite-disorder fixed point, as illustrated in the
horizontal axis of Fig.~\ref{flow_figure}.  It characterizes the
``rough'' ground state of a heterogeneous elastomer that exhibits
large sample-to-sample fluctuations relative to an idealized
homogeneous elastomer \cite{RTaerogel1,Xing-Radz2}.

\begin{figure}[!htp]
\begin{center}
\includegraphics[width=6.5cm]{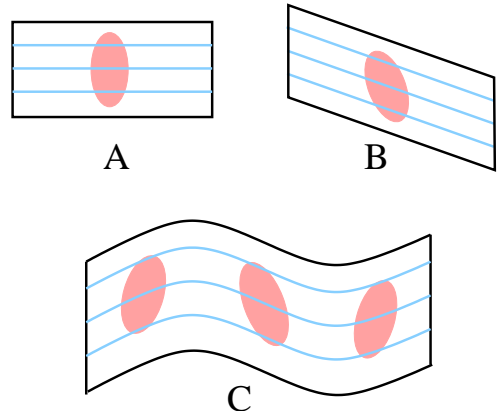}
\caption{Local fluctuations of soft deformations is the mechanism of
  anomalous elasticity in nematic elastomers. (A): the reference
  state of an ideal nematic elastomer. The ellipse denotes the
  uniaxial nematic order.  The light horizontal lines are
  visualization of local translation parallel to the director, i.e.,
  the $u_z$ phonon field.  (B): a soft volume-preserving
  deformation, which costs zero elastic energy in an ideal elastomer.
  It is a simple shear deformation combined with a uniform rotation of
  nematic order.  (C): long wavelength fluctuations of such soft
  deformations are the dominant low-energy fluctuations.  Their energy
  cost is the Frank free energy for splay distortion of nematic
  director field.}
\label{visualization}
\end{center}
\end{figure}

The physical mechanism of anomalous elasticity in nematic elastomers
can be visualized by considering the low-energy elastic excitations,
illustrated in Fig.~\ref{visualization}.  A rotation of the nematic
director and a simultaneous uniform shear deformation, shown in
Fig.~\ref{visualization}B, is a soft mode and costs zero elastic free
energy \cite{comment:similarity}.  It is important to note the total
effect of this deformation is {\em not} a simultaneous global rotation
of the solid and the nematic director.  Therefore as shown in
Fig.~\ref{visualization}C, long wavelength fluctuations of these soft
deformations are only penalized by the Frank free energy for
distortion of nematic director.  As we will show, at finite
temperature\cite{Xing-Radz1,StenLub,StenLub-2}, and/or in the
presence of elastomer's random heterogeneity\cite{Xing-Radz2}, these
fluctuations diverge with the system size, and hence necessitate a
nonperturbative treatment of relevant elastic nonlinearities.  All
modes of the elastic deformation that are coupled to the soft modes
are strongly affected by these fluctuations, leading to the
aforementioned anomalous elasticity.  By contrast, because the volume
remains fixed in soft deformations, the bulk mode is not coupled to
the soft mode.  Consequently the bulk modulus of nematic elastomers is
only weakly affected by fluctuations.

It is worth stressing that the renormalization group behavior
characteristic of nematic elastomers (and other soft solids), where,
as displayed in Fig.~\ref{flow_figure}, the low-temperature ordered
phase is controlled by a {\em non}-Gaussian fixed point, is quite
exotic from a field-theoretic point of view.  A more standard scenario
is an ordered phase controlled by a Gaussian (noninteracting) fixed
point, with fluctuations and nonlinearities only qualitatively
important when the system is fine-tuned to a critical point,
controlling a continuous transition out of the ordered state.  In
contrast, for nematic elastomers and other soft solids
\cite{GP1,RTaerogel1,RTaerogel6,NelsonPeliti,membrane_AGL,RT:MSC,RT:MSC-2},
fluctuations and nonlinearities are important throughout the ordered
phase, i.e., no fine-tuning is necessary.  Consequently, the whole
ordered phase displays the anomalous elasticity with nontrivial and
universal power-law correlations, similar to the behavior near a
critical point of more conventional systems.  It is thus appropriate
to refer to a nematic elastomer and other similar soft solids, like
the smectic and columnar liquid crystals and flat phase of tensionless
polymerized membranes, as a {\em ``critical phase''} \cite{comment:quantum}.

\subsection{Outline}

In this paper we give a detailed and pedagogical report of our recent
works on the nonlinear elasticity and fluctuations in homogeneous as
well as heterogeneous nematic elastomers.  Partial results of this
study were published earlier in brief communications
\cite{Xing-Radz1,Xing-Radz2}.  The remainder of this paper is
organized as follows.  In Sec.~\ref{Sec:Model}, we introduce basic
concepts for the development of elasticity theory for nematic
elastomers and discuss the underlying symmetries as well as the
corresponding Goldstone soft modes.  Based on these we construct a
general elastic free energy for ideal nematic elastomers, in terms of
a nonlinear strain tensor which completely encodes the soft modes.
We also discuss the relation between this elastic free energy and the
neo-classical theory.

In Sec.~\ref{Sec:Asymmetric-elasticity}, we study the elasticity of
nematic elastomers in the context of {\em micropolar elasticity
  theory}.  We demonstrate that the conservation of angular momentum
ensures that such medium is characterized by a generically {\em
  asymmetric} stress tensor, which encodes angular momentum transfer
between the translational and the internal nematic degree of freedom.
By enforcing mechanical equilibrium, we derive the stress tensor and
the couple-stress tensor of a nematic elastomer as appropriate
derivatives of the elastic free energy.  We also explicitly calculate
the Cauchy stress tensor and the couple-stress tensor for two elastic
energy models.

In Sec.~\ref{Sec:Elast-fluct}, we formulate the macroscopic elasticity
theory of nematic elastomers at a finite temperature using the
constant-strain and the constant-stress thermodynamic ensembles.
These two ensembles are related by a Legendre transformation, and
contains the same macroscopic physics.  We then proceed to construct
an effective strain-only description of a nematic elastomer by
expanding the elastic free energy in terms of the nonlinear, fully
rotationally invariant Lagrangian strain tensor of the nematic state.
We derive five Ward identities imposed by the underlying rotational
symmetry.  These identities relate the coefficients of {\em all}
anharmonic terms to that of harmonic ones, as well as require a strict
vanishing of one of the five elastic harmonic elastic moduli that
characterize a uniaxial solid.

In Sec.~\ref{Sec:Thermal-fluct}, we use this effective theory to
analyze the effects of thermal fluctuations on the long-scale elastic
properties of ideal, homogeneous nematic elastomers.  We first study the
fluctuations at the harmonic level and demonstrate that on
sufficiently long length-scales, elastic nonlinearities always become
qualitatively important for three and lower dimensional elastomers.  We
then employ the machinery of renormalization group transformation (RG)
to systematically study the long-scale effects of thermal fluctuations
in the nonlinear theory.  Incorporating the most relevant
nonlinearities, we compute the resulting long-scale, universal
anomalous elasticity of homogeneous nematic elastomers.

In Sections~\ref{Sec:Disorder-fluct} and \ref{Sec:ZetoT}, we study the
effects of network heterogeneity in nematic elastomers.  We first
identify the most relevant type of quenched disorder and show that at
a harmonic level it leads to distortions relative to the ideal nematic
reference state, that diverge for dimensions five and below.  To
characterize long-scale effects of such distortions we treat the
nonlinear elasticity using renormalization group technique and show
that the long-scale elasticity of a heterogeneous nematic elastomer is
indeed controlled by an aforementioned zero-temperature,
finite-disorder fixed point.  We calculate the renormalized
correlation functions and show that all elastic constants, except the
bulk modulus, are singular functions of probing length-scales.
Consequently, the stress-strain relation is strictly nonlinear for
arbitrary small deformation.  We further show that the spatial
correlation of the non-affine deformation field acquires anomalous
scaling at long length-scales.  Finally we conclude this paper with a
discussion of experiments and future research directions.

\section{Symmetries, Soft Modes, and Nemato-Elastic Theory}
\label{Sec:Model}
 
\subsection{Basic ingredients}

An ideal nematic elastomer is a homogeneous elastic medium with
internal orientational degree of freedom (characterized by the nematic
order parameter), which may {\em spontaneously} break the underlying,
internal rotational symmetry.  Similarly to a conventional (fluid)
liquid crystal it thus exhibits a high-temperature isotropic phase and
a low-temperature nematic phase, which are separated by a
thermodynamic phase transition \cite{comment:first-order}.  Inside the
isotropic phase and for small distortions, its elasticity is described
by the conventional elasticity theory of isotropic solids
\cite{EL:Love,EL:Landau}, characterized by two Lam\'e coefficients.

A conformation of such $d$-dimensional ordinary elastic medium is
described by a $d$-dimensional vector-valued function $\rv(\Xv)$,
which specifies the position of each mass point labeled by $\Xv$, that
is also a $d$-dimensional vector.  Through out the paper, we will
always choose the labeling $\Xv$ such that it coincides with the
average position of the corresponding mass point in the
high-temperature isotropic phase of the system.  We shall refer to
$\Xv$ as the isotropic referential, or Lagrangian, coordinate
\cite{comment:notation}.  The coordinate space where the vector $\Xv$
``lives'' shall be referred to as the {\em isotropic reference} space
and the corresponding state $\rv \equiv \Xv$ as the {\em isotropic
  reference} state (IRS).  The vector $\rv(\Xv)$, denoting the spatial
position of a mass point $\Xv$ in an arbitrary deformed state will be
referred to as the current or Eulerian coordinate and the space in
which it ``lives'' as the {\em embedding} space.  We shall always use
symbol $\vec{a}$, $\vec{b}$, etc, with a vector on the top for all
vectors, and use boldface characters such as $\mm{Q}$ and $\lm$ for
matrices and rank-two tensors, use symbols $\hat{a}$, $\hat{b}$, etc,
for unit vectors with magnitude one.

Using a {\em fixed} orthonormal coordinate system
$\{\eh_1,\eh_2,\ldots,\eh_d = \hat{z}\}$, we may decompose the vectors
as $\Xv = \sum_{a = 1}^d X_a \eh_a$, and $\rv = \sum_{i = 1}^d r_a
\eh_a$.  The elastic deformation can be locally characterized by a
$d\times d$ matrix function
\begin{equation}
\Lambda_{ia}(\Xv) = \frac{\partial r_i}{\partial X_a},
\label{Lambda_def}
\end{equation}
which is usually referred to as the {\em deformation gradient} or as
the Cauchy deformation tensor.  Note that $\Lambda_{ia}$ is a ``mixed
tensor'' with one index $a$ in the reference space and the other index
$i$ in the embedding space.  They transform differently under
rotations in each space.

Starting from the isotropic phase, the isotropic-nematic phase
transition is characterized by a development of the nematic order,
accompanied by a spontaneous uniaxial strain deformation.  The nematic
order is characterized by a second rank symmetric traceless tensor
field $\mm{Q}(\Xv)$, that naturally vanishes in the isotropic
reference state (IRS).  In a low-temperature (uniaxial) {\em nematic
  reference} state (NRS), it is given by
\begin{equation}
\mm{Q}_0 =  S \,\left( \hat{n}_0\hat{n}_0 - \frac{1}{3}\,\mm{I} \right),
\label{Q_0}
\end{equation} 
where the unit vector $\hat{n}_0$ is the nematic director that gives
the uniaxial nematic direction in the NRS and $S$ is the magnitude of
$\mm{Q}$ that characterizing the strength of nematic order; $\mm{I}$
is the identity matrix tensor.  Furthermore, the nematic reference
state is related to the isotropic reference state via a spontaneous
uniaxial deformation
\begin{equation}
\Lm_{\hat{n}_0} = \zeta_{\perp} \, \mm{I} +
(\zeta_{\parallel}-\zeta_{\perp})\, \hat{n}_0 \hat{n}_0,
\label{Lm0}
\end{equation}
where $\zeta_{\parallel}$ and $\zeta_{\perp}$ are ratios of a
deformation in the directions parallel and perpendicular to the
nematic director $\nh_0$. Because typical nematic elastomers (and more
generally rubber) are characterized by a bulk modulus that is much
larger than the shear modulus, the spontaneous deformation is nearly
volume preserving, satisfying a constraint $\det\Lm_0 \approx 1$, that
imposes a relation
\begin{equation}
\zeta_{\parallel} \,\zeta_{\perp}^{2} \approx 1. 
\end{equation}

The corresponding conformation vector $\rv^n_0(\Xv)$, defined by
\begin{equation}
(\Lambda_{\nh_0})_{ia}
	= \frac{\partial r^n_{0i}}{\partial X_a},
	\label{Lm0-1}
\end{equation}
specifying average positions of mass points in the NRS is given by
\begin{equation}
\rv_0^{\mbox{n}}(\Xv)  = \Lm_{\hat{n}_0} \cdot \Xv 
 = \left( \zeta_{\perp} \,\mm{I} 
 + (\zeta_{\parallel}-\zeta_{\perp})
 \hat{n}_0 \, \hat{n}_0 \right) \, \cdot \Xv. 
\label{Rv_0}
\end{equation}
To simplify the notations we shall always use $\xv$ to 
denote the position $\rv_0^{\mbox{n}}(\Xv)$ of mass point $\Xv$ in
NRS, i.e., 
\begin{equation}
\xv \equiv \rv_0^{\mbox{n}}(\Xv) = \Lm_{\hat{n}_0} \cdot \Xv,
\label{x-X}
\end{equation}
and will refer to it as the nematic referential (Lagrangian)
coordinates and to the space in which it lives as 
the nematic reference space.  

\begin{figure}[!htbp]
\begin{center}
\includegraphics[width=7cm]{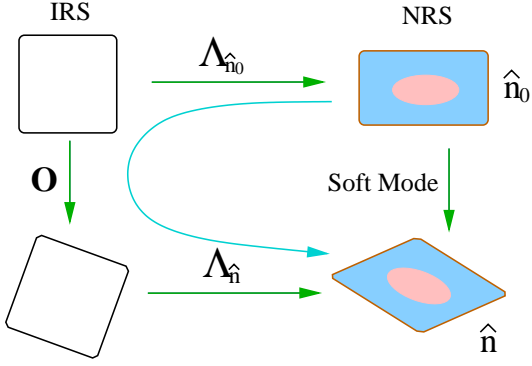}
\caption{Schematic illustration of the soft modes.  The isotropic
reference state (upper left) is spontaneously deformed into the
nematic reference state (upper right).  The ellipse in the center defines
the director $\nh_0$.  Alternatively, one may first rotate the isotropic
reference state in upper left by a global rotation ${\mathbf O}$ into
a distinct isotropic reference state in the lower left
and then have it undergo the I-N transition and the corresponding
deformation into a physically distinct nematic
reference state (lower right).  The two different nematic reference
states are energetically equivalent, and are connected 
by a soft (Goldstone mode) deformation, as given by Eq.~(\ref{soft-mode}).}
\label{fig:soft-modes}
\end{center}    
\end{figure}

An arbitrary state of a nematic elastomer is characterized by a pair
of fields $(\rv(\Xv), \Qm(\Xv))$, which specify the geometric
deformation as well as the nematic order, respectively.  Note that we
can equivalently treat $\rv$ and $\Qm$ as functions of the nematic
Lagrangian coordinate $\xv$, in which case we use the notation
$(\rv(\xv), \Qm(\xv))$ \cite{comment:abuse}.  In the nematic phase, it
is often convenient to use the deformation gradient defined relative
to the nematic reference state:
\begin{equation}
\lm_{ij} = \frac{\partial r_i}{\partial x_{j}},
\label{lambda-def}
\end{equation} 
Using Eqs.~(\ref{Lm0-1},\ref{Rv_0},\ref{x-X}), we readily find the
relation between $\Lm$ and $\lm$:
\begin{equation}
\Lambda_{ia} 
=       \frac{\partial r_i }{\partial X_a} 
=       \frac{\partial r_i }{\partial x_j}
        \frac{\partial x_j }{\partial X_a}
=       \lm_{ij}\,(\Lm_{\hat{n}_0})_{ja},
\label{Lambda-lambda}
\end{equation}
where $\Lm_{\hat{n}_0}$ is the spontaneous deformation, Eq.~(\ref{Lm0}).  
The relations between the IRS and NRS is summarized
schematically by the upper half of Fig.~\ref{fig:soft-modes}.

\subsection{Symmetries and nemato-elastic Goldstone modes}
\label{Sec:Model-II}
Since the rotational symmetry is {\em spontaneously} broken in the
nematic phase, the Goldstone theorem dictates that in the nematic
phase, the ground state must be infinitely degenerate and there must
be a soft mode associated with this degeneracy. This soft mode was
first predicted by Golubovi\'{c} and Lubensky \cite{GolLub89} on
symmetry grounds.  A thorough analysis was then given by Olmsted
\cite{Olmsted94} and in Ref. \cite{LMRX}, and so we only give a brief
discussion here.  As shown in Fig.~\ref{fig:soft-modes}, starting from
the IRS (upper left), there is a continuous manifold of different
nematic reference states available to the system.  One representative
is in the upper right, characterized by a director $\nh_0$ and a
spontaneous deformation $\Lm_{\hat{n}_0}$,
Eq.~(\ref{Lm0}). Energetically equivalent nematic reference states can
be obtained by first rotating the isotropic reference state by an
arbitrary orthogonal matrix $\mm{O}$ (lower left),
\begin{equation}
\Xv' = \mm{O}\cdot \Xv,
\end{equation}
and only then have it undergo an I-N transition to a nematic state
characterized by a distinct director $\hat{n}$ (lower right), with a
nematic order parameter
\begin{equation}
\Qm =  S \,\left( \hat{n}\hat{n}
 - \frac{1}{3}\mm{I} \right). 
\end{equation}
This transition is accompanied by a spontaneous deformation 
\begin{equation}
\Lm_{\hat{n}} =  \left( \zeta_{\perp} \, \mm{I} 
+ (\zeta_{\parallel}-\zeta_{\perp}) \, 
\hat{n} \, \hat{n} \right). 
\label{Lambda_n}
\end{equation} 
It is clear from Fig. \ref{fig:soft-modes} that starting from one NRS
with a director $\nh_0$ (upper right), the other nematic ground state
(lower right) with a director $\nh$ is accessible via a rotation of a
nematic director, combined with a deformation (right arrow in
Fig. \ref{fig:soft-modes}):
\begin{equation}
\hat{n}_0 \rightarrow \hat{n}, \hspace{3mm}
\lm_{GM} = \Lm_{\hat{n}} \, \mm{O} \,\Lm_{\hat{n}_0}^{-1}. 
\label{soft-mode}
\end{equation}
We note (as long as $\nh$ is not related to $\nh_0$ by a rotation
$\mm{O}$) this is {\em not} a simultaneous rigid body rotation of the
deformed elastomer and the nematic director in the upper right.
Instead, the above combination of a deformation and a director
rotation takes the system from one ground state to an elastically
distinct but energetically equivalent one, and therefore is the
Goldstone soft mode associated with the spontaneously broken
rotational symmetry in a solid.  Clearly then, in an ideal case the
deformation in \rfs{soft-mode} costs {\em zero} elastic free energy.
This soft mode, fully characterized by an arbitrary rotation $\mm{O}$
and the nematic director $\hat{n}$, is one of the most striking
consequence of a spontaneous symmetry breaking, and provides the key
to many unusual properties of nematic elastomers.  The main purpose of
the present paper is to study the local fluctuations of this soft mode
and their effects on the long length-scale elastic properties of
nematic elastomers.

An arbitrary uniform deformation of the system away from the NRS is
described by a triad $(\hat{n}_0, \hat{n}, \lm)$, where $\hat{n}_0$
and $\hat{n} $ are the nematic director of the initial and final state
respectively, while $\lm$, defined in Eq.~(\ref{lambda-def}), is
the deformation gradient defined relative to the NRS.  The elastic
free energy density $f(\hat{n}_0, \hat{n}, \lm)$ with strains measured
in the {\em nematic referential} coordinates $\xv$ must reflect the
symmetries discussed above.

Namely, $f$ must be invariant both under an arbitrary global rotation
of the nematic reference state and under that of the current state,
i.e., it satisfies
\begin{equation}
f(\hat{n}_0, \hat{n}, \lm) = 
	f(\mm{O}_0\cdot\hat{n}_0, 
	\mm{O}\cdot\hat{n}, \mm{O}\lm 
	\mm{O}_0^{\rm T}),
	\label{symmetry-1}
\end{equation}
where $\mm{O}_0$ and $\mm{O}$ are two arbitrary orthogonal matrices.
More importantly, the elastic energy $f$ has to vanish for an
arbitrary soft deformation $\lm_{GM}$, as given by Eq.~(\ref{soft-mode}), i.e.,
\begin{equation}
f(\hat{n}_0, \hat{n}, \Lm_{\hat{n}} \mm{O}
 \Lm_{\hat{n}_0}^{-1}) \equiv 0,
 \label{symmetry-2}
\end{equation}
with a special case of the director in the final state the same as
the initial state given by
\begin{equation}
f(\hat{n}_0, \hat{n}_0, \Lm_{\hat{n}_0} \mm{O}
 \Lm_{\hat{n}_0}^{-1}) = 0.
 \label{symmetry-21}
\end{equation}
More generally, the free energy should be independent of whether
deformation $\left( \hat{n}_0, \hat{n}, \lm \right)$ is {\em preceded}
by a soft deformation $\left( \hat{n}_0, \hat{n_0}, \Lm_{\hat{n}_0}
  \mm{O} \,\Lm_{\hat{n}_0}^{-1} \right)$.  Thus the elastic free
energy density of a nematic elastomer must satisfy
\begin{equation}
f(\hat{n}_0, \hat{n}, \lm) = f(\hat{n}_0, \hat{n}, \lm \Lm_{\hat{n}_0}
\mm{O} \,\Lm_{\hat{n}_0}^{-1} ).
 \label{symmetry-3}
\end{equation}
It clear that Eq.~(\ref{symmetry-3}) contains Eq.~(\ref{symmetry-2})
(and hence Eq.~(\ref{symmetry-21})) as a special case.

Let us examine the symmetry properties of the neo-classical theory
\cite{WarTer96,LCE:WT} for ideal, single domain nematic elastomers.
In this theory, the elastic free energy density of a uniform volume
preserving deformation $\lm$ is given by
\begin{equation}
f = \frac{1}{2} \, \mu\, \Tr \, {\mathbf l}_0\,
	\lm^{\rm T} {\mathbf l}^{-1} \lm, 
	\hspace{4mm} \det \lm \equiv 1
	\label{free-energy-neo}
\end{equation}
where the tensors ${\mathbf l}_0$ and ${\mathbf l}$, defined in
Eqs.~(\ref{lm-def}), are the so-called step length tensors and are
functions of the nematic director $\nh_0$ in the initial and $\nh$ in
the final states, respectively.  It is easy to show that indeed the
neo-classical theory, Eq.~(\ref{free-energy-neo}) satisfies properties
(\ref{symmetry-1}-\ref{symmetry-3}). However, it also exhibits another
symmetry, namely
\begin{eqnarray}
f(\hat{n}_0, \hat{n}, \lm) = 
	f(\hat{n}_0, \hat{n}, \Lm_{\hat{n}} 
		\mm{O} \,\Lm_{\hat{n}}^{-1} \lm ). 
\label{extra-symmetry}
\end{eqnarray}
Physically this corresponds to an invariance of the free energy under
a soft deformation applied {\em after} an arbitrary deformation.  This
property does not seem to follow from any underlying symmetry of a
nematic elastomer and is therefore not expected to hold in real
systems.  Thus, this unphysical constraint and associated limitations
of the neo-classical theory should be experimentally detectable.

\subsection{Invariant strains}

It is a generic feature of nonlinear elasticity theories, that there
is no unique definition of a strain tensor.  For ordinary elastic
media, where there is no independent orientational degree of freedom,
infinite number of strain tensors can be defined in terms of the
deformation gradient $\lm$ \cite{comment:Ogden}:
\begin{subequations}
\label{uv-def}
\begin{eqnarray}
\mm{u}_m &=& \frac{1}{m} \left((\lm^{\rm T} \lm)^m - \mm{I}\right), 
\\
\mm{v}_m &=& \frac{1}{m} \left((\lm \lm^{\rm T})^m - \mm{I}\right),
\end{eqnarray}
\end{subequations}
where $m$ is an arbitrary integer.  A common feature of these strain
tensors is that they vanish for an arbitrary global rotation.
Furthermore, $\mm{u}_m$'s ($\mm{v}_m$) are tensors (scalars) in the
reference space, and scalars (tensors) in the embedding space.

For isotropic solids, e.g., ordinary rubber or glass, either 
$\mm{u}$'s or $\mm{v}$'s provide an equally good description of
elasticity, with elastic free energy expressible as a function of any
of the tensors above.  For anisotropic solids, e.g., crystals, tensors
$\mm{u}_m$ provide a natural elastic description, while tensors
$\mm{v}_m$ are inadequate to capture the anisotropy of the reference
state.

In the case of nematic elastomers, a priori, none of these strain
tensor provide a complete description since they do not contain any
information about the nematic director (but see
Sec.~\ref{Sec:Effective-model}).  However, as we now show they can be
generalized to incorporate the nematic directors of the nematic
reference and deformed states.  Let us look at the following
generalized symmetric strain tensor in the embedding space:
\begin{eqnarray}
\mm{V} =  \frac{1}{2} \,\left(
	\mm{l}^{-\frac{1}{2}} \,\lm \,
	{\mathbf l}_0 \lm^{\rm T} \,
		\mm{l}^{-\frac{1}{2}} 
	-	\mm{I} \right), \label{V-def}
\end{eqnarray}
where 
\begin{subequations}
\label{lm-def}
\begin{eqnarray}
{\mathbf l}_0 &=& \Lm_{\hat{n}_0} \Lm_{\hat{n}_0}^{\rm T}
	=	\zeta_{\perp}^2 \,\mm{I} 
	+ (\zeta_{\parallel}^2 - \zeta_{\perp}^2)\,\nh_0\nh_0, 
\label{lm-def-0}\\
{\mathbf l}&=& \Lm_{\hat{n}}  \Lm_{\hat{n}}^{\rm T}
	=	\zeta_{\perp}^2 \,\mm{I} 
	+ (\zeta_{\parallel}^2 - \zeta_{\perp}^2) \,\nh\nh,
\end{eqnarray}
\end{subequations}
are the so-called ``step length tensors'' in the initial and final
state respectively.  It is easy to see that $\mm{V}$ is a
generalization of $\mm{v}_2$ \cite{comment:left-strain} as defined in
Eqs.~(\ref{uv-def}): in the isotropic limit, i.e., ${\mathbf l} =
{\mathbf l}_0 = \mm{I}$,
\begin{eqnarray}
\mm{V} \rightarrow \frac{1}{2} \left(\,
	\left(\lm \lm^{\rm T} - \mm{I} \right) \right) = \mm{v}_2.
\end{eqnarray}

The generalized strain $\mm{V}$ has many nice symmetry properties. It
is invariant with respect to arbitrary rotation in the reference
space:
\begin{eqnarray}
	\mm{V}(\mm{O}_0\cdot\hat{n}_0, 
	\hat{n}, \lm \mm{O}_0^{\rm T})
= 	\mm{V}(\hat{n}_0, \hat{n}, \lm) .	
\end{eqnarray}
It transforms as a second rank tensor with respect to rotations in the
embedding space:
\begin{eqnarray}
	\mm{V}(\hat{n}_0, \mm{O}\cdot 
	\hat{n}, \mm{O}\lm )
= \mm{O}\mm{V}(\hat{n}_0, \hat{n}, \lm)
	\mm{O}^{\rm T}.	
\end{eqnarray}
More importantly, it automatically vanishes for arbitrary soft
deformations Eq.~(\ref{soft-mode}), i.e., satisfies
Eq.~(\ref{symmetry-2}) and Eq.~(\ref{symmetry-3}), much like the 
free energy density does:
\begin{subequations}
\begin{eqnarray}
&&\mm{V}(\hat{n}_0, \hat{n}, \Lm_{\hat{n}} \mm{O}
 \Lm_{\hat{n}_0}^{-1}) \equiv 0;\\
&& \mm{V}(\hat{n}_0, \hat{n}, \lm) = 
\mm{V}(\hat{n}_0, \hat{n}, \lm \Lm_{\hat{n}_0} \mm{O} \,
\Lm_{\hat{n}_0}^{-1}  ). 
\end{eqnarray}
\end{subequations}
Because of these symmetry properties, we shall call $\mm{V}$ an {\em
invariant strain}.  A scalar function of the tensor $\mm{V}$
automatically satisfies both Eq.~(\ref{symmetry-1}) and
Eq.~(\ref{symmetry-3}).

\begin{figure}[!htbp]
\begin{center}
\includegraphics[width=6cm]{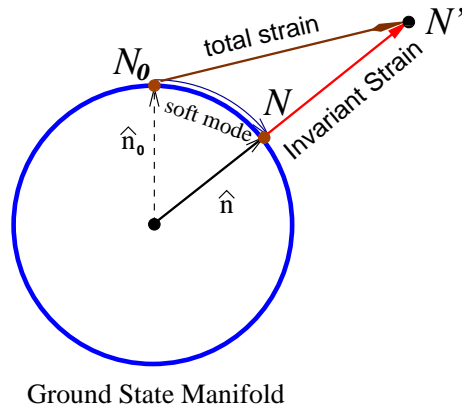}
\caption{A schematic of the invariant strain, with the blue circle
denoting the ground state manifold.  $\nh_0$ and $\nh$ are the nematic
director in the initial and final states respectively.  Point $N_0$
denotes the nematic reference state (initial state), $N$ the nematic
ground state with director $n$, and $N'$ the final strained state.
The generalized strain tensor $\mm{V}$ measures the deformation of the
system state relative to the whole ground state manifold, instead of
any particular ground state, i.e., independent of the choice of $N_0$.
This is the reason that it vanishes identically for an arbitrary soft
deformation. }
\label{invariant-strain}
\end{center}    
\end{figure}

To clarify the origin of the symmetry properties of $\mm{V}$, we
express it in terms of the deformation gradient $\Lm$ defined
relative to the {\em isotropic} reference state.  Using
Eq.~(\ref{Lambda-lambda}) and Eq.~(\ref{lm-def-0}) in
Eq.~(\ref{V-def}) we find:
\begin{eqnarray}
2\,\mm{V} = \Lm_{\hat{n}}^{-1}
	\left[ \Lm\Lm^{\rm T} 
	- \Lm_{\hat{n}} \Lm_{\hat{n}}^{\rm T}\right]
	 \Lm_{\hat{n}}^{\rm -T}. \label{V-reparam}
\end{eqnarray}
Remembering that $\Lm_{\hat{n}}$ is the energetically preferred
spontaneous deformation measured relative to the isotropic reference
state, if the nematic director of the current state is $\hat{n}$.
Therefore, the tensor inside the square bracket in
Eq.~(\ref{V-reparam}) describes the nonlinear strain $\mm{v}_2$,
measured relative to the isotropic reference state, minus its optimal
value for given nematic director $\hat{n}$.  It is same as the tensor
$\delta \mm{v}$ defined in reference \cite{LMRX}, Sec.~IV B (Theory
with strain and director).  In other word, $\mm{V}$ measures the
tensorial ``distance'' between the deformed state and the ground state
manifold, as illustrated schematically in Fig.~\ref{invariant-strain}.
This is the fundamental reason why $\mm{V}$ vanishes for an arbitrary
soft deformation.

We note in passing that all other nonlinear strain tensors $\mm{u}_m$
and $\mm{v}_m$, as defined in Eq.~(\ref{uv-def}), can be generalized
to the case of nematic elastomers.  Both properties
Eq.~(\ref{symmetry-1}) and Eq.~(\ref{symmetry-3}) are satisfied by the
generalization of $\mm{v}_m$ for arbitrary integer $m$.  On the other
hand, analogous generalization of $\mm{u}_m$ does not satisfy the
property Eq.~(\ref{symmetry-3}) and therefore does not provide an
adequate description of the elasticity of nematic elastomers.

\subsection{Modes of deformation}
For convenience of later discussion, we shall let the dimension $d$ be
an arbitrary integer larger than two.  Let $\{\hat{n},\hat{m},
\hat{l}, \ldots\}$ be an orthonormal basis of the $d$ dimensional
embedding space, where $\hat{n}$ is {\em the nematic director in the
  deformed state} \cite{comment:nh}.  This coordinate frame therefore
rotates with the nematic director $\nh$, in strong contrast with the
frame $\{\eh_1,\ldots,\eh_d = \hat{z} \}$ that we introduced earlier.
It is convenient to introduce two complementary projection tensors in
this space:
\begin{subequations}
\begin{eqnarray}
\Pn &\equiv& \hat{n}\hat{n},\\
\Pm &\equiv& \mm{I} - \Pn,
\end{eqnarray}
\end{subequations}
that naturally satisfy the following properties:
\begin{subequations}
\begin{eqnarray}
\Pn \Pn &=& \Pn, \\
\Pm \Pm &=& \Pm,\\
\Pn\Pm &=& \Pm\Pn = 0,\\
\mm{I} &=& \Pn + \Pm.
\end{eqnarray}
\end{subequations}
$\Pn$ projects out the one dimensional subspace along the nematic
director $\nh$, while $\Pm$ projects out the $d-1$ dimensional
subspace perpendicular to $\nh$.

\begin{figure}[htp!]
\begin{center}
\includegraphics[width=5cm]{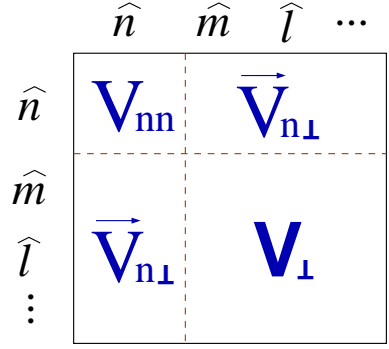}
\caption{Decomposition of the invariant strain tensor $\mm{V}$ into
three parts: $V_{nn}$, $\mm{V}_{n\perp}$, and $\mm{V}_{\perp}$. }
\label{fig:V-decomp}
\end{center}    
\end{figure}

The strain tensor $\mm{V}$ can then be decomposed into different parts
using these projectors:  
\begin{subequations} 
\begin{eqnarray}
\Pn \mm{V}\Pn &=& V_{nn}\,\Pn , \\
\Pn \mm{V} \Pm &=& \hat{n} \,\vec{V}_{n\perp} , \\
\Pm \mm{V} \Pn &=& \vec{V}_{n\perp} \, \hat{n}, \\
\Pm \mm{V} \Pm &=& \mm{V}_{\perp}.
\end{eqnarray}
\end{subequations}
This decomposition is illustrated in Fig.~\ref{fig:V-decomp}.   

Using this basis, as illustrated in Fig.~\ref{fig:V-decomp},   
we decompose the strain $\mm{V}$ into an isotropic part
and some other traceless parts according to 
\begin{eqnarray}
\mm{V} = \frac{1}{d} \,(\Tr \mm{V}) \, \mm{I} + \tilde{V}_{nn} \,
\mm{J} + 
{\begin{pmatrix} 0& \vec{V}_{n\perp}\\
\vec{V}_{n\perp}&0\end{pmatrix}}
+ {\begin{pmatrix} 0&0\\
0&\tilde{\mm{V}}_{\perp}\end{pmatrix}}, 
\nonumber\\
\label{V-parameterization}
\end{eqnarray}
where $\mm{J}$ is a $d\times d$ traceless matrix
\begin{equation}
\mm{J} = \begin{pmatrix} 
1 & 0 \\
0 & - \frac{1}{(d-1)} \mm{I}_{\perp}
\end{pmatrix},
\end{equation}
$\mm{I}_{\perp}$ a $(d-1)\times (d-1)$ identity matrix, and
a scalar $\tilde{V}_{nn}$ and a $(d-1)\times (d-1)$ traceless matrix
$\tilde{\mm{V}}_{\perp}$ are given by
\begin{subequations}
\label{V-decomps}
\begin{eqnarray}
\tilde{V}_{nn} &=& V_{nn} - \frac{1}{d}\,(\Tr\,\mm{V}),\\
 \tilde{\mm{V}}_{\perp} &=& \mm{V}_{\perp} 
- \frac{1}{(d-1)}\, (\Tr\,\mm{V}_{\perp})\, \mm{I}_{\perp}. 
\end{eqnarray}
\end{subequations}

To illustrate the geometric meaning of the different components of the
tensor $\mm{V}$ in Eq.~\ref{V-parameterization}, let us consider a
special case of small, spatially dependent deformation from the
nematic reference state.  To lowest order approximation, we need not
distinguish the Lagrangian and Eulerian coordinates.  We define a
phonon field $\uv$, a {\em linearized} symmetric strain tensor
${\boldsymbol \varepsilon}$, and a linearized antisymmetric strain
${\mathbf a}$ in a standard way:
\begin{subequations}
\label{parameterization}
\begin{eqnarray}
&&\uv(\xv)  = \rv(\xv) - \xv, \label{phonon-def}\\
&&\lm_{ia} = \frac{\partial r_i}{\partial x_a}
= \delta_{ia} + \partial_a u_i,
\label{lambda-u}\\
&&\varepsilon_{ab} = \frac{1}{2} \left(
\partial_a u_b + \partial_b u_a \right),
\label{varepsilon-def} \\
&& a_{ab} = \frac{1}{2} \left(
\partial_a u_b - \partial_b u_a \right),
\end{eqnarray}
\end{subequations}
where we have adopted the nematic Lagrangian coordinates and have used
the shorthand $\ppa$ for $\partial/\partial x_a$.  We also choose
$\nh_0 = \eh_d = \hat{z}$, and consider small fluctuation of the
nematic director:
\begin{equation}
\hat{n} = \hat{n}_0 + \delta \hat{n} 
= \hat{z}  + \delta \hat{n}, \hspace{3mm}
|\delta\hat{n}| \ll 1.
\label{n-n0}
\end{equation}

\begin{figure}
\begin{center}
\includegraphics[width=8cm]{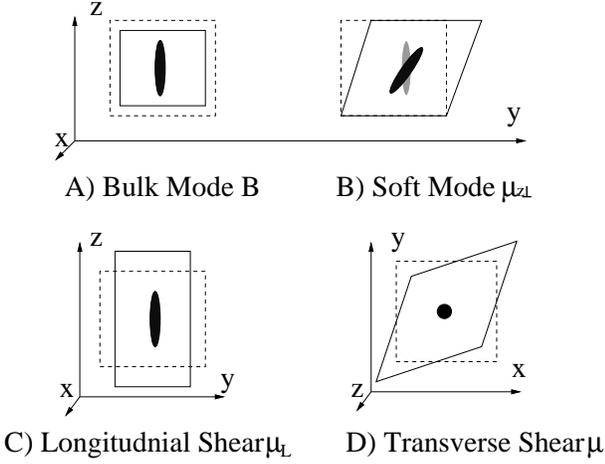}
\caption{Schematics for a bulk, soft, longitudinal shear, and
  transverse shear modes, as well as associated rigidities.  Different
  components of the strain tensors that correspond to these
  deformation modes are listed in Table~\ref{Modes}.  At lowest order,
  the bulk, longitudinal shear and transverse shear modes are the same
  as ordinary uniaxial solids. The soft mode mixes simple shear with
  rotation of the nematic director and thereby qualitatively
  distinguishes a nematic elastomer from an ordinary uniaxial solid.}
\label{deformations}
\end{center}    
\end{figure}
  
The full expressions for components of $\mm{V}$ in terms of these
variables can be obtained by using Eqs.~(\ref{parameterization}) and
Eq.~(\ref{n-n0}) in Eq.~(\ref{V-def}). Up to linear order in $\uv$ and
$\delta\hat{n}$, this leads to:
\begin{equation}
  \Tr \,\mm{V} \approx \Tr \, {\boldsymbol \varepsilon} = \nabla\cdot \uv. 
\end{equation}
Consequently $\Tr\,\mm{V}$ describes the isotropic
compression/dilation of the solid, i.e., the {\em bulk} mode,
illustrated in Fig.~\ref{deformations}A.  We also find that
\begin{eqnarray}
\tilde{V}_{nn} 
&=&	V_{nn} - \frac{1}{d} \,\Tr \,\mm{V} 
\nonumber\\
&\approx& \varepsilon_{zz} 
	- \frac{1}{d} \,\Tr \,{\boldsymbol \varepsilon}
	\nonumber\\
&=& \pz u_z -  \frac{1}{d} (\nabla\cdot \uv), 
\end{eqnarray}
where we have used the fact that $\hat{z} \approx \nh$ at the lowest order.
Therefore $\tilde{V}_{nn}$ describes a dilation (compression) along
the $\nh$ axis combined with an isotropic compression (dilation)
transverse to this axis (perpendicular subspace), with the total
volume conserved, as illustrated in Fig.~\ref{deformations}C.  We shall call this mode the {\em
  longitudinal shear} mode.  The strain component $\tilde{{\mathbf
    V}}_{\perp}$ is, up to linear order of the phonon field, given by
\begin{eqnarray}
\tilde{V}_{ij}^{\perp} &=& 
	V^{\perp}_{ij} - \frac{1}{d-1} 
	\, \sum_{k=1}^{d-1} \,V_{kk} \,\delta^{\perp}_{ij}
	\nonumber\\
	&\approx& \frac{1}{2} \left( 
	\ppi u_j + \ppj u_i \right) - \frac{1}{d-1} \,
	\left(\nabla_{\perp}\cdot \uv_{\perp} \right)
	 \,\delta^{\perp}_{ij},
\end{eqnarray}
where indices $i$ $j$ and $k$ are restricted to the $(d-1)$
dimensional transverse subspace.  Therefore $\tilde{{\mathbf
    V}}_{\perp}$ describes shear deformation in the $d-1$ dimensional
subspace perpendicular to $\hat{n}$, as illustrated in
Fig.~\ref{deformations}D.  This mode breaks the uniaxial symmetry and
we will refer to it as the {\em transverse shear} mode.  Up to a
linear order, the above three modes, i.e., the bulk, the longitudinal
shear, and the transverse shear only depend on the symmetric
linearized strain ${\boldsymbol \varepsilon}$ and are the same for
nematic elastomers as for ordinary uniaxial solids.

However, the strain operator $\vec{V}_{n\perp}$ mixes the symmetric
strain ${\boldsymbol \varepsilon}$ with the antisymmetric strain
${\mathbf a}$ and the director fluctuation $\delta\nh$.  Upto linear
order in $\uv$ and $\delta\nh$, it is given by
\begin{eqnarray}
{V}_{ni} = 
\frac{1}{2} \left(\sqrt{r}+\frac{1}{\sqrt{r}} \right) 
\left[ \varepsilon_{ni} + \left(\frac{r-1}{r+1}\right)
 (a_{ni} - \delta \nh_i)
\right], \label{V_ni}
\end{eqnarray}
where $r = \frac{\zeta_{\parallel}^2}{\zeta_{\perp}^2}$ is a
dimensionless ratio characterizing the spontaneous anisotropy of the
nematic phase.  In the limit of a vanishing nematic order,
$\zeta_{\perp} \rightarrow \zeta_{\parallel}$, $r\rightarrow 1$, and
${V}_{ni}$ consequently reduces to the linearized symmetric strain,
$\varepsilon_{ni}$.  For a non-vanishing nematic order, however, the
strain ${V}_{ni}$ involves a symmetric strain $\varepsilon_{ni}$, a
rotation of the solid $a_{ni}$, as well as a rotation of the nematic
director $\delta\nh$.  Strain components $\vec{V}_{n\perp}$
qualitatively distinguish a nematic elastomer from an ordinary
uniaxial solid. It is clear from Eq.~(\ref{V_ni}) that
$\vec{V}_{n\perp}$ describes a simple shear in a plane containing
$\nh$ and the $\hat{e}_i$ axis, compensated by an infinitesimal
rotation of the nematic director as well as a global rotation of the
solid, and vanishes for an infinitesimal {\em soft} deformation, where
the cancellation between these three contributions is complete.  This
is illustrated in Fig.~\ref{deformations}B.
 
\subsection{Nemato-elastic model}
The invariant strain tensor ${\mathbf V}$ measures deformations
relative to the uniaxial ground state manifold of an ideal nematic
elastomer, and therefore vanishes for arbitrary soft deformation.
Thus, for a uniform deformation, the elastic energy of an ideal
nematic elastomer can be expressed as a scalar function of ${\mathbf
V}$.  Indeed, in the neo-classical theory, the elastic free energy per
unit volume \cite{comment:volume-measure}, 
Eq.~(\ref{free-energy-neo}), can be written as
\begin{equation}
f_{\rm neo} = \frac{1}{2}\,\mu\,
	\Tr \, \mm{l}_0 \lm^{\rm T} \,\mm{l}^{-1}\lm
	= \mu\, \Tr \, \mm{V} + \frac{d}{2} \,\mu, 
	\label{f-neo}
\end{equation}
where we have used the cyclic property of the trace operation and
 $\det\mm{l} = \det \mm{l}_0$ that follows from Eq.~(\ref{lm-def}).
 Using Eq.~(\ref{V-def}), the incompressibility constraint $\det \lm
 \equiv 1$ can also be written in terms of ${\mathbf V}$ as
\begin{equation}
\det (\mm{I} + 2\,{\mathbf V}) \equiv 1.
\label{incomp-constraint}
\end{equation}

Because of the highly nonlinear incompressibility constraint
\rfs{incomp-constraint}, the neo-classical theory Eq.~(\ref{f-neo}) is
not particularly convenient for our goal of studying long wavelength
fluctuations of a nematic elastomer.  Furthermore, as we have noted at
the end of Sec.\ref{Sec:Model-II}, the neo-classical theory exhibits
extra unphysical symmetry, \rfs{extra-symmetry}, and is thus not the
most general theory satisfying physical elastomer symmetries
Eqs.~(\ref{symmetry-1}), (\ref{symmetry-3}).  We therefore relax the
incompressible constraint and expand the elastic free energy in powers
of the invariant strain $\mm{V}$.  This removes the aforementioned
unphysical symmetry of the neo-classical theory.  Linear terms cannot
appear since the nematic reference state is a minimum of the elastic
free energy.  In accordance with the uniaxial symmetry in the nematic
phase, five different quadratic terms can appear.  Ignoring all higher
order terms, the elastic energy density can be expressed 
as \cite{comment:energy}
\begin{eqnarray}
f_e  &=& \frac{1}{2}\,B_z \, V_{nn}^2 + \lambda_{z\perp} 
	\, V_{nn}(\Tr \, \mm{V}_{\perp})  	
	+ \frac{1}{2}\,\lambda \, (\Tr \, \mm{V}_{\perp})^2
	\nonumber\\
&+& \mu \, \Tr \, \mm{V}_{\perp}^2 
	+ \mu_{n\perp}\, \vec{V}_{n\perp}^2
		 \label{elast-energy-1} \\
&=& \frac{1}{2}\,B\,(\Tr \,\mm{V})^2 
		+ C\, (\Tr \,\mm{V})\, \tilde{V}_{nn}
		+ \frac{1}{2}\, \mu_{\rm L} \, \tilde{V}_{nn}^2 
	\nonumber\\
	&+& \mu \,\Tr\, \tilde{\mm{V}}_{\perp}^2 
	 + \mu_{n\perp}\, (\vec{V}_{n\perp})^2,
	 \label{elast-energy-2}
\end{eqnarray}
where we have used two different decompositions of $\mm{V}$, related
to each other by Eqs.~(\ref{V-decomps}).  Note that we have used the
coordinate system $\{\hat{n},\hat{m},\hat{l},\ldots\}$ as illustrated
in Fig.~\ref{fig:V-decomp}.
The total elastic free energy is then given by $f_e$
integrated over the nematic referential volume element 
$d^d x$ \cite{comment:f-e}
\begin{equation}
F_e = \int f_e \,d^d x = \int J^{-1} \,f_e\,d^d r ,
\end{equation}
where 
$$J = \det \lm = \det \frac{\partial r_i}{\partial x_j}$$ is the
Jacobian factor relating the Eulerian volume element to the Lagrangian
volume element.  For most rubbery materials the Jacobian factor is
very close to unity (up to ~$10^{-5}$).

The two sets of parameter in Eq.~(\ref{elast-energy-1}) and
Eq.~(\ref{elast-energy-2}) are related via:
\begin{subequations}
\label{def_moduli}
\begin{eqnarray}
B_z &=& B + 2\,(1-\frac{1}{d}) \, C 
	+ (1-\frac{1}{d})^2\,\mu_{\rm L} ,\\
\lambda_{n\perp} &=& B + (1-\frac{2}{d})\,C
 - \frac{1}{d^2} (d-1) \mu_{L},\\
\lambda &=& B  - \frac{2}{d} \, C + \frac{1}{d^2} \,\mu_{\rm L}
 - \frac{2}{(d-1)} \,\mu.
\end{eqnarray}
\end{subequations}
%
Inverting the relations (\ref{def_moduli}), we find
\begin{subequations}
\label{def_moduli_2}
\begin{eqnarray}
\hspace{-5mm}
B &=& \frac{1}{d^2}\,\left(
	B_z+(d-1) (2 C+(d-1) \lambda +2 \mu ) \right),\\
\mu_{\rm L} &=& B_z - 2 C+\lambda +\frac{2 \mu }{d-1},\\
C &=&\frac{1 }{d} \left(
B_z+(d-2)C-(d-1) \lambda -2 \mu\right).
\end{eqnarray}
\end{subequations} 
%

Following our discussions of different deformation modes in the
preceding subsection, we shall call the elastic constants $B$,
$\mu_{\rm L}$ and $\mu$ in Eq.~(\ref{elast-energy-2}) the bulk
modulus, the longitudinal shear modulus, and the transverse shear
modulus, respectively.
Generically, rubber is characterized by a bulk modulus $B$ that is
four to five orders of magnitude larger than all other elastic moduli.
Furthermore, as we shall see in Sec.~\ref{Sec:Thermal-fluct} and
Sec.~\ref{Sec:Disorder-fluct}, ideal nematic elastomers are actually
{\em strictly incompressible} due to strong fluctuations that
singularly renormalize elastic moduli, driving all ($\mu_{\rm L}$,
$\mu$, and $C$) but the bulk modulus $B$ to zero at long
length-scales.  This essential difference between the bulk modulus $B$
and other moduli ($\mu_{\rm L}$, $\mu$, $C$) indicates that
Eq.~(\ref{V-parameterization}) and Eq.~(\ref{elast-energy-2}) provide
a more natural parameterization to the invariant strain $\mm{V}$ and
the elastic energy, respectively.  By contrast, if one insists on
using the parameterization in Eq.~(\ref{elast-energy-1}) with
independent parameters $B_z$, $\lambda_{z\perp}$, $\lambda$, $\mu$,
one would find, from Eqs.~(\ref{def_moduli}) that all renormalized
elastic constants except $\mu$ flow to the same limiting value, i.e.,
to the renormalized non-universal bulk modulus $B$.  The essential
physics (i.e., strict incompressibility, universal Poisson ratios,
etc.) is thus missed in this latter parameterization, hidden in the
{\em differences} between these renormalized constants and $B$, which
go to zero in the long wavelength limit.

We note that by virtue of the invariant strain $\mm{V}$, the elastic
energy (\ref{elast-energy-2}) vanishes identically for an arbitrary
soft deformation, i.e., it satisfies
Eqs.~(\ref{symmetry-2}-(\ref{symmetry-3}).  On the other hand, since
the elastic energy is expanded up to the second order of the strain
$\mm{V}$, it is valid for small value of $\mm{V}$, i.e., the strained
state should be not far away from the ground state manifold.  Because
of the uniaxial nematic order, the elastic energy
(\ref{elast-energy-2}) of a nematic elastomer bears a resemblance to a
conventional uniaxial solid, that to a quadratic order in the
nonlinear Lagrange strain tensor $\mm{u}$ (using the decomposition
introduced above) is given by \cite{EL:Landau}
\begin{eqnarray}
f_e &=& \frac{1}{2} \,C_1\, u_{nn}^2 
	+  C_4\, u_{nn} \Tr\,\vec{u}_{\perp}
 + 	\frac{1}{2} \,  C_2 \, (\Tr\,\vec{u}_{\perp})^2 		
				\nonumber\\
&+&	  C_3 \, \Tr\,\vec{u}_{\perp}^2
		+ C_5\, (\vec{u}_{\perp n})^2.
	\label{H-uniaxial}
\end{eqnarray}
Although both elastic energies, Eq.~(\ref{elast-energy-1}) and
Eq.~(\ref{H-uniaxial}) are characterized by five independent moduli,
there is an essential qualitative difference between a conventional
uniaxial solid and an ideal nematic elastomer: while the Lagrange
strain $\mm{u}$ is a tensor in the reference space and only depends on
the geometric deformation $\rv(\xv)$, the invariant strain $\mm{V}$ is
a tensor in the embedding space and depends both on the geometric
deformation and on the nematic director rotation, and more importantly, 
vanishes for arbitrary soft deformations.

For non-uniform deformations, the elastic energy
Eq.~(\ref{elast-energy-1}) and Eq.~(\ref{elast-energy-2}) must be
augmented by the free energy cost for distortion of the nematic order, that
in 3d is given by the Frank free energy
\begin{eqnarray}
F_d[\hat{n}] &=& \int f_d \,  d^3 x = \int f_d \, J^{-1} d^3 r,
\nonumber\\
f_d
&=& \frac{1}{2}\, K_1\, (\nabla \cdot \hat{n})^2 
+\frac{1}{2}\, K_2 \,(\hat{n} \cdot \nabla \times \hat{n})^2 
\nonumber\\
&+& \frac{1}{2}\,K_3 \,(\hat{n} \times \nabla 
\times \hat{n})^2,
\label{f_Frank}
\end{eqnarray}
where $K_1$, $K_2$ and $K_3$ are the splay, twist, and bending
constants, respectively.  Again, we have defined the free energy
density $f_d$ as the Frank free energy per unit volume measured in the
nematic referential coordinate system.  Here it is important to note
that the gradient operator $\nabla$ is defined in the Eulerian
coordinates $\rv$, i.e.,  $\nabla_i = {\partial}/{\partial r_i}$,
because the nematic director interaction is dominated by the liquid
fraction \cite{comment:Frank}. 

For small deformation, we can take $\nh\approx\hat{z}+\delta\nh$ and
make linear approximation of the invariant strain $\mm{V}$ in
Eq.~(\ref{elast-energy-2}), in terms of linearized strains
${\boldsymbol \varepsilon}$, $\mm{a}$ and the director fluctuation
$\delta \nh$.  The elastic energy Eq.~(\ref{elast-energy-2}) then
reduces to
\begin{eqnarray}
f_e &=& \frac{1}{2}\,B\,(\Tr \,{\boldsymbol \varepsilon})^2 
		+ C\, (\Tr \,{\boldsymbol \varepsilon})\, \tilde{\varepsilon}_{zz}
		+ \frac{1}{2}\, \mu_{\rm L} \, \tilde{\varepsilon}_{zz}^2 
	 \label{H-quadratic}	\\
	&+& \mu \,\Tr\, \tilde{{\boldsymbol \varepsilon}}_{\perp}^2 
	 + \tilde{\mu}_{n\perp}\, \left[ \varepsilon_{zi} 
	 + \left(\frac{r-1}{r+1}\right)  (a_{zi} - \delta \nh_i) \right]^2,
\nonumber
\end{eqnarray}
where
\begin{eqnarray}
\Tr \,{\boldsymbol \varepsilon} &=& \varepsilon_{aa} =  \varepsilon_{zz} 
+ \sum_{i = 1}^{d-1}\varepsilon_{ii} ,\\
\tilde{\varepsilon}_{zz} &=& \varepsilon_{zz} - \frac{1}{d} \Tr {\boldsymbol \varepsilon},\\ 
\tilde{\varepsilon}_{ij} &=&  \varepsilon_{ij}
 - \frac{1}{d-1} \varepsilon_{kk} \, \delta_{ij},\\
\tilde{\mu}_{n\perp} &=& \frac{1}{2} \left(\sqrt{r}
	+\frac{1}{\sqrt{r}} \right)^2{\mu}_{n\perp}.
\end{eqnarray}
The last term in Eq.~(\ref{H-quadratic}) is the most interesting one.
It explicitly demonstrates our earlier general symmetry assertion that
a simple shear $\varepsilon_{ni}$ can be completely compensated by a
combination of the elastomer and the nematic director rotations,
$a_{ni}$ and $\delta\nh$ respectively, such that the net elastic
energy cost is zero.  An infinite set of such soft elastic distortions
is specified by a constraint
\begin{equation}
\varepsilon_{ni} + \left(\frac{r-1}{r+1}\right)  (a_{ni} - \delta \nh_i) 
 = 0. 
\label{softconstraint}
\end{equation} 
This complete compensation is a consequence of the spontaneous symmetry
broken nematic elastomer state and is the lowest order manifestation
of the soft deformation Eq.~(\ref{soft-mode}).  

If one is primarily interested in the elastic degrees of freedom, then
director fluctuations $\delta\nh$ can be integrated out of the
\rfs{H-quadratic}, thereby obtaining an effective elastic description
purely in terms of the strain ${\boldsymbol \varepsilon}$.  It is easy
to show that at the quadratic order this leads to a rotation tensor
$\mm{a}$ and $\delta\nh$ that automatically adjust to ${\boldsymbol
  \varepsilon}$ to satisfy the soft mode constraint
\rfs{softconstraint}. Thus it is quite clear that the resulting
effective elastic free energy density has the form \rfs{H-quadratic},
but with the last $\mu_{n\perp}$ shear term identically vanishing, a
result that was first derived (in a quite a different way) by
Golubovi\'{c} and Lubensky \cite{GolLub89,LMRX}.  We thus observe that
the elasticity of uniaxial nematic elastomers resembles that of
ordinary uniaxial solids, but with key distinction that the transverse
shear elastic modulus $C_5$ strictly vanishes, which is guaranteed by
the underlying rotational symmetry spontaneously broken in the nematic
state.

\begin{figure}  
\begin{center}
\includegraphics[width=7.5cm]{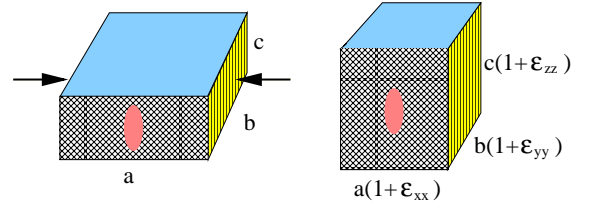}
\caption{A simple experiment on a uniaxial nematic elastomer.}
\label{poissonratio}
\end{center}    
\end{figure}

To illustrate the basic structure of the harmonic theory
Eq.~(\ref{H-quadratic}), let us consider a simple mechanical experiment
illustrated in Fig.~\ref{poissonratio}.
With the initial nematic order (uniaxial axis) along $\hat{z}$, we
study the elastic response to a small strain deformation
$\varepsilon_{xx} = \varepsilon$.  To avoid the elastic stripe
instability \cite{Finkelmann-95-stripe}, we take $\varepsilon$ to be
negative, as shown in Fig.~\ref{poissonratio}.  To find the
equilibrium state we minimize the elastic free energy
Eq.~(\ref{H-quadratic}) with respect to all strain components except
$\varepsilon_{xx}$.  In strong contrast to a stretch along $\hat{e}_x$
axis, a compression along $\hat{e}_x$ is stable against an
inhomogeneous soft deformation, and therefore $\delta \nh$ and strain
components $\varepsilon_{ni}$ as well as $a_{ni}$ remain zero. We can
therefore ignore the $\tilde{\mu}_{n\perp}$ term in
Eq.~(\ref{H-quadratic}).

For concreteness we focus on three dimensions, and simplify the
analysis by taking the bulk modulus $B$ to infinity, a reasonable
approximation for elastomers that are essentially incompressible.
Minimizing the elastic energy Eq.~(\ref{H-quadratic}) over
$\varepsilon_{yy}$, $\varepsilon_{xy}$, and $\varepsilon_{zz}$ for
fixed $\varepsilon_{xx} = \varepsilon$, we obtain:
\begin{subequations}
\label{e-results}
\begin{eqnarray}
\varepsilon_{yy}  &\rightarrow&  \frac{\mu - \mu_{\rm L}}
        { \mu + \mu_{\rm L} } \,\varepsilon,\\
\varepsilon_{zz}  &\rightarrow&  \frac{ - 2\,\mu }
        { \mu + \mu_{\rm L}  } \,\varepsilon,\\
\varepsilon_{xy} &=& 0,
\end{eqnarray} 
\end{subequations}
which satisfy the traceless incompressibility condition
\begin{equation}
\Tr \,{\boldsymbol \varepsilon} = \sum_{i}^{3}\varepsilon_{ii} + \varepsilon_{zz}
\equiv 0.
\end{equation}

As expected, in this simplest (fluctuation-free) harmonic treatment
the compressional elastic response coincides with that of an ordinary
uniaxial solid.  Because in the incompressible limit the
cross-coupling $C$ drops out, the ratios between different strain
components (i.e., Poisson ratios) only depend on a single ratio
$\mu_{\rm L}/\mu$.  As we shall see later in this paper, long
wavelength fluctuations drive this ratio to a {\em universal}
constant, set by the ratio between the renormalized shear moduli
$\mu^R$ and $\mu_{\rm L}^R$.  The simple experiment illustrated in
Fig.~\ref{poissonratio} allows a direct test of the universal Poisson
ratios predicted by our theory.

Although our focus here (and throughout the manuscript) is on the most
experimentally relevant uniaxial nematic state, our phenomenological
model, Eq.~(\ref{elast-energy-1}) or Eq.~(\ref{elast-energy-2}), can
be easily generalized to a biaxial nematic elastomer.  Indeed we can
define the invariant strain tensor $\mm{V}$ by Eq.~(\ref{V-def}) for
arbitrary $m$-axial nematic order simply by using corresponding
$m$-axial step-length tensors $\mm{l}_0$ and $\mm{l}$.  However, such
non-uniaxial nematic elastomer generically has many more elastic
constants in its free energy.  For example, a three-dimensional
biaxial nematic elastomer is characterized by biaxial step length
tensors $\mm{l}_0$ and $\mm{l}$.  All rotational symmetries are broken
and the corresponding quadratic elastic free energy density has $9$
independent parameters:
\begin{eqnarray}
f^{3d}_{biaxial} &=& C_1\, V_{nn}^2 + C_2 \,V_{mm}^2 + C_3 \, V_{ll}^2
		\nonumber\\
	&+&  C_4 \,V_{nn}\,V_{mm} + C_5 \,V_{nn}\,V_{ll} 
	+ C_6 \,V_{mm}\,V_{ll} \nonumber\\
	&+&  C_7 \,V_{nm}^2 +  C_8 \,V_{nl}^2 
	 + C_9 \,V_{ml}^2 ,
		 \label{elast-energy-biaxial}
\end{eqnarray}
where $\{\hat{n}, \hat{m},\hat{l}\}$ are three mutually orthogonal
principle axes of the biaxial nematic order in the deformed state.
Similarly a two-dimensional ideal nematic elastomer is characterized
by $4$ elastic constants and the following elastic free energy
density:
\begin{equation}
f^{2d}_{biaxial} = C_1\, V_{nn}^2 + C_2 \, V_{mm}^2 
	+ C_3 \,V_{nn}  V_{mm} 
	+ C_4\, V_{mn}^2, 
		 \label{elast-energy-2d}
\end{equation}
where $\{\hat{m},\hat{n}\}$ are the principle axes of the nematic order. 

We conclude this section by contrasting the differences between the
neo-classical elastic model, Eq.~(\ref{free-energy-neo}) and the
nemato-elastic model, Eq.~(\ref{elast-energy-2}).  While both theories
can be formulated in terms of the invariant strain $\mm{V}$ and both
capture the full nematic elastomer symmetry, the neo-classical theory
has the advantage of simplicity (characterized by only two independent
parameters, the shear modulus $\mu$ and the dimensionless ratio
$\zeta_{\parallel}/\zeta_{\perp}$), and can be easily used to describe
large deformations.  On the down side, as mentioned earlier, the 
neo-classical model is over-constrained, containing a larger
symmetry, Eq.~(\ref{extra-symmetry}) than that exhibited by a generic
nematic elastomer. The neo-classical theory is also not well-suited for
studying long wavelength fluctuations.

On the other hand, although the Laudau theory
Eq.~(\ref{elast-energy-2}) does not have this redundant symmetry,
characterized by five elastic moduli, it can only be used for small
deformations near the ground state manifold.  In the limit of infinite
bulk modulus, the moduli $B$ and $C$ drop out and the Landau theory
considerably simplifies to three independent parameters. Furthermore,
it allows for a systematic study of long wavelength fluctuations and
heterogeneities as well universal properties of nematic elastomers.


\section{Cosserat, micropolar, and asymmetric elasticity}
\label{Sec:Asymmetric-elasticity}

\subsection{Micropolar elastic media}
Nematic, as well as other liquid crystalline elastomers, are examples
of a wider class of micropolar elastic media \cite{EL:Nowacki} that
are elastic solids with both translational and rotational degrees of
freedom.  A proper description of micropolar elasticity therefore
falls outside the domain of the classical elasticity theory
\cite{EL:Landau}.  Related to the concept of micropolar elastic media,
there is another concept of micropolar fluids, i.e., fluids with broken
rotational symmetry.  Examples of micropolar fluids include nematic,
hexatic, and ferroelectric fluid phases, all characterized by an
asymmetric stress tensors.  In between micropolar solids and
micropolar fluids, there are also systems with orientational order and
{\em partial} translational order.  For example, in a smectic liquid
crystal, in addition to the mass-density wave position, whose
fluctuations are specified by a phonon field $u(\xv)$, the orientation
of the nematic director field $\nh(\xv)$ must be specified to fully
characterized the state. In the limit of infinitely strong anchoring
(which can be shown to be always the case in the long wavelength
limit), the director $\nh(\xv)$ becomes locked to the layer normal,
thereby allowing an effective description in terms of purely
translational elastic degrees of freedom.  Although a
three-dimensional material, a smectic liquid crystal is not a
three-dimensional solid, since it has rigidity only in one direction
alone the layer normal.
%

Historically \cite{comment:history}, the elasticity of micropolar
elastic solids was first studied by W. Voigt and the Cosserat brothers
back in early 1900's, and is often referred to as Cosserat media.  In
the standard micropolar elasticity theory, a deformation of the solid
is described by the displacement field $\xv(\Xv)$ as well as an
independent rotation vector $\vec{\phi}(\Xv)$, with latter specifying
the orientation of the ``molecules'' (which are usually modeled as
isotropic rigid bodies) relative to some reference configuration.  A
distinguishing macroscopic property of such media is that they can
transmit torque {\em independent} of force.  Not being widely noticed
for a long time, this subject was revived about half a century later,
most notably by Nowacki, Eringen and others \cite{EL:Nowacki}, in the
mechanical engineering and applied mathematics communities, and was
renamed as micropolar elastic media.  One major motivation underlying
these efforts seems to study the elasticity of solids with
microstructures.  The theories developed by these authors are quite
formal and with presentation that parallels the classic work of
elasticity by Love \cite{EL:Love}.  They focus on conservation laws,
as well as general equations of motion, without paying much attention
to the microscopic physical mechanism that leads to the unconventional
elasticity.  The success of these theoretical efforts on micropolar
elasticity is limited by the lacking of well controlled experiments on
relevant materials.  We note that only quite recently did {\em
  physicists} start to successfully synthesize these elastic solids
with internal degree of freedom.  Furthermore, most of the research on
the micropolar elasticity was concentrated on the {\em isotropic}
state of the micropolar media, where the coupling between the relative
rotation and shear deformation, e.g., the last term in
Eq.~(\ref{H-quadratic}) or Eq.~(\ref{elast-energy-2}), does not
appear.  As we will show in subsequent sections, this nemato-elastic
coupling is in fact the most interesting and most important feature of
micropolar elasticity.

For nematic elastomers, the internal orientational degree of freedom
is characterized by a unit vector field $\nh(\xv)$ with two degree of
freedom, not a rotation vector $\vec{\phi}(\xv)$ with three degree of
freedom.  In this section, we shall examine the elasticity of nematic
elastomers from the view point of micropolar elasticity theory, with
the conservation of angular momentum as the central principle.  Our
discussion shall thereby clarify various aspects of qualitative
difference between ordinary solids and nematic elastomers.  We will
also derive equations that must be satisfied by the stress tensor and
the coupled-stress tensor in elastic equilibrium, obtained from the
variation of the elastic energy with respect to a virtual deformation
and rotation.  From these equations (working in 3d) we will find
explicit expressions for the true stress tensor and the couple-stress
tensor for a uniformly deformed nematic elastomer.

Following the standard convention of the nonlinear elasticity
literature, in this section we shall use a slightly different notation
than the rest of the paper.  We will use $\Xv$ to denote the
referential (Lagrangian) coordinate and $\xv$ the target space
(Eulerian) coordinate \cite{comment:not-original}.  A state of an
ordinary elastic media is then given by a function $\xv(\Xv,t)$ where
$t$ is the time.  Let $d\Xv$, $d\vec{A}$, $dV$ be the line, surface,
and volume elements of the Lagrangian coordinates. Then the
corresponding quantities in the Eulerian coordinates $d\xv$,
$d\vec{a}$, $dv$ are given by standard relations \cite{EL:Ogden}:
\begin{subequations}
\begin{eqnarray}
d\xv &=& \lm \cdot d\Xv, \\
d\vec{a} &=& J \,\lm^{\rm -T} \cdot d\vec{A},
\label{da-dA}\\
dv &=& J \,dV,
\label{dv-dV}
\end{eqnarray}
\end{subequations}
where $J = |\partial \xv/\partial \Xv|$ is the corresponding Jacobian
factor.

A local physical quantity $g$, e.g., a certain extensive quantity per
unit mass, can be treated as a function of either the Lagrangian
or the Eulerian coordinates, i.e.,
\begin{equation}
g(\xv,t) = g(\xv(\Xv,t),t). 
\end{equation}
We use $\dot{g}$ to denote a material time derivative, i.e., the time
derivative of $g$ with the Lagrangian coordinate $\Xv$ fixed, and use
$\partial g/\partial t$ for the time derivative with the Eulerian
coordinate $\xv$ fixed.  Clearly these two derivatives are related to
each other in a standard way
\begin{equation}
\dot{g} = \frac{\partial g}{\partial t} + \vec{v} \cdot \nabla g,
\end{equation}
where 
\begin{equation}
\vec{v} = \frac{d}{dt} \, \xv(\Xv,t) = \dot{\xv}(\Xv,t);
\end{equation}
is the velocity field.  

As the elastic medium deforms, the region $\Omega(t)$ that it occupies
in the Eulerian coordinate space changes with time.  The corresponding
region $\Omega_0$ in the Lagrangian coordinate space remains constant
in time.  Taking the mass density $\rho_0$, measured in the Lagrangian
coordinates to be spatially uniform, we have:
\begin{equation}
\frac{d}{dt} \int_{\Omega_0} g \,\rho_0\,dV
 =\int_{\Omega_0} \dot{g} \,\rho_0\,dV,  
  \label{transport-theorem-0}
\end{equation}
since the material time derivative commuting with integral over the
Lagrangian coordinates.  The mass density, $\rho$, measured using the
Eulerian coordinates is related to $\rho_0$ by
\begin{equation}
\rho\,dv = \rho\, J\,dV = \rho_0\,dV 
\longrightarrow
\rho = J^{-1} \,\rho_0.
\end{equation}
We can use this to express Eq.~(\ref{transport-theorem-0}) in terms of
the Eulerian coordinates:
\begin{equation}
\frac{d}{dt} \int_{\Omega(t)} g \,\rho\,dv
 = \int_{\Omega(t)} \dot{g} \,\rho\,dv
 = \int_{\Omega(t)} \left( \frac{\partial g}{\partial t}
 + \vec{v} \cdot \nabla g
 \right)\,\rho\,dv.  
 \label{transport-theorem}
\end{equation}
Eqs.~(\ref{transport-theorem}) and (\ref{transport-theorem-0}) are
referred to as the transport theorem \cite{EL:Ogden} in elasticity
literature.

\subsection{Conservation of angular momentum in conventional elastic
  media}

Let $\xv$ (Eulerian coordinate) be an arbitrary point inside an
elastic media and $d\vec{a} = \hat{a} da $ a surface element passing
through $\xv$, where $\hat{a}$ is the unit normal vector.  The {\em
  stress vector} field $\vec{t}(\xv, d\vec{a})$ is defined as the
elastic force that is exerted on one side of the surface element
$d\vec{a}$ from the other side.  The famous Cauchy's theorem
\cite{EL:Love,EL:Ogden} then dictates that the stress vector $\vec{t}$
depends linearly on the vector $d\vec{a}$ and there exists a stress
tensor field $\mm{T}(\xv)$, referred to as the {\em Cauchy stress}, or
the {\em true stress}, given by
\begin{equation}
\vec{t}(\xv,d\vec{a}) = \mm{T}(\xv)\cdot d\vec{a},
\hspace{4mm} t_i = T_{ij} {a}_j.
\end{equation}
As we will see, for ordinary liquids or solids, conservation of
angular momentum requires the Cauchy stress tensor to be
symmetric, i.e., $T_{ij} = T_{ji}$.  We can also express the stress
vector in terms of the Lagrangian surface element $d\vec{A}$:
\begin{equation}
\vec{t}(\xv, d\vec{a}) 
= \mm{T}(\xv)\cdot d\vec{a} = \mm{S}(\Xv) \cdot d\vec{A},
\end{equation}
through the so-called {\em nominal stress} tensor $\mm{S}$ \cite{EL:Ogden},
related by Eq.~(\ref{da-dA}) to $\mm{T}$ via:
\begin{equation}
\mm{S} = J\, \mm{T} \lm^{\rm -T}, \hspace{4mm}
\mm{T} = J^{-1} \, \mm{S} \lm^{\rm T}. 
\label{relation:S-T}
\end{equation}
The nominal stress is generally {\em not} symmetric.  

For a deformed body, Newton's second law dictates that the rate of
change of total linear momentum is given by the total force acting on
the body:
\begin{equation}
\frac{d}{dt} \int_{\Omega(t)} \vec{v}\, \rho dv = 
  \int_{\Omega(t)} \vec{f} \,\rho dv
  + \int_{\partial \Omega(t)} \mm{T}(\xv)\cdot d\vec{a},
\end{equation}
which consists of the bulk and surface contributions, with $\vec{f}$
the body force per unit of mass acting at a point $\xv$.  Applying
the transport theorem on the left hand side, as well as the Gauss
theorem on the right hand side, we find the following equation of
motion for an (ordinary or micropolar) elastic solid:
\begin{eqnarray}
\int_{\Omega(t)}\rho \, \dot{\vec{v}} \,dv = 
\int_{\Omega(t)} \left(  \rho \, \vec{f}(\xv) + \nabla\cdot\mm{T} \right) \,dv.
\label{eomInt}
\end{eqnarray}
where $\nabla\cdot\mm{T}$ is a vector with components $\partial
T_{ij}/\partial x_j$.  Since this must be true for an arbitrary volume
element, \rfs{eomInt} must be satisfied by equality of integrands and
thus leads to a local field equation of motion
\begin{eqnarray}
\rho \, \dot{\vec{v}} &=& \rho \, \vec{f}(\xv) + \nabla\cdot\mm{T},
\label{Cauchy-1}
\end{eqnarray}
referred to as Cauchy's first law of motion.  For an elastic body in
equilibrium and under a vanishing body force, we recover the well
known result that static stress field is divergenceless in the absence
of external forces, i.e., $\partial_j T_{ij} = 0$.

The angular momentum of an ordinary solid is due to the center of mass
motion of its constituents (atoms, molecules, colloids) and is given
by
\begin{equation}
\vec{L}_e = \int_{\Omega} \rho dv \,\vec{x} \times \vec{v},
\label{Lm-def}
\end{equation}
where volume $v$ is not to be confused with velocity $\vec{v}$.  The
time derivative of the angular momentum is given by the total torque
acting on the elastic body:
\begin{eqnarray}
\frac{d}{dt} \vec{L}_e &=& 
	\frac{d}{dt} \int_{\Omega} \rho dv \,\, \vec{x} \times \vec{v} 
 = 	 \int_{\Omega} \rho dv \,\, \vec{x} \times \dot{\vec{v} }
	\nonumber\\
&=& \int dv \,\, \vec{x} \times (\rho \,  \vec{f} + \nabla\cdot\mm{T}),
\label{dLdt}
\end{eqnarray} 
where we used the transport theorem, Eq.~(\ref{transport-theorem})
and Cauchy's first law of motion, Eq.~(\ref{Cauchy-1}).  Defining a
pseudo-vector $\vec{T}_{A}$ associated with the antisymmetric part of
$\mm{T}$ by:
\begin{equation}
(T_{A})_i \equiv \epsilon_{ijk} T_{jk},
\end{equation}
it is easy to see that
\begin{equation}
\left[ \vec{x} \times (\nabla\cdot\mm{T}) \right]_i
=		\partial_l \left( \epsilon_{ijk} x_j T_{kl} \right)
+ (T_{A})_i. 
\end{equation}
Using this identity together with Gauss's theorem in Eq.~(\ref{dLdt})
we find
\begin{eqnarray}
\frac{d}{dt} \vec{L}_e = 
\int dv \left( \rho\,\vec{x} \times \vec{f} 
+ \vec{T}_{A}  \right)
+ \int_{\partial \Omega} 
\vec{x} \times \mm{T} \cdot d \vec{a}. 
\label{dLe-dt}
\end{eqnarray}
For a vanishing body force $\vec{f}$, the bulk contribution to angular
momentum $\vec{L}_e$ must be locally conserved, with only an external
surface stress contributing to a change in $\vec{L}_e$.  From above
expression this is clearly true if and only if $\vec{T}_{A} =0$.  We
therefore find Cauchy's second law of elasticity, namely, that in an
ordinary solid the Cauchy stress tensor $\mm{T}$ must be symmetric.

\subsection{Conservation of angular momentum in nematic elastomers}

A nematic elastomer is an anisotropic micropolar elastic medium, and
its complete description is characterized by the mapping $\xv(\Xv)$
and the nematic director field $\nh(\Xv)$. The time dependent
variation of the nematic director gives an additional {\em intrinsic}
contribution, $\vec{L}_i$ to the total angular momentum,
\begin{equation}
\vec{L}_{tot}=\vec{L}_e+\vec{L}_i.
\end{equation}
The extrinsic part $\vec{L}_e$, defined in Eq.~(\ref{Lm-def}), gives a
contribution associated with the motion of the center of mass of the
molecules.  The interaction between the translational and
orientational degrees of freedom allows an exchange between the
intrinsic and extrinsic angular momenta, and thus, neither is
separately conserved \cite{comment:observation}.   
The total angular momentum, $\vec{L}_{tot}$ is of course always conserved.
Going back to Eq.~(\ref{dLe-dt}), we see that for nematic elastomers,
the antisymmetric part of the Cauchy stress tensor, $\vec{T}_A$ is
generally nonzero, describing the production of the extrinsic angular
momentum due to the coupling between translational and orientational
degrees of freedom.

To study the intrinsic angular momentum, we need to introduce two
physical quantities special to micropolar media.  Adopting the
terminology from the elasticity literature \cite{EL:Nowacki}, the body
torque, $\vec{G}$, is the external torque (e.g., due to an external
magnetic field) per unit of volume acting on the nematic degrees of
freedom.
The couple-stress vector $\vec{c}(\xv,d\vec{a})$ is defined as the
torque transmitted through an infinitesimal surface element
$d\vec{a}$, solely due to the distortion of nematic director across
the surface.  Clearly this torque acts on the orientational degrees of
freedom, i.e., the nematic director \cite{comment:torque}.  
  One can extend the Cauchy's theorem and show that the
couple-stress vector also depends linearly on $d\vec{a}$ and therefore
can be written as the contraction between $d\vec{a}$ and the {\em
  couple-stress tensor} $\mm{C}(\xv)$:
\begin{equation}
\vec{c}(\xv,d\vec{a}) = \mm{C}(\xv) \cdot d\vec{a}.
\label{couplestress-def}
\end{equation}
The couple-stress tensor describes the torque transmitted through a
unit surface area due to a distortion of the nematic order, instead of
a geometric deformation \cite{comment:torque-2}. 

From a general principle, the rate of change of the total angular
momentum is given by the total torque, which is given by combination 
of the body force and body torque acting on the bulk, and by the stress 
and couple-stress vectors acting on the boundary:
\begin{eqnarray}
\frac{d}{dt} \vec{L}_{tot} &=& 
 \int \rho dv\, \left(	
\xv \times \vec{f} + \vec{G}  \right)
+ \int_{\partial \Omega} \,\left(
\xv \times \mm{T} + \mm{C} \right)
\cdot d\vec{a}
\nonumber\\
&&
\label{dLdt-2}
\end{eqnarray}
Subtracting Eq.~(\ref{dLe-dt}) from Eq.~(\ref{dLdt-2}), we find the
rate of change of the intrinsic angular momentum
\begin{equation}
\frac{d\vec{L}_i}{dt} 	 = 
 \int \rho dv\, \left( \vec{G} - \vec{T}_{A} \right)
+ \int_{\partial \Omega}  \mm{C} \cdot d\vec{a}.
\label{dLidt}
\end{equation}
In the absence of body force and torque as well as the boundary stress
and couple-stress, Eqs.~(\ref{dLe-dt}) and (\ref{dLidt}) reduce to
\begin{equation}
\frac{d\vec{L}_e}{dt} = - \frac{d\vec{L}_i}{dt} 
=  \int \rho dv \,\, \vec{T}_{A}
= \int  \rho dv \,\, \hat{e}_i\epsilon_{ijk}T_{jk},
\end{equation}
demonstrating that the antisymmetric part of the Cauchy stress tensor
describes angular momentum transfer between translational and
rotational degrees of freedom of a micropolar elastic medium.

\subsection{Equilibrium conditions, stress and couple-stress}

The total free energy of an ideal nematic elastomer consists of the
elastic energy as a function of $\lm$ and $\nh$, as well as the Frank
energy Eq.~(\ref{f_Frank}).  External body forces, such as those due
to e.g., a gravitational or magnetic fields, can also be easily
included. For example the free energy associated with the interaction
a magnetic field $\vec{h}$ and the nematic director field $\nh(\xv)$
is given by
\begin{eqnarray}
F_{m} = \int_{\Omega} f_m J^{-1}dv 
= - \frac{1}{2} \,\gamma_a\int_{\Omega} (\nh(\xv) \cdot \vec{h})^2\,J^{-1}dv, 
\label{F_m}
\end{eqnarray}
that induces body torque.  We note that the energy density $f_m$ is 
defined with respect to unit volume in Lagrangian coordinate and 
the integration is over the Eulerian coordinates $\xv$. 

To derive the equilibrium conditions, let us apply external force
$\hat{e}_i\,\ct_{ij}(\xv) \,da_{j}$ and torque $\hat{e}_i\,
\cc_{ij}(\Xv)\,da_{j}$ on the surface element $d\vec{a}$ at the
boundary.  We consider an infinitesimal variation of the equilibrium
state, i.e., translation of the center-of-mass Eulerian coordinate
as well as rotation of the nematic director field as follows,
\begin{subequations}
\label{variation}
\begin{eqnarray}
\xv &\rightarrow& \xv'= \xv + \delta \xv,
\label{variation-1}\\
\nh(\xv) &\rightarrow& \nh'(\xv') = \nh(\xv) + \delta\nh(\xv),
\label{variation-2}\\
\delta \nh(\xv) &=& \delta \vec{\omega}(\xv) \times \nh(\xv).
\label{variation-3}
\end{eqnarray}
\end{subequations}
Operationally, we first rotate the nematic director by
$\delta\omega(\xv)$ without deformation, and then translate each point
$\xv$ (together with the rotated director $\nh + \delta \nh$) to
$\xv'$ without rotation of the director field.  The external traction
\cite{comment:traction} on the boundary introduces two following terms
in the total free energy:
\begin{eqnarray}
\delta F_{boundary} = - \int_{{\partial \Omega}} \left(
\ct_{ij}(\xv) \,  \delta x_i + \cc_{ij}(\xv)\, 
 \delta \omega_i(\xv) \right) \,da_{j}, 
 \label{dF-b}
\end{eqnarray}
The force couples to the translational degree of freedom $\delta\xv$
while the torque couples to the rotational degree of freedom $\delta
\nh$.  The total free energy of nematic elastomer under external
boundary traction is then given by
\begin{eqnarray}
F_{tot} = \int_{\Omega}\left( f_e + f_d + f_m - \nu(\xv)\nh^2\right)J^{-1} dv
+ \delta F_{boundary},\nonumber\\
&&\label{F_tot}
\end{eqnarray} 
where for convenience, we included a Lagrange multiplier term to
ensure that the constraint that $\nh$ is a unit vector field is
satisfied. 

In equilibrium, the first-order variation of the total free energy
must vanish. Since the referential volume element $J^{-1}dv = dV$ in
Eq.~(\ref{F_tot}) is invariant under the variation of the current
coordinates $\xv \rightarrow \xv'$, the variation of the total elastic
energy is given by
\begin{eqnarray}
\delta F_{tot} &=& 0\nonumber\\
&=& \int_{\Omega} \left( 
\delta f_e + \delta f_d + \delta f_m 
- \nu(\xv) \, \nh_i \delta \nh_i\right) \, J^{-1} dv\nonumber\\
&+&\delta F_{boundary}.
\label{deltaF}
\end{eqnarray} 

Let us look at the variation of the different part of the bulk free
energy separately.  The elastic free energy density $f_e[\lm,\nh]$ is
a function of the deformation gradient $\lm_{i\alpha} = \partial
x_i/\partial X_{\alpha}$ and the nematic director $\nh$.  (It is also
a function of the initial director $\nh_0$, which does not change in
the variation.)  Under the variation Eqs.~(\ref{variation}) the
deformation gradient changes as follows:
\begin{eqnarray}
\lm_{i\alpha} &\rightarrow& \lm'_{i\alpha} 
 = \frac{\partial x'_i}{\partial X_\alpha} 
 = \lm_{i\alpha} + \partial_{\alpha} \delta x_i,
 \\
\delta \lm_{i\alpha} &=& \partial_{\alpha} \delta x_i 
= \frac{\partial x_j}{\partial X_{\alpha}} 
\frac{\partial \delta x_i}{\partial x_j}
=  \lm_{j\alpha} \partial_j \delta x_i,
\end{eqnarray}
where to avoid confusion, we use Roman indices $i$, $j$, $k$, etc.,  to
label Eulerian coordinates and use Greek subscripts $\alpha$, $\beta$,
etc., to label Lagrangian coordinates.  The first-order variation of
the elastic free energy is therefore given by
\begin{eqnarray}
\delta F_e &=& \int_{\Omega} \left( 
\frac{\partial f_e}{\partial \nh_i} \,\delta \nh_i
+ \frac{\partial f_e}{\partial \lm_{i\alpha}} \, \lm_{j\alpha}
\partial_j \delta x_i \right)  J^{-1} dv
\nonumber\\
&=& \int_{\Omega} \left[ J^{-1} 
\frac{\partial f_e}{\partial \nh_i} \,\delta \nh_i
- \frac{\partial}{\partial x_j} \left( 
J^{-1} \frac{\partial f_e}{\partial \lm_{i\alpha}} 
\, \lm_{j\alpha} \right)  \delta x_i \right]  dv 
\nonumber\\
&+& \oint_{\partial \Omega} 
J^{-1} \frac{\partial f_e}{\partial \lm_{i\alpha}} \, \lm_{j\alpha}
\delta x_i \, da_j . 
\label{dF-e}
\end{eqnarray}

The Frank free energy density $f_d [\nh, \nabla\nh]$ is a function of
both the director $\nh$ and its gradient $\nabla \nh$.  Under the
variation \rf{variation}, the gradient $\nabla \nh$ changes as
follows:
\begin{eqnarray}
\frac{\partial \nh_i} {\partial x_j} \rightarrow 
\frac{\partial \nh'_i}{\partial x'_j} 
= \frac{\partial x^k}{\partial x'_j} 
\frac{\partial }{\partial x_k} \left(
\nh_i + \delta \nh_i \right),
\label{deltanh}
\end{eqnarray}
which, using Eqs.~(\ref{variation}), to the lowest
order gives
\begin{eqnarray}
\delta \partial_j \nh_i = \partial_j \delta\nh_i 
- \partial_j \delta x_k \partial_k \nh_i
\end{eqnarray}
The first-order variation of the Frank free energy is therefore 
\begin{eqnarray}
\delta F_d &= & \int_{\Omega}  \left(
\frac{\partial f_d}{ \partial \nh_i} \delta \nh_i 
+ \frac{\partial f_d}{\partial \eta_{ji}}\,
\partial_j \delta\nh_i 
\right)J^{-1} dv, \nonumber\\
&-&  \int_{\Omega} 
  \frac{\partial f_d}{\partial \eta_{ji} }
 \eta_{ki}\,\partial_j \delta x_k \, J^{-1} dv, 
\label{dF-d}
\end{eqnarray}
where 
\begin{equation}
\eta_{ji} = \frac{\partial \nh_i}{\partial x_j}.
\end{equation}

Finally, the magnetic free energy density $f_m[\nh]$, as given by
Eq.~(\ref{F_m}), is a function of the nematic director only.  The
variation of $F_m$ is therefore given by
\begin{eqnarray}
\delta F_m &=& \int_{\Omega} \frac{\partial f_m}
{\partial \nh_i} \delta \nh_i \, J^{-1}dv  
\nonumber\\
&=& - \gamma_a H_i \int_{\Omega} 
(\nh\cdot\vec{h}) \delta \nh_i \,J^{-1}dv.
\label{dF-m} 
\end{eqnarray}

Substituting Eq.~(\ref{dF-e}), Eq.~(\ref{dF-d}), Eq.~(\ref{dF-m}) and
Eq.~(\ref{dF-b}) into Eq.~(\ref{deltaF}), and performing integration
by parts appropriately, we arrive at
\begin{eqnarray}
\delta F_{tot}&=&
\oint_{\partial \Omega} da_j \left[
J^{-1}\left(  
\frac{\partial f_e}{\partial \lm_{i\alpha}} \lm_{j\alpha} 
- \frac{\partial f_d}{\partial \eta_{jk} } \eta_{ik}
\right)  - \ct_{ij}(\xv)
\right]\delta x_i 
\nonumber\\
\label{deltaF-2-1}\nonumber\\
 &-& \int_{\Omega}  \partial_j \left[ J^{-1}\left(  
\frac{\partial f_e}{\partial \lm_{i\alpha}} \lm_{j\alpha} 
- \frac{\partial f_d}{\partial \eta_{jk} } \eta_{ik}
\right) \right]
\delta x_i\, dv
  \label{deltaF-2-2}\nonumber\\
&+&  \oint_{\partial \Omega} da_j 
\left[ J^{-1}\frac{\partial f_d}{\partial \eta_{ji} }  \delta \nh_i
- \cc_{ij}(\xv) \delta \omega_i(\xv)
\right]  \label{deltaF-2-3}\nonumber\\
&+&  \int_{\Omega} J^{-1}\left[
\frac{\partial f_e}{\partial \nh_i} 
+ \frac{\partial f_d}{\partial \nh_i}
- J\, \partial_j \left( J^{-1}\frac{\partial f_d}
{\partial \eta_{ji}}\right) 
\right. \nonumber\\
&+& \left.\frac{\partial f_m}{\partial \nh_i}
- \nu(\xv) \nh_i \right] \delta \nh_i \,dv 
  \label{deltaF-2-4}\nonumber\\
 &=& 0.
\end{eqnarray}
At mechanical equilibrium, each volume and surface integral in the left
hand side must vanish separately.

Let us first look at the surface integral Eq.~(\ref{deltaF-2-1}),
which is linear in $\delta \xv$.  Since by definition $\ct_{ij}$ is
the $i-$th components of the external force per unit area acting on
the surface element with normal $\hat{e}_j$, it has to be balanced by
the corresponding component of elastic force, i.e., the Cauchy stress
tensor.  We immediately see that the remaining terms in the surface
integrand must be the Cauchy stress tensor $T_{ij}$, giving us an
important constitutive relation for the stress tensor field:
\begin{eqnarray}
T_{ij}(\xv) = J^{-1} \left( \frac{\partial f_e}
{ \partial \lm_{ia}} \lm_{ja} 
- \frac{\partial f_d}{\partial \eta_{jk}} \eta_{ik}  \right). 
\label{Cauchy-stress}
\end{eqnarray}
For the convenience of later discussion, we separate the stress tensor
into two parts,
\begin{eqnarray}
T_{ij} = T^e_{ij} + T^d_{ij}, 
\end{eqnarray}
where $T^e_{ij}$ due to the elastic deformation, and
$T^d_{ij}$ due to a distortion of the nematic director field are given by
\begin{subequations}
\begin{eqnarray}
T^e_{ij} &=&   J^{-1} \frac{\partial f_e}
{ \partial \lm_{ia}} \lm_{ja} ,\label{Te-fe}\\
T^d_{ij} &=& - J^{-1} \frac{\partial f_d}
{\partial \eta_{jk}} \eta_{ik}. \label{Td-fd}
\end{eqnarray}
\end{subequations}

For a conventional solid, there is no Frank free energy $f_d$, and
therefore the stress tensor is solely given by $T^e$ in
Eq.~(\ref{Te-fe}).  Furthermore, rotational symmetry in the embedding
space implies that the elastic energy must be a function of the metric
tensor defined as
\begin{equation}
g_{\alpha\beta} = \frac{\partial \xv}{\partial X_{\alpha}}\cdot
        \frac{\partial \xv}{\partial X_{\beta}} 
      = \lm_{i\alpha}\lm_{i\beta}. \label{metric_def}
\end{equation}
which then gives for the Cauchy stress tensor of an ordinary solid
\begin{equation}
T_{ij} = T^e_{ij} =2\, J^{-1} \frac{\partial f}{\partial
g_{\alpha\beta}} \lm_{i\alpha}\lm_{j\beta} ,
\end{equation}
which (as expected on general grounds discussed above) is obviously
symmetric.

Let us now calculate the stress tensor $T^e_{ij}$ for an ideal nematic
elastomer.  For the neo-classical elastic energy, Eq.~(\ref{f-neo}),
to impose the incompressibility constraint, we introduce a Lagrange
multiplier $p$, physically corresponding to the {\em pressure}.  We
then find that the elastic stress tensor is given by
\begin{equation}
\mm{T}^e =  \mu \, \lm \, {\mathbf l}_0 \, 
	\lm^{\rm T}\,  {\mathbf l}^{-1} 
	- p \, \mm{I}. 
\end{equation}
which is clearly asymmetric, unless ${\mathbf l}$ commutes with $\lm
\, {\mathbf l}_0 \, \lm^{\rm T}$.  The pressure term is always
isotropic and symmetric regardless of the deformation, as one would
expect.

A different form of the stress tensor ${\mathbf T}^e_{ij}$ is obtained
for our phenomenological elastic model, Eq.~(\ref{elast-energy-2}).
A standard but somewhat tedious calculation gives
\begin{widetext}
\begin{eqnarray}
{\mathbf T}^e = (2\,\mm{V} + \mm{l}) \,\left[ 
	(B \,\Tr \,\mm{V} + C\, \tilde{V}_{nn} ) \,\mm{I}
	+ (C\,\Tr \,\mm{V} +  \mu_{\rm L}\, \tilde{V}_{nn})
	(\nh\nh - \frac{1}{3} \mm{I}) 
+  \mu \,\tilde{\mm{V}}_{\perp} 
	+ \mu_{n\perp} \mm{V}_{n\perp}
	\right],
\end{eqnarray}
\end{widetext}
which is also generically asymmetric.

Let us next turn to $T^d_{ij}$ in Eq.~(\ref{Td-fd}), derived from the
Frank free energy.  It is non-vanishing only for an inhomogeneous
configuration of the director field.  Except for the isotropic
pressure term, which can be associated instead with $T^e_{ij}$,
$T^d_{ij}$ is identical to the Ericksen stress tensor
\cite{LC:deGennes} studied extensively in nematic liquid crystals
literature.  Using Frank free energy, Eq.~(\ref{f_Frank}) inside
Eq.~(\ref{Td-fd}) we find
\begin{eqnarray}
T^d_{ij} = &-& K_1 (\nabla\cdot\nh) \ppi n_j 
- K_2 (\nh \cdot \nabla \times \nh) 
 \epsilon_{jkl} n_l \ppi n_k 
 \nonumber\\
 &-& K_3 n_j (\ppi n_k) (\nh \cdot \nabla n_k),
\end{eqnarray}
that is also generically asymmetric, unless all Frank elastic constants
are the same.

Let us now look at the bulk term (volume integral) in
Eq.~(\ref{deltaF-2-2}) associated with the variation $\delta \xv$ .
It is clear that the integrand describes the total force per unit
volume acting on the element $dv$. Indeed the integrand is identical
to $- \partial_j T_{ij}/\delta x_i$, with $T_{ij}$ the Cauchy stress
tensor given by Eq.~(\ref{Cauchy-stress}).  Mechanical equilibrium
then implies
\begin{equation}
\partial_j T_{ij} \equiv\nabla\cdot{\mathbf T} = 0,
\end{equation} 
corresponding to a force balance equation of ordinary solids, in the
case of a vanishing body force.

Similarly, the integrand in the surface integral of
Eq.~(\ref{deltaF-2-3}) involving orientational degrees of freedom must
also vanish.  Since the tensor $\cc_{ij}$ is the $i-$th component of
the external torque per unit area acting on the the surface element
with normal $\hat{e}_j$, it must be balanced by the corresponding
component of the couple-stress term, defined in
Eq.~(\ref{couplestress-def}).  Using Eq.~(\ref{variation-3}) to relate
$\delta\nh$ to $\delta \vec{\omega}$, we see that the couple-stress
tensor $C_{ij}$ must be given by
\begin{eqnarray}
C_{ij}(\xv) = J^{-1} \epsilon_{ikl} \frac{\partial f_d}{\partial
\eta_{jl}} \nh_k. \label{C-fd}
\end{eqnarray}
Substituting the Frank free energy Eq.~(\ref{f_Frank}) into
Eq.~(\ref{C-fd}) we find
\begin{eqnarray}
C_{ij} &=& - K_1 (\nabla \cdot \nh) \epsilon_{ijk}n_k 
- K_2 (\nh\cdot\nabla \times\nh) (\delta_{ij} - n_i n_j) 
\nonumber\\
&+& K_3 \epsilon_{ikl} n_j n_k (\nh \cdot \nabla) n_l. 
\end{eqnarray}
Similar to the stress tensor $T^d_{ij}$, the couple-stress tensor
$C_{ij}$ is nonzero only if the nematic director field is non-uniform.

Finally, we look at the integrand of the bulk term in
Eq.~(\ref{deltaF-2-4}) linear in $\delta \nh$.  Defining a local
molecular field acting on the nematic director by
\begin{eqnarray} 
h_i (\xv) = -\frac{\partial f_e}{\partial \nh_i} - \frac{\partial
  f_d}{\partial \nh_i} - \frac{\partial f_m}{\partial \nh_i} +
J\,\partial_j \left( J^{-1}\frac{\partial f_d}{\partial \eta_{ji}}
\right),
\end{eqnarray}
bulk torque balance requires a vanishing of this integrand.
This then gives a torque balance equation
\begin{eqnarray}
  h_i(\xv) + \nu(\xv) \nh_i  = 0,
\end{eqnarray}
which imposes an equilibrium orientational constraint that the local
molecular field must be parallel to the nematic director.  We do not
analyze the molecular field $\vec{h}$ in detail in this work.

\section{Nonlinear elastic model of  {\em homogeneous} nematic 
elastomers} 
\label{Sec:Elast-fluct}

\subsection{Stress versus strain ensembles}
The main objective of this paper is to study the effects of long
wavelength fluctuations on the macroscopic elasticity, e.g., the
stress-strain relation, of an ideal nematic elastomer.  To formulate
the problem, we need to consider the relation between fluctuations and
finite temperature elasticity in some detail.  At finite temperature,
mass points in an elastic media fluctuate around their equilibrium
positions, regardless of whether or not the system is coupled to an
external traction.  On the other hand, an external traction causes a
macroscopic deformation, which changes equilibrium positions of mass
points.  Furthermore, the spectrum of the fluctuations may also be
modified by a macroscopic deformation.  

The central quantity we want to calculate is the elastic free energy,
as usual obtained as a sum over all elastic configurations with an
appropriate Gibbs-Boltzmann weight, determined by the elastic energy
discussed in the previous section. It then determines a macroscopic
elasticity, renormalized by long wavelength fluctuations. The free
energy can be computed either as a function of the imposed macroscopic strain
tensor and the nematic director, or as a function of the external
traction.  The choice of the macroscopic control variable (strain or
traction) amounts to a choice of a statistical ensembles within which
to study thermal fluctuations.  As usual, we expect that in the
thermodynamic limit, macroscopic quantities are independent of the
choice of the ensemble.

Let us first look at the constant strain ensemble by fixing the
macroscopic deformation gradient $\lm_0$ and the nematic director
$\hat{n}_0$.  For simplicity, we shall only study the case where
$\lm_0$ and $\nh_0$ are uniform.  We shall switch back to the
notations we introduced in Sec.~\ref{Sec:Model}, using $\xv$ for the
nematic referential coordinates and $\rv(\xv)$ for the deformed
(target) coordinates.  To compute the thermodynamics (e.g., the
partition function) we need to sum over all fluctuations of the
position field $\rv(\xv)$ and the nematic director field $\nh(\xv)$
around their equilibrium values.  Naturally, we parameterize these two
fields in terms of the phonon displacement field, $\uv(\xv)$ and the
nematic director fluctuation, $\delta\nh(\xv)$, defined by
\begin{eqnarray}
\rv(\xv) &=& \rv_0(\xv) + \uv(\xv) =  \lm_0\cdot \xv + \uv(\xv) ,
\label{r-l0-u} \\
\nh(\xv) &=& \nh_0 + \delta\nh(\xv).
\end{eqnarray}
Expressing the Landau elastic energy, Eq.~(\ref{elast-energy-2}) in
terms of these fields, we obtain the elastic ``Hamiltonian''
functional $H[\lm_0,\nh_0,\uv,\delta\nh]$.  Since $\uv(\xv)$ and
$\delta\nh(\xv)$ are local fluctuations, they have well-defined
Fourier transformations \cite{comment:phonon-linear}.  The
corresponding thermodynamic potential, given by
\begin{eqnarray}
F[\lm_0,\nh_0] &=&  -T\,\log \int D\uv(\xv) D\delta\nh(\xv)\,
e^{-H[\lm_0,\nh_0,\uv,\delta\nh]/T},\nonumber\\
&&\label{F-functintegral}
\end{eqnarray}
is the elastic free energy, as a function of the macroscopic
deformation $\lm_0$ and the average nematic director $\nh_0$. In above
functional integrals a short distance (large momentum) cutoff is
implicit, set by the lattice constant in crystalline materials and
polymer network mesh size in elastomer materials and we are using
units in which the Boltzmann constant $k_B$ is $1$.

To calculate the functional integral in Eq.~(\ref{F-functintegral}),
we first expand the elastic Hamiltonian around its minimum for the given
$\lm_0$ and $\nh_0$, which, baring the possibility of an elastic
instability, is given by $\uv = 0$ and $\delta\nh=0$, i.e., a
uniformly deformed reference state.  The zero-th order (tree level)
approximation to the functional integral is then simply given by the
saddle point approximation, i.e., by setting $\uv=0$ and $\delta \nh =
0$ in the integrand of Eq.~(\ref{F-functintegral}), i.e.,
\begin{equation}
F[\lm_0,\nh_0] \approx H[\lm_0,\nh_0,\uv=0,\delta\nh=0] 
\equiv H_{mft}[\lm_0,\nh_0].
\end{equation}
This is just the Landau theory, that does not take into account any long
wavelength fluctuations.

To systematically address the effects of fluctuations on the
macroscopic elasticity beyond the Landau (mean-field) approximation,
we expand the elastic energy in terms of the phonon field $\uv(\xv)$
and the director fluctuation $\delta\nh$, and calculate the functional
integral Eq.~(\ref{F-functintegral}) order by order in corresponding
nonlinearities.  These fluctuations are generically important for all
soft solids.  As we discussed in Sec.~\ref{Sec:Intro}, for the
particular case of ideal nematic elastomers, the fluctuations of
nemato-elastic Goldstone modes lead to qualitative breakdown of
perturbation theory in the thermodynamic limit, as well as necessity
of a renormalization group analysis.
 
We can also study the constant traction ensemble, by introducing a
boundary traction term, expressed in terms of the Lagrangian
coordinates and the external traction tensor $S_{i\alpha}=J (\ct
\lm^{-T})_{i\alpha}$ using Eq.~(\ref{da-dA}):
\begin{equation}
\delta H = - \int_{\partial \Omega_0} \delta r_i(\xv) \,
S_{i\alpha}(\xv)\,dA_{\alpha}, \label{boundary-traction}
\end{equation} 
For simplicity we shall focus on the case of a vanishing external
torque traction, $\cc_{ij}=0$.

In the case where the external traction tensor $\mm{S}$ is independent
of $\xv$, we can use Gauss's theorem to rewrite $\delta H$ \cite{comment:S} as
\begin{equation}
\delta H \rightarrow - \int_{\Omega_0} dV\, S_{i\alpha}
\partial_{\alpha} r_i = - \int_{\Omega_0} dV\, S_{i\alpha}
\lm_{i\alpha},
\label{S-lambda}
\end{equation}
a form that confirms that the deformation gradient
$\lm_0$ and the traction tensor $\mm{S}$ are conjugate variables.
Therefore, in equilibrium, $\mm{S}$ is equal to the nominal stress
tensor, as introduced in \rfs{relation:S-T}.    In fact this explains 
the notation we used in Eq.~(\ref{boundary-traction}).   

The thermodynamic potential for this ensemble can then be calculated by
summing over {\em all} elastic configurations with the modified
Gibbs-Boltzmann weight:
\begin{eqnarray}
G[\mm{S}] = -T \log \int D\rv(\xv) D\nh(\xv)e^{-(H+\delta H)/T},
\label{G-functintegral}
\end{eqnarray}
As usual the thermodynamic potential $G[\mm{S}]$,
Eq.~(\ref{G-functintegral}) and the elastic free energy
Eq.~(\ref{F-functintegral}) are related by a Legendre transformation
\cite{comment:analogue}:
\begin{equation}
G[\mm{S}] = (F[\lm] - V\,\Tr \, \mm{S} \lm )|_{\lm_0},
\end{equation}
where $V$ is the system volume and $\lm_0$ is the solution to the
following equation
\begin{equation}
\frac{\partial}{\partial \lm} \left(F(\lm) - V\, \Tr\, \mm{S} \lm \right)
= \frac{\partial}{\partial \lm} F(\lm) - V\,\mm{S} = 0.
\end{equation}
This is precisely the relation between the macroscopic nominal stress
tensor and the macroscopic deformation gradient \cite{EL:Ogden}, both
of which can be directly measured in experiments.

\subsection{Referential coordinates at finite temperature}
\label{Sec:reference-state}
As defined in Sec.~\ref{Sec:Model}, the nematic reference state
minimizes the elastic Hamiltonian and thus precludes the appearance of
terms linear in the invariant strain tensor $\mm{V}$.  However, in the
presence of thermal fluctuations and elastic nonlinearities, the
average position $\langle \rv(\xv) \rangle$ of a mass point is not the
same as its nematic referential coordinate $\xv$.  Since the nematic referential coordinate $\xv$ and the average position $\langle \rv(\xv) \rangle$ differ only at finite temperature, their difference (to lowest order) must be proportional to $T$, and inversely proportional to an elastic modulus.  

For an {\em isotropic} solid, symmetry dictates that $\xv$ and $\langle \rv(\xv) \rangle$ can only differ by a scalar factor and therefore the associated modulus must be the bulk modulus, $B$: 
\begin{eqnarray}
\langle \rv(\xv) \rangle - \xv \approx \mbox{const.} \frac{T\,\xi^{-d}}{B} \xv,
\end{eqnarray}
where $\xi$ is the microscopic cutoff length-scale.  For ordinary
isotropic rubber we have ${T\,\xi^{-d}}/{B}\ll 1 $ and therefore the
difference between the referential coordinate and the average position
of a mass point is negligible.  In contrast, for an {\em anisotropic}
elastomers such as a nematic elastomer, the relation between $\langle
\rv(\xv) \rangle$ and $\xv$ involves other, significantly smaller
elastic constants, which are comparable with $T\,\xi^{-d}$.  Therefore,
generically thermally induced deformation of the average position
$\rv(\xv)$ is non-negligible, particularly near the I-N transition.
For a given elastic Hamiltonian, this relation can be systematically
computed and in principle is experimentally accessible.

Because of this non-negligible difference between $\xv$ and
$\langle\rv(\xv)\rangle$ there are two (in principle equivalent)
complementary descriptions of the elastic theory of a nematic
elastomer.  One, most conceptually clear formulation is in terms of
phonon modes expanded about a coordinate system defined by
$\langle\rv(\xv)\rangle$.  The advantage of such approach is that the
computed renormalized elastic free energy is by construction minimized
by a vanishing phonon displacement, i.e., precludes any linear terms
in the nonlinear strain tensor $\mm{V}$. A technical disadvantage is
that the elastic Hamiltonian (that does not include long wavelength
fluctuations) will then necessarily admit terms {\em linear} in
$\mm{V}$, whose coefficients are adjusted order by order in
perturbation theory so as to ensure that the renormalized elastic free
energy is free of such linear terms \cite{comment:counter-term}.  The
latter is thus quite cumbersome to work with.  An alternative,
technically more convenient approach is to choose the nematic
reference coordinate system $\xv$, which ensures that instead the
elastic Hamiltonian is free of terms linear in the strain. The
unavoidable price of this is that such linear terms will then be
generated in the renormalized elastic free energy, as a signal of a
shift in the elastic minimum configuration.  These terms will be
discussed in much more detail when we carry out the renormalization
group analysis in later sections \cite{comment:linear-terms}.

Thus, for convenience we
will always choose the nematic referential coordinates $\xv$ to label
mass points and about which to expand phonon fluctuations.
Nevertheless, our final results in this work will of course be {\em
  independent} of the choice of the reference state.\cite{LC:deGennes}

\subsection{Strain-only nonlinear elastic model}
\label{Sec:Effective-model}

We are interested in the long wavelength elastic fluctuations in
nematic elastomers, i.e., fluctuations of the phonon and the nematic
director fields around their equilibrium values.  In thermal
equilibrium, the Gibbs-Boltzmann distribution of these fluctuations is
controlled by the coarse-grained elastic Hamiltonian functional,
Eq.~(\ref{elast-energy-2}), augmented by the Frank energy
Eq.~(\ref{f_Frank}) for the distortion of nematic order.  Being fully
invariant under rotations as well as arbitrary soft deformations, this
energy functional is rather complicated, containing a large number of
nonlinearities, most of which are irrelevant at long length-scales.
To analyze effects of fluctuations, we therefore look for a simpler,
minimal model formulation, which from the start contains only those
nonlinearities that survive at long length-scales.  Furthermore,
focusing on the elastic degrees of freedom, we will integrate out the
nematic director fluctuation in the partition function and obtain an
effective theory in terms of only the translational degree of freedom
$\rv(\xv)$.  This operation is well justified if the nematic director
is tightly coupled to the elastic media, such that the fluctuaftions
of the nematic director from its energetically favored direction
(determined by the elastic degrees of freedom) are small.  This
condition is always satisfied \cite{comment:condition} and is quite
similar to the locking (at long length-scales) of the nematic director
fluctuations to the local layer normal fluctuations in a smectic
liquid crystal, thereby allowing a well-known purely elastic
description of smectic liquid crystals.

A ``strain-only'' elastic model can be obtained by integrating out the
nematic director fluctuations from the Hamiltonian
(\ref{elast-energy-2}), formulated in the nematic state. However, to
clarify the role of the spontaneous symmetry breaking and to simplify
the analysis it is more convenient to do this in a model formulated in
the isotropic phase, i.e., expressed in terms of conformational
degrees of freedom relative to the isotropic reference state and the
nematic order parameter $\Qm$. Integrating out the nematic order
parameter, $\Qm$ leads to an effective elastic energy functional in
which the effective shear modulus changes sign at the I-N transition
\cite{GolLub89,LMRX}, thereby inducing a uniaxial distortion of the
elastomer matrix. Expansion about the resulting nematic reference
state then gives the sought-after purely elastic description of the
nematic elastomer.

Our goal is to deduce the most general form of the resulting elastic
Hamiltonian, carefully ensuring its underlying rotational and
translational invariance.  To ensure these conditions we take it to be
the most general scalar function of the metric tensor $\mm{g}$,
defined in Eq.~(\ref{metric_def}).  Without loss of generality, we can
express it as a function of the following scalar quantities
\begin{eqnarray}
S_n = \Tr \,\mm{g}^n = \Tr\,(\Lm^{\rm T} \Lm)^n, \hspace{5mm}
\hspace{3mm} n = 1, 2, 3,\ldots
\end{eqnarray}
which are rotationally invariant both in the reference space and in the
embedding space.  In three dimensions, the metric tensor has only
three eigenvalues, ensuring that first three invariants $S_n$ are
independent \cite{comment:dimension}. 

Hence the elastic Hamiltonian is a function of $S_1$, $S_2$, and
$S_3$:
\begin{equation}
H[\mm{g}] = H[S_1,S_2,S_3].  \label{elastic_energy}
\end{equation} 

In the high-temperature isotropic phase, the elastic Hamiltonian is
minimized by $\mm{g} = \mm{I}$, for which $S_1=S_2=S_3=3$, obviously
corresponding to the isotropic reference state.  In the
low-temperature broken-symmetry phase, the elastic energy is minimized
by the nematic reference state (NRS), characterized by a uniaxial
metric tensor
\begin{equation}
\mm{g}_0 = \Lm^{\rm T}_0 \Lm_0, 
\end{equation}
where $\Lm_0$, given by Eq.~(\ref{Lm0}), is the spontaneous
deformation accompanying the I-N transition.  Inside the nematic
phase, we can expand the elastic energy Eq.~(\ref{elastic_energy})
around the NRS:
\begin{eqnarray}
\calH_{el} [\mm{g}] &=&  f[S_1, S_2, S_3] \nonumber\\
&=& \calH_{el}[S_1^0+ \delta S_1, S_2^0 + \delta S_2, 
S_3^0 + \delta S_3] \nonumber\\
&=&  f[S_1^0, S_2^0, S_3^0]
 + \sum_{n} \Phi_n \delta S_n  \nonumber\\
&+& \frac{1}{2} \sum_{m,n} \Phi_{mn}
 \delta S_m \delta S_n 
+ O[\delta S_n^3], \label{f_expansion1}
\end{eqnarray}
where 
\begin{subequations}
\begin{eqnarray}
\delta S_n &=& S_n - S_n^0 
=\Tr (\mm{g}^n - \mm{g}_0^n), \\
\Phi_n &=& \frac{\partial f}{\partial S_n} \rvert_0,\\
\Phi_{nm} &=& \frac{\partial f}{\partial S_n \partial S_m} |_0,
\end{eqnarray}
\end{subequations}
and we have truncated the series at the second order in $\delta S_n$.
Since, by definition $\delta S_n$'s are fully rotationally invariant,
the rotational symmetry as well as the associated soft mode of the
nematic phase are guaranteed by every term in Eq.~(\ref{f_expansion1})
and are not compromised by the truncation.

To make contact with elasticity theory we express the scalars $\delta
S_n$ in terms of the more familiar {\em nonlinear} Lagrange strain tensor
defined relative to the nematic reference state:
\begin{eqnarray}
\mm{e}&=& \frac{1}{2} \left( \lm^{\rm T} \lm - \mm{I}
	\right),\label{e-def}\\
e_{ab} &=& \frac{1}{2} \left( 
\frac{\partial \rv}{\partial x_a} \cdot
\frac{\partial \rv}{\partial x_b} 
-\delta_{ab} \right) \nonumber\\
&=& \frac{1}{2} \left(  
\partial_a u_b + \partial_b u_a 
+ \partial_a \uv \cdot \partial_b \uv
\right), \label{eu-relation}
\end{eqnarray}
where in the last expression we used Eq.~(\ref{phonon-def}) to 
express $\mm{e}$ in terms of the phonon field $\vec{u}$.
To this end we make use of the relation Eq.~(\ref{Lambda-lambda})
between deformation tensors in the isotropic and nematic reference
states, finding:
\begin{subequations}
  \label{def_Sn}
\begin{eqnarray}
\hspace{-6mm}
\delta S_1 &=& 	2 \, \Tr\, \mm{l}_0\, \mm{e},
	\label{S1}\\
\delta S_2 &=& 	4 \,\Tr \, \mm{l}_0^2 \,\mm{e} 
+ 4 \,\Tr \,(\mm{l}_0 \, \mm{e})^2, 
	\label{S2}\\
\delta S_3 &=& 	6 \,\Tr \, \mm{l}_0^3 \, \mm{e} 
	 + 12 \,\Tr \, ( \mm{l}_0^2 \, \mm{e} 
 	\, \mm{l}_0 \,\mm{e}) 
	+ 8\, \Tr \, ( \mm{l}_0 \,\mm{e})^3. \nonumber
	\label{S3}
\end{eqnarray}
\end{subequations}
where we defined 
\begin{equation}
\mm{l}_0 = \Lm_0\Lm_0^{\rm T}.
\end{equation}

Choosing the nematic director of the NRS to be along the $z$ axis, we may
express the tensor $ \mm{l}_0$ as a matrix:
\begin{equation}
\mm{l}_0 = \mm{g}_0
 =\left[ \begin{array}{ccc}
\zeta_{\perp}^2&0&0\\0&\zeta_{\perp}^2&0\\
0&0&\zeta_{\parallel}^2 \end{array} \right].
\label{g0}
\end{equation}
Substituting Eqs.~(\ref{def_Sn}) into Eq.~(\ref{f_expansion1}) and
rearranging the series in the order of increasing powers of the
Lagrange strain tensor $\mm{e}$, we obtain an effective elastic free
energy for nematic elastomers with spontaneously broken rotational
symmetry:
\begin{eqnarray}
\calH_{el} &=& a_{\perp} \, e_{ii} + a_{z} \, e_{zz} 
+ \mu_{z\perp} \, e_{zi}^2 \label{f_expansion2}\\
&+& \frac{1}{2} \,\left( B_z \, e_{zz}^2
 + 2 \,\lambda_{z\perp} \, e_{zz} \, e_{ii}
 + \lambda \, e_{ii} \, e_{jj} + 2 \, \mu \, e_{ij}^2 \right)
\nonumber\\ 
&+& b_1 \, e_{zz} \, e_{iz}^2 + b_2 \, e_{kk} \, e_{iz}^2
 + b_3 \, e_{ij} \, e_{iz} e_{jz}
 + c \, e_{iz}^2 \, e_{jz}^2+\ldots, \nonumber
\end{eqnarray}
where repeated subscripts $i$, $j$, $k$ are summed over transverse
components, $x,y$ only.  In above, we have only kept the most
important terms, i.e., those that are relevant in the renormalization
group sense, as we will verify a posteriori. The values of the eleven
coefficients appearing in Eq.~(\ref{f_expansion2}) are listed in
Appendix \ref{App:Coefficients}.

Not all of the eleven coefficients in Eq.~(\ref{f_expansion2}) are
independent to each other.  It is straightforward, although tedious,
to check that these coefficients, as given by
Eqs.~(\ref{Coefficients}), satisfy the following five Ward identities:
\begin{subequations}
\label{ward_identities}
\begin{eqnarray}
\mu_{z\perp} &=& 2 \,(\alpha\, a_z - (1+\alpha)\, a_{\perp}),
\label{ward1}\\
b_1 &=& 2\, \alpha\, B_z - 2\,(1+ \alpha)\,
 (\lambda_{z\perp} + \mu_{z\perp}),
\label{ward2}\\
b_2 &=& 2\,\alpha\,\lambda_{z\perp} - 2 \,(1+\alpha)\, \lambda,
\label{ward3}\\
b_3 &=&2\, \alpha \,\mu_{z\perp} - 4 \,(1+ \alpha )\, \mu,
\label{ward4}\\
c &=& 2 \,\alpha^2 \,B_z + 2 \,(1+\alpha)^2 \,(\lambda + 2 \,\mu) 
        \nonumber\\
 &-& 4 \,\alpha \,(1+\alpha)\, (\lambda_{z\perp} + \mu_{z\perp}),
\label{ward5}
\end{eqnarray}
\end{subequations}
where 
\begin{equation}
\alpha = \frac{\zeta_{\perp}^2}{\zeta_{\parallel}^2 
 - \zeta_{\perp}^2}  =  \frac{1}{r-1}, 
 \hspace{4mm} r = \frac{\zeta_{\parallel}^2}{\zeta_{\perp}^2}
 \label{alpha}
\end{equation}
are two dimensionless ratios characterizing the anisotropy of the
nematic phase.  Therefore out of eleven there are only six independent
parameters in Eq.~(\ref{f_expansion2}).  The five Ward identities,
Eqs.~(\ref{ward_identities}) reflect the underlying rotational
symmetry in the isotropic reference state, {\em spontaneously} broken
by the nematic state.

We note that although we have derived these identities in three
dimensions, as we explicitly show in Appendix
\ref{App:Ward-identities}, they actually hold in arbitrary dimensions
$d\geq 3$, generalized by simply allowing indices $i,j,k,\ldots$ to
range over all $d-1$ transverse directions.  The elastic Hamiltonian,
Eq.~(\ref{f_expansion2}) and the set of Ward identities
Eqs.~(\ref{ward_identities}) agree with Eq.~(4.3) and Eqs.~(4.4) of
Ref. \cite{StenLub-2}, after appropriate redefinition of various
constants.  The correspondence is established in Table \ref{WT-mapping}.
\begin{widetext}
\begin{center}
\begin{table}[!hbt]
\vspace{5mm}
\begin{tabular}{|c|c|c|c|c|c|c|c|c|c|c|c|}
\hline\hline This work&
$a_z$ & $a_{\perp}$ & $\mu_{z\perp}$ & $B_z$ & $\lambda_{z\perp}$ & $\lambda$ & $\mu$ & $b_1$ & $b_2$ & $b_3$ & $c$ \\\hline Reference \cite{StenLub-2}&
$\zp^2\,a_1$ &$ \zt^2\,a_2$ & $\zp^2\zt^2\,b_5$ & $2\,\zp^4\,b_1$
& $\zp^2\,\zt^2\,b_2$ & $2\,\zt^4\,b_3$ & $\zt^4\,b_4$ & $\zp^4\,\zt^2\,c_1$ & $\zp^2\,\zt^4\,c_2$ & $\zp^2\,\zt^4\,c_3$ & $\zp^4\,\zt^4\,d_1$
\\\hline
\end{tabular}
\caption{Correspondence between the Ward identities in this paper and those in reference \cite{StenLub-2}.  The parameter $s$ used in reference \cite{StenLub-2} corresponds to $(\zp^2-\zt^2)/2$ in our notations. }
\label{WT-mapping}
\end{table}
\end{center}
\end{widetext}

As we have already discussed in Sec.~\ref{Sec:reference-state}, 
it is convenient to choose $\zeta_{\perp}$ and $\zeta_{\parallel}$ in
Eq.~(\ref{g0}) such that the nematic reference state $\mm{e} = 0$
minimizes the elastic Hamiltonian Eq.~(\ref{f_expansion2}).  
This ensures that the coefficients $a_z$ and $a_{\perp}$ of linear terms
are exactly zero, which further reduces the number of independent
parameters in Eq.~(\ref{f_expansion2}) to four.  
More importantly, this choice, when combined with the first Ward identity
Eq.~(\ref{ward1}), dictates that the shear modulus $\mu_{z\perp}$
strictly vanishes in the nematic phase, as we have shown in a
complementary way in Sec.~\ref{Sec:Model}.  As emphasized there, a
vanishing of $\mu_{z\perp}$ without any fine-tuning is a consequence
of the underlying rotational invariance of the isotropic reference
state and is the lowest order manifestation of the corresponding soft
Goldstone mode. This feature qualitatively distinguishes a nematic
elastomer from an ordinary uniaxial solid, whose corresponding modulus
$C_5$ in Eq.~(\ref{H-uniaxial}) is generically nonzero.  

The other four Ward identities, Eqs.~(\ref{ward2}-\ref{ward5}),
dictate that the coefficients of all anharmonic terms in
Eq.~(\ref{f_expansion2}), $b_1$, $b_2$, $b_3$ and $c$, are completely
determined by the remaining $4$ quadratic coefficients, the elastic
moduli $B_z,\lambda_{z\perp},\lambda,\mu$.  As we will see in the next
section and as is clear from general symmetry principles from which
these arise, these relations are preserved in the presence of
fluctuations.  Using these relations, we can define an ``effective''
nonlinear strain tensor $\mm{w}$ with components:
\begin{subequations}
\label{w-def}
\begin{eqnarray}
w_{zz} &=& e_{zz} + 2 \alpha \, e_{zi}^2 ,\\
w_{ij}  &=& e_{ij} - 2 (1+\alpha)\,  e_{zi}\, e_{zj},
\end{eqnarray}
\end{subequations}
where $\alpha$ is defined in \rfs{alpha}, and indices $i$ and $j$ are
limited to the perpendicular subspace $1,2,\ldots,d-1$.  The elastic
Hamiltonian Eq.~(\ref{f_expansion2}) then reduces to
\begin{eqnarray}
\calH_{el} = \frac{1}{2} \,B_z\,w_{zz}^2
	 + \frac{1}{2} \,\lambda \, w_{ii}^2
+     \lambda_{z\perp}\, w_{zz}\,w_{ii} 
+      \mu\, w_{ij}^2, 
\label{f_expansion3}
\end{eqnarray}
which, after supplemented with the curvature energy corresponding
to the Frank free energy of the nematic director, will be the form we
will use for further study of thermal fluctuations and network
heterogeneities.

\subsubsection{Two-dimensional nematic elastomer}
\label{Sec:2d-NE}
The same analysis can be carried out straightforwardly for a
two-dimensional nematic elastomer.  An important property special to
the two-dimensional case, is that the subspace perpendicular to the
nematic order is one-dimensional, and thus has no shear mode
(characterized in higher dimensions by the modulus $\mu$) associated
with the transverse subspace.

Consequently, the effective elastic Hamiltonian for a 2d nematic elastomer is 
given by:
\begin{eqnarray}
\calH_{el}^{\rm 2d} = \frac{1}{2}\,\lambda_{xx}\, w_{xx}^2 +
     \frac{1}{2}\,\lambda_{yy}\,w_{yy}^2 + \lambda_{xy}\,
     w_{xx}\,w_{yy}, \label{f_expansion_2d}
\end{eqnarray}
where we chose the nematic director $\hat{n}_0$ to be along the $x$-axis.
The nonlinear strain components $w_{xx}$ and $w_{yy}$ are
\begin{subequations}
\begin{eqnarray}
w_{xx} &=& e_{xx} + 2 \alpha \, e_{xy}^2 ,\\
w_{yy}  &=& e_{yy} - 2 (1+\alpha)\,  e_{xy}^2,
\end{eqnarray}
\end{subequations}
where $\alpha = {\zeta_y^2}/{(\zeta_x^2-\zeta_y^2)}$.
We will leave the analysis of this very interesting two-dimensional 
model for a future investigation.

\subsection{Relevant nonlinearities and the minimal strain-only 
elastic model}
As we shall see later in this work, the nonlinear elastic model,
Eq.~(\ref{f_expansion3}) is an interesting coupled combination of a
smectic-like and columnar-like nonlinear elasticities, involving $u_z$
and $u_i^\perp$, respectively.  Experience with the nonlinearities in
the presence of fluctuations in these two systems
\cite{GP1,GP2,RT:MSC-2} suggests that it is the former that are more
relevant.  We will verify this rigorously a posteriori in
Sec.~\ref{Sec:har-fluct}.  Thus, anticipating a stronger relevance of
$u_z$ elastic nonlinearities to long length-scale fluctuations, we
substitute Eq.~(\ref{eu-relation}) into Eq.~(\ref{w-def}) and
Eq.~(\ref{f_expansion3}), and only keep those anharmonic terms
(smectic-like nonlinearities) that are proportional to $(u_{z}^3,
u_{z}^4, \uv_{\perp} \uv_{z}^2)$, ignoring all others.  It is
straightforward to show that this amounts to making the following
replacements in Eq.~(\ref{f_expansion3}):
\begin{subequations}
\begin{eqnarray}
 w_{zz} &\rightarrow& 
   \ppz u_z + \frac{\alpha}{2} \,
    (\nabla_{\perp}u_z)^2,\\
 w_{ij} &\rightarrow&
   \frac{1}{2} (\ppi u_j + \ppj u_i ) 
 - \frac{\alpha}{2}\, (\ppi u_z \,\ppj u_z),
\end{eqnarray}
\end{subequations}
where $\alpha$  is defined in Eq.~(\ref{alpha}).  

In order to streamline the notations, we further rescale the phonon
fields by
\begin{subequations}
\begin{eqnarray}
u_z &\rightarrow& \alpha^{-1} u_z,\\
\uv_{\perp} &\rightarrow&  \alpha^{-1} \uv_{\perp},
\end{eqnarray}
\end{subequations}
and absorb an overall factor $\alpha^{-2}$ of the elastic Hamiltonian
by redefining all the elastic constants:
\begin{equation}
(B_z, \lambda_{z\perp}, \lambda, \mu) \rightarrow
\alpha^2 \, (B_z, \lambda_{z\perp}, \lambda, \mu).
\end{equation}

In order to stabilize soft-mode phonon fluctuations, the Hamiltonian
in Eq.~(\ref{f_expansion3}) must be augmented by the the Frank free
energy, \rfs{f_Frank}, using the relation \rfs{softconstraint} to
eliminate director fluctuations $\delta\nh$ in favor of the phonon
fields.  Furthermore, since we are most interested in the fluctuations
of the $u_z$ phonon field, which in the momentum space is controlled
by the pole $q_z \sim q_{\perp}^2$, it can be easily shown that both
the bend ($K_3$) and twist ($K_2$) term are (dangerously) irrelevant
in the RG sense \cite{comment:bending}.   We will therefore ignore
these two terms for the purpose of the RG analysis, and to simplify
the notation will use $K$ to denote the $K_1$ splay modulus.

The resulting minimal elastic Hamiltonian density has the following form:
\begin{widetext}
\begin{subequations}
\label{f_expansion5}
\begin{eqnarray}
\hspace{-20mm}
{\mathcal H}_{el} &=&  \frac{1}{2} B_z \,w_{zz}^2 
+ \lambda_{z\perp}\, w_{zz}\, w_{ii} 
+ \frac{1}{2}\, \lambda\, w_{ii}^2  +  \mu\, w_{ij}^2 
+ \frac{K}{2}\,(\nabla_{\perp}^2 u_z)^2, 
\label{f_expansion4}\\
&=& \frac{B}{2} \,(\Tr \, \mm{w})^2
 + C\, (\Tr \, \mm{w}) \, \tilde{w}_{zz} 
+ \frac{\mu_{\rm L}}{2}  \,\tilde{w}_{zz}^2
+ \mu \, \tilde{w}_{ij} \,\tilde{w}_{ij}
+\frac{K}{2}\,(\nabla_{\perp}^2 u_z)^2, 
\label{f_expansion41}
\end{eqnarray}
\end{subequations}
\end{widetext}
where the components of the (rescaled) effective strain tensor
$\mm{w}$ now become
\begin{subequations}
\label{effective_strain}
\begin{eqnarray}
w_{zz} &=& \partial_z u_z + \frac{1}{2}
 \, (\nabla_{\perp} u_z)^2,\\
w_{ij} &=& \frac{1}{2} \,\left(\ppi u_j + \ppj u_i
-  \,\ppi u_z \,\ppj u_z\right),
\end{eqnarray}
\end{subequations}
and 
\begin{subequations}
\label{modes:w}
\begin{eqnarray}
\mbox{Bulk} \hspace{4mm}
\Tr\,\mm{w} &=& w_{zz} + w_{ii},\\
\mbox{Longitudinal Shear} \hspace{4mm}
\tilde{w}_{zz} &=& w_{zz} - \frac{1}{d} \, \Tr\,\mm{w},\\
\mbox{Transverse Shear} \hspace{4mm}
\tilde{w}_{ij} &=& w_{ij} -\frac{w_{kk}}{(d-1)}\,\delta_{ij}\hspace{1cm}.
\end{eqnarray}
\end{subequations}
The elastic constants $B$ (bulk modulus), $\mu_{\rm L}$ (longitudinal
shear modulus), $\mu$ (transverse shear modulus) and $C$ have already
been defined in Eqs.~(\ref{def_moduli}).

The elastic nonlinearities contained in different modes of deformation
are particularly interesting.  Qualitatively speaking, the relevant
elastic nonlinearities arise from the coupling between various modes
of deformation with the soft mode.  It is the fluctuations of the soft
mode that renormalize various elastic constants and lead to anomalous
elasticity.  We observe that the bulk mode
\begin{equation}
\Tr \mm{w} =  w_{zz}+w_{ii} = \ppz u_z + \ppi u_i = \nabla \cdot \uv
\end{equation}
\begin{widetext}
\begin{center}
\begin{table}[!hbt]
\vspace{5mm}
\begin{tabular}{|c|c|c|c|c|}
\hline\hline
Moduli & $B$ & $\mu_{\rm L}$ & $\mu$ & $\mu_{z\perp}$ \\\hline
 Descriptions & Bulk Mode
& Longitudinal Shear & Transverse shear & Soft Mode \\\hline
$\mm{V}$ & $\Tr\,\mm{V}$ & $\tilde{V}_{nn} = V_{nn} - d^{-1}\,(\Tr \,\mm{V})$ &
$\tilde{\mm{V}}_{\perp}$ & $\vec{V}_{n\perp}$ \\\hline
$\mm{w}$ & $ \Tr \, \mm{w}$ & $\tilde{w}_{zz} =w_{zz} - d^{-1}\,(\Tr\,\mm{w})$ & $\tilde{w}_{ij}$ & N/A
\\\hline
${\boldsymbol \varepsilon}$ & $\Tr {\boldsymbol \varepsilon}$ & $\tilde{\varepsilon}_{zz} = \varepsilon_{zz} - d^{-1}\,(\Tr \, {\boldsymbol \varepsilon})$ &
$\tilde{\varepsilon}_{ij}$& Eq.~(\ref{V_ni})\\ \hline
\end{tabular}
\caption{Various modes of deformation and their operators in different descriptions}
\label{Modes}
\end{table}
\end{center}
\end{widetext}
does not contain any relevant nonlinearities.  This is due to the fact
that the soft deformation Eq.~(\ref{soft-mode}) preserves the volume
and therefore its local fluctuations do not couple to the
bulk mode.  Consequently we expect that renormalization of the bulk
modulus is qualitatively unimportant.  On the other hand, from
Eqs.~(\ref{modes:w}) and Eqs.~(\ref{effective_strain}) we observe that 
longitudinal shear mode
\begin{equation}
\tilde{w}_{zz} = (1-d^{-1}) \, \ppz u_z - d^{-1}\,\ppi u_i 
+ (\nabla_{\perp} u_z)^2
\end{equation} 
and the transverse shear mode 
\begin{eqnarray}
\tilde{w}_{ij} &=& \frac{1}{2} \left(
\partial_i u_j + \partial_j u_i 
- \frac{\nabla_{\perp} \cdot \uv_{\perp}}{(d-1)} \delta_{ij}
\right) \nonumber\\
&+&  \frac{1}{2} \left(
\partial_i u_z \partial_j u_z 
- \frac{(\nabla_{\perp} u_z)^2}{(d-1)} \delta_{ij}
\right),
\end{eqnarray}
do involve relevant elastic nonlinearities.  We therefore expect that
the renormalization the longitudinal and transverse shear moduli by
fluctuations to be qualitatively important.  These expectations will
be verified by explicit RG calculations in the next section.

We emphasize the structural similarity between
Eq.~(\ref{f_expansion41}), Eq.~(\ref{elast-energy-2}), and
Eq.~(\ref{H-quadratic}), with the only exception that the term with
the coefficient $\mu_{n\perp}$ in Eq.~(\ref{elast-energy-2}) and
Eq.~(\ref{H-quadratic}) is replaced by the term with the coefficient
$K$ in Eq.~(\ref{f_expansion41}).  Indeed the elastic free energy
Eq.~(\ref{f_expansion41}) can be obtained from
Eq.~(\ref{elast-energy-2}), augmented by the Frank free energy
Eq.~(\ref{f_Frank}), by integrating out the fluctuations of nematic
director, and dropping irrelevant nonlinearities.  This procedure
replaces the $\mu_{n\perp}$ term by the curvature $K$ term.

Finally we note that we have represented the bulk, longitudinal shear,
and transverse shear modes in terms of the invariant strain $\mm{V}$,
the linearized strain tensors $\epsilon$ and $\mm{a}$
(Sec.~\ref{Sec:Model}), as well as the effective strain tensor
$\mm{w}$ (the current section).  At the linear order in the phonon
field, they all agree with each other.  In contrast, because the soft
mode involves both strain deformation and rotations of nematic
director, it cannot be expressed in terms of $\mm{w}$ alone since the
latter does not contain $\delta\nh$. Thus, the soft mode only appears
indirectly in the current $\mm{w}$ description, through an exact
vanishing of $\mu_{z\perp}$ and the form of the nonlinear strain
tensor $\mm{w}$. The relations are summarized by Table.~\ref{Modes}.

The effective strain-only model, Eq.~(\ref{f_expansion5}) will be the
starting point for all of our further analysis of thermal fluctuations
in the presence of elastic nonlinearities.

\section{Long-scale thermal elasticity of a {\em homogeneous} 
nematic elastomer}
\label{Sec:Thermal-fluct}

In this section we use the effective elastic model derived in the 
previous section, \rfs{f_expansion5} to study the long-scale
properties of a nematic elastomer in the presence of thermal
fluctuations.

\subsection{Harmonic phonon fluctuations}
\label{Sec:har-fluct}
We begin by studying thermal fluctuations within a harmonic
approximation, ignoring all elastic nonlinearities in the model,
\rf{f_expansion41}. As mentioned above, this is equivalent to going
back to the harmonic theory, Eq.~(\ref{H-quadratic}), supplemented by
Frank free energy, and integrating out the director fluctuation
$\delta \nh$.  To the lowest order at long wavelength this amounts to
the nemato-elastic coupling in Eq.~(\ref{H-quadratic}) simply enforcing
the replacement
\begin{equation}
\delta \nh_i \rightarrow \frac{r+1}{r-1} \varepsilon_{ni} + a_{ni}
\label{replacement}
\end{equation} 
in the Frank free energy Eq.~(\ref{f_Frank}).  
The resulting quadratic effective elastic energy is then given by
\begin{eqnarray}
H^0_{el} = \int d^d x
   \, &\frac{1}{2} &  \left[   B_z\, (\ppz u_z)^2 
    + 2\, \lambda_{z\perp}\, (\ppz u_z)\, (\ppi u_i )
 \right.   \nonumber\\
   &+& (\lambda + \mu)\, (\ppi u_i)^2 
   + \mu \,(\ppi u_j)^2
   \nonumber\\
 &+&
   \left. K_1\, (\nabla_{\perp}^2 u_z)^2 
    + K_3\, (\partial_z^2 \uv_{\perp})^2 
    \right], 
    \label{H_0}
\end{eqnarray} 
where we have absorbed some constant factors into the Frank elastic
constants $K_1$ (splay) and $K_3$ (bending).  The twist term ($K_2$)
turns out to be less relevant than $K_1$ and $K_3$, and therefore is
ignored in Eq.~(\ref{H_0}).  Except for the last two terms (from Frank
free energy), Eq.~(\ref{H_0}) is identical to the quadratic (in phonon
field $\uv$) parts of the nonlinear elastic Hamiltonian
Eq.~(\ref{f_expansion3}).
 
In terms of the Fourier transform of the phonon field $\uv(\xv)$
\begin{eqnarray}
\uv(\qv) &=& \int d^d x \,\uv(\xv)\, e^{-i\,\qv\cdot \xv}, 
\end{eqnarray}
the harmonic elastic Hamiltonian $H_0$ takes the form
\begin{equation}
H_{el}^0  = \frac{1}{2} \, \int d^d q \,\,
\Gamma_{ab}(\qv)\,u_a(\qv)\,u_b(-\qv),
\end{equation}
where indices $a$ and $b$ are summed over all $d$ dimensions, and the
kernel matrix $\Gamma_{ab}(\qv)$ is given by
\begin{subequations}
\label{gamma_0}
\begin{eqnarray}
\Gamma_{ij}  &=&  (\lambda + \mu)\, q_i\,q_j +
         (\mu \,q_{\perp}^2 + K_3\, q_z^4 )\,\delta_{ij},\\
\Gamma_{zi}  &=& \lambda_{z\perp}\, q_z\,q_i,\\
\Gamma_{zz} &=&  B_z\, q_z^2+K_1\,q_{\perp}^4. 
\end{eqnarray}
\end{subequations}

The harmonic phonon correlation functions 
\begin{equation}
\langle u_a(\qv)\,u_b(-\qv')\rangle_0 
= \frac{1}{Z}\int{\mathcal D}\uv \,
        u_a(\qv)\,u_b(-\qv')\,e^{-{\mathcal H}_0[\uv]/T}. 
\end{equation}
can be calculated using the equipartition theorem or equivalently by
performing above simple Gaussian integral.  They are
given by
\begin{equation}
\langle u_a(\qv) \, u_b(-\qv') \rangle_0  
        =  (2\,\pi)^d \, \delta^d(\qv-\qv') \,G_{ab}(\qv),
\end{equation}
where as usual $G_{ab}(\qv)$ is related to $\Gamma_{ab}(\qv)$ via:
\begin{equation}
G_{ab}(\qv)\, \Gamma_{bc}(\qv) = T\,\delta_{ac}.
\label{G-Gamma-0}
\end{equation}

The components of the propagator matrix $G_{ab}(\qv)$ are fairly
complicated.  For example, $G_{zz}(\qv)$ is given by
\begin{equation}
G_{zz}(\qv) = \frac{T}{C_z(\qv) \,q_z^2+K_1 \,q_{\perp}^4},
\label{G0zz}
\end{equation}
where  $C_z(\qv)$ is a wavevector-dependent constant
\begin{equation}
C_z(\qv) = B_z - \frac{\lambda_{z\perp}^2}
{\lambda+2\,\mu + K_3\,q_z^4/q_{\perp}^2 }. 
\label{Cq-def}
\end{equation}
Since we are most interested in small $\qv$ (long length-scale)
properties, the $\qv$ dependent term $K_3\, q_z^4/q_{\perp}^2$ in the
denominator of Eq.~(\ref{Cq-def}) is generically much smaller than the
$\qv$ independent term $\lambda+2\mu$, and therefore can be ignored
\cite{comment:small-region}.  Therefore we may approximate
$G_{zz}(\qv)$ as
\begin{eqnarray}
G_{zz}(\qv) &\approx& \frac{T}{\hat{\mu} \, q_z^2
 + K_1 \, q_{\perp}^4 }, \label{G0zz-1}
\end{eqnarray}
where 
\begin{equation}
\hat{\mu} = B_z - \frac{\lambda_{z\perp}^2}{\lambda+2\,\mu}
= C_z(\qv)|_{q_z=0} 
\label{hatmu-def}
\end{equation}

The correlator Eq.~(\ref{G0zz-1}) is identical to the harmonic phonon
correlation function of a conventional smectic liquid crystal
\cite{LC:deGennes} with $\hat{\mu}$ and $K_1$ the modulus for layer
compression and layer bending, respectively.  In three dimensions, the
real space mean-squared fluctuations of the $u_z$ phonon field are
given by:
\begin{eqnarray}
\langle u_z(\rv)^2\rangle_0 &=&
	 \int \frac{d^{2} \vec{q}_{\perp}dq_z}{(2\pi)^{3}}
        \frac{T} {\hat{\mu}\,q_z^2 + K_1 q_{\perp}^4}
        \nonumber\\        
 &\propto&
 	\frac{T}{\sqrt{\hat{\mu}\,K_1}} \,\log \frac{L}{a}.
        \label{uz_fluct}
\end{eqnarray}
where $L$ is the system size and $a$ the small cutoff length-scale,
set by the molecular size.  The fact that $\langle
u_z(\rv)^2\rangle_0$ diverges with the system size suggests a
breakdown of the harmonic elasticity theory and the qualitative
importance of elastic nonlinearities that have so far been ignored.
As in the case of smectic liquid crystal, we expect that the
elasticity of three- (and two-) dimensional nematic elastomers is
dominated by long wavelength fluctuations of the $u_z$ phonon field.
 
We can use Eqs.~(\ref{def_moduli}) to express $\hat{\mu}$ as
function of $B$, $C$, $\mu_{\rm L}$ and $\mu$.  In the limit of
an infinite bulk modulus, i.e., $B \rightarrow \infty$ with $C$, $\mu_{\rm
L}$ and $\mu$ fixed, we find $\hat{\mu}$ approaching a finite limit
\begin{equation}
\hat{\mu} \longrightarrow
	 \frac{2 (d-2)}{(d-1)}\, \mu + \mu_{\rm L},
	\hspace{3mm}  \mbox{as $B \rightarrow \infty$}.
	\label{muhat-limit}
\end{equation}
Therefore in this limit, the correlation function of $u_z$ field
simplify considerably, becoming independent of the bulk modulus $B$.
We will see that the same result holds for other correlation functions
as well.  Physically, for a large $B$, the bulk mode is essentially
frozen out, and as a result does not play any role in long wavelength
fluctuations.

Matrix inversion of $\Gamma_{ab}$ also gives the harmonic correlations
of $\uv_{\perp}(\qv)$:
\begin{eqnarray}
 G_{ij}(\qv) =  G_{\rm L}(\qv) P^{\rm L}_{ij}(\vec{q}_{\perp})
 + G_{\rm T}(\qv) P^{\rm T}_{ij}(\vec{q}_{\perp}),
        \label{G0ij}
\end{eqnarray}
where
\begin{subequations}  
\begin{eqnarray}
 G_{\rm L}(\qv)  &=& 
 \frac{T}{ C_{\perp}(\qv)\,q_{\perp}^2 
        + K_3\, q_z^4 } ,  
 \\
 G_{\rm T}(\qv) &=&            
 		\frac{T}{\mu\, q_{\perp}^2 + K_3\, q_z^4} ,\\
C_{\perp}(\qv) &=& \lambda+ 2\,\mu - 
\frac{\lambda_{z\perp}^2}{B_z + K_1\,q_{\perp}^4/q_z^2},
\end{eqnarray}
\end{subequations}
and 
\begin{equation}
P^{\rm L}_{ij}(\vec{q}_{\perp}) 
	= \frac{q_i\, q_j} {q_{\perp}^2},
\hspace{5mm}
P^{\rm T}_{ij}(\vec{q}_{\perp}) = 
	 \delta_{ij} - \frac{q_i\, q_j}
       	 {q_{\perp}^2}
\end{equation}
are the longitudinal and transverse projection operators (with respect
to $q_i$) in the $(d-1)$ dimensional subspace perpendicular to
$\hat{z}$.  Eq.~(\ref{G0ij}) is similar in structure to the harmonic
phonon correlations of a columnar liquid crystal
\cite{RT:MSC,RT:MSC-2}.  A straightforward calculation shows that real
space fluctuation of $\uv_{\perp}$ remains finite in three dimensions.
Therefore the long wavelength fluctuations of the phonon fields,
$\uv_{\perp}$ are qualitatively unimportant as compared with those of
$u_z$.  This qualitative difference between the phonon fields $u_z$
and $\uv_{\perp}$ is a result of spontaneous broken rotational
symmetry in an originally isotropic system.  By contrast, in
ordinary uniaxial solids, all phonon fluctuations remain finite in
three dimensions, thus leading only to quantitative corrections to
properties of conventional solids.

The cross correlation functions between $u_z$ and $\uv_{\perp}$ are
given by
\begin{eqnarray}
G_{zi}(\qv) &=& 
V^{-1}\langle u_z(\qv) u_i (-\qv) \rangle_0,\\
 &=& - \frac{\Gamma_{zi}(\qv)} {\Gamma_{zz}(\qv)\Gamma_{\rm L}(\qv) 
- (\Gamma_{zi}(\qv))^2}, 
\label{G0zi}
\end{eqnarray}
where 
\begin{equation}
\Gamma_{\rm L}(\qv)
 = (\lambda+2\mu)\,q_{\perp}^2+ K_3\,q_z^4,
\end{equation}
and $V$ is the volume of the system.
For a sufficiently small $\qv$, $G_{zi}(\qv)$ can be approximated by
\begin{eqnarray}
&&V^{-1}\langle u_z(\qv) u_i(-\qv) \rangle_0 =\\
&&- \frac{\lambda_{z\perp}q_z q_i}
{(B_z(\lambda+2\mu)-\lambda_{z\perp}^2) 
(q_z^2 + \frac{K_1}{\hat{\mu}}q_{\perp}^4)
(q_{\perp}^2 + \frac{K_3}{\tilde{\mu}}q_{z}^4)},
\nonumber
\end{eqnarray}
where 
\begin{equation}
\tilde{\mu} = (\lambda+2\mu) - \frac{\lambda_{z\perp}^2}{B_z}.
\end{equation}

\subsection{Naive scaling and critical dimension}

To study fluctuations in the nonlinear elastic theory beyond the
harmonic approximation of the previous subsection, one might naively
hope to perform a perturbative expansion in the nonlinear elastic
terms. However, a standard analysis, which is relegated to Appendix
\ref{App:RG-general} shows that such direct perturbation theory is
hopelessly divergent at long length scales, as already suggested by
divergent phonon fluctuations in e.g., Eq.~(\ref{uz_fluct}).

Thus, similarly to systems near a critical point (but here applied
throughout the nematic phase), to treat elastic nonlinearities we need
to employ the machinery of the renormalization group transformation
(RG) \cite{RGreview-Wilson,RGreview-Fisher,RGreview-Ma} which
establishes relations between physical quantities (e.g., correlation
functions) at different length-scales.  These relations then allow us
to extract universal long-scale nonperturbative (in nonlinearities)
properties of the system from their perturbatively computable
short-scale versions.

To this end, we need to study the property of the system under
rescaling of length-scales and coarse-graining (thinning) of degrees
of freedom. This procedure, executed explicitly in Appendix
\ref{App:RG-general}, is quite nontrivial in the presence of
nonlinearities. However, it simplifies considerably to zeroth order in
nonlinear terms, becoming equivalent to a rescaling transformation on
the ``bare'' Hamiltonian.  

Applying this to the nematic elastomer model,
Eq.~(\ref{f_expansion41}) it is not difficult to see that it is
invariant under each of the following two rescaling operations, with
$b$ an arbitrary rescaling factor:
\begin{enumerate}
\item Rescaling of $z$ axis: $R_{\parallel}(b)$
\begin{subequations}
\label{rescaling_z}
\begin{eqnarray} 
(\xv_{\perp}, z, \uv_{\perp}, u_z)  &=&
(\xv_{\perp}',b\,z', b^{-2} \,\uv_{\perp}',b^{-1}\, u_z'),\\
(B_z, \lambda_{z\perp}, \lambda, \mu) 
&=& b^{3} (B_z', \lambda_{z\perp}', \lambda', \mu'), \\
K &=& b^1\, K',\\
(Q_{\perp},Q_{\parallel})  &=& 
(Q_{\perp}', b^{-1}\,Q_{\parallel}').
\end{eqnarray}
\end{subequations}
\item Rescaling of $\xv_{\perp}$ plane: $R_{\perp}(b)$ 
\begin{subequations}
\label{rescaling_perp}
\begin{eqnarray}
(\xv_{\perp}, z, \uv_{\perp}, u_z)  &=&
(b\,\xv_{\perp}', z', b^{3} \,\uv_{\perp}',b^{2}\, u_z'),\\
(B_z, \lambda_{z\perp}, \lambda, \mu) 
&=& b^{-(d+3)} (B_z', \lambda_{z\perp}', \lambda', \mu'), \\
K &=& b^{1-d}\, K',\\
(Q_{\perp},Q_{\parallel})  &=& 
( b^{-1}\,Q_{\perp}', Q_{\parallel}').
\end{eqnarray}
\end{subequations}
\end{enumerate}
We note that in above transformations, in addition to system's
coordinates, phonon fields and elastic moduli, we have introduced two
large wavevector cutoffs $Q=\{Q_{\parallel},Q_{\perp}\}$ beyond which
our coarse-grained model is inapplicable.  Here $Q_{\parallel}$ and
$Q_{\perp}$ are cutoffs in the directions parallel and perpendicular
to $\nh_0=\hat{z}$, respectively, roughly set by the inverse of the 
mesh-size of the polymer network.   They also provide an
ultraviolet regularization to the path integral representation of the
partition function
\begin{equation}
{\mathcal Z} = \int D\uv \,e^{-H_{el}/T}, \label{partition_function}
\end{equation}
where $H_{el} = \int d^d x\,\, {\mathcal H}_{el}$, with ${\mathcal
  H}_{el}$ given by Eq.~(\ref{f_expansion41}) and $d^d x \equiv
dx_\perp^{d-1} dz$.  Thus it is understood throughout that
functional integrals (and wavevector mode sums that result from these)
are over phonon fields $\uv(\xv)$ whose Fourier amplitudes have
support $|\qv_{\perp}| \leq Q_{\perp}$ and $|q_{z}| \leq
Q_{\parallel}$.  To simplify the subsequent analysis, it is convenient
to choose a cylindrical cutoff, in which $Q_{\parallel} = \infty$
(i.e., $q_z$ is not cutoff at all) and to denote $Q_{\perp}$ simply by
$Q$.  Universality principle guarantees that the long length-scale physics 
is independent of the choice of the short-scale cutoff.

It is convenient to consider a special combination of these two
rescaling operations $R_{\parallel}(b^2) \, R_{\perp}(b)$:
\begin{subequations}
\label{rescaling_perpz}
\begin{eqnarray}
\hspace{-30mm}
(\xv_{\perp}, z, \uv_{\perp}, u_z)  &=&
(b\,\xv_{\perp}', b^2\, z', b^{-1}\,\uv_{\perp}',\, u_z'),
\nonumber\\
\\
(B_z, \lambda_{z\perp}, \lambda, \mu, K) 
&=& b^{3-d} (B_z', \lambda_{z\perp}', \lambda', \mu',K'), 
\label{rescaling-2}
\nonumber\\
\\
(Q_{\perp},Q_{\parallel})  &=& 
( b^{-1}\,Q_{\perp}', b^{-2} \,Q_{\parallel}').
\label{rescaling-Q}
\end{eqnarray}
\end{subequations}
that changes all elastic moduli by a common factor $b^{3-d}$ and
therefore preserves their relative ratios and the form of the
Hamiltonian. 
Let $c$ and $c'$ be the shorthands for the sets of all elastic
constants before and after rescaling, as shown in each side of
\rfs{rescaling-2}.  Since $H_{el}$ is linear in the set of couplings
$c$, it transforms according to
\begin{equation}
H_{el}[\uv,c] = H_{el}[\uv',c'] = H_{el}[\uv',b^{d-3}c] = b^{d-3} H_{el}[\uv',c].
\label{H-H'}
\end{equation}
Substituting this result into \rfs{partition_function}, we find that
such a rescaling transformation \rfs{H-H'} is equivalent to a
rescaling of the temperature according to
\begin{equation}
T' = b^{3-d} \,T. 
\end{equation}
Thus we find that for $d < d_c=3$ the effective temperature grows
under rescaling, demonstrating increased importance of thermal
fluctuations.  In contrast, above the critical dimension $d_c=3$,
fluctuation corrections to harmonic theory are small at low $T$.

To simplify the notation, in subsequent calculations we rescale all
elastic moduli by temperature $T$ so that it does not explicitly
appear in the functional integral, but can be easily restored by
undoing this rescaling.

\subsection{Renormalization-group analysis}

To treat effects of nonlinearities beyond the zeroth order analysis
of the previous subsection we employ the momentum-shell
renormalization group, detailed in Appendix
\ref{App:RG-general}.  
\begin{figure}
\begin{center}
  \includegraphics[width=4.5cm]{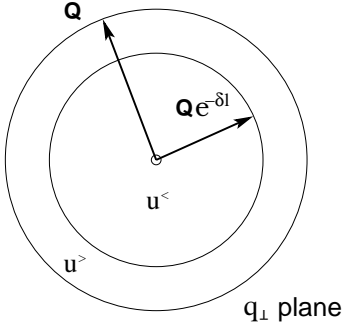}
  \caption{Basic scheme of momentum shell RG, defining the short
    ($\uv_{\perp,z}^>$) and long ($\uv_{\perp,z}^<$) scale phonon fields. }
\label{shell}
\end{center}    
\end{figure}
To summarize, we coarse-grain the system by separating the phonon field 
into high- and low-wavevector components, as shown in Fig.~\ref{shell}:
\begin{subequations}
\begin{eqnarray}
\uv_{\perp}(\rv)&=&\uv_{\perp}^>(\rv)+\uv_{\perp}^<(\rv), \\
u_z(\rv)&=&u_z^>(\rv)+u_z^<(\rv),
\end{eqnarray}
\end{subequations}
where $\uv_{\perp}^>$ and $u_z^>$ have support in the momentum shell
$Q \, e^{-\delta l} < q_{\perp} \leq Q$, while $\uv_{\perp}^<$ and
$u_z^<$ have support in the inner cylinder $0 \leq q_{\perp} \leq Q \,
e^{-\delta l}$.  We then integrate out the high-wavevector parts
$\uv_{\perp}^>$, $u_z^>$, perturbatively in the anharmonic terms in
$H_{el}$, \rfs{f_expansion5}, thereby obtaining a coarse-grained elastic
Hamiltonian in terms of fields $u_{\perp}^<$ and $u_z^<$, with all
elastic constants renormalized by fluctuations of the
$\uv_{\perp,z}^>$ fields.  

\begin{figure}
 \vspace{0.5cm}
\begin{center}
  \includegraphics[width=5cm]{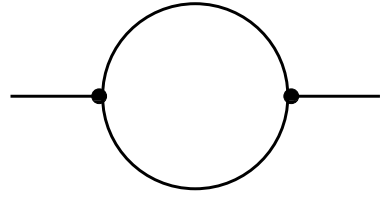}
\caption{Feynman diagrams renormalizing $B$, $C$, $\lambda$, $\mu$,
and $K$. }
\label{quadradiagram}
\end{center}    
\end{figure}

As we have shown in the preceding section, the coefficients of the
nonlinear terms in the elastic Hamiltonian Eq.~(\ref{f_expansion5})
are completely determined by those of the quadratic terms, as
summarized by the Ward identities, Eqs.~(\ref{ward_identities}) and
the form of $H_{el}$, Eq.~(\ref{f_expansion5}).  Since these
identities are enforced by the underlying rotational symmetry, they
are preserved by the above coarse-graining procedure, i.e., the
perturbative corrections to various elastic constant must satisfy the
same Ward identities.  This observation considerably simplifies our
calculations, as it allows us to focus on the renormalization of the
harmonic terms. These are summarized by the Feynman diagram in
Fig.~\ref{quadradiagram}, with correlators given by the harmonic
theory, Eqs.~(\ref{G0zz},\ref{G0ij},\ref{G0zi}), that for $K_3=0$
reduce to:
\begin{subequations}
\label{simplified_propagator}
\begin{eqnarray}
G_{zz}(\qv) &=& \left(\hat{\mu}\,q_z^2 
        + K\, q_{\perp}^4 \right)^{-1},
        \\
G_{zi}(\qv) &=& -\frac{\lambda_{z\perp}} {(\lambda + 2\, \mu) }
	\,\frac{q_z\, q_i}{(\hat{\mu}\,q_z^2 + K\, q_{\perp}^4)\,q_{\perp}^2 } ,\\
G_{ij}(\qv) &=& \frac{1}
        {(\lambda + 2\, \mu) } \,
        \frac{(B_z\, q_z^2+K\, q_{\perp}^4)\,q_i\, q_j}
        { (\hat{\mu}\,q_z^2 
        + K\, q_{\perp}^4)\,q_{\perp}^4}
        \nonumber\\
     &+& \frac{1}{\mu\, q_{\perp}^2} \,\left(\delta_{ij} 
        - \frac{q_i\, q_j}{q_{\perp}^2}\right),
\end{eqnarray}
\end{subequations}
with $\hat{\mu}$ defined in Eq.~(\ref{hatmu-def}).  The underlying
symmetry then enforces that the corrections to anharmonic terms are
then completely determined by the Ward identities, as we have
explicitly verified via detailed calculations presented in Appendix
\ref{App:RG-general}.  

In order to relate the resulting coarse-grained Hamiltonian with
the ``bare'' one, it is convenient to apply a combination of two
rescaling transformation $R_{\parallel}(e^{\omega\,\delta
  l})\,R_{\perp}(e^{\delta l})$,
Eqs.~(\ref{rescaling_z},\ref{rescaling_perp}), to the field theory of
low-wavevector fields $\uv^<(\xv)$, so as to restore the ultraviolet
cutoff for $\qv_{\perp}$ back to $Q$.  Here $\omega$ is an arbitrary
constant that can be chosen by convenience.  The transformation of the
spatial coordinates, the phonon fields, and various elastic constants
can be read off from Eq.~(\ref{rescaling_z}) and
Eq.~(\ref{rescaling_perp}):
\begin{subequations}
\label{RGrescaling}
\begin{eqnarray}
(\xv_{\perp},z) &=& (e^{\delta l} \xv'_{\perp},
   e^{\omega \delta l}z'),\\
  \uv_{\perp}^<(\xv) &=& e^{ (3-2\,\omega)\delta l}\,
 \uv_{\perp}'(\xv'),\\
u_z^<(\xv) &=& e^{(2-\omega) \delta l}u_z'(\xv'),\\
(B_z, \lambda_{z\perp}, \lambda, \mu) 
&=& b^{(-d-3+3\omega)\delta l} (B_z', \lambda_{z\perp}', 
\lambda', \mu'),\hspace{1cm}\\
K &=& b^{(-d+1+\omega)\delta l} \, K'. 
\end{eqnarray}
\end{subequations}
where the choice of the phonon field rescalings are dictated by the
convenience of keeping the form of the nonlinear strain tensor,
$\mm{w}$ unchanged.

The end result of these two operations (partial tracing and rescaling)
is a nematic elastomer Hamiltonian, identical in form to that in
\rfs{f_expansion5}, expressed in terms of phonon fields $\uv'(\xv')$,
the original momentum cutoff $Q$, and effective elastic moduli $\{ B_z
+ \delta B_z, \lambda_{z\perp} + \delta\lambda_{z\perp}, \lambda +
\delta\lambda, \mu + \delta\mu,K+\delta K\}$.  Assembling corrections
to the elastic constants both from rescaling and from tracing out of
short-scale fluctuations $\uv^>$, we obtain their one-loop flow
equations with the scale parameter $l$:
\begin{subequations}
\label{RGcorrections}
\begin{eqnarray}
\frac{d \,B_z }{d\,l} &=&  (d+3-3\,\omega)\, B_z 
  	 -\frac{ \psi_d}{4 \sqrt{\hat{\mu}K^3}}
	  \,(B_z - \lambda_{z\perp})^2,
		 \label{dB}\\
\frac{d\, \lambda_{z\perp}}{d\,l} &=& 
	(d+3-3\,\omega)\, \lambda_{z\perp}
\nonumber\\
 	&-&  \frac{\psi_d}{4 \sqrt{\hat{\mu}K^3}}
        \,(B_z - \lambda_{z\perp}) 
        \left(\lambda_{z\perp} - \lambda -  \mu \right),
        	\label{dC}\\
\frac{d\,\lambda }{d\,l} &=& 
		(d+3-3\,\omega)\, \lambda
		\label{dBp}\\
	&-& \frac{\psi_d}{4 \sqrt{\hat{\mu}K^3}}\,
	\left[ (\lambda - \lambda_{z\perp})^2
	+ 2\, (\lambda-\lambda_{z\perp})\, \mu
	+ \frac{1}{2}\,\mu^2 \right],
			\nonumber\\
\frac{d\, \mu}{d\,l} &=&
	 (d+3-3\,\omega)\,\mu 
 		- \frac{\psi_d}{8 \sqrt{\hat{\mu}K^3}}\,\mu^2,
 		\label{dmu}\\
\frac{d\, K}{d\,l} &=& 
	 (d-1-\omega)\, K 
 	  +  \frac{\psi_d}{8 \sqrt{\hat{\mu}K^3}} 
	  \frac{1}{(\lambda +2\, \mu)}\times
	 	\label{dK}\\
	&&  \left[ B_z(\lambda+2\,\mu) 
	+ 12 \,\mu\,(\lambda+\mu)\,K
- 4\,C\,\mu - C^2 \right], 
	  \nonumber
\end{eqnarray}
\end{subequations}
where 
\begin{equation}
\psi_d = \frac{\Omega_{d-1} Q^{d-3}}
{2\,(2 \, \pi)^{d-1}},
\label{psid_def}
\end{equation}
and $\Omega_{d-1}$ is the surface area of a $d-1$-dimensional sphere.
We note that (keeping consistent with a one-loop calculation) in above
we have set $d=3$ in the calculation of all diagrammatic corrections.

The above flow equations simplify considerably when formulated in terms
of parameters $(B , C , \mu_{\rm L} , \mu )$.  Using
Eqs.~(\ref{def_moduli}), we find:
\begin{subequations}
\label{RGflows}
\begin{eqnarray}
\frac{d\, B }{d\,l} &=& (d+3-3\, \omega - \eta_B )\, B  ,\\
\frac{d\, C }{d\,l} &=&  (d+3-3\, \omega- \eta_{L} )\,  C  ,\\
\frac{d\, \mu_{\rm L} }{d\,l} &=&  (d+3-3\, \omega- \eta_{\rm L} ) \,\mu_{\rm L} ,\\
\frac{d\, \mu }{d\,l} &=& (d+3-3\, \omega - \eta_{\perp} )\, \mu ,\\
\frac{d\, K }{d\,l} &=& (d-1-\omega+\eta_K ) \,K , 
\end{eqnarray}
\end{subequations}
where various $\eta $ exponents encode diagrammatic corrections and
are given by
\vspace{5mm}
\begin{subequations}
\label{eta_exponents}
\begin{eqnarray}
\eta_B  &=& \frac{1}{4}\, y ^2\, g_{\rm L} , \hspace{3mm}
\eta_{L}  = \frac{1}{4}\,g_{\rm L} ,  \hspace{3mm}
\eta_{\perp}  = \frac{1}{8}\, g_{\perp} ,\label{etaL-etaP}\\
\eta_K  &=& \frac{g_{\rm L} \left(g_{\perp} \left(8 x-24 y
   \sqrt{x}+27\right)-3 g_{\rm L} \left(y^2-1\right)\right)}
   {8\left( 9 g_{\perp} x+g_{\rm L} \left(x-6 y \sqrt{x}+9\right) \right)}.
        \nonumber\\
        \label{eta_K}
\end{eqnarray}
\end{subequations}
In above, the $g_{\rm L} $, $g_{\perp} $, and two ratios $x $ and $y $ are all
dimensionless and are given by
\vspace{-3mm}
\begin{subequations}
\label{def_couplings}
\begin{eqnarray}
g_{\rm L}  &=& \frac{\psi_d \,\mu_{\rm L} }
{ \sqrt{K ^3\, \hat{\mu}}},
  \hspace{4mm}
g_{\perp}  = \frac{\psi_d \,\mu }
        {\sqrt{K ^3\, \hat{\mu}}},
        \label{def_gLT}\\
 x  &=& \frac{{\mu}_L }{B },   \hspace{15mm}
 y  = \frac{C }{\sqrt{B  \mu_{\rm L} }}. \label{def_xy}
\end{eqnarray}
\end{subequations}

The flow equations for these four dimensionless coupling constants can
be calculated from Eqs.~(\ref{RGcorrections}) and
Eqs.~(\ref{def_gLT}-\ref{def_xy}):
\begin{widetext}
\begin{subequations}
\label{flow_couplings}
\begin{eqnarray}
\frac{d \,g_{\rm L}}{d \,l} = \epsilon \, g_{\rm L} &+&
\frac{g_{\rm L}^2} {16 \left(9 g_{\perp} x+g_{\rm L} \left(x-6 y
   \sqrt{x}+9\right)\right) \left(9 g_{\rm L}
     \left(y^2-1\right)-g_{\perp} \left(4 x+12 y
   \sqrt{x}+9\right)\right)}  \times
   \nonumber\\
   \vspace{2mm}
&& \big[9  g_{\rm L}^2 (y-1) (y+1) \left(-4 x+24 y \sqrt{x}+9
   \left(y^2-5\right)\right)
   \nonumber\\   \vspace{2mm}
   &&+ 2  g_{\rm L}  g_{\perp}  \left(972
   \sqrt{x} y^3-9 (64 x+135) y^2-24 \sqrt{x} (x+36) y+8 x^2+522
   x+1377\right)
   \nonumber\\
   \vspace{2mm}
   &&+ g_{\perp}^2 \left(-2763 x y^2+6
   \sqrt{x} (38 x+153) y+4 x (89 x+495)+2106\right) \big] , 
      		        \label{flow_gL}\\
		           \vspace{5mm}
\frac{d \, g_{\perp}}{d \, l} = \epsilon\, g_{\perp} &+&
\frac{g_{\perp}} {16 \left(9 g_{\perp}
   x+g_{\rm L} \left(x-6 y \sqrt{x}+9\right)\right)
   \left(g_{\perp} \left(4 x+12 y \sqrt{x}+9\right)-9 g_{\rm L}
   \left(y^2-1\right)\right)} \times
 \nonumber\\
&&\big[ -81 g_{\rm L}^3 \left(y^2-1\right)^2 +18
  g_{\rm L}^2  g_{\perp} (y-1) (y+1) \left(31 x-114 y \sqrt{x}+144\right) 
   \nonumber\\
  &&+ g_{\rm L} g_{\perp}^2 \left(3069 x y^2+114 \sqrt{x} (2 x-9)
   y-4 (x (55 x+477)+567)\right) -18 g_{\perp}^3 x
   \left(4 x+12 y \sqrt{x}+9\right)\big],\hspace{1cm}
          \label{flow_gT} 
 \end{eqnarray}
\end{subequations}
\end{widetext}     
\begin{subequations}
\begin{eqnarray}
\frac{d \, x} {d \, l} &=& - \frac{1}{4}\,g_{\rm L} \,x \,(1 - y^2),
        \label{flow_x}\\
\frac{d \, y} {d \, l} &=& -\frac{1}{8}\, g_{\rm L}\, y\, (1-y^2),
\label{flow_y}
\end{eqnarray}
\label{flow_xy}
\end{subequations}
where $\epsilon = 3-d$ is the small parameter controlling the flow
equations expansion, and as before the $l$ dependence of the coupling
constants is implicit.  The initial conditions for the flow equations,
(\ref{flow_couplings}-\ref{flow_xy}) are determined by the bare
values of corresponding elastic moduli:
\begin{subequations}
\label{def_couplings-2}
\begin{eqnarray}
g_{\rm L}(l=0)  &=& \frac{\psi_d \,\mu_{\rm L} }
{ \sqrt{K ^3\, \hat{\mu}}},
  \hspace{4mm}
g_{\perp}(l=0)  = \frac{\psi_d \,\mu }
        {\sqrt{K ^3\, \hat{\mu}}},
        \nonumber\\
        \label{def_gLT-2}\\
 x(l=0)  &=& \frac{{\mu}_L }{B },   \hspace{10mm}
 y(l=0)  = \frac{C }{\sqrt{B  \mu_{\rm L} }}. 
 \nonumber\\
 \label{def_xy-2}
\end{eqnarray}
\end{subequations}

\subsection{Solution of renormalization-group equations in 
three dimensions}

Naturally, we are most interested in three-dimensional nematic
elastomers, corresponding to $\epsilon = 0$. Deferring the analysis of
the $d<3$ regime to Appendix \ref{RGsolution-d<3}, we set $\epsilon =
0$ in the right hand sides of Eqs.~(\ref{flow_gL}-\ref{flow_gT}),
which leads to a vanishing of all linear (in $g_{\rm L}$ and
$g_{\perp}$) terms.  Also, as we discussed earlier, rubber is nearly
incompressible and is therefore characterized by a bare value of the
bulk modulus $B$ that is much larger than other moduli, $C$, $\mu_{\rm
  L}$, and $\mu$.  Thus, using Eqs.~(\ref{def_couplings}) we see that
the initial values $x(l=0)$ and $y(l=0)$ are much less than one.  Also,
as we will prove later in this section, small $x(l)$ and $y(l)$ both flow to
zero as $l \rightarrow \infty$.  Consequently it becomes
asymptotically exact to directly set $x$ and $y$ to zero in the flow
equations for the coupling constants $g_{\rm L}(l)$ and $g_{\perp}(l)$,
Eqs.~(\ref{flow_gL}-\ref{flow_gT}), which greatly 
simplifies these equations:
\begin{subequations}
\label{flow-gs}
\begin{eqnarray}
\frac{d \,g_{\rm L} }{d \,l} &=& 
-\frac{g_{\rm L} \left(5 g_{\rm L}^2+34 g_{\perp} g_{\rm L}+26
   g_{\perp}^2\right)}{16 (g_{\rm L}+g_{\perp})},
        \label{flow_gL2}\\
 \frac{d \, g_{\perp} }{d \, l} &=& 
-\frac{g_{\perp} \left(g_{\rm L}^2+32 g_{\perp} g_{\rm L}+28
   g_{\perp}^2\right)}{16 (g_{\rm L}+g_{\perp})}.\qquad
        \label{flow_gperp}
\end{eqnarray}
\end{subequations}
To solve for these two equations, we define a new variable $\sigma(l)$:
\begin{equation}
\sigma(l) = \frac{g_{\rm L}(l)}{g_{\perp}(l)} - \frac{1}{2}.
\label{sigma-def}
\end{equation}
From Eqs.~(\ref{flow-gs}) we can then derive the flow equations for
$g_{\rm L}$ and $\sigma$:
\begin{eqnarray}
\frac{d\sigma}{dl} &=& -\frac{g_{\rm L}}{4}  \, \sigma; 
\label{flow-sigma}\\
\frac{dg_{\rm L}}{dl} &=& 
-\frac{g_{\rm L}^2 \left(20 \sigma ^2+156 \sigma +177\right)}{16
   \left(4 \sigma ^2+8 \sigma +3\right)}.
   \label{flow-gL-3}
\end{eqnarray}
Anticipating that, for large $l$, $g_{\rm L}(l)$ asymptotically flows
to zero as
\begin{equation}
g_{\rm L}(l) \approx \frac{\Upsilon}{l}, 
\label{assump-gL}
\end{equation}
with $\Upsilon$ a positive constant (which we shall determine below), we
easily see from Eq.~(\ref{flow-sigma}) that $\sigma(l)$ also monotonically
flows to zero:
\begin{equation}
\sigma(l) \approx  l^{-\frac{\Upsilon}{4}}. 
\end{equation}
Going back to Eq.~(\ref{sigma-def}), we find that $\sigma \rightarrow
0$ implies that the ratio between the transverse and longitudinal shear
moduli flows to a {\em universal} value $2$:
\begin{equation}
\frac{g_{\perp}(l)}{g_{\rm L}(l)} = 
\frac{\mu(l)}{\mu_{\rm L}(l)} \rightarrow 2, 
\hspace{5mm} \mbox{as} \hspace{3mm}l\rightarrow \infty.
        \label{ratio_limit}
\end{equation}
Because of the slowness of $\sigma(l)$ decay (power-law in $l$ and
logarithmic in a length-scale) we expect that it will be difficult to
observe this universal ratio in a real experiment.

In the asymptotic regime where $\sigma(l)$ is small, we may simplify the
flow equation for $g_{\rm L}$, Eq.~(\ref{flow-gL-3}), by setting
$\sigma$ to zero:
\begin{equation}
\frac{dg_{\rm L}}{dl} = - \frac{59}{16} \, g_{\rm L}^2,
\end{equation}
which is solved by
\begin{equation}
g_{\rm L}(l) \approx \frac{16}{59\,l}. 
\label{solution_gL}
\end{equation}
We have therefore verified the assumption (\ref{assump-gL}), with
\begin{equation}
\Upsilon = 16/59.
\label{Upsilon}
\end{equation}
Combining this with $\sigma(l)\rightarrow 0$, we find the asymptotic
flow of $g_{\perp}(l)$ to be
\begin{equation}
g_{\perp}(l) \approx \frac{32}{59\,l}.\label{solution_gT}
\end{equation}

We can now explicitly show that two dimensionless ratios $x(l)$ and $y(l)$
indeed both flow to zero for large $l$.  Substituting the asymptotic solution
Eq.~(\ref{solution_gL}) into Eqs.~(\ref{flow_x}-\ref{flow_y}), we
find:
\begin{subequations}
\label{flow_xy2}
\begin{eqnarray}
\frac{d \, x} {d \, l} &=&
 - \frac{\Upsilon}{4\,l}  \,x \,(1 - y^2)
 \approx  - \frac{\Upsilon}{4\,l}  \,x ,
        \\
\frac{d \, y} {d \, l} &=& 
-\frac{\Upsilon}{8\,l}\, y\, (1-y^2)
\approx -\frac{\Upsilon}{8\,l}\, y.
\end{eqnarray} 
\end{subequations}
with approximate asymptotic solutions
\begin{subequations}
\label{solution_xy}
\begin{eqnarray}
x(l) &\approx& x_0\,l^{-\Upsilon/4},\\
y(l) &\approx& y_0\,l^{-\Upsilon/8}
\end{eqnarray}
\end{subequations}
indeed flowing to $0$ as anticipated.

Armed with the results in
Eqs.~(\ref{solution_gL},\ref{solution_gT},\ref{solution_xy}), we
then obtain the asymptotic behaviors of the $\eta(l)$ exponents as
defined in Eqs.~(\ref{eta_exponents}):
\begin{subequations}
\label{eta_exponents-2}
\begin{eqnarray}
\eta_B(l) &=& \frac{4\,y_0^2}{59} \,l^{-(1+\Upsilon/4)},\label{eta_B-2}\\
 \eta_{\rm L}(l) &=&  \eta_{\perp}(l)  =\frac{4}{59\,l},\\
\eta_K(l)  &=& \frac{38}{59\,l}. 
        \label{eta_K-2}
\end{eqnarray}
\end{subequations}
Even though all $\eta(l)$ exponents vanish as $l \rightarrow \infty$,
$\eta_B(l)$ does so qualitatively faster than all others.  As we shall
discuss next, this leads to a qualitatively different scaling behavior
of the renormalized bulk modulus as compared to other renormalized
elastic moduli, and is responsible for an exact asymptotic
incompressibility of three-dimensional nematic elastomers.

\subsection{Anomalous elasticity of a thermal (homogeneous) nematic 
elastomer} 
\label{Sec:matching-thermal}

The renormalization group analysis of the previous subsection allows
us now to relate the (difficult to compute) long-scale correlation
functions, computed with bare elastic moduli to the corresponding
functions at short-scales, but with renormalized elastic moduli. The
latter are determined by the solution of the flow equations
\rf{RGcorrections}, that are related to the microscopic moduli through
the initial conditions. In contrast to the pure thermodynamics, which
follows from the partition function and thus only affected by
transformation of the elastic moduli, correlation functions are also
affected by the RG rescaling of the coordinates and the phonon
fields. In reciprocal space, these transform according to:
\begin{subequations}
\label{RGrescaling2}
\begin{eqnarray}
\qv_{\perp}&=&\qv_{\perp}(l)e^{-l},\\
  q_z       &=& q_z(l) e^{-\omega\,l},\\
\uv_{\perp} (\qv)&=&e^{ (d+2-\,\omega)\, l}
 \uv_{\perp}^l(\qv(l)),\\
u_z(\qv)&=&e^{(d+1) \,l}u_z^l(\qv(l)),
\end{eqnarray}
\end{subequations}
where at this point the anisotropy exponent $\omega$ is arbitrary.

As an example, let us explicitly consider the RG transformation of the
two-point correlation function of the $u_z$ phonon.  Using the
transformations of fields and wavevectors above, we find:
\begin{eqnarray}
\hspace{-0.2cm}
\langle u_z(\qv) u_z(\qv') \rangle
&=& (2\,\pi)^d \delta^d(\qv+\qv') \,
 G_{zz}(\qv, c) \nonumber\\
&=& e^{2(d+1)l}\,\langle u_z^l(\qv(l)) u_z^l(\qv'(l)) \rangle\nonumber\\
&=&  e^{2(d+1)l} \,(2\,\pi)^d \delta^d(\qv(l)+\qv'(l))\,
  G_{zz}(\qv(l),c(l)),\nonumber\\ 
&&
\end{eqnarray}
where again we have used $c$ as a shorthand for all parameters of the
elastic Hamiltonian, that also enter the correlator $G_{zz}(\qv, c)$.

Utilizing a multiplicative transformation of $\delta^d(\vec{q}) =
\delta(q_z) \delta^{(d-1)}(\vec{q}_{\perp})$ under rescaling \rf{RGrescaling2}
we find:
\begin{equation}
G_{zz}(\qv, c) = e^{(d+3-\omega)l}\, G_{zz}(\qv(l),c(l)). 
\label{RG_correlation1}
\end{equation}
Similarly analysis gives the transformations of other phonon correlators:
\begin{subequations}
\label{RG_correlation2}
\begin{eqnarray}
G_{zi}(\qv, c) &=& e^{(d+4-2\,\omega)l} \,G_{zi}(\qv(l),c(l)),\\
G_{ij}(\qv, c) &=& e^{(d+5-3\,\omega)l}\,G_{ij}(\qv(l),c(l)),
\end{eqnarray}
\end{subequations}
that then give the renormalized two-point vertex functions
$\Gamma_{ab}$, the inverse of $G_{ab}$:
\begin{subequations}
\label{RG_vertex}
\begin{eqnarray}
\Gamma_{zz}(\qv, c) &=& 
 e^{-(d+3-\omega)l} \, \Gamma_{zz}(\qv(l),c(l)),\\
\Gamma_{zi}(\qv, c) &=& 
 e^{-(d+4-2\,\omega)l} \, \Gamma_{zi}(\qv(l),c(l)),\\ 
\Gamma_{ij}(\qv, c) &=& 
 e^{-(d+5-3\,\omega)l} \, \Gamma_{ij}(\qv(l),c(l)). 
\end{eqnarray}
\end{subequations}

Now we may use the asymptotic solutions Eqs.~(\ref{eta_exponents-2})
for $\eta$ exponents in the flow equations (\ref{RGflows}) that
control the large $l$ behavior of $c(l)$:
\begin{subequations}
\label{RGflows2}
\begin{eqnarray}
\frac{d\, B}{d\,l} &=& - \frac{4\,y_0^2}{59}
\,\frac{B}{l^{(1+\Upsilon/4)}},
\label{flow-B-2}\\
\frac{d}{d\,l} \left( C, \mu_{\rm L}, \mu \right)
 &=& - \frac{4}{59\,l} \,  \left( C, \mu_{\rm L}, \mu \right) ,\\
\frac{d\, K}{d\,l} &=&  \frac{38}{59\,l} \, K. 
\end{eqnarray}
\end{subequations}
where for later convenience we chose the arbitrary rescaling parameter
$\omega$ to be $\omega =2$.

These flow equations are easily solved to give:
\begin{equation}
(C(l), \mu_{\rm L}(l), \mu(l), K(l)) 
 =  (\frac{C^*}{l^{\Upsilon/4}},
 \frac{\mu_{L}^*}{l^{\Upsilon/4}}, 
\frac{\mu^*}{l^{\Upsilon/4}}, 
K^* \, l^{\frac{19\,\Upsilon}{8}} ),
\label{moduli-solutions}
\end{equation}
where $\mu^*= 2 \mu_{\rm L}^*$ due to Eq.~(\ref{ratio_limit}). The
slow decay of the effective $C,\mu_L,\mu$ and growth of $K$ with
length-scale contrasts with the long-scale behavior of the bulk
modulus that flows to a nonzero and nonuniversal value $B^*$
determined by $y_0$ as well as the initial (bare) value of $B$:
\begin{equation}
B(l) \longrightarrow B^*, \hspace{4mm} \mbox{as $l\rightarrow \infty$} 
\label{B-solution}
\end{equation}

We can now use these relations to determine the wavevector dependence
of these renormalized vertex functions by choosing the flow parameter
$l$ such that
\begin{subequations}
\label{l-choice}
\begin{eqnarray}
q_{\perp}(l) &=& q_{\perp} \, e^l = Q,\\
q_z(l) &=& q_z \, e^{\omega\,l} = q_z\frac{Q^2}{q_{\perp}^2}. 
\end{eqnarray}
\end{subequations}
This choice ensures that the rescaled vertex functions on the right
hand side of Eqs.\rf{RG_vertex} are easy to compute as they are
evaluated at the large wavevector $Q$ and with small nonlinear couplings
$g_{\rm L,\perp}(l)$, as latter flow to $0$ for large $l(q_\perp)=\log
Q/q_\perp$ (small $\qv$).  Namely, in this regime the effect of
anharmonic fluctuations is negligible and we can replace the right hand
sides of Eqs.~(\ref{RG_vertex}) by harmonic vertex functions.
Focusing
on $\Gamma_{zz}(\qv, c)$ as an explicit example, we have:
\begin{widetext}
\begin{eqnarray}
\Gamma_{zz}(\qv,c) 
&=& e^{-4\,l}\, \Gamma_{zz}(\qv(l),c(l)) =
 e^{-4\,l}\,\left( B_z(l)\,q_z(l)^2+K(l)\,q_{\perp}(l)^4\right)
\nonumber\\
&=& e^{-4\,l}\,\left[
 \big(B(l)+\frac{4}{3}C(l) 
+ \frac{4}{9} \mu_{\rm L}(l)\big) q_{z}(l)^2
  + K(l)\,q_{\perp}(l)^4 \right]
\end{eqnarray}
where we have used Eq.~(\ref{def_moduli}) (with $d=3$) to express
$B_z$ in terms of $B$, $C$, and $\mu_{\rm L}$.  Using
Eqs.~(\ref{moduli-solutions},\ref{B-solution},\ref{l-choice}), we
finally find the renormalized vertex function $\Gamma_{zz}(\qv,c)$ to
be given by
\begin{eqnarray}
\Gamma_{zz}(\qv) &=& \left(\frac{q_{\perp}}{Q}\right)^4 
 \left[ \left(B^* + ( \frac{4}{3}\, C^* + \frac{4}{9}\, \mu_{L}^* ) 
\left|\log \frac{Q}{q_{\perp}}\right|^{-\frac{\Upsilon}{4}}\right) \,
q_z^2\,q_{\perp}^{-4}Q^4
+ K^* \,\left|\log \frac{Q}{q_{\perp}}\right|^{\frac{19\,\Upsilon}{8}}
Q^4\right]. \nonumber\\
&=&  \left[ B^* + \left( \frac{4}{3}\, C^* 
+ \frac{4}{9}\, \mu_{L}^*  \right) 
\left|\log \frac{Q}{q_{\perp}}\right|^{-\frac{4}{59}} \right] q_{z}^2
+ K_1^* \left|\log \frac{Q}{q_{\perp}}\right|^{\frac{38}{59}} q_{\perp}^4.
\end{eqnarray}

The other two vertex functions can be similarly calculated,
\begin{eqnarray}
\Gamma_{zi}(\qv) &=& 
  \left[ B^* + \left( \frac{1}{3}\, C^* - \frac{2}{9}\, \mu_{L}^*  \right) 
\left|\log \frac{Q}{q_{\perp}}\right|^{-\frac{4}{59}} \right] q_{z}\,q_i
	\nonumber\\     
 \Gamma_{ij}(\qv) &=&
 \left[ B^* + \left( -\frac{2}{3}\, C^* + \frac{1}{9}\, \mu_{L}^*  \right) 
\left|\log \frac{Q}{q_{\perp}}\right|^{-\frac{4}{59}} \right] q_i q_j 
+\left[\mu^*\left|\log \frac{Q}{q_{\perp}}\right|^{-\frac{4}{59}} q_{\perp}^2 
+ K_3 \, q_z^4 \right] \delta_{ij},
\end{eqnarray}
\end{widetext}
where we have restored $K_3$ in $\Gamma_{ij}$.  Comparing these
renormalized vertex functions with their bare forms, Eqs.\rf{gamma_0},
we can interpret the effects of long wavelength elastic fluctuations
as {\em wavevector-dependent} renormalized elastic moduli:
\begin{subequations}
\label{renormalized_moduli}
\begin{eqnarray}
B(\qv) &\cong& B^*,\\
\mu_{\rm L}(\qv), \mu(\qv), C(\qv) &\propto&
\left|\log \frac{Q}{q_{\perp}}\right|^{-\frac{4}{59}},\\
K_1(\qv)  &\propto&\left|\log \frac{Q}{q_{\perp}}\right|^{\frac{38}{59}}.
\end{eqnarray}
\end{subequations} 
We observe that the renormalized shear moduli $\mu$ and $\mu_{\rm L}$
are singular functions of the wavevector and vanish in the long
wavelength limit.  In contrast, the renormalized bulk modulus remains
finite, while the renormalized splay constant diverges in the long
wavelength limit.  As advertised, these results show that in the
thermodynamic limit an ideal nematic elastomer is effectively {\em
  strictly incompressible}. They also imply an absence of a linear
stress-strain response (i.e., a breakdown of Hooke's law) even for a
{\em nonsoft} shear deformation, as (due to a vanishing of $\mu(\qv)$
at long wavelengths) the shear stress vanishes faster than linearly
with a vanishing strain. Finally due to Eq.~(\ref{ratio_limit}), the
ratio of two renormalized shear moduli goes to a universal number:
\begin{equation}
\lim_{q\rightarrow 0} \frac{\mu(\qv)}{\mu_{\rm L}(\qv)}
 = \lim_{l\rightarrow \infty} \frac{g_{\perp}(l)}{g_{\rm L}(l)}
        = 2. \label{ratio}
\end{equation}

The above anomalous behavior begins to manifest itself roughly when
the asymptotic values of coupling constants (as given by
Eq.~(\ref{solution_gL}) and Eq.~(\ref{solution_gT})) become comparable
with their bare values, given by Eqs.~(\ref{def_couplings-2}). This
happens at $l = l^*$, which satisfies
\begin{equation}
\frac{16}{59\,l^*} = \frac{\psi_d\,\mu_{\rm L}\,k_BT}{\sqrt{\hat{\mu} K^3}},  
\end{equation}
where we restored the temperature dependence.  This value of $l^*$ in
turn determines the two crossover length-scales $\xi_{\perp}$ and
$\xi_{\parallel}$ via
\begin{eqnarray}
e^{l^*} = Q\,\xi_{\perp},\hspace{3mm}
e^{2\,l^*} = Q\,\xi_{\parallel}, 
\end{eqnarray}
which are thus given by
\begin{subequations}
\label{xi-thermal}
\begin{eqnarray}
\xi_{\perp} &=&    Q^{-1} \, 
	\exp\left(\frac{16 \sqrt{\hat{\mu}K_1^3}} {59\psi_d\,\mu_{\rm L}\,k_B\,T} \right),\\
\xi_{\parallel} &=&  Q^{-1}\,
 \exp\left(\frac{32 \sqrt{\hat{\mu}K_1^3}} {59\psi_d\,\mu_{\rm L}\,k_B\,T} \right),
\end{eqnarray}
\end{subequations}
with $\hat{\mu}$ defined by Eq.~(\ref{muhat-limit}).  

It is interesting to estimate the order of magnitude of these two
important crossover length-scales $\xi_{\perp}$ and $\xi_{\parallel}$.
In three dimensions, $\psi_d = 1/4\pi $. The typical value of $Q^{-1}$
(mesh-size of the polymer network) is a few nanometers.  At room
temperature, $k_B T$ is roughly $4\times 10^{-21}J$.  The splay
constant $K_1$ is usually $2-4 \times 10^{-12}N$ for low molecular
weight nematic liquid crystals, but may be several times larger than
this for liquid crystalline polymers.  The shear moduli $\hat{\mu}$
and $\mu_{\rm L}$ may vary significantly near the vulcanization
transition, with a typical range of $10^4 Pa - 10^6 Pa$.  Substituting
these numbers into Eq.~(\ref{xi-thermal}), we find a wild range of
these two crossover length-scales:
\begin{eqnarray}
10^{-9} m < \xi_{\perp} < 10^2 m, \hspace{5mm}
10^{-6} m < \xi_{\parallel} < 10^{12} m, 
\end{eqnarray}
where the lower limit corresponds to large shear moduli.  Therefore
for soft elastomers, the anomalous effects of thermal fluctuations may
be impossible to observe, while for hard elastomers, thermal
fluctuations may become important even at the scale of the polymer network
mesh size.  The ratio between two length-scales $\xi_{\perp}$ and
$\xi_{\parallel}$ also exhibit a wide range and depends sensitively on
the shear moduli.  This extreme sensitivity of $\xi_{\perp}$ and
$\xi_{\parallel}$ to shear moduli is of course due to the exponential
functional form in Eqs.~(\ref{xi-thermal}), which is a consequence of
the marginal irrelevance of temperature in three dimensions.

\section{Long-scale elasticity of a {\em heterogeneous} nematic 
elastomer}
\label{Sec:Disorder-fluct}

\subsection{Elastomer heterogeneity}

In all preceding analyses we have treated nematic elastomers as
homogeneous continuous elastic media.  However, as all rubber and
gels, nematic elastomers are random polymer networks that are only
statistically homogeneous and isotropic at the macroscopic level.  At
the microscopical level, their elasticity is due to a heterogeneous
polymer network, made by random crosslinking of a polymer melt, a
process called ``vulcanization'' or ``gelation''.  Thus, for a
complete understanding of the elasticity of nematic elastomers, it is
essential to examine the nature and consequences of the network
heterogeneity.

Experimental evidence unambiguously shows the importance of network
heterogeneity.  Elastomers crosslinked in the isotropic phase, upon
lowering temperature below the I-N transition point of the
uncrosslinked system, are observed to exhibit a poly-domain ``nematic
phase'', where a nematic order freezes locally into a micron-size
domain structure, with no long-range orientational order.  When the
system is uniaxially stretched, these randomly orientated nematic
domains align as the strain deformation exceeds a threshold value
\cite{PhysRevE.60.1847}.  When the strain is removed, the original
poly-domain pattern is restored \cite{comment:however}.  On the other
hand, nematic elastomers crosslinked under anisotropic condition,
i.e., inside the nematic phase or under a uniaxial stretch, do exhibit
a long-range nematic order, but are characterized by {\em semi}-soft
elasticity, with a small but finite shear modulus $\mu_{z\perp}$
\cite{KupferFinkelmann94,Warner99}.  When heated back to high
temperature, the system exhibits a para-nematic phase with a weak
remnant nematic order. Thus, such network permanently breaks the
rotational symmetry with the isotropic phase no longer accessible.
That is, the system ``remembers'' the anisotropic crosslinking
conditions, i.e., its history of formation.  The mechanism underlying
this memory effect is poorly understood.

To understand how the history of formation affects the properties of
nematic elastomers (or more generally speaking, vulcanized solids), a
detailed characterization of elastic heterogeneities is needed.  A
semi-microscopic level of understanding of network heterogeneity may
be obtained from the vulcanization theory.  As pointed out by de
Gennes \cite{polymer:deGennes} and Edwards \cite{Deam-Edwards},
elastomers (gels) are ``frozen'' random systems.  Their physical
properties depend on both the state within which the measurement is
taken, and the preparation state of the system.  A complete study of
these systems necessarily involves two ensembles: a preparation
ensemble and a measurement ensemble.  The vulcanization theory
\cite{Deam-Edwards,Vulcan_Goldbart} captures both of these and
therefore has the potential to yield a better statistical
characterization of elastic heterogeneities in elastomeric materials,
as well as their connections to the crosslinking process and elastic
properties of the resulting amorphous solid.  One recent theoretical
work \cite{Vulcan:MGXZ} along this direction has indeed showed some
success.  At this stage, however, it is beyond the Landau approach
that we take in this work and will therefore not be discussed further.

Within the framework of continuous elasticity theory, two types of
quenched disorders can be identified: random elastic constants and
random internal, or residual stress.  Internal stress is a stress
configuration of solids that does not vanish even if all external
tractions are removed.  It is a generic property of random solids and
systems with topological defects.  At the equilibrium state, the
internal stress has to satisfy
\begin{eqnarray}
\partial_i \sigma^{\rm I}_{ij} = 0, \label{equi-sigma-I}
\end{eqnarray}
which is essentially the condition for mechanical equilibrium.  
This implies that random internal stress heterogeneity only couples to
the linearized strain at the quadratic order:
\begin{eqnarray}
\int d^d x \,\,\sigma^{\rm I}_{ij}e_{ij} &=& 
\int d^d x \,\, \sigma^{\rm I}_{ij} \frac{1}{2} \left( 
\partial_i u_j + \partial_j u_i + \partial_i \uv \cdot \partial_j \uv \right)
\nonumber\\
&\rightarrow&\int d^d x \,\,  \frac{1}{2} \,\sigma^{\rm I}_{ij} 
\partial_i \uv \cdot \partial_j \uv,
\label{sigma-uu}
\end{eqnarray}
where $e_{ij}$ is the nonlinear Lagrange strain tensor defined
relative to the true equilibrium state.

DiDonna and Lubensky \cite{Didonna-Lub-nonaffine} have recently
studied the effects of random internal stress and random elastic
constants in macroscopically deformed conventional solids.  They found
that random elastic constants, but not random internal stress, induce
nonaffine components of the deformation field.  These authors have
also introduced {\em short-range} correlated random stress field into
an otherwise homogeneous solid, by, e.g., starting from a triangular
lattice of particles connected by central force springs, and randomly
change the equilibrium length of each spring.  They found that, after
re-expanding the elastic energy around the true equilibrium state, the
system exhibited both random elastic constants and random internal
stress.

In this paper we shall take a phenomenological approach similar to
that of DiDonna and Lubensky \cite{Didonna-Lub-nonaffine}.  We start
from an ideal homogeneous nematic elastomer and introduce {\em
short-range} (uncorrelated) random tensor fields that couple to strain
and nematic order parameter.  By re-expanding the elastic Hamiltonian
around the ideal nematic reference state (i.e., around the true ground
state for the disorder-{\em free} system) and integrating out the
fluctuations of the nematic director field, we obtain an effective
strain-only model for heterogeneous nematic elastomers, where the
nonlinear Lagrange strain tensor is coupled to a random symmetric
tensor field, which we shall refer to as the {\em random initial
stress}.  Such a stress tensor field, ${\boldsymbol \sigma}$
generically contains both the longitudinal part (satisfying $\nabla
\times \ {\boldsymbol \sigma}^L = 0$) and a transverse part
(satisfying $\nabla \cdot \ {\boldsymbol \sigma}^T = 0$).  However,
since we shall only consider the linear coupling between the random
initial stress and the {\em linearized} strain tensor (see
\rfs{sigma-uu}), our model ultimately only captures the fluctuations
in the {\em longitudinal} part of the initial stress, but {\em not}
the transverse one. Although we might argue that the latter is less
important as it only couples to the nonlinear part of the strain, a
more detailed analysis and justification of this approximation is
necessary, but will not be studied here.

Our analysis below shows that these initial random stress fluctuations
qualitatively modify macroscopic elastic properties of heterogeneous
nematic elastomers, leading to a zero-temperature analog of anomalous
elasticity discussed in the previous sections for thermal homogeneous
elastomers.  In fact, we will find that these random initial stress
effects strongly dominate over purely thermal fluctuations and already
below five (and therefore in the physically most interesting three)
dimensions lead to shear moduli that vanish as power-laws of a probing
wavevector, and to a breakdown of Hooke's law, replaced by a strictly
nonlinear stress-strain relation for arbitrarily small strain
deformation. These anomalous elastic properties are consequences of
the subtle interplay between the elastic heterogeneities and the long
wavelength fluctuations of soft deformations.  Nevertheless, we find
the the long-range nematic order is stable against these random
initial stress fluctuations in three dimensions, at least within the
approximation of ignoring transverse internal stress fluctuations, and
within the approximation of one-loop renormalization group analysis.
We emphasize that our analysis of the ordered state provides a
necessary stability condition for the existence of a mono-domain
nematic elastomer in the presence of {\em weak} quenched disorder.
Finally, our way of modeling nematic elastomer heterogeneity, though
simple, is far from well justified. A justification of this approach
must come from a more detailed heterogeneous model of rubber
elasticity and ultimately from comparison with experiments.

In the following sections we present a detailed analysis of the
fluctuations of the initial stress in nematic elastomers and their
effects on the long-scale elastic properties of the systems at low
temperature.  A summary of these results have already appeared in
Ref. \cite{Xing-Radz2}.  We leave other important and challenging
questions such as for example the nature of the isotropic-nematic
transition in heterogeneous elastomers (i.e., whether it survives or
is replaced by a crossover) to another publication \cite{Xing-Radz3}.

\subsection{Heterogeneous strain-only elastic model}
\label{Sec:disorder-Hamiltonian}

\begin{figure}[!htbp]
\begin{center}
 \includegraphics[width=6.5cm]{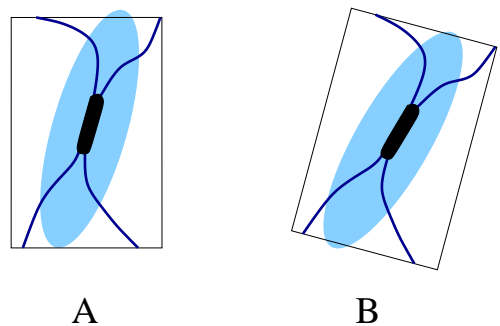}
  \caption{A cartoon for a random tilting field coupled to a
    nematic order of a nematic elastomer. In (A), a crosslink (the black
    rod) connects two polymer chains. Neighboring liquid crystalline
    mesogenic units tend to align along this rod-like
    crosslink, an effect that can be modeled as a random tilting field
    coupling to nematic order (shown as blue ellipses).  In (B), this random
    tilting field rotates with the polymer network.  Therefore elastic
    deformation of the network also changes this random tilting field seen
    by local nematic order. This property is captured by
    Eq.~(\ref{model:disorder}).}
\label{randomtilt}
\end{center}    
\end{figure}

We are interested in nematic elastomers crosslinked under isotropic
conditions, i.e., inside the isotropic phase and in the absence of
shear strain.  Under these conditions, the lowest order elastic energy
terms that involve network heterogeneities, elastic deformations, and
nematic order, and at the same time satisfy all relevant symmetries
are:
\begin{eqnarray}
H_{\mbox{d}}&=& -  \int d^d X
\left[\Tr\bigg(\mm{F}_1(\Xv) \Lm^{\rm T} \Lm\bigg)\right.
\label{model:disorder}\\
&& \left.
+\Tr\bigg(\mm{F}_2(\Xv) \Lm^{\rm T} \mm{Q}(\Xv) \Lm\bigg)\right],\nonumber
\end{eqnarray}
where, as defined in Sec.~\ref{Sec:Model}, $\Xv$ is the isotropic
referential coordinate and $\Lm$ the deformation gradient defined
relative to the isotropic reference state.  $\mm{F}_1(\Xv)$ and
$\mm{F}_2(\Xv)$, encoding network heterogeneity, are two random tensor
fields in the {\em isotropic reference space}.  They transform as
scalars with respect to rotations in the embedding space, in agreement
with the fact that they describe the intrinsic properties of random
network.  Since the nematic order parameter $\mm{Q}$ is a tensor in
the embedding space, it can couple to the tensor field $\mm{F}_2$ only
through the deformation gradient $\Lm$, hence the structure of the
second term in Eq.~(\ref{model:disorder}).  Since the system is
isotropic at the macroscopic level, the spatial correlation of the
tensor fields $\mm{F}_1(\Xv)$ and $\mm{F}_2(\Xv)$ must be isotropic.
To simplify the analysis, we shall also assume that they are Gaussian
and short-range correlated.  We may also say something about the
microscopic origin of these two terms.  $\mm{F}_1(\xv)$, directly
coupled to the Lagrange strain tensor, may be simply due to the
fluctuations of crosslink density.  $\mm{F}_2(\xv)$, coupled to the
nematic order through the deformation gradient, may be due to the
orientational effects of rod-like crosslinkers on local nematic order,
as is illustrated in Fig.~\ref{randomtilt}.

For a given realization of quenched disorder fields $\mm{F}_1(\Xv)$
and $\mm{F}_2(\Xv)$, in contrast to the homogeneous case of
Sec.~\ref{Sec:Model}, the isotropic reference state (IRS) defined by
$$\rv(\Xv) = \Xv, \hspace{4mm} \mm{Q} = 0 $$ is generically {\em not}
the true ground state, i.e., it does not minimize the total
heterogeneous elastic energy, which is the sum of
Eq.~(\ref{model:disorder}) and the elastic energy for ideal nematic
elastomers Eq.~(\ref{elast-energy-2}).  This raises the question of
physical significance of the IRS in a realistic heterogeneous
elastomer.  Indeed IRS has no physical significance; it just provides
a conceptually convenient but arbitrary reference point around which
we expand the elastomers free energy.  In such an expansion, the
lowest order two terms involving random heterogeneity are shown in
Eq.~(\ref{model:disorder}) \cite{comment:elastomer-crystal}. 

The true ground state of a disordered
nematic elastomer can always be found by minimizing the total elastic
energy.  Hence for a given realization of the disorder, there is a
well-defined relation between the IRS and the true ground state.  One
could in principle re-expand the elastic energy around the true ground
state.  However, due to the nonlinear nature of the elastic system,
this analysis is rather messy and the final result is not
illuminating.

We re-expand the disorder Hamiltonian in
Eq.~(\ref{model:disorder}) around the ideal nematic reference state
(NRS), which is the ground state for a disorder-free system as defined
by Eq.~(\ref{elast-energy-1}), and is related to the IRS by
Eq.~(\ref{Rv_0}).  As usual we choose the (uniform) nematic order in
the NRS to be along $\hat{z}$ axis so that $\nh_0 = \hat{z}$ in
Eq.~(\ref{Rv_0}).  As in Sec.~\ref{Sec:Model}, we shall use the
notation $\xv$ for the position of mass point in the ideal nematic
reference state, and define the ``phonon field'' $\uv(\xv)$ to be the
displacement from this NRS, i.e.,
\begin{equation}
\rv(\xv) = \xv + \uv(\xv).  
\end{equation} 

To further simplify the analysis, we integrate out the fluctuations of
nematic director $\delta \nh$ around the ideal nematic reference
state.  After a tedious but conceptually straightforward calculation,
which we relegate to Appendix \ref{App:Disorder-deriv}, we find that,
to a {\em linear} order in the phonon field $\uv$, the most relevant
part of the disorder Hamiltonian Eq.~(\ref{model:disorder}) reduces to
\begin{equation}
H_{\rm d} =  - \int d^d x \, \sigma_{ab}(\xv) \,
   \varepsilon_{ab}(\xv),
        \label{H_disorder}
\end{equation}
where
\begin{eqnarray}
 \varepsilon_{ab}(\xv) =  \frac{1}{2}\left(\partial_a u_b(\xv)
        + \partial_b u_a(\xv) \right), 
      \label{varep-def}
\end{eqnarray}
are the components of the linearized symmetric strain tensor, defined
relative to the ideal nematic reference state.  The quenched random
tensor field ${\boldsymbol \sigma(\xv)}$ as a functional of $\mm{F}_1$
and $\mm{F}_2$ is given by Eqs.~(\ref{sigma-F}).  Since it is coupled
to the {\em linearized} strain, we shall call it the {\em random
  initial stress}, to make a distinction with an internal stress field
in the equilibrium state which satisfies the mechanical equilibrium
condition $\partial_a \sigma^I_{ab} = 0$. Obviously, as mentioned
above, only the longitudinal part of the random initial stress
$\sigma_{ab}(\xv)$ contributes to Eq.~(\ref{H_disorder}).

It is further shown in Appendix \ref{App:Disorder-deriv} that, inside
Eq.~(\ref{H_disorder}), the most relevant terms are $\sigma_{iz}
\varepsilon_{iz}$, where $i$ and $j$ only take values $x$ and $y$.
Other terms, such as $\sigma_{ij}\varepsilon_{ij}$ and
$\sigma_{zz}\varepsilon_{zz}$, are less relevant (in the RG sense) and
therefore can and will be neglected.  The statistics of the disorder
fields $\sigma_{zi}(\xv)$ is assumed to be Gaussian with short-range
correlations:
\begin{equation}
\overline{\sigma_{zi}(\xv)\,\sigma_{zi}(\xv')}
 =  4\, \Delta \, \delta_{ij}\,\delta^d(\xv-\xv').
        \label{disorder_variance-0}
\end{equation}

In the framework of the strain-only description, the total elastic 
Hamiltonian is then given by Eq.~(\ref{f_expansion4}) for the pure
system, augmented by the leading disorder terms in Eq.~(\ref{H_disorder}):
\begin{eqnarray} 
 \label{disorder_Hamiltonian}
\hspace{-2mm}
H &=& \int d^d x
\left[  \frac{1}{2}B_z w_{zz}^2 
        + \lambda_{z\perp} w_{zz}w_{ii}
                + \frac{1}{2}\lambda w_{ii}^2
                     \right.
       + \mu w_{ij} w_{ij}\nonumber\\
&+& \left. \frac{K_1}{2} (\nabla_{\perp}^2 u_z)^2 
+ \frac{K_3}{2} (\partial_z^2 \uv_{\perp})^2 
        - {\mathbf \sigma}_{zi}(\xv) \varepsilon_{zi}(\xv)
        \right],
  \end{eqnarray}
where the effective strain tensor $\mm{w}$ is defined in
Eqs.~(\ref{effective_strain}).

Since the Hamiltonian (\ref{disorder_Hamiltonian}) contains random
stress terms linear in $\varepsilon_{iz}$, it is clear that the NRS
with $\rv = \xv$ is not the true ground state.  Rather the latter is
given by
\begin{equation}
\rv_0(\xv,\sigma) = \xv + \uv_0(\xv, \sigma),
\label{rv0}
\end{equation}
where $\uv_0(\xv)$ describes the deformation of the ground relative to
the ideal nematic reference state due to the presence of quenched
disorder (random stresses). The physical phonon field, which we denote
by $\delta \rv(\xv)$, is defined as the displacement from the true
ground state \cite{comment:natural}.  
It is related to the fictitious ``phonon field''
$\uv(\xv)$ by
\begin{eqnarray}
\rv(\xv) &=& \rv_0(\xv,\sigma)  + \delta \rv(\xv) 
 = \xv + \uv(\xv) ,\\
 \delta\rv(\xv) &=& \uv(\xv) - \uv_0(\xv,\sigma),
\end{eqnarray}

The harmonic thermal fluctuations around the disorder-dependent ground
state $\rv_0(\xv,\sigma)$ can be characterized by the disorder
averaged {\em thermal} correlation function of the physical phonon
field $\delta\rv(\xv)$,
\begin{eqnarray}
\mm{G}^{\rm T}(\xv-\vec{y}) 
= \overline{ \langle \delta\rv(\xv) \delta \rv(\vec{y}) \rangle}
= \overline{(\uv - \langle \uv \rangle)
(\uv - \langle \uv \rangle)},
\label{GT-def}
\end{eqnarray}
where we use angular brackets and over-bar to denote thermal and
disorder averages, respectively. Note that $\mm{G}^{\rm T}$ is a
tensor with components $G^{\rm T}_{ab}$.  On the other hand, the
quenched fluctuations, i.e., sample-to-sample variations of the ground
state relative to the ideal NRS can be characterized by the quenched
correlator
\begin{eqnarray}
\mm{G}^{\Delta}(\xv-\vec{y}) = \overline{ \uv_0(\xv, \sigma) \uv_0(\vec{y}, \sigma) }
=  \overline{ \langle \uv \rangle \langle \uv \rangle},
\end{eqnarray}
where $\mm{G}^{\Delta}$ is also a tensor with components
$G^{\Delta}_{ab}$.  As shown in Appendix \ref{App:Replica},
$\mm{G}^{T}$ and $\mm{G}^{\Delta}$ can be calculated using the replica
method \cite{comment:replica}.

Even though the choice of the ideal isotropic and nematic reference
states are rather arbitrary, the distortion field $\uv_0(\xv)$ has
important physical significance, as it is closely related to the local
nonaffine deformation field for a macroscopically strained elastomer.
To see this, let us consider applying a macroscopic deformation $\Lm$,
with $\Lm$ a constant matrix.  Because of the network heterogeneities,
the elastomer does not deform affinely as in the idealized homogeneous
(disorder-free) case.  The new ground state configuration subject to
the macroscopic strain can be parameterized by
\begin{eqnarray}
\rv_0(\xv, \sigma,\Lm) = \Lm \cdot \xv + \uv_0(\xv,\sigma,\Lm), 
\label{rv0-lm}
\end{eqnarray}
where we have shown explicitly the dependence of the ground state on
$\Lm$.  $\uv_0(\xv,\sigma,\Lm)$ can be found by minimizing the total
elastic energy Eq.~(\ref{disorder_Hamiltonian}) subject to the
constraint of macroscopic deformation $\Lm$.  The local nonaffine part
of the deformation field can then be characterized by:
\begin{eqnarray}
\vec{t}(\xv,\sigma,\Lm) &=& 
\rv_0(\xv, \sigma,\Lm) - \Lm \cdot \rv_0(\xv, \sigma)
\nonumber\\
&=&  \uv_0(\xv,\sigma,\Lm) - \Lm \cdot  \uv_0(\xv,\sigma),
\label{tv-def}
\end{eqnarray}
where we have used Eq.~(\ref{rv0}) and Eq.~(\ref{rv0-lm}) in the last
equality.  $\vec{t}(\xv,\sigma,\Lm) $ is generically nonzero for a
system with quenched disorder.  The correlation function of the
nonaffine deformation field $\vec{t}$ can also be calculated using the
replica method and is found \cite{Xing-Radz3} to be linearly
proportional to the quenched correlator $\mm{G}^{\Delta}$:
\begin{eqnarray}
\overline{\vec{t}(\xv,\sigma) \vec{t}(\vec{y},\sigma)} 
\propto \overline{\uv_0(\xv,\sigma) \uv_0(\vec{y},\sigma)} 
\equiv \mm{G}^{\Delta}(\xv- \vec{y}).
\end{eqnarray}
As we will demonstrate shortly, the {\em renormalized} quenched
correlator $\mm{G}_R^{\Delta}$ (and hence the nonaffinity correlation
function) will exhibit scaling that is distinct from that of the
thermal correlator $\mm{G}_R^{T}$.  Physically, this is a reflection
of the strong sensitivity of the nonaffine response of the ground
state to an external traction. The resulting macroscopic strain
deformation is a nonlinear function of the stress even for
an infinitesimal value of the stress.

\subsection{Replica trick and harmonic theory} 

The fields $\sigma_{zi}(\xv)$ are quenched random variables, i.e.,
their values are fixed for a particular sample by the network
heterogeneities, and in contrast to the phonon and nematic director
fields do not fluctuate on experimental time scales.  We make a
standard assumption that the system is self-averaging, namely, that
physical quantities for a typical system coincide with corresponding
disorder averaged quantities.

The disorder average can be performed using the standard replica
trick, which we review in Appendix \ref{App:Replica}.  As explained
there, for an arbitrary nonzero integer $n$, we define an
$n$-replicated Hamiltonian $H_n[\uv^1,\uv^2,\ldots,\uv^n]$ by
\begin{eqnarray}
\exp{\left(-H_n[\uv^1,\uv^2,\ldots,\uv^n]/T\right)} 
&=&\overline{\,\,\prod_{\alpha=1}^n
\exp{\left(-H[\uv^{\alpha}]/T\right)} \,\, }
\nonumber\\
&=& \overline{ \,\,\exp{\left(-\sum_{\alpha=1}^n
      H[\uv^{\alpha}]\right)} \,\, }.
\hspace{0.7cm}
\label{replicated}
\end{eqnarray}
To compute physical quantities, at the end of their calculation we
will analytically continue all replicated $n$-dependent quantities
from integer $n$ to positive real values $0<n<1$, and will take the
replica limit $n \rightarrow 0$.

To calculate the disorder average in Eq.~(\ref{replicated}), 
we use Gaussian, zero-mean field identity
\begin{equation}
\overline{e^{\sigma}} = e^{\frac{1}{2}\overline{\sigma^2}},
\end{equation}
together with Eqs.~(\ref{H_disorder},\ref{disorder_variance-0}), we find
\begin{eqnarray}
&&\overline{\exp\big[-\frac{1}{T}\sum_{\alpha}H_d[\uv^{\alpha}]\big]\,} =
\\ 
&&\exp\big[\frac{\Delta}{2\,T^2} 
\int d^d x \sum_{\alpha \beta}
\,\,(\partial_i u_z^{\alpha} + \partial_z u_i^{\alpha})
(\partial_i u_z^{\beta} + \partial_z u_i^{\beta})\big].\nonumber
\end{eqnarray}

Consequently, we find that the replicated Hamiltonian is given by
\begin{widetext}
\begin{eqnarray}
H_n[\uv^1,\uv^2,\ldots,\uv^n] &=& \int d^d x \sum_{\alpha} 
 \left[ \frac{B}{2} \,(\Tr \, \mm{w}^{\alpha})^2
 + C\, (\Tr \, \mm{w}^{\alpha}) \, \tilde{w}_{zz}^{\alpha} 
+ \frac{\mu_{\rm L}}{2}  \,(\tilde{w}_{zz}^{\alpha})^2
+ \mu \, \tilde{w}_{ij}^{\alpha} \,\tilde{w}_{ij}^{\alpha} \right.
		\nonumber\\
 &+&  \left.  \frac{1}{2} \, K_1
 (\nabla_{\perp}^2 u^{\alpha}_z)^2 
 + \frac{1}{2} K_3(\partial_z^2 \uv_{\perp}^{\alpha})^2 \right]
 -  \frac{2\Delta}{T}\,\int d^d x  \sum_{\alpha, \beta }
 \varepsilon^{\alpha}_{iz}  \varepsilon^{\beta}_{iz} 
         \label{replicated_Hamiltonian}
\end{eqnarray}
where $\alpha$ and $\beta$ are replica indices summed over $1, \ldots,
n$.

We first study the harmonic part of the replicated elastic Hamiltonian
Eq.~(\ref{replicated_Hamiltonian}), which is given by:
\begin{equation}
{\mathcal H}_{n}^{0} = \frac{1}{2} \,
	\sum_{\alpha} {\mathcal H}_0[\uv^{\alpha}] 
           -    \frac{2\Delta}{T}  \sum_{\alpha, \beta }
 \varepsilon^{\alpha}_{iz}  \varepsilon^{\beta}_{iz} ,
        \label{har_disorder}
\end{equation}
where ${\mathcal H}_0$ is defined in Eq.~(\ref{H_0}).  In Fourier
space, it can be written as
\begin{equation}
{\mathcal H}_{n}^0(\qv) 
 = \frac{1}{2} \, \sum_{\alpha\beta}\sum_{ab}
        \,\big[ \Gamma_{ab}(\qv)\,\delta_{\alpha\beta} 
        - J_{ab}(\qv) \big]\, 
        u^{\alpha}_a(\qv)\,u^{\beta}_b(-\qv),
        \label{har_disorder-2}
\end{equation}
\end{widetext}
where indices $a$ and $b$ are summed over all $d$ possible values
$x,y,\ldots,z$, $\Gamma_{ab}(\qv)$ is given by Eq.~(\ref{gamma_0}), and
$\mm{J}$ is a $d\times d$ matrix with components $J_{ab}$:
\begin{eqnarray}
J_{ij} &=& \frac{1}{T}\,\Delta \,q_z^2\,\delta_{ij},\\
J_{iz} &=& J_{zi} = \frac{1}{T}\,\Delta\,q_z\,q_i,\\
J_{zz} &=& \frac{1}{T}\,\Delta\,q_{\perp}^2.
\end{eqnarray}
In replica space one can think of $\mm{J}$ as either an $n\times n$
matrix with all elements equal to $1$ or simply as a scalar equal to
$1$.

The harmonic correlation functions (propagators) of the replicated
model can be easily calculated:
\begin{equation}
{G}_{ab}^{\alpha\beta}(\qv) =
        \frac{1}{V}\langle u^{\alpha}_a(\qv)
        \,u^{\beta}_b(-\qv) \rangle_0 
   =  G^{T}_{ab}(\qv) \, \delta_{\alpha\beta} 
  + G^{\Delta}_{ab}(\qv),  
        \label{replicated_propagators}
\end{equation}
where the thermal correlators $G_{ab}^T$ are defined by
Eq.~(\ref{G-Gamma-0}) and are given by
Eqs.~(\ref{G0zz},\ref{G0ij},\ref{G0zi}).  The harmonic thermal
correlators are linear in temperature and independent of random stress
variance $\Delta$.  The harmonic quenched correlators
$G^{\Delta}_{ab}$ are given by
\begin{equation}
G^{\Delta}_{ab} = T^{-1} \left(\mm{G}\,\mm{J}\,\mm{G} \right)_{ab},
\end{equation}
and are linear in the disorder variance $\Delta$ and independent of
$T$.  It is easy to check that in the $n\to 0$ limit
${G}_{ab}^{\alpha\beta}(\qv)$ is indeed the inverse matrix of the
kernel in Eq.~(\ref{har_disorder-2}):
\begin{equation}
\sum_{\beta} \sum_{b}{G}_{ab}^{\alpha\beta}\,
 \big[ \Gamma_{bc}(\qv)\,\delta_{\beta\gamma} 
        - J_{bc}(\qv) \big]
  =    T\, \delta_{ac}\,\delta_{\alpha\gamma}.
\end{equation}

Let us first look at the quenched fluctuations of the $u_z$ phonon
around the ideal nematic reference state in real space:
\begin{subequations}
\begin{eqnarray}
\hspace{-7mm}
\overline{\langle u_z(\rv)\rangle_0^2}    &=& 
  {G}^{\alpha\neq\beta}_{zz}(\rv=0)
  =  \int\frac{d^d q}{(2\,\pi)^d} 
{G}^{\Delta}_{zz}(\qv) 
  \nonumber\\
 &=&  \int\frac{d^d q}{(2\,\pi)^d} 
    \Delta \left[q_i G_{zz}(\qv) + q_z G_{zi}(\qv)\right]^2
 \label{u-quenched-1}\\
 &\approx&  \int \frac{d^d q}{(2\,\pi)^d} 
  \frac{\Delta\,q_{\perp}^2} {(\hat{\mu} \,q_z^2+K\,q_{\perp}^4)^2}
 \label{u-quenched-2}\\
  &\propto& \frac{\Delta}{\sqrt{\hat{\mu}K^3} } \,L^{(5-d)}, 
  \hspace{5mm} \mbox{for $d<5$},
 \label{u-quenched}
\end{eqnarray}
\end{subequations}
where in going from Eq.~(\ref{u-quenched-1}) to
Eq.~(\ref{u-quenched-2}), we have only kept the most (infra-)divergent
term, proportional to $G_{zz}^2$.  The result in
Eq.~(\ref{u-quenched}) shows that quenched fluctuations of $u_z$ field
diverge with the system size for $d<5$, suggesting a break down of the
harmonic theory at long length-scale.  Therefore our model of a
heterogeneous nematic elastomer, Eq.~(\ref{replicated_Hamiltonian})
has an upper-critical dimension $d_c^\Delta=5$.  This should be
contrasted with a homogeneous thermal elastomer that we studied in
Sec.~\ref{Sec:Thermal-fluct}, where the critical dimension is $d_c=3$.

On the other hand, {\em thermal} fluctuations of the phonon $u_z$
about the ground state for a specific realization of quenched disorder
are given by:
\begin{eqnarray}
\overline{\langle u_z^2 \rangle_0}
 - \overline{\langle u_z \rangle_0^2}
&=&  \int\frac{d^d q}{(2\,\pi)^d} 
 ({G}^{\alpha\alpha}_{zz}(\qv)
 - {G}^{\alpha\beta}_{zz}(\qv))
\nonumber\\
&\approx& \int\frac{d^d q}{(2\,\pi)^d} 
\frac{T}{\hat{\mu}\,q_z^2 + K\,q_{\perp}^4},
\end{eqnarray}
and as found earlier are far weaker, finite for $d > 3$.  However, as
we will see shortly, thermal fluctuations of $u_z$ around the
disorder-renormalized state are in fact finite at and even below three
dimension.

Similarly, we can show that the most divergent part of the quenched
fluctuations of $\uv_{\perp}$ about the ideal NRS are given by
\begin{eqnarray}
&&\overline{\langle \uv_{\perp}(\rv)\rangle_0^2}   =
  {G}^{\alpha\neq\beta}_{ii}(\rv=0)
  =  \int\frac{d^d q}{(2\,\pi)^d} 
{G}^{\Delta}_{ii}(\qv) 
  \nonumber\\
 &\approx&  \int \frac{d^d q}{(2\,\pi)^d} 
  \frac{\Delta\,q_{\perp}^2}{T^2} 
  [G_{\rm L}^2 + (d-1) G_{\rm T}^2]
 \nonumber\\
 &=&  \Delta \int \frac{d^d q}{(2\,\pi)^d} 
 \left[   \frac{q_z^2} {(\tilde{\mu} \,q_{\perp}^2
  + K_3\,q_{z}^4)^2} 
  +  \frac{(d-1) q_z^2} {( \mu \,q_{\perp}^2
  + K_3\,q_{z}^4)^2} \right],
   \nonumber\\
 \label{up-quenched}
\end{eqnarray}
where, again, we have only kept the most infra-divergent terms.  One
can readily see that this integral diverges with the system size only
when $d<7/2$, similarly to a randomly pinned columnar liquid crystal
and vortex lattice in a disordered magnetic superconductor
\cite{RT:discotic,RT:MSC-2}.  As in the case of thermal fluctuations
of a homogeneous elastomer, we again find that the fluctuations of
$u_z$ phonon field are qualitatively more important than those of
$\uv_{\perp}$.  The difference in critical dimensions for these two
phonon fields justifies our model in Eq.~(\ref{disorder_Hamiltonian}),
where we only kept the more relevant smectic (as opposed to columnar)
nonlinearities.

\subsection{Renormalization group analysis}

We will therefore focus on the larger quenched phonon $u_z$
fluctuations. To simplify the analysis, again we drop the dangerously
irrelevant term $K_3 (\partial_z^2 \uv_{\perp})^2$ in the elastic
Hamiltonian, \rfs{disorder_Hamiltonian}.  Furthermore, we will neglect
the coupling between the quenched random stress field $\sigma_{zi}$
and the $\uv_{\perp}$ phonon fields, as we have just seen that
fluctuations of the latter are subdominant to those of the $u_z$
field.  These simplifications amount to making the following
replacements in the elastic Hamiltonian
(\ref{replicated_Hamiltonian}):
\begin{subequations}
\label{irrelevant-ignore}
\begin{eqnarray}
  K_3 &\longrightarrow& 0, \\
\varepsilon^{\alpha}_{iz} \varepsilon^{\beta}_{iz} 
 &\longrightarrow& \frac{1}{4}
 (\nabla_{\perp}u_z^{\alpha})
 \cdot (\nabla_{\perp}u_z^{\beta}). 
\end{eqnarray}
\end{subequations}

Similar to Sec.~\ref{Sec:Thermal-fluct}, we rescale elastic constants
$B$, $C$, $\mu_{\rm L}$, $\mu$ and $K\equiv K_1$ by temperature $T$,
and at the same time also scale the disorder variance $\Delta$ by
$T^2$, such that $T$ does not appear in the formalism.  We shall
restore the $T$ dependence after the technical RG analysis.
 
To study the effects of the network heterogeneity and the elastic
nonlinearities beyond harmonic approximation, we perform a
momentum-shell RG analysis.  We use the same cylindrical ultraviolet
cutoff scheme $(Q_{\perp} = Q, Q_{\parallel}=\infty)$ and rescaling
transformations Eqs.~(\ref{RGrescaling}) as in
Sec.~\ref{Sec:Thermal-fluct}.  After ignoring the irrelevant terms
according to Eqs.~(\ref{irrelevant-ignore}), the harmonic correlators
of the replicated elastic Hamiltonian,
Eq.~(\ref{replicated_propagators}) reduce to:
\begin{subequations}
\label{replicated_propagators-2}
\begin{eqnarray}
{G}^{\alpha \beta}_{zz}(\qv) &=& G_{zz}(\qv) 
        \left( \delta_{\alpha \beta}
        + \Delta \, q_{\perp}^2 \, G_{zz}(\qv) \right),
        \label{Gzz_disorder}\\
{G}^{\alpha \beta}_{zi}(\qv) &=& G_{zi}(\qv) 
\left( \delta_{\alpha \beta} + \Delta \,
         q_{\perp}^2 \, G_{zz}(\qv) \right),\\
{G}^{\alpha \beta}_{ij}(\qv) &=& 
        G_{ij}(\qv) \, \delta_{\alpha \beta} 
        + \Delta \, q_{\perp}^2 \, G_{zi}(\qv) \, G_{zj}(\qv),
\hspace{1cm}
\end{eqnarray}
\end{subequations}
with $G_{ab}(\qv)$ the harmonic thermal correlators given by
Eqs.~(\ref{simplified_propagator}).

Detailed calculations of the diagrammatic corrections to various model
parameters $\{B_z, \lambda_{z\perp},\lambda,\mu,K, \Delta \}$ are
quite involved and are relegated to Appendix \ref{App:RG-disorder}.
Their flow equations are listed in Eqs.~(\ref{RGcorrectionsdisorder}).
Similarly to the case of thermal fluctuations, we find that the RG
flows of these parameters are controlled by four dimensionless
coupling constants, defined as follows:
\begin{subequations}
\label{coupling_replica}
\begin{eqnarray}
\mathsf{g}_{\rm L} &=& \frac{\psi_d \,\Delta\,\mu_{\rm L}}
{ \sqrt{K^5\,\hat{\mu}}}, \hspace{4mm}
\mathsf{g}_{\perp} = \frac{\psi_d \,\Delta\,\mu}
{ \sqrt{K^5\,\hat{\mu}}} , \\
x &=& \frac{\mu_{\rm L}}{B},\hspace{12mm}
y = \frac{C}{\sqrt{B \mu_{\rm L}}},
\end{eqnarray}\end{subequations}
where $\psi_d$ is defined in Eq.~(\ref{psid_def}).  In contrast with
the case of thermal fluctuations discussed in
Sec.~\ref{Sec:Thermal-fluct} (Eqs.~(\ref{def_couplings})), however,
here the two coupling constants $\mathsf{g}_{\rm L}$ and
$\mathsf{g}_{\perp}$ are proportional to the disorder variance,
$\Delta$, instead of temperature, $T$.  The flow equations of all
elastic constants are given by:
\begin{subequations}
\label{flow_2}
\begin{eqnarray}
\frac{d B}{d\,l} &=& (d+3-3\, \omega - \eta_B)\, B ,\\
\frac{d C}{d\,l} &=& (d+3-3 \,\omega- \eta_{L})\, C ,\\
\frac{d \mu_{\rm L}}{d\,l} &=&  (d+3-3 \,\omega- \eta_{\rm L})\, \mu_{\rm L},\\
\frac{d \mu}{d\,l} &=& (d+3-3\, \omega - \eta_{\perp})\, \mu,\\
\frac{d K}{d\,l} &=& (d-1-\omega+\eta_K)\, K, \\
\frac{d \Delta}{d\,l} &=& (d+1-\omega+\eta_{\Delta})\,\Delta,
\end{eqnarray}
\end{subequations}
where anomalous $\eta$ exponents are functions of dimensionless
coupling constants $\mathsf{g}_{\rm L}$, $\mathsf{g}_{\perp}$, $x$,
and $y$, given in Eqs.~(\ref{eta_exponents-3},\ref{eta_exponents-4})
of Appendix \ref{App:RG-disorder}.

As discussed in the context of homogeneous elastomers, the bare values
of two dimensionless ratios $x$ and $y$ are much smaller than one.
Furthermore, as we shall show in Appendix \ref{App:RG-disorder}, they
flow to zero {\em exponentially} as long as $d<5$.  Therefore to a
good approximation, which becomes asymptotically exact, we can set
them to zero in all flow equations, thereby considerably simplifying
calculations.  In this limit, the $\eta$ exponents reduce to:
\begin{subequations}
\label{def_eta}
\begin{eqnarray}
\eta_B &=& \frac{3}{8} \mathsf{g}_{\rm L} y^2
\longrightarrow 0, \label{def_etaB}\\
\eta_{L} &=& \frac{3}{8}\, \mathsf{g}_{\rm L},\\
\eta_{\perp} &=& \frac{1}{16} \, \mathsf{g}_{\perp},\\
\eta_K &=& \frac{1}{16} \, \mathsf{g}_{\rm L} + \frac{35}{32} \,\mathsf{g}_{\perp},\\
%
\eta_{\Delta} &=& \frac{1}{32} \, \mathsf{g}_{\rm L} 
	+ \frac{3}{64}\,\mathsf{g}_{\perp}. 
%
\end{eqnarray}
\end{subequations}

The RG flows of the dimensionless coupling constants, $\mathsf{g}_{\rm
  L}$, $\mathsf{g}_{\perp}$, $x$ and $y$ are also calculated in
Appendix \ref{App:RG-disorder}, and in the limit that both $x$ and $y$
approach zero reduce to
\begin{subequations}
\label{flow_gs}
\begin{eqnarray}
\frac{d \,\mathsf{g}_{\rm L}}{d \,l} &=& \epsilon \,{\mathsf{g}_{\rm L}} 
 - \frac{5\, {\mathsf{g}_{\rm L}}\left( 4\,{{\mathsf{g}_{\rm L}}}^2 
 + 44\,{\mathsf{g}_{\rm L}}{\mathsf{g}_{\perp}} + 
      51\,{{\mathsf{g}_{\perp}}}^2 \right) }{32
    \left( 2\,{\mathsf{g}_{\rm L}}
  + 3\,{\mathsf{g}_{\perp}} \right) },
  \nonumber\\
  \label{flow_gL3}\\
 \frac{d \, \mathsf{g}_{\perp}}{d \, l} &=&
  \epsilon \,\mathsf{g}_{\perp} 
 - \frac{ {\mathsf{g}_{\perp}}\left( -4\,{{\mathsf{g}_{\rm L}}}^2
  + 188\,{\mathsf{g}_{\rm L}}{\mathsf{g}_{\perp}} + 
        261\,{{\mathsf{g}_{\perp}}}^2 \right) }{32
    \left( 2\,{\mathsf{g}_{\rm L}} + 3\,{\mathsf{g}_{\perp}} \right) },
  \qquad
    \nonumber\\
    \label{flow_gperp3}
\end{eqnarray} \end{subequations}
where $\epsilon = 5 - d$.   

\begin{figure}[!htbp]
\begin{center}
  \includegraphics[width=8cm,height=3.5cm]{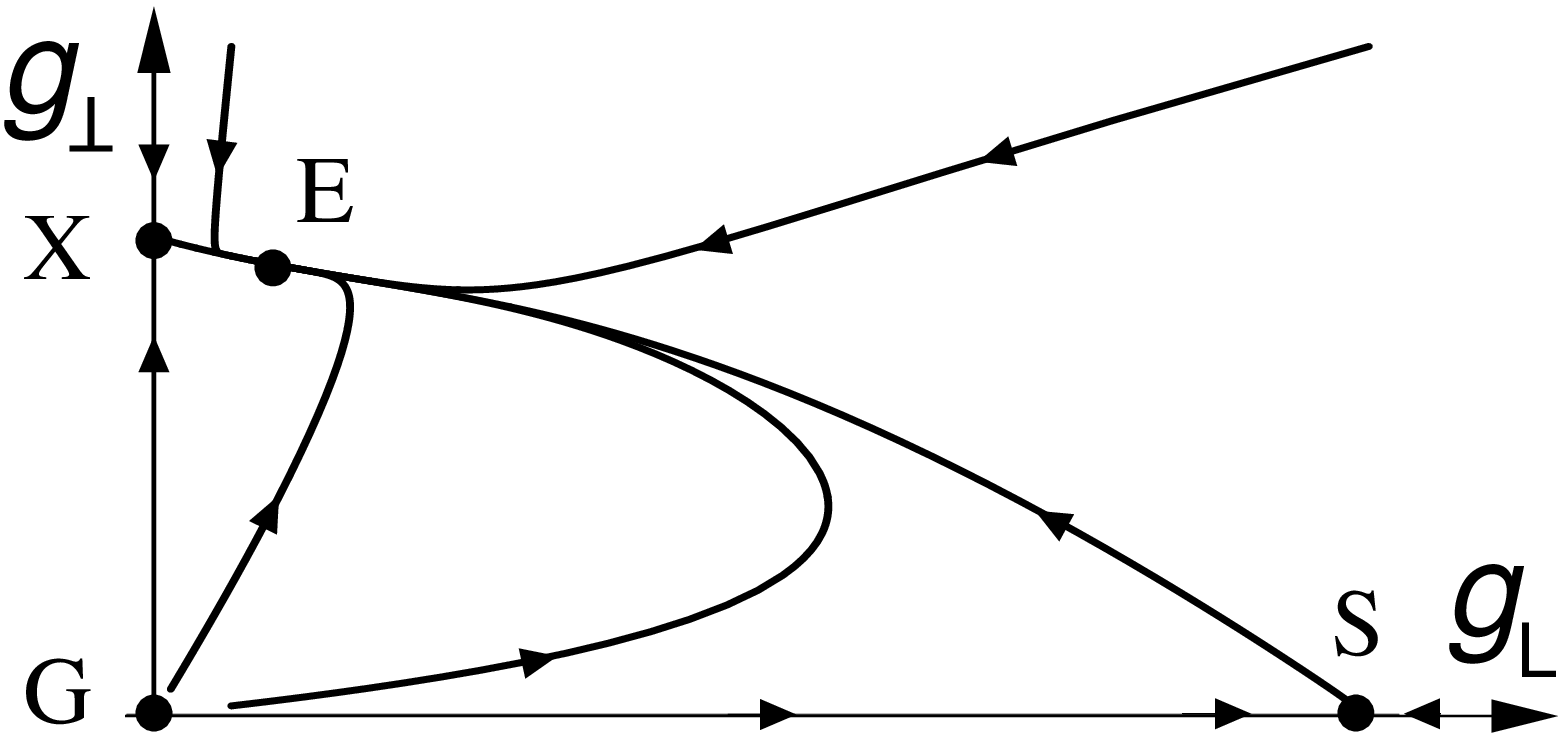}
  \caption{Renormalization group flow diagram for dimensionless
    couplings $\mathsf{g}_{\rm L}$ and $\mathsf{g}_{\perp}$ of a
    heterogeneous nematic elastomer. Gaussian fixed point G of a
    harmonic elastomer is unstable to elastic nonlinearities, flowing
    to a globally stable fixed point E that controls long-scale
    properties of a heterogeneous nematic elastomer. The fixed point S
    at $\mathsf{g}_{\perp}$ is identical to that of a randomly pinned
    smectic studied by Radzihovsky and Toner
    \cite{RTaerogel1,RTaerogel2}.}
\label{flowdiagram}
\end{center}    
\end{figure}

Below five dimensions, $\epsilon > 0$, Eqs.~(\ref{flow_gs}) admit four
fixed points, Gaussian (G), Smectic (S), X, and Elastomer (E), which
we list in Table \ref{table-fixpoints}.  Also shown in Table
\ref{table-fixpoints} are the $\eta$ exponents for all elastic
constants, as defined in Eqs.~(\ref{def_eta}).  The flow pattern of
$\mathsf{g}_{\rm L}$ and $\mathsf{g}_{\perp}$ under a renormalization
group transformation for $d<5$ is shown in Fig.~\ref{flowdiagram}.

It is particularly interesting to note that at fixed point S, where
$\mathsf{g}_{\perp}$ vanishes, various $\eta$ exponents are identical
to those of a smectic liquid crystal confined in random environment of
e.g., an aerogel matrix, discovered by Radzihovsky and Toner
\cite{RTaerogel1,RTaerogel2}.  This is not merely a coincidence.  It
is clear from Eq.~(\ref{disorder_Hamiltonian}) that if the transverse
shear modulus $\mu=0$ vanishes (that is, $\mathsf{g}_{\perp}$), then
$\uv_{\perp}$ phonon fields can be integrated out completely.  The
resulting effective model is identical to that of a smectic liquid
crystal with one-dimensional phonon field $u_z$ and a shifted
compressional modulus, as well as a random tilting field with variance
$\Delta$.  This is precisely the model extensively studied by
Radzihovsky and Toner \cite{RTaerogel1,RTaerogel2,RTaerogel6}, and a
recovery of their randomly-pinned smectic flows and exponents here is
a nontrivial check on our calculations.
  
We are not aware of any physical system that is described by the fixed
point X, characterized by $\mathsf{g}_{\rm L}=0$.  We note that
$\eta_B$ vanishes at all four fixed points.  Consequently the bulk
modulus $B$ does not acquire any anomalous dimension, with the
underlying physical reason for this already discussed in
Sec. \ref{Sec:Thermal-fluct}.

As shown in Fig.~\ref{flowdiagram} only the fixed point E is stable in
both $\mathsf{g}_{\rm L}$ and $\mathsf{g}_{\perp}$ directions.  It
therefore controls the long length-scale elasticity of a heterogeneous
nematic elastomer, which, based on this fixed point shall be analyzed
in detail in the next section.

\begin{table}[!hbt]
\begin{center}
\begin{tabular}{|c|c|c|c|c|c|c|c|c|}
\hline\hline \,\, F.P. \,\, & \,\,$\mathsf{g}_{\rm L}^*$ \,\, & \,\,$\mathsf{g}_{\perp}^*$ \,\,
&  \,\,$\eta_B$  \,\,& \,\, $\eta_{\rm L} $  \,\,
& \,\,$\eta_{\perp}$ \,\,& \,\,$\eta_K$ \,\,&$ \,\,\eta_{\Delta}$ \,\,\\
\hline G & 0 & 0 & 0 & 0&0&0&0 \\
S&$ \frac{16 \epsilon}{5}$&0& $0$ &
 $\frac{6 \epsilon}{5}$&0& $\frac{\epsilon}{5}$
&$\frac{\epsilon}{10}$\\
X & 0 & $\frac{32 \epsilon}{87}$ &0
&0&$\frac{2 \epsilon}{87}$
&$\frac{35 \epsilon}{87}$
&$\frac{\epsilon}{58}$\\
E &$\frac{16 \epsilon}{263}$& 
$\frac{96 \epsilon}{263}$
 & 0 & $\frac{6 \epsilon}{263}$
&$\frac{6 \epsilon}{263}$
&$\frac{106 \epsilon}{263}$
&$\frac{5 \epsilon}{263}$\\
&&&&&&&\\
\hline
\end{tabular}
\caption{Fixed point couplings and $\eta$ exponents for heterogeneous 
nematic elastomer. $x^*=y^*=0$ at each of these fixed points.}
\label{table-fixpoints}
\end{center}
\end{table}

\section{Anomalous elasticity of {\em heterogeneous} nematic
  elastomers}
\label{Sec:ZetoT}

As we discussed in previous section (and is true more generally),
fluctuations associated with local network heterogeneity are dominant
over thermal fluctuations.  This is also supported by the difference
between the critical dimensions of the homogeneous and heterogeneous
nematic elastomers.  We further note that the coupling constants
$\mathsf{g}_{\rm L}$ and $\mathsf{g}_{\perp}$ defined in
Eqs.~(\ref{coupling_replica}) depend on quenched disorder $\Delta$ but
not on temperature $T$.  On the other hand the thermal coupling
constants $g_{\rm T}$ and $g_{\perp}$, whose flow we studied in
Sec. \ref{Sec:Thermal-fluct} for homogeneous case, are less relevant.
In fact we shall see below that the fixed points we have identified in
the preceding section are finite-disorder, zero-temperature fixed
points.  Qualitatively speaking, the universal, long-scale elastic
properties of a heterogeneous nematic elastomer are determined by its
ground state, which sensitively depends on the particular realization
of disorders and external traction, i.e., macroscopic strain
deformation.  Thermal fluctuations around the ground state remain
finite and are qualitatively unimportant.  Similar scenario also
appears in the equilibrium physics of random field Ising model and
smectic liquid crystal confined in random geometry.

\subsection{Zero-temperature fixed point}
To illustrate the physics of the new fixed point E that we found in
the preceding section, let us first summarize the RG analysis for the
disordered model in a slightly different notations.  Let us first
restore the temperature dependence in the replicated Hamiltonian
Eq.~(\ref{replicated_Hamiltonian}) by some appropriate rescaling
transformation:
\begin{widetext}
\begin{eqnarray}
\frac{1}{T}{\mathcal H}_n[\uv^1\ldots\uv^n] &=& 
\sum_{\alpha} \left[ \frac{1}{2}B \,(\Tr \, \mm{w})^2
 + C\, (\Tr \, \mm{w}^{\alpha}) \, \tilde{w}_{zz}^{\alpha} 
+ \frac{1}{2}  \mu_{\rm L}\,\tilde{w}_{zz}^{\alpha}\tilde{w}_{zz}^{\alpha} \right.
\nonumber\\
&+&  \left. \mu \, \tilde{w}_{ij}^{\alpha} \,\tilde{w}_{ij}^{\alpha}
+\frac{1}{2}K\,(\nabla_{\perp}^2 u_z^{\alpha})^2 
- S_{ia} \partial_a u_i \right]
-\Delta \,\sum_{\alpha\beta} (\partial_i u_z^{\alpha})^2 
\label{replicated_Hamiltonian-1} \\
&\rightarrow& \frac{1}{T} \sum_{\alpha}
\left[ \frac{1}{2} \chi_B \,(\Tr \, \mm{w})^2
 + \chi_C \, (\Tr \, \mm{w}^{\alpha}) \, \tilde{w}_{zz}^{\alpha} 
+ \frac{1}{2}  \,\tilde{w}_{zz}^{\alpha}\tilde{w}_{zz}^{\alpha} \right.
\nonumber\\
&+&  \left. \nu \, \tilde{w}_{ij}^{\alpha} \,\tilde{w}_{ij}^{\alpha}
+\frac{1}{2}\,(\nabla_{\perp}^2 u_z^{\alpha})^2 
- \tilde{S}_{ia} \partial_a u_i \right]
- \frac{\tilde{\Delta}}{T^2} \sum_{\alpha\beta} (\partial_i u_z^{\alpha})^2 , 
\label{replicated_Hamiltonian-2}
\end{eqnarray}
\end{widetext}
where, to facilitate later discussion on renormalized stress-strain
relation, we have also introduced an external nominal stress $S_{ia}$
coupled linearly to the deformation gradient, which we first discussed
in Sec.~\ref{Sec:Elast-fluct} (c.f. Eq.~(\ref{S-lambda}) ).  The
relations between the two set of coefficients appearing in
Eq.~(\ref{replicated_Hamiltonian-1}) and
Eq.~(\ref{replicated_Hamiltonian-2}) are shown in Table
\ref{Coefficients-mapping}.
\begin{center}
\begin{table}[!hbt]
\vspace{5mm}
\begin{tabular}{|c|c|c|c|c|c|c|c|}
\hline\hline Eq.~(\ref{replicated_Hamiltonian-1}) 
& $B$ & $C$ & $\mu_L$ & $\mu$ & $K$ & $\Delta$ & 
$S_{i\alpha}$
\\\hline 
Eq.~(\ref{replicated_Hamiltonian-2}) 
& $T^{-1} \chi_B$ & $T^{-1} \chi_C$ & $T^{-1}$ &
 $T^{-1} \nu$ & $T^{-1}$ & $T^{-2}\tilde{\Delta}$ 
 & $T^{-1} \tilde{S}_{i\alpha}$
\\\hline
\end{tabular}
\caption{Correspondence between the coefficients in
Eq.~(\ref{replicated_Hamiltonian-1}) and
Eq.~(\ref{replicated_Hamiltonian-2}). }
\label{Coefficients-mapping}
\end{table}
\end{center}

The form of $H_n$ in \rfs{replicated_Hamiltonian-2} leads to form of
model parameters that are slightly more convenient for a study of
consequences of the flow equations.  Firstly, we observe that since
both $\mu_L$ and $K$ are mapped into $T^{-1}$ in
Eq.~(\ref{replicated_Hamiltonian-2}), in order to preserve the
Hamiltonian's functional form under the RG flow (a convenience for
later analysis), the anisotropy scaling exponents $\omega$ in
Eqs.~(\ref{flow_2}) must be chosen such that $\mu_{\rm L}$ and $K$
flow exactly the same.  This condition gives
\begin{equation}
\omega = 2 - \frac{1}{2} \left(
\eta_{\rm L} + \eta_K \right). 
\label{omega-value}
\end{equation}
We shall see that this $\omega$ is precisely the physical anisotropy
exponent which appears in all renormalized correlation
functions. Since $\eta_L$ and $\eta_K$ are positive we observe that at
long-scales the $2:1$ anisotropy between $\xv_\perp$ and $z$ of the
harmonic theory is {\em reduced} below $2$ for a fully nonlinear
heterogeneous nematic elastomer.  Using Eqs.~(\ref{flow_2}) and Table
\ref{Coefficients-mapping} we can readily derive the flow equations
for the new parameters of the rescaled Hamiltonian in
Eq.~(\ref{replicated_Hamiltonian-2}):
\begin{subequations}
\label{flow_3}
\begin{eqnarray}
\frac{dT}{dl} &=& (3-d -\frac{1}{2}(\eta_{\rm L} + 3\eta_K)) T = y_T T ,
\label{T-flow}\\
\frac{d \chi_B}{dl} &=& (\eta_{\rm L} - \eta_B) \chi_B
= \eta_L \chi_B,\\
\frac{d \chi_C}{dl} &=& 0,\hspace{5mm}
\frac{d \nu}{d l} = 0,\\
\frac{d\tilde{\Delta}}{d l} &=& 
 (5-d - \frac{1}{2} \eta_L - \frac{5}{2} \eta_K + \eta_{\Delta})
 \tilde{\Delta}= 0,\hspace{8mm}
\end{eqnarray}
In above we have used the property of $\eta$ exponents deduced in the
previous section, such that $\eta_B\rightarrow 0$, and the fact that
at the critical point $E$ the following relation holds:
\begin{eqnarray}
(5-d - \frac{1}{2} \eta_L - \frac{5}{2} \eta_K + \eta_{\Delta})= 0, 
\label{WardExp}
\end{eqnarray}
\end{subequations}
guaranteed by the underlying rotational invariance and the form of the
nonlinear strain tensor $\mm{w}$, \rfs{effective_strain}.  The flow of
the external traction $S_{i\alpha}$, which we have not derived, can be
inferred from other flows using equilibrium fluctuation-dissipation
relation, as we shall shortly demonstrate.

From the flow equations (\ref{flow_3}) we can draw several important
conclusions about the long wavelength physics of the elastic
Hamiltonian Eq.~(\ref{replicated_Hamiltonian-2}).  
\begin{enumerate}
\item The crossover exponent $y_T$ for temperature is given by
\begin{eqnarray}
y_T =  3-d -\frac{1}{2}(\eta_{\rm L} + 3\eta_K).  
\label{y-T}
\end{eqnarray}
Since both $\eta_L$ and $\eta_K$ are positive for $d<5$, in three
dimensions, $y_T$ is always negative.  Therefore $T$ flows to zero as
$l \rightarrow \infty$, and thermal fluctuations are irrelevant at
long length-scales in physical 3d elastomers.
\item $\chi_B$ flows to infinity, indicating that in the thermodynamic
  limit the bulk mode freezes out at long length-scales.  For the same
  reason the cross coupling term $\chi_C$ is irrelevant at
  long-scales.
\item $\nu = \mu/\mu_L$ flows to a constant value as $l
  \rightarrow \infty$.  Thus a nematic elastomer exhibits universal
  Poisson ratios at long length-scales.
\end{enumerate}

With $T$ flowing to zero and disorder variance $\tilde{\Delta}$
flowing to a constant, the long wavelength physics is controlled the
zero-temperature, finite-disorder fixed point E.  That is, long-scale
properties of heterogeneous nematic elastomers are dominated by the
ground state energy for a given random initial stress $\sigma(\xv)$, a
nontrivial problem due to the elastic nonlinearities of the model.  In
the scaling regime, $\nu$ and $\tilde{\Delta}$ flow to constants,
while $\chi_B$ and $\chi_C$ play no role since the bulk mode is frozen
out.  Consequently, all physical quantities become functions of only
two parameters temperature $T$ and the external stress $S_{ia}$.

\subsection{Correlation functions}

We now proceed to combine RG flow equations with a matching scale
analysis to deduce long-scale properties of a nematic elastomer
through the behavior of renormalized correlation functions. To this
end we look at the RG transformation of the renormalized (i.e., with
all effects of fluctuations and disorders included) correlation
functions.  For simplicity, we shall consider the zero external stress
case and take $G^{\alpha\beta}_{zz}(\qv)$ as an example.  Using
Eq.~(\ref{RG_correlation1}) we have
\begin{equation}
G_{zz}^{\alpha\beta}(\qv_{\perp}, q_z, T) = e^{(d+3-\omega)l}
G_{zz}^{\alpha\beta} (\qv \,e^l, q_z\,e^{\omega l},T e^{y_T\,l}),
\label{G-scaling}
\end{equation}
where the exponent $y_T$ is given by Eq.~(\ref{y-T}) and the
anisotropy exponent $\omega$ fixed by Eq.~(\ref{omega-value}).  Using
the results in Appendix \ref{App:Replica}, we can separate the
replicated correlation into a quenched part $G^{\Delta}$ and a thermal
part $G^T$:
\begin{eqnarray}
G^{\alpha\beta}_{ab}(\qv) &=& 
	G^{\Delta}_{ab}(\qv)  	
+ G^T_{ab}(\qv) \,\delta_{\alpha\beta} 
	\nonumber\\
 &=&  \overline{\langle u_a \rangle \langle u_b \rangle}  + 	
 \left[ \, \overline{\langle u_a u_b\rangle} 
   -    \overline{\langle u_a \rangle \langle u_b \rangle} \, \right]
	\delta_{\alpha\beta}.\hspace{8mm} \label{G-TDelta}
\end{eqnarray} 
The thermal correlator is expected to be proportional to temperature
$T$,
with all other $T$ dependence
subdominant due to the irrelevance of $T$ at the fixed point E. The
quenched correlator $G^{\Delta}$ is expected to be {\em independent}
of temperature, and as we have discussed in
Sec.~\ref{Sec:Disorder-fluct}, is directly related to the nonaffinity
correlation function.

Combining Eq.~(\ref{G-scaling}) with Eq.~(\ref{G-TDelta}) and
proportionality of $G^T$ with $T$
we obtain
\begin{eqnarray}
G^{\Delta}_{zz}(\qv_{\perp}, q_z) &=& e^{(d+3-\omega)l} 
G^{\Delta}_{zz}(\qv_{\perp}\,e^l, q_z\,e^{\omega \,l}) ,
\nonumber\\
\label{GDelta-scaling}\\
G^{T}_{zz}(\qv_{\perp}, q_z, T) &=& e^{(d+3-\omega+y_T)l} 
G^{T}_{zz}(\qv_{\perp}\,e^l, q_z\,e^{\omega \,l},T)
\nonumber\\
&=& e^{(4 - \eta_K)l} 
G^{T}_{zz}(\qv_{\perp}\,e^l, q_z\,e^{\omega \,l},T), 
\label{GT-scaling}
\nonumber\\
\end{eqnarray}
where we used Eq.~(\ref{omega-value}) and Eq.~(\ref{T-flow}).  From
above, it is already clear that it is the dangerously irrelevant
temperature, $T$ that leads to different scaling behaviors of $G^T$
and $G^{\Delta}$.

Now by taking $e^l = Q/q_{\perp}$ in Eq.~(\ref{GT-scaling}) we
find
\begin{eqnarray}
G^{T}_{zz}(\qv_{\perp}, q_z) &=&
q_{\perp}^{-4+\eta_K} T\Phi_{z}^T\left(
\frac{q_z}{q_{\perp}^{\omega}} \right).  
\end{eqnarray}
where 
\begin{equation}
\Phi_{z}^T(x) = T^{-1}G^T_{zz}(1,x,T),
\end{equation}
and for simplicity of notation we have set $Q=1$ (used units of the
uv-cutoff).  In the limit $q_{z}\rightarrow 0$, $G^{T}_{zz}$ must be a
function of $q_{\perp}$ only.  Therefore we have
\begin{equation}
G^{T}_{zz}(\qv_{\perp}, q_z = 0) =
q_{\perp}^{-4+\eta_K} T\Phi^T_z\left( 0 \right)  
\propto q_{\perp}^{-(4-\eta_k)}. 
\end{equation}
Likewise, in the limit $q_{\perp} \rightarrow 0$, $G^{T}_{zz}$ must be
a function of $q_{z}$ only.  This requires that as $x \rightarrow
\infty$,
\begin{eqnarray}
\Phi^T_z(x)\sim x^{-\frac{4-\eta_k}{\omega}},
\end{eqnarray}
which leads to 
\begin{eqnarray}
G^{T}_{zz}(\qv_{\perp} =0, q_z )\propto
q_z^{-\frac{4-\eta_k}{\omega}}.  \label{GT-z-scaling}
\end{eqnarray}
 
We can use the renormalized correlation function to calculate the real
space fluctuations of the $u_z$ phonon:
\begin{eqnarray}
\langle (u_z - \langle u_z \rangle)^2 \rangle  &= &
\int \frac{dq_z d^{d-1} {q}_{\perp}}{(2\pi)^d} 
G^{T}_{zz}(\qv_{\perp}, q_z) \nonumber\\
&=& T\int \frac{dq_z d^{d-1} {q}_{\perp}}{(2\pi)^d} 
q_{\perp}^{-4+\eta_K} \Phi^T_z \left(
\frac{q_z}{q_{\perp}^{\omega}} \right).\hspace{1cm}
\label{uz-RG-theraml}
\end{eqnarray} 
To determine whether thermal fluctuation are divergent in a
macroscopic limit, power-counting is sufficient.  To this end we
observe that the integral is dominated by the region where $q_z \sim
q_{\perp}^{\omega}$, which gives
\begin{eqnarray}
&&\langle (u_z - \langle u_z \rangle)^2 \rangle  \sim q_{\perp}^{\delta},
\\
&& \delta = (d-3) + \frac{1}{2} (\eta_K - \eta_L) = \frac{100}{263} >0, 
\,\,\,\,\mbox{for}\,\,\, d = 3.\nonumber
\end{eqnarray}
Therefore thermal fluctuations are indeed finite in three dimensions
for the renormalized heterogeneous elastomer, even though they are
divergent with system size for a homogeneous one.

Similarly, we can deduce the renormalized quenched correlator
$G^{\Delta}_{zz}$.  Setting $e^l = q_{\perp}^{-1}$ in
Eq.~(\ref{GDelta-scaling}) and using the Ward exponent identity,
\rfs{WardExp}, we find
\begin{eqnarray}
G^{\Delta}_{zz}(\qv_{\perp},q_z) &=& q_{\perp}^{-(6-2\eta_K+\eta_{\Delta})}
\Phi^{\Delta}_z (\frac{q_z}{q_{\perp}^{\omega}}).   
\nonumber\\
&\rightarrow& \left\{ \begin{array}{ll}
q_{\perp}^{-(6-2\eta_K+\eta_{\Delta})}, & q_z \ll q_{\perp}^{\omega}\\
\vspace{3mm}
q_{z}^{-\frac{1}{\omega}(6-2\eta_K+\eta_{\Delta})}, &
 q_z \gg q_{\perp}^{\omega}.\end{array} \right.\hspace{8mm}
\end{eqnarray}
As argued above, the nonaffinity correlation function
$\overline{{t}_z(\qv){t}_z(-\qv)}$ (defined in Eq.~(\ref{tv-def})) is
expected to exhibit the same scaling behavior.

\subsection{Breakdown of linear response theory: 
nonlinear stress-strain relation}

The effective renormalized shear (and other elastic) moduli can be
extracted from the phonon renormalized correlation functions.  To this
end, we introduce a uniaxial traction $S \ppz u_z$ in the
elastic Hamiltonian Eq.~(\ref{replicated_Hamiltonian}).  It can then
be shown that, for an incompressible and uniaxial solid, the
renormalized longitudinal shear modulus $\mu_L$ is related to the
renormalized correlation function $G^T_{zz}$ via the following
Kubo-like formula:
\begin{equation}
\frac{1}{\mu_{L}(S)} = T^{-1} \lim_{\qv \rightarrow 0} 
q_z^2 \,G^T_{zz}(\qv,S). \label{Kubo-formula}
\end{equation}

Using Eq.~(\ref{GT-z-scaling}) and Eq.~(\ref{omega-value}), we find
that the linear shear longitudinal modulus (i.e., $S \rightarrow 0$) vanishes
in the long-scale limit according to
\begin{equation}
\mu_{L}(S\rightarrow0) \propto \lim_{q_z \rightarrow 0} 
q_z^{\eta_L/\omega} = 0.
\end{equation}
Analogous result can be derived for the linear transverse shear
modulus.  This implies that the elasticity of our system is strictly
nonlinear for arbitrary small stress.

\begin{figure}[!htbp]
\begin{center}
  \includegraphics[width=8cm]{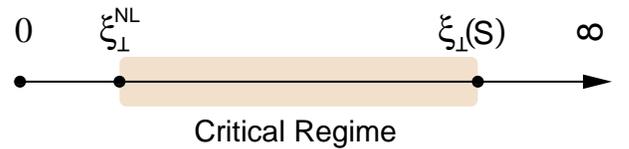}
  \caption{External stress introduces new stress-dependent
    length-scales $\xi_{\perp,z}(S)$, that diverge with a vanishing
    stress according to Eq.~(\ref{new-lengthscales}). On longer scales
    the critical scaling of elastic moduli and the associated strictly
    nonlinear anomalous elasticity is cut off by $\xi_{\perp,z}$.  The
    short-scale cutoffs $\xi^{NL}_{\perp,z}$ for the critical regime
    are set by a combination of network mesh size (uv-cutoff) and the
    bare elastic constants.}
\label{qregime}
\end{center}    
\end{figure}

The origin of nonlinear elasticity is straightforward to understand
within the RG formulation.  The external traction $S_{ia}$ is a
relevant coupling whose value grows under RG rescaling, thereby
flowing away from the nematic elastomer fixed point at $S_{ia}=0$.
The scale at which a weak external stress rescales to a large, maximum
(as defined by elastic model's uv cutoff and bare elastic moduli)
effective stress defines a new length-scales $\xi_{\perp,z}(S)$,
beyond which the RG flow is cut off and the associated
critical scaling breaks down.  For a nonzero $S$, the scaling form of
the thermal correlation function, Eq.~(\ref{GT-scaling}) is extended
to be
\begin{equation}
G^{T}_{zz}(\qv_{\perp}, q_z, S) = e^{(4 - \eta_K)l} 
G^{T}_{zz}(\qv_{\perp}\,e^l, q_z\,e^{\omega \,l},S\,e^{y_S\,l}),
\end{equation}
where $y_S$ is a {\em positive} exponent (calculable within our
$\epsilon$ RG expansion; see below) governing the flow of the
dimensionless stress $S$ (measured in units of a maximum stress for
which the elastic model remains valid) under the RG transformation.
Choosing $S\,e^{y_S\,l} = 1$, we obtain
\begin{equation}
G^{T}_{zz}(\qv_{\perp}, q_z, S) = 
S^{-\frac{4-\eta_K}{y_S}} \Phi\left(
q_{\perp}S^{-\frac{1}{y_s}}, q_{z}S^{-\frac{\omega}{y_s}}
\right).  \label{GTS-scaling}
\end{equation}
This automatically gives us length-scales 
\begin{equation}
\xi_{\perp}(S) = S^{-\frac{1}{y_s}},
\hspace{3mm}
\xi_z(S) = S^{-\frac{\omega}{y_s}}, \label{new-lengthscales}
\end{equation}
relative to which all other scales (e.g., $1/q_{\perp,z}$) are
measured.  Setting $\qv_{\perp}$ to zero in Eq.~(\ref{GTS-scaling}),
we find that the correlation function behaves very differently in the
large and small $q_z\xi_z$ regimes.  For $q_z \xi_z(S) \gg1$, we must
recover the critical scaling critical scaling form
(\ref{GT-z-scaling}) valid at $S=0$
\begin{equation}
G^{T}_{zz}(0, q_z, S) = q_{z}^{-\frac{4-\eta_K}{\omega}},\,\,\,
\mbox{for}\,\, q_z \xi_z(S) \gg1
\label{G-S-scaling-1}
\end{equation}
On the other hand, for $q_z \xi_z(S) \ll1$, we must have
\begin{eqnarray}
G^{T}_{zz}(0, q_z, S) &=&q_{z}^{-2}S^{\frac{1}{y_s}(2\,\omega-4+\eta_K)}
,\,\,\,\mbox{for}\,\, q_z \xi_z(S)\ll 1 \nonumber\\
&=& q_{z}^{-2} \,S^{-\eta_L/y_s}, 
\label{G-S-scaling-2}
\end{eqnarray}
whose form is dictated by the requirement of a $q_z$ independent shear
modulus in \rfs{Kubo-formula}.  Using this inside
Eq.~(\ref{Kubo-formula}), we find
\begin{equation}
\mu_{\rm L}(S) \propto S^{\eta_L/y_S} 
\sim \xi_{\perp}(S)^{-\eta_L}, \label{mu-scaling}
\end{equation}
that vanishes with a vanishing stress, $S \rightarrow 0$.

To calculate the unknown exponent $y_S$, we may study RG flow equation
for $S$ near the fixed point $E$.  However, we can also obtain the
value of $y_S$ from the scaling form of the elastic free energy
density, which is a function of $T$ and $S$ near the fixed point:
\begin{equation}
f(T,S) = e^{-(d-1+\omega )l - y_T\,l} \, f(T e^{y_T l}, S e^{y_S l}). 
\label{fTS-scaling}
\end{equation}
This scaling form is different from the corresponding form of thermal
critical phenomena by a factor of $e^{-y_T\,l}$.  The difference is
again due to the fact that long-scale properties of the elastomer are
controlled by a zero-temperature fixed point, E, with $T$ a {\em
  dangerously} irrelevant coupling.

As we have discussed above, the fact that the thermal exponent $y_T$
is negative implies that thermal fluctuations are not important at low
temperature and thus the singular part of the free energy, to the
leading order, is independent of temperature, reducing simply to the
ground state energy for given disorder realization and external
traction.  Thus neglecting this subdominant $T$ dependence in
Eq.~(\ref{fTS-scaling}) and choosing $l$ such that $e^l = S^{-1/y_s}$,
we find
\begin{equation}
f(S) \propto S^{\frac{1}{y_S} \left(
d-1+\omega +y_T \right)}.
\end{equation}
Using the fundamental definition of the inverse (nonlinear) shear
modulus as the second derivative of the free energy with respect to
the external stress $S$ we obtain:
\begin{equation}
\frac{1}{\mu_{\rm L}(S)} = \frac{\partial^2 f}{\partial S^2}
\propto S^{\frac{1}{y_S} \left(
d-1+\omega +y_T -2 \,y_S\right)}. 
\end{equation}
Comparing this with Eq.~(\ref{mu-scaling}) we find 
\begin{eqnarray}
y_S &=& 2 - \eta_K,
\end{eqnarray}
which then gives
\begin{eqnarray}
\mu_{\rm L}(S) &\propto& S^{\frac{\eta_L}{2-\eta_K}}. 
\end{eqnarray}
This agrees with the result we obtained from a complementary argument
in Ref.\,\onlinecite{Xing-Radz1}, based on the fact that the external
stress $S$ experiences only a ``trivial'' (not independent of other
couplings) renormalization, analogous to an external magnetic field in
an Ising model ($\phi^4$ theory).

To the leading order in the $\epsilon$-expansion, we can read off the
values of various exponents from Table \ref{table-fixpoints}:
\begin{subequations}
\label{exponents-value}
\begin{eqnarray}
\eta_{\rm L} &=& \eta_{\perp} = \eta= \frac{6 \epsilon}{263},\\
\eta_K   &=& \frac{106\epsilon}{263},\\
\eta_{\Delta} &=& \frac{5\epsilon}{263}. 
\end{eqnarray}
\end{subequations}
which in three dimensions lead to 
\begin{equation}
\mu_{\rm L}(S) \propto S^{\frac{2 \epsilon}{157}} 
= S^{\frac{6}{157}}. 
\end{equation}
While this violation of linear strain-stress relation is quite weak in
three dimensions (the exponent is small), it is qualitatively quite
significant.

As we have just seen from Eq.~(\ref{G-S-scaling-1}) and
Eq.~(\ref{G-S-scaling-2}), critical fluctuations of a nematic
elastomer are suppressed by the external traction beyond the stress
dependent crossover length-scales $\xi_{\perp}(S)$ and $\xi_z(S)$.
Similar effect also arises from the application of a magnetic field
along the nematic director, which adds a new term into the
nematic elastomer Hamiltonian:
\begin{eqnarray}
\delta H_m = - \frac{1}{2}\gamma_a 
\int d^d x \, (\nh \cdot \vec{h})^2
\approx \frac{\gamma_a}{4} h^2 \int d^d x \,\delta \nh(\xv)^2.\hspace{6mm}  
\end{eqnarray}
After integrating out the nematic director fluctuations using
Eq.~(\ref{replacement}) as we discussed earlier, $\delta H_m$ reduces
to
\begin{eqnarray}
\delta H_m =  \frac{\gamma_a}{4} h^2 \int d^d x \,(\nabla_{\perp} u_z)^2
+ \mbox{less relevant} .  
\end{eqnarray}
This can be further transformed (using $w_{zz}=\partial_z u_z 
+ \frac{1}{2}\, (\nabla_{\perp} u_z)^2$) to reduce $\delta H_m$ to
\begin{eqnarray}
\delta H_m = - \frac{1}{2}\gamma_a h^2 
\int d^d x \,(\ppz u_z) +
\mbox{less relevant},
\end{eqnarray}
after shifting the reference state appropriately.  Therefore $\gamma_a
h^2$ is equivalent to the stress component $S$ along the director, which
we considered in Eq.~(\ref{Kubo-formula}).  With this mapping in hand,
the following results straightforwardly follow:
\begin{enumerate}
\item A magnetic field along the director introduces two crossover
length-scales
\begin{equation}
\xi_{\perp}(h) = h^{-\frac{2}{y_s}},
\hspace{3mm}
\xi_z(h) = h^{-\frac{2\omega}{y_s}},
\end{equation}
beyond which the critical fluctuations are suppressed. 
\item A magnetic field along the director induces spontaneous strain
along the director, which scale as
\begin{eqnarray}
e_{zz} \propto h^{2(1+\frac{\eta}{2-\eta_K})}.  
\end{eqnarray}
\item If we apply both a magnetic field $h$ and an external traction
$S$ along the director, then the stress-strain relation is linear for
$S \ll \gamma_a h^2$ and is nonlinear for $S \gg \gamma_a h^2$.
\end{enumerate}
These predictions should in principle be experimentally testable.

\subsection{Stability of the anomalous nematic phase}

The validity of above analysis and the associated stability of the
critical phase requires that the nematic order is long-ranged.  This
translates into a constraint that elastic uniaxial anisotropy survives
thermal fluctuations and network heterogeneity, a condition given by
\begin{equation}
\omega = 2 - \frac{1}{2} \left( \eta+\eta_K\right) >1. 
\label{nematic-stability}
\end{equation}
Using our one-loop exponents approximation in Table
\ref{table-fixpoints}, we find that the condition
(\ref{nematic-stability}) is satisfied as long as
\begin{equation}
d>d_{lc} = 17/56.
\label{dlc}
\end{equation}
This is clearly satisfied by the physical case of $d=3$, in which
\begin{equation}
\omega_{d=3} \approx 1.574,
\end{equation}
is significantly above unity, as required for nematic state stability.

The same condition, Eq.~(\ref{nematic-stability}) can be obtained in a
complementary way by looking at the real space fluctuations of the
nematic director.  Recalling that a director fluctuation $\delta\nh$
is massively tied to the asymmetric linearized strain through
Eq.~(\ref{V_ni}), the fluctuations of $\delta \nh$ can be estimated by
\begin{eqnarray}
\overline{\langle \delta\hat{n}(\qv)^2 \rangle}
&\approx&  (r-1)^{-2}  \left(  q_{\perp}^2 
        \overline{\langle |u_z(\qv)|^2\rangle} \right.\\
&+&     \left.  2\,r\,  q_z \,q_i\,
        \overline{\langle u_z(\qv)\,u_i(-\qv)\rangle}
+      r^2\,q_z^2\,
        \overline{\langle |\uv_{\perp}(\qv)|^2\rangle}
        \right).
\nonumber
\end{eqnarray} 

Given the scaling RG analysis above, the dominant contribution on the
right hand side is given by quenched fluctuations of the $u_z$ phonon:
\begin{eqnarray}
\overline{\langle \delta\hat{n}(\qv)^2 \rangle} &\propto &
q_{\perp}^2 \overline{\, \langle u_{z}(\qv)\rangle
\langle u_{z}( - \qv)\rangle \,}
= q_{\perp}^2 G^{\Delta}_{zz}(\qv)
\nonumber\\
&=& q_{\perp}^{-(d+1-\omega)} \,\Phi_z^\Delta
\left(\frac{q_z}{q_{\perp}^{\omega}}\right), 
\end{eqnarray}
where we have used the scaling form of $G^{\Delta}_{zz}$,
Eq.~(\ref{GDelta-scaling}).  In real space, we have
\begin{equation}
\overline{\langle \delta\hat{n}(\xv)^2 \rangle} \propto 
\int d^{d-1} {q}_{\perp} dq_z 
q_{\perp}^{-(d+1-\omega)} \,\Phi_z^\Delta
\left(\frac{q_z}{q_{\perp}^{\omega}}\right).
\end{equation}
A straightforward power-counting shows that the above integral scales
as $q_{\perp}^{2\omega - 2}$, and therefore converges as long as
$\omega > 1$, in agreement with the result in Eq.~(\ref{nematic-stability}).

We have so far been ignoring the nonlinearities associated with
$\uv_{\perp}$ phonon, since they are less relevant at the Gaussian
fixed point than the $u_z$ nonlinearities.  However, we have also seen
in Eq.~(\ref{up-quenched}) that the harmonic quenched fluctuations of
$\uv_{\perp}$ are actually divergent, where the unrenormalized
correlator is used.  To check whether these quenched fluctuations are
still divergent in the renormalized theory, i.e., at fixed point E, we
need to use the renormalized correlator $G^R_{L/T}$ in
Eq.~(\ref{up-quenched}).  Without detailed calculation, we observe
that the renormalized disorder variance $\Delta$ diverges and the
shear moduli $\tilde{\mu}$, $\mu$ vanish as $q\rightarrow 0$,
according to our RG analysis.  On the other hand, the bending constant
$K_3$ does not acquire anomalous dimension from $u_z$ fluctuations.
Therefore we deduce that the renormalized fluctuations of
$\uv_{\perp}$ becomes stronger, i.e., more divergent than the naive
harmonic analysis at the Gaussian fixed point.  This suggests that we
may not be able to ignore the columnar elastic nonlinearities
associated with $\uv_{\perp}$.  However, a simultaneous treatment of
both smectic and columnar nonlinearities is a challenging open problem
that we leave to the future research.

\subsection{Universal Poisson ratios}

We conclude our analysis of long-scale anomalous elasticity, by
observing from Table \ref{table-fixpoints} that the fixed point values of
two coupling constants $\mathsf{g}_{\rm L}$ and $\mathsf{g}_{\perp}$
satisfy a relation:
\begin{eqnarray}
\frac{\mathsf{g}^*_{\perp}}{\mathsf{g}^*_L } =  6.
\end{eqnarray}
Combing this observation with the definitions of $\mathsf{g}_{\rm L}$
and $\mathsf{g}_{\perp}$, Eqs.~(\ref{coupling_replica}) we find that
the ratio between the long-scale effective transverse and longitudinal
shear moduli approaches a universal constant:
\begin{eqnarray}
\lim_{l\rightarrow \infty} \frac{\mu(l)}{\mu_{\rm L}(l)} = 6.  
\label{Poisson-ratio}
\end{eqnarray}
On the other hand, since $\eta_B = 0$ at the fixed point E, we also have
\begin{eqnarray}
\lim_{l \rightarrow\infty} \frac{\mu(l)}{B(l)} =  0.
\end{eqnarray}
That is, nematic elastomers are {\em strictly} (not just
quantitatively) incompressible at long length-scales. The existence of
these universal ratios characterizing a critical nematic elastomer
phase is an analog of a well-known universal amplitude ratios at a
continuous phase transition.

Let us now revisit the experiment shown in Fig.~\ref{poissonratio},
where we impose a fixed strain deformation $\varepsilon_{xx}$ in the
$x$ direction and let the sample relax without any macroscopic
reorientation of the nematic director.  The other strain components
are still given by Eqs.~(\ref{e-results}), if we replace all bare
elastic constants by the renormalized ones.  Using
Eq.~(\ref{Poisson-ratio}), we find two universal Poisson ratios:
\begin{eqnarray}
\varepsilon_{yy} &=& \frac{5}{7} \varepsilon_{xx},\label{uyy}\\ 
\varepsilon_{zz} &=& -\frac{12}{7} \varepsilon_{xx}.\label{uzz}
\end{eqnarray} 
This prediction should be experimentally testable.

\section{Summary and Conclusions}
\label{Sec:Conclusion}

In this paper, we have developed a theoretical framework for the
nonlinear elasticity of uniaxial nematic liquid crystalline
elastomers, a fascinating class of materials which has an internal
orientational order as a consequence of a spontaneous symmetry breaking.
We have analyzed the soft Goldstone modes associated with the
spontaneously broken rotational symmetry.  In a strain-only
description, nematic elastomers resemble conventional uniaxial elastic
solids, but with a strictly vanishing shear modulus $\mu_{zi}$. We have
developed a model elastic free energy, in terms of an invariant strain
tensor, which completely encodes the soft modes.  We have also
discussed its connection and difference with the popular neo-classical
theory of nematic elastomers.

We have discussed the angular momentum transfer between the
translational and orientational degrees of freedom in nematic
elastomers, characterized the antisymmetric part of the Cauchy stress
tensor.  We have also established the connection between stress
tensor, couple-stress tensor, and the elastic free energy, and have
calculated these quantities explicitly using two model elastic free
energies.

We have also developed a complementary strain-only elastic model of
nematic elastomers, and have derived a set of (rotational
symmetry-enforced) Ward identities relating the coefficients of all
anharmonic terms to those of quadratic ones.  Using this minimal
model, we analyzed the long-scale properties of ideal nematic
elastomers and found that due to soft modes, the long-scale elastic
properties are qualitatively modified by thermal fluctuations and
polymer network heterogeneities, with the latter dominating over the
former.

Our key finding is that thermal and quenched fluctuations lead to the
``anomalous elasticity'' phenomena, familiar from a number of other
soft-matter contexts, such as smectic and columnar liquid crystals,
polymerized membranes and putative spontaneous vortex lattices in
ferromagnetic superconductors. Specifically, as a result at long
scales nematic elastomers are characterized by singular, length-scale
dependent shear elastic moduli, a divergent splay elastic constant,
long-scale incompressibility, universal Poisson ratios and, a
non-Hookean (nonlinear) stress-strain relation down to arbitrary small
strains.  Furthermore, we show that these properties are {\em
  universal}, and are controlled by a nontrivial zero-temperature
fixed point, constituting a qualitative breakdown of classical
elasticity theory. Thus, nematic elastomers constitute a stable
``critical phase'', characterized by universal power-law correlations
akin to a critical point of a continuous phase transition, but
extending over an entire phase, as illustrated in Fig.~\ref{qregime}.
We have also found that for weak disorder and low thermal fluctuations
the anomalous elasticity {\em stabilizes} long-range nematic elastomer
order in three dimensions.  Such orientational order in a 3d
disordered system with purely orientational degrees of freedom, e.g.,
spin system or a nematic liquid crystal, is known to be impossible.
Thus a stable orientational order in nematic elastomers is a
consequence of stabilizing interplay between orientational and elastic
degrees of freedom.

However, experimentally, at low temperature, nematic elastomers
crosslinked under isotropic conditions always exhibit a polydomain
nematic director pattern with short-range nematic order, in apparent
contradiction with our $\epsilon$-expansion results.  There are a
number of possible explanation for this disagreement.  Firstly, and
most likely in our view, real nematic elastomers are quite likely
characterized by {\em strong} heterogeneity, which may not be captured
by our perturbative RG analysis.  It is also possible that
$\epsilon=2$ is too large to trust the $\epsilon$-expansion.  Although
qualitatively subdominant, columnar nonlinearities may play an
important quantitative role.  Also our analysis neglected the
transverse part of the random stress.  This naturally couples to the
nematic topological defects, and therefore for strong disorder may
lead to a proliferation of the latter, thereby destroying long-range
orientational order.  This problem is being actively studied by the
authors.  Finally, it is also quite possible that nematic order is
indeed stable in real weakly-heterogeneous elastomers, but (as in the
case of the random-field spin systems \cite{BirgeneauRFIMexp})
requires long equilibration times or ``field-cooling'' to realize.  If
the nematic orientational order is indeed unstable to disorder, but
with a long orientational correlation length, our analysis and
predictions are expected to be valid in the resulting polydomain
nematic for a range of length-scales between the nonlinear elasticity
scale $\xi_{NL}$ and the orientational correlation length-scale
$\xi_{nem}$, with latter diverging in the weak heterogeneity limit.
Further theoretical and experimental studies are clearly needed in
order to clarify the true nature of the nematic elastomer ground
state.

Our present work leaves open a number of interesting questions. One
very important one is the fate of the isotropic-nematic {\em
  transition} in a heterogeneous elastomers. Also, we exclusively
focussed on the ideal case, ignoring the fact that all mono-domain
elastomers are crosslinked under pre-stretched conditions, that
imprints a weak uniaxial anisotropy, that was recently shown to lead
semi-soft elasticity. Thus for a direct comparison with experiments
our theory must be extended to a detailed treatment of semi-soft
elasticity.  Interesting work on this subject was recently carried out
by Ye and collaborators.\cite{SemiSoftLubensky} It is quite clear from
their and our analyses that imprinted anisotropy cuts off the critical
phase (anomalous elasticity) phenomenology at long scales and weak
external traction.  We leave further detailed work on these subjects
to future investigations.

\begin{acknowledgments}
  We acknowledge the financial support from the American Chemical
  Society under grant PRF 44689-G7 (XX) and the National Science
  Foundation under grants MRSEC DMR-0213918 (LR), DMR-0321848 (LR), as
  well as from David and Lucile Packard Foundations (LR).  We thank
  the hospitality of the Harvard Physics Department, KITP at Santa
  Barbara and University of Illinois Physics Department during the
  past five years, where a fraction of this research was conducted.
  We are grateful to Tom Lubensky for introducing us to the
  fascinating problem of liquid crystal elastomers and for earlier
  fruitful collaborations without which much of this work would not be
  possible. We also thank John Toner and Pierre Le Doussal for a
  number of helpful discussions.
\end{acknowledgments}

\appendix

\section{Elastic Moduli in the Effective Strain-Only Model}
\label{App:Coefficients}
Here, for completeness we list all eleven coefficients appearing in
the effective elastic Hamiltonian Eq.~(\ref{f_expansion2}):
\begin{subequations}
\label{Coefficients}
\begin{eqnarray}
a_{\perp} &=& 2\,\zeta_{\perp}^2\,\left( \Phi_{1} 
  + 2\,\zeta_{\perp}^2\,\Phi_{2} + 
    3\,\zeta_{\perp}^4\,\Phi_{3} \right),\nonumber\\
a_z &=& 2\,\zeta_{\parallel}^2\,\left( \Phi_{1}
   + 2\,\zeta_{\parallel}^2\,\Phi_{2} + 
    3\,\zeta_{\parallel}^4\,\Phi_{3} \right),\nonumber\\
\mu_{z\perp} &=& 4\,\zeta_{\perp}^2\,\zeta_{\parallel}^2
   \,\left(2\, \Phi_{2} 
  +  3\,\left( \zeta_{\perp}^2 
   + \zeta_{\parallel}^2 \right) \,\Phi_{3} \right),\nonumber\\
B_z &=& 4 \sum_{i,j=1}^3 (i \cdot j)\, \Phi_{ij}\,
  \zeta_{\parallel}^{2 (i+j)}
 + 8 \zeta_{\parallel}^4 \Phi_2 
+ 24 \zeta_{\parallel}^6 \Phi_3,\nonumber\\
\lambda_{z\perp} &=& 4  \sum_{i,j=1}^3 
 (i \cdot j) \,\Phi_{ij} \,
 \zeta_{\parallel}^{2 i}\,  \zeta_{\perp}^{2 j},\nonumber\\
\lambda &=&  4 \sum_{i,j=1}^3 (i \cdot j) 
\,\Phi_{ij} \,\zeta_{\perp}^{2 (i+j)},\nonumber\\
\mu &=& 4\,\zeta_{\perp}^4\,\left( \Phi_{2}
 + 3 \,\zeta_{\perp}^2\,\Phi_{3} \right),\nonumber\\
b_1 &=& 8\,\zeta_{\perp}^2\,\zeta_{\parallel}^4
  \,\left(3\, \Phi_3 + 2\, \Phi_{12} 
+  3 \,\left( \zeta_{\perp}^2 
+ \zeta_{\parallel}^2 \right) \Phi_{13}
\right.  \nonumber\\ 
 &+& 4 \,\zeta_{\parallel}^2 \, \Phi_{22}  
+  6 \, \zeta_{\parallel}^2 \,\left(
 \zeta_{\perp}^2 + 2 \, \zeta_{\parallel}^2 \right)\,\Phi_{23}
\nonumber\\
 &+& \left. 9 \,\zeta_{\parallel}^4\, \left( \zeta_{\perp}^2
   + \zeta_{\parallel}^2 \right)
    \, \Phi_{33} \right),\nonumber\\
b_2 &=&8\,\zeta_{\perp}^4\,\zeta_{\parallel}^2
 \,\left(2\, \Phi_{12}
 +  3 \,\left( \zeta_{\perp}^2 
 + \zeta_{\parallel}^2 \right) \Phi_{13} 
 + 4 \,\zeta_{\perp}^2 \, \Phi_{22}  \right.
\nonumber\\ 
&+&  \left. 6 \, \zeta_{\perp}^2 \,\left(
    2 \, \zeta_{\perp}^2 + \zeta_{\parallel}^2 
     \right)\,\Phi_{23}
    +  9 \,\zeta_{\perp}^4\, \left( \zeta_{\perp}^2 
  + \zeta_{\parallel}^2 \right)
    \, \Phi_{33} \right), \nonumber\\
b_3 &=& 24\,\zeta_{\perp}^4\,
 \zeta_{\parallel}^2\,\Phi_{3},\nonumber\\
c &=& 8\,\zeta_{\perp}^4\,\zeta_{\parallel}^4
   \,\left(4\, \Phi_{22} + 
    12\,\left( \zeta_{\perp}^2 
 + \zeta_{\parallel}^2 \right) \,\Phi_{23} 
\right. \nonumber\\
 &+& \left. 9 \,\left( \zeta_{\perp}^2
 + \zeta_{\parallel}^2 \right)^2 \,\Phi_{33}\right).\nonumber
\end{eqnarray}
\end{subequations}

\section{Ward Identities revisited} 
\label{App:Ward-identities}

In this appendix we re-derive the Ward identities
Eqs.~(\ref{ward_identities}) via an alternative, more straightforward
approach, whose advantage is that for $d \geq 3$ it is independent of
space dimensions.

To this end, let us consider a $d$-dimensional ideal uniaxial nematic
elastomer.  Using Eq.~(\ref{soft-mode}), it is easy to see that the
most general expression for soft deformations with constraint
$\nh=\nh_0$ is given by
\begin{equation}
\hat{n}_0 \rightarrow \hat{n}_0, \hspace{3mm}
\lm = \Lm_{\hat{n}_0} \, \mm{O} \,\Lm_{\hat{n}_0}^{-1},
\label{soft-mode-1}
\end{equation}
where $\mm{O}$ is an arbitrary rotation and $\Lm_{\hat{n}_0}$ is given
by Eq.~(\ref{Lambda_n}) with $\nh$ replaced by $\nh_0$:
\begin{equation}
\Lm_{\nh_0} = \zeta_{\perp}^2 \mm{I} 
+ (\zeta_{\parallel}^2 - \zeta_{\perp}^2) \hat{n}_0\hat{n}_0, 
\label{Lambda_n0}
\end{equation}

Let us choose the coordinate system such that $\nh_0$ is parallel to
the $z$ axis and consider a rotation $\mm{O}$ in the $xz$ plane.  All
other $d-2$ axes are unchanged by the rotation and therefore do not
need to be considered.  Therefore we only need to keep track of
$(xx)$, $(xz)$, $(zx)$, and $(zz)$ components of the tensors $\mm{O}$
and $\Lm_{\nh_0}$, and can express them as $2\times 2$ matrices:
\begin{eqnarray}
\mm{O} = \begin{pmatrix} \cos \theta& -\sin \theta\\
\sin\theta &\cos\theta\end{pmatrix}, 
\hspace{3mm}
\Lm_{\nh_0} =  \begin{pmatrix} \zeta_{\perp}^2&0\\
0&\zeta_{\parallel}^2\end{pmatrix} 
\label{OLm}
\end{eqnarray}
acting in this 2d subspace.  Using Eq.~(\ref{Lambda_n0}) and
Eq.~(\ref{OLm}), we find the Lagrange strain tensor corresponding to
this soft deformation is given by
\begin{eqnarray}
\mm{e} &=& \begin{pmatrix}
e_{xx}&e_{xz}\\e_{zx}&e_{zz}\end{pmatrix}
 = \frac{(r-1)}{2}
\begin{pmatrix}\sin^2\theta&-\frac{1}{2\sqrt{r}}\sin 2\theta\\
-\frac{1}{2\sqrt{r}}\sin 2\theta&-\frac{\sin^2\theta}{r}\end{pmatrix}
\nonumber\\
&\approx& \frac{1}{2}(r-1)\begin{pmatrix}
\theta^2 &-\frac{\theta}{\sqrt{r}}\\
-\frac{\theta}{\sqrt{r}}&-\frac{\theta^2}{r}
\end{pmatrix}. \label{soft_strain_d}
\end{eqnarray}
where we have expanded up to order of $\theta^2$.   

For this particular soft deformation (\ref{soft_strain_d}), the first
three terms of the elastic free energy, Eq.~(\ref{f_expansion2}) are
given by
\begin{subequations}
\label{soft-theta}
\begin{eqnarray} 
e_{zz} &=& \frac{1}{2}(\frac{1}{r}-1) \theta^2,\\
e_{ii} &=& e_{xx} =  \frac{1}{2}(r-1)\theta^2, \\
e_{iz}^2 &=& e_{xz}^2 =  \frac{1}{4\,r}(r-1)^2\theta^2,
\end{eqnarray}
\end{subequations}
which are all at order of $\theta^2$.  It is also clear that all other
terms explicitly shown in Eq.~(\ref{f_expansion2}) are at least of
order $\theta^4$, and with others (not shown) are even of higher, at
least $\theta^6$ order.  Therefore, in order for the elastic energy
Eq.~(\ref{f_expansion2}) to vanish at order of $\theta^2$, we must
have
\begin{eqnarray}
0 &=& a_{z} \, e_{zz}  + a_{\perp} \, e_{ii} 
+ \mu_{z\perp} \, e_{zi}^2
\\
&=& \frac{(r-1)^2 \theta^2}{4\,r} \left( 
- \frac{2}{r-1} a_z + \frac{2\,r}{r-1} a_z 
+ \mu_{z\perp}\right), \nonumber
\end{eqnarray}
which can be seen to be identical to the first Ward identity,
Eq.~(\ref{ward1}), after using the relation Eq.~(\ref{alpha}) between
$r$ and $\alpha$.

We can then make an appropriate choice of the nematic reference state
such that the first three terms in Eq.~(\ref{f_expansion2}) vanish
simultaneously.  The remaining eight terms, all of order of
$\theta^4$, can be reformulated into
\begin{eqnarray}
{\mathcal H} &=& \frac{1}{2} \,B_z\,\left( e_{zz} 
        + 2\alpha_z \, e_{zi}^2 \right)^2
        + \frac{1}{2} \,\lambda \,  
        \left( e_{ii} + 2\alpha_{\perp} 
        \, e_{zi}^2 \right)\nonumber\\
&+&     \lambda_{z\perp}\, 
        \left( e_{zz} +2 \alpha_z \, e_{zi}^2 \right)
        \left( e_{ii} +2 \alpha_{\perp} \,e_{zi}^2 \right)
        \nonumber\\
&+&     \,\mu\, \left(e_{ij} +2 \alpha_{\perp}'\,
        e_{zi}\, e_{zj}\right)^2 
        + \bar{c} \, e_{zi}^2\,e_{zj}^2\,.
\label{elastic_energy2}
\end{eqnarray}
The new coefficients $\alpha_z$, $\alpha_{\perp}$, $\alpha'_{\perp}$
and $\bar{c}$ can be determined by comparing coefficients of every
terms between Eq.~(\ref{elastic_energy2}) and
Eq.~(\ref{f_expansion2}):
\begin{subequations}
\label{ward_identities-2}
\begin{eqnarray}
b_1 &=& 2 \, \alpha_z \, B_z 
+ 2 \,\alpha_{\perp} \, \lambda_{z\perp},\\
b_2 &=& 2\,\alpha_z\, \lambda_{z\perp} 
+ 2\,\alpha_{\perp} \, \lambda,\\
b_3 &=& 4 \, \alpha_{\perp}' \, \mu,\\
c &=& 2 \, \alpha_z^2 \, B_z 
 +  4 \,\alpha_z\, \alpha_{\perp} \, \lambda_{z\perp}
 + 2\,\alpha_{\perp}^2 \, \lambda \nonumber\\
 &+& 4 \, (\alpha_{\perp}')^2 \, \mu 
 + \bar{c}
\end{eqnarray}
\end{subequations}

Given that $e_{ij}$ is a soft deformation, Eq.~(\ref{soft-theta}), all
terms in Eq.~(\ref{elastic_energy2}) order by order in $\theta$ and in
particular to order $\theta^4$ must vanish identically.  Furthermore,
the elastic energy must be nonnegative for arbitrary strain
deformation.  The only way to satisfy both conditions is by requiring
that {\em every single term} in Eq.~(\ref{elastic_energy2}) {\em
vanishes} for the soft deformation Eq.~(\ref{soft_strain_d}).  This
imposes stringent constraints on coefficients $\alpha_z$,
$\alpha_{\perp}$, $\alpha'_{\perp}$ and $\bar{c}$ as follow:
\begin{eqnarray}
\alpha_z &=& \frac{1}{r-1} = \alpha,\\
\alpha_{\perp} &=& \alpha'_{\perp}= -\frac{r}{r-1}= - (1+\alpha),\\
\bar{c} &=& 0.
\end{eqnarray}
Under these conditions, Eqs.~(\ref{ward_identities-2}) reduce to the
last four Ward identities in Eqs.~(\ref{ward_identities}) with
$\mu_{z\perp}$ set to zero, while the elastic energy
Eq.~(\ref{elastic_energy2}) reduces to Eq.~(\ref{f_expansion3}).

\section{Derivation of RG flow equations for an ideal homogeneous elastomer}
\label{App:RG-general}

In this appendix we present a detailed derivation of renormalization
group flow equations (\ref{RGcorrections}) near three dimensions for
elastic constants characterizing a nematic elastomer.  We will also
show explicitly that the form of Hamiltonian Eq.~(\ref{f_expansion5})
is preserved by the renormalization group transformation, as we have
already argued in Sec.\ref{Sec:Thermal-fluct} based on symmetry
grounds.

\subsection{Momentum shell RG}
\label{app:MSRG}
The principle of renormalization group transformation (RG) is to
transform a Hamiltonian $H$ for a fluctuating field $\uv(\xv)$, and
characterized by a set of coupling constants $c$ and a uv wavevector
cutoff $Q$ to a Hamiltonian of the same form for a coarse-grained
field $\uv'(\xv')$, the same momentum cutoff $Q$, and characterized
by new set of couplings constants $c'$. The goal is to establish a
relation (RG flow) between coupling constants $c'$ and $c$ under such
successive RG transformations.  Armed with the RG flow of the coupling
set $c$, as well as the relation between the original field $\uv(\xv)$
and the coarse-grained one $\uv'(\xv')$ allows one to extract the
scaling behavior of correlation functions of $\uv(\xv)$.

A momentum-shell RG (MSRG) consists of two steps: (i) an integration
of short wavelength fluctuations of the field $\uv(\xv)$ (i.e.,
coarse-graining), whose Fourier transform has support in the momentum
shell $Q\,e^{-\delta l} \leq q \leq Q$, giving a coarse-grained field
theory with uv wavevector cutoff $Q\,e^{-\delta l}$, and a set of
modified coupling constants; (ii) a rescaling of coordinates $\xv$ and
field $\uv(\xv)$ so as to restore the uv cutoff to the original value
of $Q$. The rescaling transformation (ii) leads to further
modification of the coupling constants $c$. Although this step is in
principle unnecessary, it facilitates the establishing of the
relation between coupling constants $c$ and $c'$, characterizing the
two Hamiltonians.

For a uniaxial system such as a nematic elastomer, there are generally
two different momentum cutoff $Q_{\parallel}$ and $Q_{\perp}$, where
subscripts $\parallel$ and $\perp$ denote directions parallel and
perpendicular to the nematic director, respectively. However, the
principle of universality guarantees that the long length-scale
physics of the system is independent of different choices of momentum
cutoff. It is convenient to choose a cylindrical cutoff scheme,
where $(Q_{\parallel} = \infty, Q_{\perp} = Q)$.  

For most applications it is sufficient to focus on the partition
function $Z[c, Q]$, given by
\begin{equation}
Z[c, Q] = \int^Q {\mathcal D} \uv \,e^{ -H[\uv,C]},
\end{equation}
where the functional integral is over field $\uv(\xv)$, whose Fourier
transform has support in the region $|\qv_{\perp}| \leq Q$, and we
have explicitly shown the dependence of the partition function on the
uv cutoff $Q$ and the set of coupling constants $c$.

To carry out the coarse-graining procedure, as illustrated in
Fig.~\ref{shell}, we decompose the field $\uv(\qv)$ into a sum of high-
and low-wavevector parts $\uv^>(\qv)$ and $\uv^<(\qv)$,
\begin{equation}
\uv(\qv) = \uv^>(\qv) + \uv^<(\qv),
\end{equation}
where $\uv^>(\qv)$ has support in the outer cylinder shell of the
$\qv$ space:
\begin{equation}
\uv^>(\qv) \left\{ \begin{array}{ll} \neq 0, 
  & \mbox{   if}\hspace{5mm} Q e^{-l} 
< |\qv_{\perp}| < Q,\\
= 0, & \mbox{   if} \hspace{5mm}0 
< |\qv_{\perp}| < Q e^{-l},
\end{array}\right.
\end{equation}
while $\uv^<(\qv)$ has support in the inner cylinder of the $\qv$ space:
\begin{equation}
\uv^<(\qv) \left\{ \begin{array}{ll} = 0, 
  & \mbox{   if}\hspace{5mm} Q e^{-l} 
< |\qv_{\perp}| < Q,\\
\neq 0, & \mbox{   if} \hspace{5mm} 0 
< |\qv_{\perp}| < Q e^{-l}.\end{array}\right.
\end{equation}
Because of the translational invariance, the harmonic part of the
Hamiltonian can be separated into higher and lower momentum pieces,
giving
\begin{eqnarray}
H[\uv] &=& H_0[\uv^<+\uv^>] 
+ H_I[\uv]  \nonumber\\
&=& H_0[\uv^<] + H_0[\uv^>] + H_I[\uv],
\end{eqnarray}
where $H_I[\uv]$ contains all the anharmonic terms in $\uv(\rv)$.

This allows one to separate the functional integral into a product of
two parts,
\begin{eqnarray}
\int^Q D\uv = \int^{Q\,e^{-dl}} D\uv^< \int_{Q\,e^{-dl}}^Q D\uv^>
= \int^<\int^<,
\end{eqnarray}
where we have also introduced a short hand for functional integral
over $\uv^<$ and $\uv^>$.  Carrying out the functional integral over
the high-wavevector field $\uv^>$, we obtain another field theory with
a wavevector cutoff $Q\,e^{-\delta l}$ and a modified Hamiltonian
defined by:
\begin{eqnarray}
Z[c, Q] &=& \int^< e^{-H_0[\uv^<]} \int^> e^{-H_0[\uv^>] -
H_I[\uv^<+\uv^>]} \nonumber\\ &=& Z_0^> \int^< e^{-H_0[\uv^<]} \langle
e^{-H_I[\uv^<+\uv^>]}\rangle_0^>, \nonumber\\
\label{partition-function}
\end{eqnarray}
where 
\begin{eqnarray}
  Z_0^> &=& \int^> e^{-H_0[\uv^>, \{\lambda\}]},\\
  \langle A \rangle_0^> &=& \frac{1}{Z_0^>}
  \int^> e^{-H_0[\uv^>]} A.
\end{eqnarray}

This allows us to define a coarse-grained Hamiltonian as a functional
of $\uv^<$ and new coupling constants 
$c + \delta_g c$ \cite{comment:subscripts}: 
\begin{equation}
H[\uv^<, c+\delta_g c ]
 = H_0[\uv^>] - \log  \langle 
e^{-H_I[\uv^<+\uv^>, c]} \rangle_0^>,
\label{HRG}
\end{equation}
in terms of which the partition function (\ref{partition-function})
can now be written as 
\begin{equation}
Z[c, Q]  = Z_0^>\int^< e^{-H[\uv^<, c+\delta_g c]}
= Z_0^> Z[c+\delta_g c, Q e^{-\delta l}]
\end{equation}
This thereby transforms a computation of $Z$ into a functional
integral over a coarse-grained field $\uv^<(\xv)$, momentum cutoff $Q
e^{-\delta l}$ and coupling constants $c+\delta_g c$.

For convenience, we furthermore rescale the spatial coordinates 
\cite{comment:rescaling} such that the wavevector cutoff is restored to $Q$: 
\begin{equation}
\xv = \xv' \, e^{\delta l},
\hspace{3mm}
\qv = \qv' \, e^{-\delta l}.
\end{equation}
This leads to further transformation of the coupling constants, which
we denote as $\delta_r c$ \cite{comment:rescaling-2}: 
\begin{equation}
Z[c, Q] = Z_0^> Z[c+\delta_g c + \delta_r c, Q]. 
\end{equation}
This allows us to summarize the effect of the RG transformation in terms of 
a transformation between coupling constants, given by
\begin{equation}
c + \delta c = c + \delta_g c  +  \delta_r c, \label{correction}
\end{equation}
with $\delta C = \delta_g c + \delta_r C$ proportional to an
infinitesimal $\delta l$. The RG transformation can therefore be
summarized by ordinary differential equations for $c(l)$, describing
the RG flow of coupling constants under successive infinitesimal
renormalization group transformations.

\subsection{Derivation of RG flow equations, Eqs.~(\ref{RGcorrections})}
\label{flow_equations}

The key non-trivial step in the RG transformation is the calculation
of the second term in the right hand side of Eq.~(\ref{HRG}). Near the
upper-critical dimension, $d_{uc}$ (above which nonlinearities and
fluctuations are no longer important), this step can be carried out
perturbatively in nonlinearities (which amounts to be perturbatively
in $\epsilon=d_{uc}-d$) using cumulant expansion:
\begin{widetext}
\begin{equation}
- \log  \langle e^{-H_I[\uv^<+\uv^>, \{\lambda\}]} \rangle_0^>
 = \langle H_I \rangle_0^> - \frac{1}{2} \left(
\langle H_I^2\rangle_0^> - (\langle H_I\rangle_0^>)^2\right)
+ \cdots.
\label{cumulants_expansion}
\end{equation}

The total nematic elastomer Hamiltonian is given by
Eq.~(\ref{f_expansion4}), with the effective strain tensor $\mm{w}$
defined in Eqs.~(\ref{effective_strain}).  The harmonic part
$H_0[\uv]$ is given by:
\begin{equation}
H_0 = \frac{1}{2} \left[ B_z (\pz u_z)^2 
       + 2 \lambda_{z\perp} (\pz u_z) (\ppi u_i )
       + ( \lambda + \mu) (\ppi u_i)^2  
       + \mu (\nabla_{\perp}\cdot \uv_{\perp})^2
       + K (\nabla_{\perp} u_z)^2 \right].
\end{equation}
The nonlinear part $H_I[\uv]$ contains all cubic and quartic terms given by
\begin{subequations}
\label{H-anhar}
\begin{eqnarray}
H_I[\uv] &=& H_{\rm cubic} + H_{\rm quartic},\\
H_{\rm cubic} &=& A_{ij}(\uv) \,(\ppi u_z) (\ppj u_z) ,\\
H_{\rm quartic} &=& B_{ijkl} \,
(\ppi u_z) (\ppj u_z)  (\ppk u_z) (\ppl u_z) ,
\end{eqnarray}
\end{subequations}
where 
\begin{subequations}
\label{AB_vertex}
\begin{eqnarray}
A_{ij}(\uv) &=& \frac{1}{2}\, \left[ (B_z - \lambda_{z\perp}) (\pz
  u_z) \delta_{ij} + (\lambda_{z\perp}-\lambda) (\nabla_{\perp} \cdot
  \uv_{\perp})\delta_{ij} - \mu (\ppi u_j + \ppj u_i) \right] ,\\
  B_{ijkl} &=&\frac{1}{24}\, \left(\lambda + 2\,\mu + B_z
  -2\,\lambda_{z\perp}\right) \left(\delta_{ij} \delta_{kl}+
  \delta_{ik} \delta_{jl} + \delta_{il} \delta_{jk}\right).
\end{eqnarray}
\end{subequations}
These can be conveniently represented diagrammatically, as illustrated
in Fig.~\ref{vertex}.
\begin{figure}[!htbp]
\begin{center}
\includegraphics[width=6.5cm]{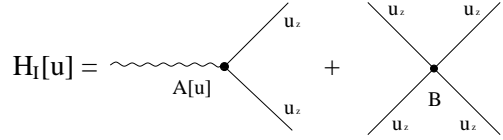}
\caption{Feynman diagrams for cubic and quartic nonlinearity,
respectively. The solid lines represent field $\ppi u_z$, while the
wiggly line represents field $A_{ij}[\uv]$.}
\label{vertex}
\end{center}    
\end{figure}

We calculate the cumulant expansion (\ref{cumulants_expansion}) up to
1-loop order, with every term in this expansion representable by a
Feynman diagram.  We refer the reader to reference \cite{CMP:CL} for a
introduction of graphical representation of a cumulant
expansion. Since we are after a correction that is a functional of
coarse-grained fields, with only high-wavevector fields integrated
out, all Feynman diagrams have harmonic propagators of $\uv^>$ as
internal lines and $\uv^<$ as external lines.

A basic property of a cumulant expansion, is that no disconnected
diagrams appear in the expansion.  Furthermore, momentum conservation
requires that all so-called one-particle-{\em reducible} diagram,
which can be separated into two pieces by cutting one internal line,
vanish identically.  An example is shown in Fig.~\ref{1PI}.
\begin{figure}[!htbp]
\begin{center}
\includegraphics[width=7cm]{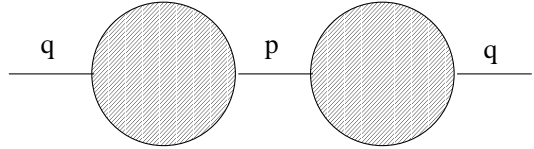}
\caption{A one-particle-{\em reducible} diagram. Because the internal
line carries a high momentum $p>Q\,e^{-\delta l}$ while the external
line carries lower momentum $q<Q\,e^{-\delta l}$, the diagram vanishes
identically under MSRG by momentum conservation.}
\label{1PI}
\end{center}    
\end{figure}
Therefore we only have to keep track all one-particle-{\em
irreducible} diagrams in our calculation.

The first-order cumulant in Eq.~(\ref{cumulants_expansion}) can be
represented as the sum of four Feynman diagrams shown in
Fig.~\ref{cumulant1}.
\begin{figure}[!htbp]
\begin{center}
\includegraphics[width=6cm]{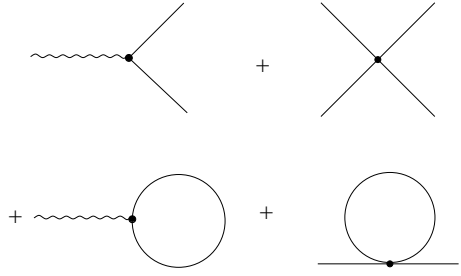}
\caption{Non-vanishing Feynman diagrams for the first cumulant in
Eq.~\ref{cumulants_expansion}. Other diagrams either vanish or are
only constants independent of fields $\uv^<$. External lines represent
field $u_z^<$ (solid lines) or $A_{ij}[\uv^<]$ (wiggly
lines). Internal lines represent $u_z^>$ or $A_{ij}[\uv^>]$.}
\label{cumulant1}
\end{center}    
\end{figure}
It is easy to see that the first two diagrams are just $H_I[\uv^<]$.
The last two diagrams generate terms such as
$$\int d^d x \,\, A_{ij}[\uv^<] \hspace{3mm} \mbox{and} \hspace{3mm}
\int d^d x \,\,(\nabla_{\perp}\uv^<_z)^2$$ which do not appear in the
original Hamiltonian Eq.~(\ref{f_expansion5}). However, they can be
expressed as a linear combination of $w_{zz}$ and $w_{ii}$ and
therefore can be eliminated by an appropriate shift of the nematic
reference state, as we will show explicitly later in this appendix.

By calculating the second order cumulants we find corrections to all
quadratic terms in the Hamiltonian, Eq.~(\ref{f_expansion4}), which
can be represented by the diagram in Fig.~\ref{quadradiagram}.
Because the functional form of the elastic Hamiltonian
Eq.~(\ref{f_expansion5}) is preserved by the renormalization group
transformation, these are only terms that we need to calculate.
Rotational symmetry (Ward identities) ensure that from these
corrections to quadratic terms we can infer corrections to all
anharmonic (cubic and quartic) terms appearing in
Eq.~\ref{f_expansion5}.

The part of the second order cumulant that renormalizes elastic
constants $B_z$, $\lambda_{z\perp}$, $\lambda$ and $\mu$ is given by
\begin{equation}
- A_{ij} [\uv^<] A_{kl} [\uv^<] 
\int^> \frac{d^d p}{(2\,\pi)^d} \,\,
 p_i\,p_j\,p_k\,p_l\,G_{zz}(\pv)^2,
 \label{dH_011}
\end{equation}
where 
\begin{equation}
\int^> \frac{d^d p}{(2\,\pi)^d} = 
\frac{Q^{d-1}}{(2\,\pi)^d}\delta l
\int d \Omega_{d-1} \int d p_z  
\end{equation}
is the integral over the momentum shell $(-\infty < p_z < \infty,
Q\,e^{-\delta l} < p_{\perp} < Q)$.  $d \Omega_{d-1}$ is the
differential surface element of a unit sphere in a $d-1$ dimensional
space. The integral of the product $p_i\,p_j\,p_k\,p_l$ over the $d-2$
sphere can be easily calculated to give:
\begin{equation}
\int d \Omega_{d-1} \, p_i\,p_j\,p_k\,p_l
= \frac{\Omega_{d-1} Q^4}{(d^2-1)} \left( \delta_{ij}\delta_{kl}
+ \delta_{ik}\delta_{jl} + \delta_{il}\delta_{jk} \right).
\end{equation}
Therefore Eq.~(\ref{dH_011}) reduces to 
\begin{eqnarray}
&& - \frac{\Omega_{d+3} Q^{d-1}\delta l}{(2\,\pi)^d}
\frac{1}{(d^2-1)}\left( \delta_{ij}\delta_{kl}
+ \delta_{ik}\delta_{jl} + \delta_{il}\delta_{jk} \right)
\,\,
 A_{ij} [\uv^<] A_{kl} [\uv^<] 
\int_{-\infty}^{\infty} \frac{d p_z}{\left(
\hat{\mu}p_z^2+K\,Q^4 \right)^2}
\nonumber\\ 
&=& - \frac{\Omega_{d-1}Q^{d-3}}{2\,(2\,\pi)^{d-1}}
 \frac{ \delta l}{(d^2-1) \sqrt{K^3 \hat{\mu}}}
\left(\frac{}{} (B_z-\lambda_{z\perp})^2 (\pz u_z^<)^2 \right.
+ 2 (B_z - \lambda_{z\perp})(\lambda+\mu - \lambda_{z\perp})
 (\pz u_z^<) (\nabla_{\perp} \cdot \uv_{\perp}^<) \nonumber\\
 &+& \left. (\lambda+\mu - \lambda_{z\perp})^2 
 (\nabla_{\perp} \cdot \uv_{\perp}^<)^2
 + \frac{1}{2} \mu^2 (\ppi u_j^<)^2\right).
\end{eqnarray}
According to the momentum shell RG scheme outlined above, the
preceding set of diagrammatic correction is to be equated to:
\begin{equation}
\frac{1}{2} \{ \delta_g B_z (\pz u_z^<)^2 
+ 2 \delta_g \lambda_{z\perp} (\pz u_z^<) (\ppi u_i^< )
           + (\delta_g \lambda + \delta_g \mu) (\ppi u_i^<)^2 
       + \delta_g \mu (\ppi u_j^<)^2 \},\label{dH_01}
\end{equation}
which gives graphic corrections to elastic constants $B_z$,
$\lambda_{\perp}$, $B_{\perp}$ and $\mu$ as shown in the second terms,
on the right hand sides of Eqs.~(\ref{dB}-\ref{dmu}).

The part of the second order cumulant that renormalizes the splay
constant $K$ is represented by the wavevector dependent part of the
Feynman diagram shown in Fig.\ref{quadradiagram}.  The corresponding
correction to Hamiltonian Eq.~(\ref{H_0}) is given by
\begin{eqnarray}
&&\frac{1}{2} \delta_g K \, q_{\perp}^4 |u_z(\qv)|^2 
= - \frac{1}{2}\cdot 2\cdot 2 \, q_i q_j q_k q_l |u_z(\qv)|^2 
\frac{1}{2} \frac{\partial^2}{\partial q_k \partial q_l}
\int^> \frac{d^d p}{(2\,\pi)^d}       \times \nonumber\\
&& \left[ \langle A_{im}[\uv(\pv+\qv)] 
A_{jn}[\uv(-\pv-\qv)] \rangle_0^>
 \langle p_m p_n |u_z(\pv)|^2\rangle_0^> \right. \nonumber\\
&+& \left. \langle A_{im}[\uv(\pv+\qv)] 
(p_n + q_n) u_z(\pv+\qv) \rangle_0^>
 \langle p_m u_z(\pv) A_{jn}(\pv)|^2
\rangle_0^> \right] \nonumber\\
&=& \frac{\Omega_{d-1}Q^{d-3}\delta l}{2(2\,\pi)^{d-1}}
\frac{\left( B_z (\lambda+2\,\mu) + 12\,\mu(\lambda+\mu)
-4\,\mu\,C-C^2 \right)} {16 (B_{\perp}+ \mu) 
\sqrt{K \hat{\mu}}} q_{\perp}^4 |u_z(\qv)|^2, 
\end{eqnarray}
and can be readily computed using Mathematica.  Since we are after a
result that is lowest order in $\epsilon$, the graphical correction
can be approximated by its value in $d=3$.  From above we obtain the
graphic correction to $K$ used in the second term on the right hand
side of Eq.~(\ref{dK}) of the main text.

Finally, we apply the rescaling transformation
$R_{\parallel}(e^{\omega\,\delta l})R_{\perp}(e^{\delta l})$ to
restore the momentum cutoff to $Q$. From Eqs.~(\ref{rescaling_z}) and
Eqs.~(\ref{rescaling_perp}) it is easy to see that the corrections to
various elastic constants from this rescaling are
\begin{subequations}
\begin{eqnarray}
(\delta_r B_z, \delta_r \lambda_{z\perp}, \delta_r \lambda, \delta_r \mu)
&=& (d+3-3\,\omega) (B_z, \lambda_{z\perp},\lambda,\mu) \delta l,
\nonumber\\
\\
\delta_r K &=& (d-1-\omega) K \delta l.  
\end{eqnarray}
\end{subequations}
Assembling these graphical and rescaling corrections to all elastic
constants, we obtain the RG flow equations (\ref{RGcorrections}) of
the main text.

\subsection{Generation of linear strain terms}
\label{RG-linear}
The vanishing of linear (in $\mm{w}$) terms in the ``bare'' elastic
Hamiltonian Eq.~(\ref{f_expansion4}) is a consequence of our choice of
the nematic reference state.  However, for this choice of the bare
Hamiltonian, due to thermal fluctuations and elastic nonlinearities,
thermal averages $\langle \ppz u_z\rangle$ and $\langle \ppi
u_i\rangle$ do not vanish in equilibrium.  This implies that the
nematic reference state, around which we expand the elastic
Hamiltonian, is not the true equilibrium state in the presence of
thermal fluctuations.  This manifests itself through a generation of a
nonzero linear terms $a_{z} w_{zz} + a_{\perp} w_{ii}$ by the
coarse-graining.

However, Ward identities, Eqs.~(\ref{ward1}), dictate that we must
also generate a term $\mu_{z\perp} (\nabla_{\perp}u_z)^2,$ with
$\mu_{z\perp}$ satisfying Eq.~{\ref{ward1}}.  This guarantees that
such terms assemble into a nonlinear strain $\mm{w}$ term, and can
therefore be shifted away by re-expanding the coarse-grained
Hamiltonian around the true (thermal fluctuations corrected) ground
state. Thus around the new state, both linear coefficients
$a_{z,\perp}$, as well as the shear modulus $\mu_{z\perp}$ are
guaranteed to vanish.  In this section, we explicitly verify that this
is indeed the case.

In the process of renormalization group transformation, cubic
nonlinearities $A_{ij}[\uv] \ppi u_z \ppj u_z$ generate terms linear
in the phonon field $\uv$, the so-called tadpole diagrams. At one-loop
order, this is represented by the third diagram in
Fig.~\ref{cumulant1} and is given by
\begin{equation}
\frac{\Omega_{d-1}Q^{d-1}\delta l}{2(2\,\pi)^{d-1}}
\frac{1}{8\,\sqrt{K^3\hat{\mu}}}
\left( - (\lambda+\mu-\lambda_{z\perp})\nabla_{\perp}\cdot \uv_{\perp}
+ (B_z - \lambda_{z\perp}) \ppz u_z \right).
\end{equation}
\end{widetext}

At the same time, the fourth diagram in Fig.~\ref{cumulant1} as well
as a part from the diagram in Fig.~\ref{quadradiagram} also generate a
quadratic contribution
\begin{equation}
\frac{\Omega_{d-1}Q^{d-1}\delta l}{2(2\,\pi)^{d-1}}
\frac{1}{16\,\sqrt{K^3\hat{\mu}}}
(\lambda+\mu+B_z-2\,\lambda_{z\perp}) (\nabla_{\perp}u_z)^2.
\end{equation}

It is straightforward to see that the sum of these two corrections can
be written in the form of
\begin{equation}
\delta a_z w_{zz} + \delta a_{\perp} w_{ii}
\label{linear_correction}
\end{equation}
with 
\begin{eqnarray}
\delta a_z &=& \frac{\Omega_{d-1}Q^{d-1}\delta l}{2(2\,\pi)^{d-1}}
 \frac{(B_z - \lambda_{z\perp})}{8\,\sqrt{K^3\hat{\mu}}},\\
\delta a_{\perp} &=&
 - \frac{\Omega_{d-1}Q^{d-1}\delta l}{2(2\,\pi)^{d-1}}
 \frac{(\lambda+\mu-\lambda_{z\perp})}{8\,\sqrt{K^3\hat{\mu}}}.
\end{eqnarray}

These linear terms in Eq.~(\ref{linear_correction}) can then be
eliminated by an appropriate shift of the nematic reference state.

\subsection{Renormalization of cubic and quartic elastic nonlinearities}

As we discussed in the main text, rotational invariance requires that
perturbative corrections to cubic and quartic nonlinearities be
related to those of the quadratic terms found above, so as to preserve
the form of the nonlinear elastic Hamiltonian.

Diagrammatically, corrections to cubic terms $A_{ij}[\uv]\ppi u_z \ppj
u_z$ can be represented by Feynman diagrams in
Fig.~\ref{forkdiagram}. Detailed, somewhat technically involved
calculations, that we omit here give:
\begin{widetext}
\begin{eqnarray}
\delta_g H_{cubic} &=&  
\frac{\Omega_{d-1}Q^{d-1}\delta l}{2(2\,\pi)^{d-1}}
\frac{1}{64 \,\sqrt{K^3\,\hat{\mu}}}
\left( \frac{}{} - 2(B_z - \lambda_{z\perp})
(B_z + \lambda +\mu-2\,\lambda_{z\perp}) 
(\ppz u_z) \delta_{ij}\right.,\nonumber\\
 &+& ( 2\,\left( \lambda - \lambda_{z\perp}  \right) \,
   \left( B_z - 2\,\lambda_{z\perp} + \lambda  \right)  + 
  2\,\left( B_z - 3\,\lambda_{z\perp} 
 + 2\,\lambda  \right) \,\mu   
 +  {\mu }^2) (\nabla_{\perp}\cdot\uv_{\perp}) \delta_{ij}
\nonumber\\
&+& \left.\mu^2 (\ppi u_j + \ppj u_i)  \frac{}{}\right)
 (\ppi u_z) (\ppj u_z).
\end{eqnarray}
\begin{figure}
\begin{center}
  \includegraphics[width=8cm,height=6cm]{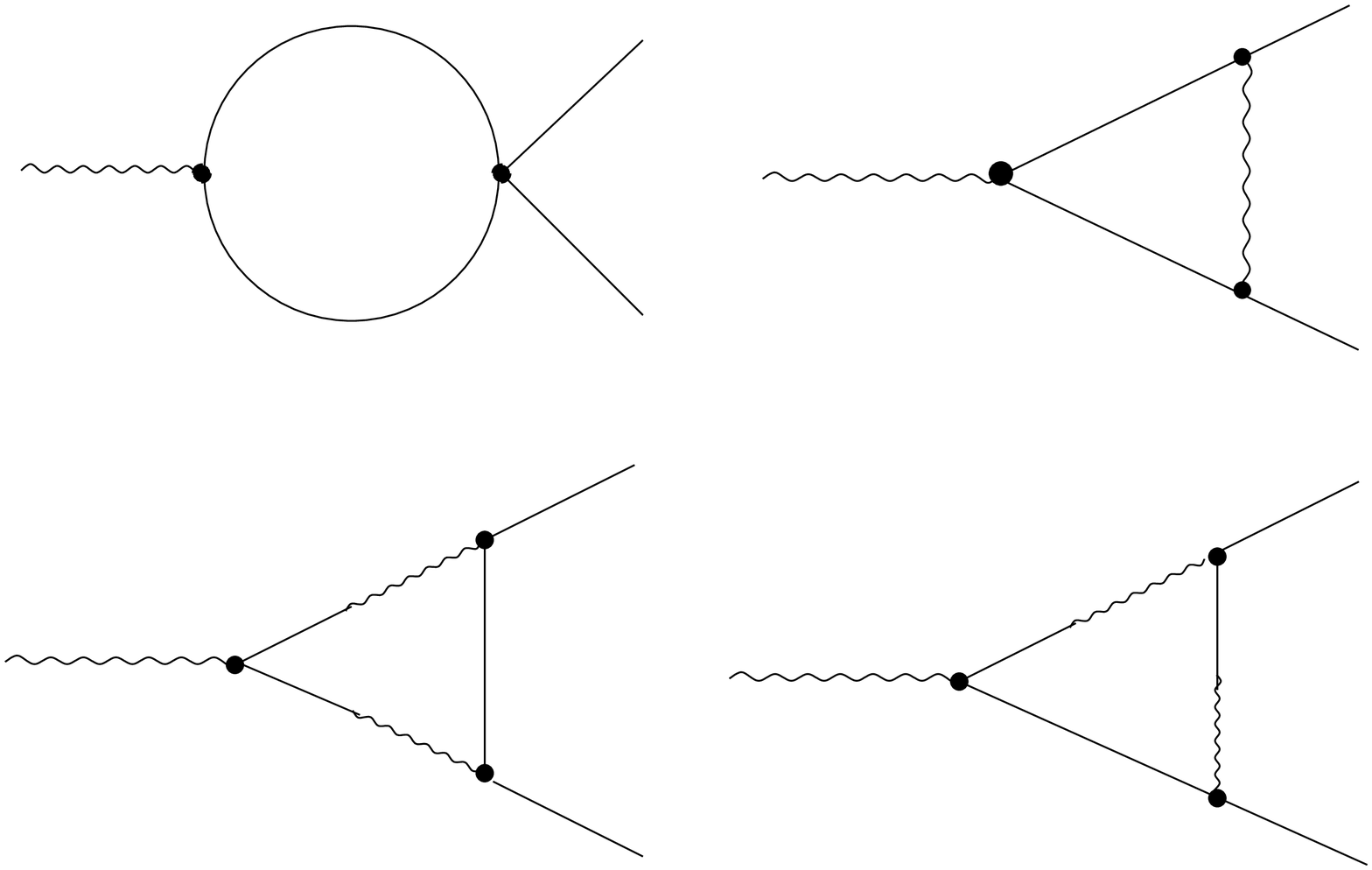}
  \vspace{0.5cm}
\caption{Feynman diagrams renormalizing the cubic elastic terms.}
\label{forkdiagram}
\end{center}    
\end{figure}
Interpreting this as a correction to elastic constants of a
coarse-grained cubic terms from Eqs.~(\ref{H-anhar}) and
Eq.~(\ref{AB_vertex})
\begin{equation}
\frac{1}{2} [ (\delta_g B_z - \delta_g \lambda_{z\perp})
  (\pz u_z) \delta_{ij} 
  + (\delta_g \lambda_{z\perp} - \delta_g \lambda) (\nabla_{\perp}
  \cdot \uv_{\perp})\delta_{ij} 
   - \delta_g  \mu (\ppi u_j + \ppj u_i) ]  (\ppi u_z \ppj u_z),
\end{equation}
we find that these corrections, $\delta_g B$,
$\delta_g\lambda_{z\perp}$,\ldots are identical to those obtained from 
the quadratic term above.

Similarly, fluctuation correction to the quartic term, represented by
diagrams in Fig.~\ref{crossdiagram}, is given by
\begin{equation}
\delta_g H_{quartic} =
\frac{\Omega_{d-1}Q^{d-1}\delta l}{2(2\,\pi)^{d-1}}
 \frac{\left( 2\,{\left( B_z - 2\,\lambda_{z\perp} 
  + \lambda  \right) }^2 
  +   4\,\left( B_z - 2\,\lambda_{z\perp}
  + \lambda  \right) \,\mu  
  + 3\,{\mu }^2\right)} 
 {256 \,\sqrt{K^3\,\hat{\mu}}}
 (\nabla_{\perp}\cdot u_z)^4 ,
\end{equation}
which when identified with
\begin{equation}
\frac{1}{8} \left( \delta_g \lambda + 2 \,\delta_g \mu + \delta_g B_z
- 2\,\delta_g \lambda_{z\perp} \right) (\nabla_{\perp} u_z)^4,
\end{equation} 
again gives the same corrections to elastic moduli as quadratic and
cubic terms. 

\begin{figure}
\begin{center}
  \includegraphics[width=8cm,height=6cm]{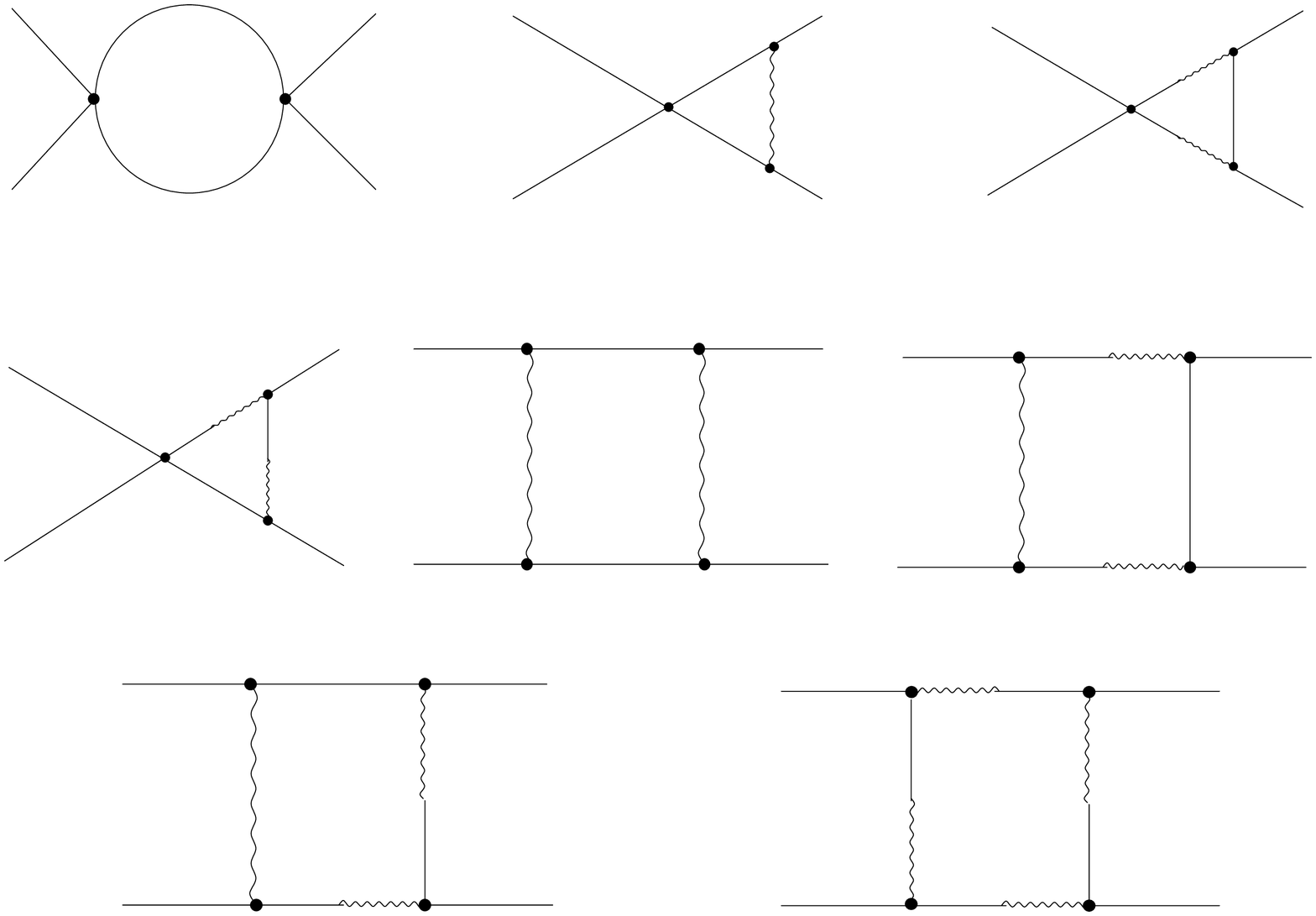}
  \vspace{0.5cm}
\caption{Feynman diagrams renormalizing quartic term}
\label{crossdiagram}
\end{center}    
\end{figure}
        
Thus, as required by symmetry we have demonstrated explicitly that the
functional form of the elastic Hamiltonian Eq.~(\ref{f_expansion5})
(encoded through the nonlinear form of the strain tensor $\mm{w}$) is
preserved by thermal fluctuations and therefore by the coarse-graining
RG transformation.
\end{widetext}

\subsection{Solution of RG flow equations for $d\neq3$}
\label{RGsolution-d<3}

In this appendix, we solve the renormalization group equations 
(\ref{flow_couplings}) for $d\neq 3$.

For $d>3$, $\epsilon < 0$ and it is clear that only the Gaussian fixed
point $g_{\rm L}^*=g_{\perp}^* =0$ is stable.  This means that elastic
nonlinearities are irrelevant, and the harmonic approximation becomes
asymptotically exact, with subdominant corrections (from elastic
nonlinearities), that can be computed in a controlled perturbative
expansion in elastic nonlinearities.

For $d<3$, it is clear that elastic nonlinearities destabilize the
Gaussian fixed point, with two nonlinear, dimensionless couplings
$g_{\rm L}(l)$ and $g_{\perp}(l)$ expected to flow to a finite fixed
point.  We first look at the flow equations for the dimensionless
ratios $x(l)$ and $y(l)$, Eqs.~(\ref{flow_xy}).  As discussed in
Sec.~\ref{Sec:Thermal-fluct}, for physical elastomers, the bare values
of $x$ and $y$ are much less than unity.  Furthermore, for a finite
and positive $g_{\rm L}(l)$ it is easy to see from
Eqs.~(\ref{flow_xy}) that both $x(l)$ and $y(l)$ flow to zero as
$l\rightarrow \infty$.  Therefore to a very good approximation, which
becomes asymptotically exact, we can set $x(l), y(l)$ to zero in the
flow equations for $g_{\rm L}$ and $g_{\perp}$, Eq.~(\ref{flow_gL})
and Eq.~(\ref{flow_gT}), which then drastically simplify to:
\begin{subequations}
\begin{eqnarray}
\frac{d \,g_{\rm L}}{d \,l} &=& \epsilon \, {g_{\rm L}} 
-\frac{g_{\rm L} \left(5 g_{\rm L}^2+34 g_{\perp} g_{\rm L}+26
   g_{\perp}^2\right)}{16 (g_{\rm L}+g_{\perp})},
   \nonumber\\
   \\
 \frac{d \, g_{\perp}}{d \, l} &=& \epsilon g_{\perp} 
-\frac{g_{\perp} \left(g_{\rm L}^2+32 g_{\perp} g_{\rm L}+28
   g_{\perp}^2\right)}{16 (g_{\rm L}+g_{\perp})},
   \nonumber\\
\end{eqnarray}
\end{subequations}
Solving these two equations we find 4 fixed points, that we list along
with the corresponding set of $\eta$ exponents in
Table~\ref{fixedpoints}.

\begin{table}[!hbt]
\begin{center}
\begin{tabular}{|c|c|c|c|c|c|c|c|}
\hline\hline Fixed point & $g_{\rm L}^*$ & $g_{\perp}^*$
& $\eta_B$ & $\eta_{\rm L} = \eta_{C}$ &$\eta_{\perp}$&$\eta_K$\\
\hline G & 0 & 0 & 0 & 0&0&0 \\
S &$ \frac{16 \epsilon}{5}$&0& $0$ &
 $\frac{4 \epsilon}{5}$&0& $\frac{2 \epsilon}{5}$\\
X & 0 & $\frac{4 \epsilon}{7}$ &0&0&$\frac{\epsilon}{14}$&0\\
E &$\frac{16 \epsilon}{59}$& $\frac{32 \epsilon}{59}$
 & 0 & $\frac{4 \epsilon}{59}$&$\frac{4 \epsilon}{59}$
&$\frac{38 \epsilon}{59}$
\\\hline
\end{tabular}
\caption{Fixed point and corresponding dimensionless couplings and
$\eta$ exponents, with $x^*=y^*=0$ at all fixed point.}
\label{fixedpoints}
\end{center}
\end{table}

\begin{figure}[bhtp]
\begin{center}
  \includegraphics[width=7cm]{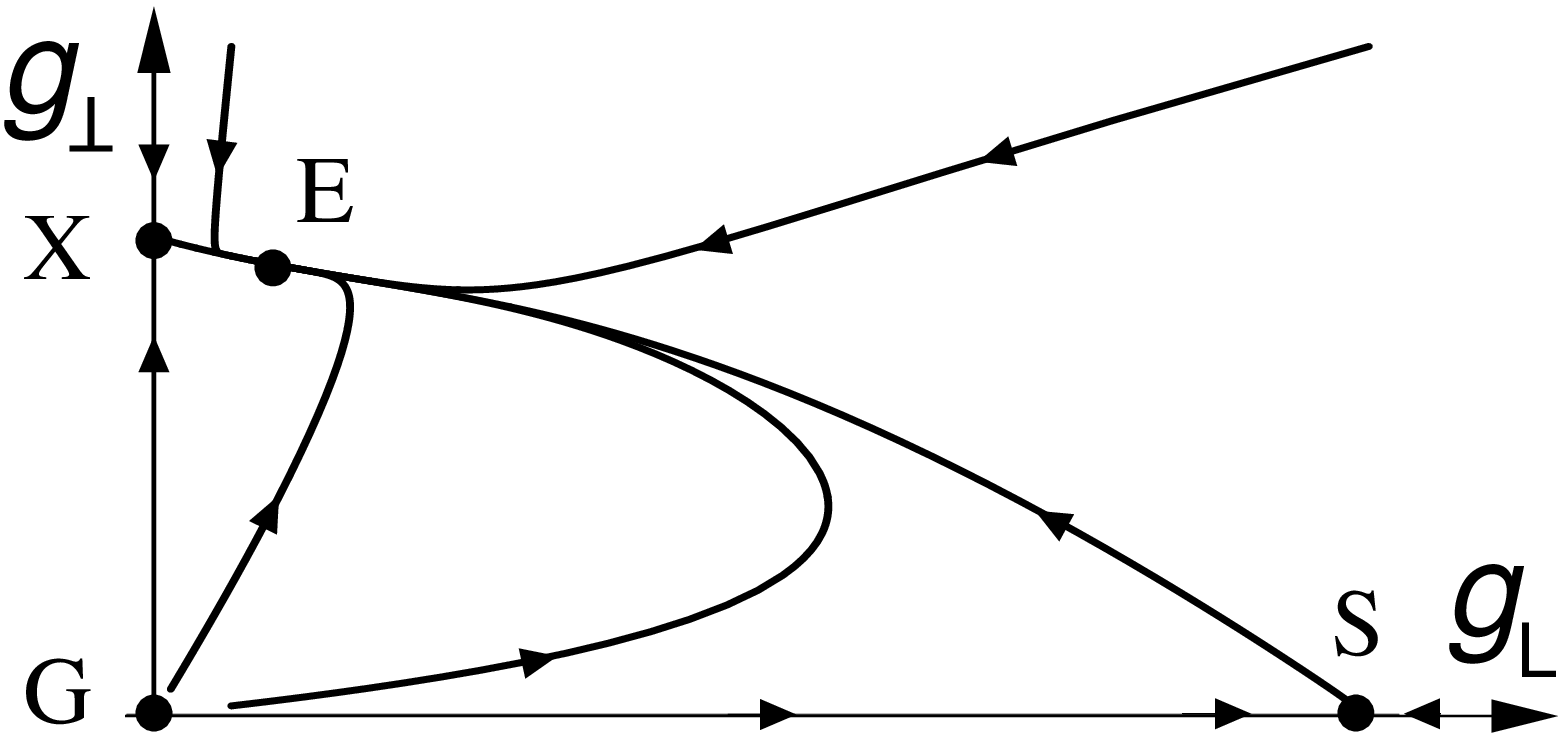}
\caption{Flow diagram for coupling constants $g_{\rm L}$ and
$g_{\perp}$ of an ideal {\em homogeneous} nematic elastomer.  $S$ is
the smectic fixed point discovered by Grinstein and Pelcovits
\cite{GP1}. The globally attractive fixed point E controls the
long-scale properties of an ideal homogeneous nematic elastomer,
characterizing its universal anomalous elasticity.}
\label{flowdiagram-2}
\end{center}    
\end{figure}

The flow pattern of $g_{\rm L}$ and $g_{\perp}$ under renormalization
group transformation for $d<3$ is shown in
Fig.~\ref{flowdiagram-2}. It is interesting to note that the critical
exponents $\eta_{\rm L}$ and $\eta_K$ at fixed point S are identical
to those of smectic liquid crystal in $3-\epsilon$ dimension, first
studied by Grinstein and Pelcovits \cite{GP1}.  Since this fixed point
S is attractive for the bare value of the coupling $g_{\perp}=0$
(proportional to the transverse shear modulus $\mu$) a nematic
elastomer with $\mu=0$ shares the same long wavelength ``anomalous
elasticity'' with a smectic liquid crystal.  This is not merely a
coincidence, as it is clear from Eq.~(\ref{f_expansion5}) that for a
vanishing $\mu$, the phonon field $\uv_{\perp}$ can be integrated out.
The resulting effective model is exactly a smectic liquid crystal with
a one dimensional phonon field $u_z$ and a shifted compressional
modulus.  Another way to understand this result is to note that with a
vanishing in-plane transverse shear modulus $\mu$, our elastomer model
Eq.~(\ref{f_expansion5}) is essentially a stack of liquid membranes
with $u_z$ the inter-plane displacement.  Physically, this is clearly
equivalent to a smectic liquid crystal.

Fig.~\ref{flowdiagram-2} shows that this smectic fixed point, S is
unstable to a turning on a finite in-plane shear modulus, with the
flowing toward a globally stable fixed point E at finite couplings
$g_{\perp}$ and $g_{rm L}$. It thus controls the long length-scale
physics of ideal uniaxial nematic elastomers below three dimensions.

Although as noted above, physical elastomers are usually characterized
by small $x$ and $y$ parameters, it is interesting to examine a
special case of $y(0) = \pm 1$.  As is clear from Eqs.~(\ref{flow_xy})
for $y=\pm 1$ neither $x(l)$ nor $y(l)$ flow, giving two unstable
fixed lines $(0\leq x \leq \infty,y=\pm1)$ that correspond to very
special systems.  Setting $y=\pm 1$ in Eqs.~(\ref{flow_couplings}), we
obtain considerably simplified flow equations for $g_{\rm L}$ and
$g_{\perp}$ for this special case:
\begin{subequations}
\begin{eqnarray}
\frac{d \, g_{\rm L}} {d \, l} 
  &=&   \epsilon \, g_{\rm L}
-\frac{g_{\rm L}^2 \left(4 g_{\rm L}
   \left(3 \mp \sqrt{x}\right)^2+g_{\perp} \left(89 x \mp 210
   \sqrt{x}+234\right)\right)}{16 \left(g_{\rm L}
   \left(3 \mp \sqrt{x} \right)^2+9 g_{\perp} x\right)} , \nonumber
       \\
       \\
\frac{d \, g_{\perp}} {d \, l}
 &=&    \epsilon \, g_{\perp}
 -\frac{g_{\perp}^2 \left(18 g_{\perp} x+g_{\rm L} \left(55 x \mp 222
   \sqrt{x}+252\right)\right)}{16 \left(g_{\rm L}
   \left(3 \mp \sqrt{x}\right)^2+9 g_{\perp} x\right)} . 
   \nonumber
   \\
\end{eqnarray}
\end{subequations}
\begin{widetext}
Solving these two equations, we find following fixed lines
parameterized by $x$:
\begin{enumerate}
\item Line of unstable fixed points:
$$(g_{\rm L}^*=g_{\perp}^*=0, y^*=\pm 1)$$ 
 These two lines are unstable with respect to all directions.
\item Line of mixed fixed points:
 $$(g_{\rm L}^*=4 \epsilon,
  g_{\perp}^* = 0,y^*=\pm 1)$$ These two lines are unstable in the 
  $g_\perp$ direction and stable in the $g_{\rm L}$ direction.
\item Line of mixed fixed points:
 $$(g_{\rm L}^*=0, g_{\perp}^* =
8\,\epsilon,y^* = \pm 1)$$ These two lines are unstable in the $g_{\rm
L}$ direction and stable in the $g_{\perp}$ direction.
\item Line of stable fixed points:
$$(g_{\rm L}^* =\frac{8 \left(19 x \mp 6 \sqrt{x}+9\right)}{91 x \mp
222 \sqrt{x}+252} ,g_{\perp}^* = 2\, g_{\rm L}^*,y^*=\pm 1 )$$ These
two lines are stable with respect to both $g_{\rm L}$ and
$g_{\perp}$. $\eta$ exponents for various elastic constants explicitly
depend on a continuous parameter $x$:
\begin{subequations}
\begin{eqnarray}
 \eta_B &=& \eta_{\rm L} = \eta_C = \eta_{\perp}     
=\frac{2 \left(19 x \mp 6 \sqrt{x}+9\right)}{91 x \mp 222 \sqrt{x}+252} ,
         \nonumber\\
         \\
 \eta_K &=& \frac{16 \left(8 x \mp 24 \sqrt{x}+27\right)}
 {91 x \mp 222 \sqrt{x}+252}.
 \end{eqnarray} 
 \end{subequations}
\end{enumerate}

Therefore systems characterized by $y^2 = C^2/B\,\mu_{\rm L} =1$
exhibit a continuous family of universal long length-scale elastic
properties. It is not clear to us at the moment what physical system
is characterized by $y=1$ and therefore shares properties given by
these fixed points.
\end{widetext}

\section{Model of heterogeneity in nematic elastomer}
\label{App:Disorder-deriv}

In this appendix we derive the disorder Hamiltonian
Eq.~(\ref{H_disorder}) starting with a phenomenological, symmetry-based 
disorder model in Eq.~(\ref{model:disorder}):
\begin{eqnarray}
H_{\mbox{d}} 
 = -  \int d^d X \, \left[ \Tr \,\mm{F}_1(\Xv) \Lm^{\rm T} \Lm 
 + \Tr \,\mm{F}_2(\Xv) \Lm^{\rm T} \mm{Q}(\Xv) \Lm 
 \right], \nonumber\\
  \label{model:disorder-2}
\end{eqnarray}
where $\Lm$ is the deformation gradient relative to the isotropic
reference state and is defined in Eq.~(\ref{Lambda_def}).  $\mm{F}_1$
and $\mm{F}_2$ are random tensor fields in the reference space.

Using Eq.~(\ref{Lambda-lambda}) to express $\Lm$ in terms of the
deformation gradient relative to the nematic reference state, $\lm$,
and defining two new tensor fields $\tilde{\mm{F}}_1(\Xv)$ and
$\tilde{\mm{F}}_2(\Xv)$ by
\begin{eqnarray}
\tilde{\mm{F}}_i = J_0^{-1} \Lm_0 \mm{F}_i \Lm_0^{\rm T},
\end{eqnarray} 
where $J_0$ is a Jacobian factor 
\begin{equation}
J_0 = \det \frac{\partial \xv}{\partial \Xv},
\end{equation}
we can express the disorder Hamiltonian Eq.~(\ref{model:disorder-2}) as
\begin{equation}
H_{\mbox{d}} 
 = -  \int d^d x \, 
 \left[ \Tr \,\tilde{\mm{F}}_1 \lm^{\rm T} \lm 
 + \Tr \,\tilde{\mm{F}}_2 \lm^{\rm T} \mm{Q} \lm 
 \right]. \label{model:disorder-3}
\end{equation}

Choosing the direction of the nematic order in NRS to be the $z$ axis,
$\hat{n}_0 = \hat{z}$, to first-order in $\delta \nh $ (where
$\delta\nh \cdot \hat{z} = 0$), the nematic order parameter $Q$ for
the current state is given by
\begin{eqnarray}
\mm{Q} &=& \mm{Q}_0 + S\,(\hat{z}\,\delta \hat{n}
        + \delta\hat{n}\,\hat{z}), 
       \label{Q-deltan}
\end{eqnarray}
where
\begin{eqnarray}
 \mm{Q}_0 &=& S\,(\hat{z}\hat{z} - \frac{1}{d} \mm{I}),       
\end{eqnarray}

Using Eq.~(\ref{lambda-u}) and Eq.~(\ref{Q-deltan}), we find that
up to linear order in $\uv$ and $\delta \nh$,
\begin{equation}
\Tr \,\tilde{\mm{F}}_1 \lm^{\rm T} \lm = 
\mbox{const.} + 2 (\hat{F}_1)_{ab} \varepsilon_{ab},
\end{equation}
and 
\begin{widetext}
\begin{equation}
\Tr \,\tilde{\mm{F}}_2 \lm^{\rm T} \mm{Q} \lm = 
\mbox{const. } + 2 S (\tilde{F}_2)_{iz} \left(
\frac{d-1}{d} \,(\partial_i u_z) -\frac{1}{d} \partial_z u_i 
+ \delta \nh_i \right)
+ \frac{2(d-1)}{d} S (\tilde{F}_2)_{zz} \partial_z u_z
-  \frac{1}{d} S (\tilde{F}_2)_{ij} \partial_i u_j , 
\label{tildeF2}
\end{equation}
\end{widetext}
where, as usual, indices $i$ and $j$ are limited to the $d-1$
dimensional subspace perpendicular to $z$ axis.

We proceed to integrate out the fluctuations of the nematic director
$\delta\nh$.  To the lowest order this amounts to making the
replacement Eq.~(\ref{replacement}) in Eq.~(\ref{tildeF2}).  It is
interesting to see that this replacement exactly cancels the
antisymmetric strain components $a_{zi}$ in Eq.~(\ref{tildeF2}), so
that as the final result, the disorder Hamiltonian only depends on the
linearized symmetric strain $\varepsilon_{ab}$:
\begin{eqnarray}
H_{\rm d} \rightarrow
- \int d^d x \, \sum_{a,b} \,\sigma_{ab}(\xv)\varepsilon_{ab},   
	\label{H_d-sigma}
\end{eqnarray}
where the random stress field ${\boldsymbol \sigma}(\xv)$ is given by
\begin{subequations}
\label{sigma-F}
\begin{eqnarray}
\sigma_{zz} &=& 2 (\hat{F}_1)_{zz}+ 
\frac{2(d-1)}{d}S ( \tilde{\mm{F}}_2 )_{zz},\\
\sigma_{ij} &=& 2 (\hat{F}_1)_{ij}
- \frac{2}{d}S ( \tilde{\mm{F}}_2 )_{ij},\\
\sigma_{iz} &=& \sigma_{zi} = 
2 (\hat{F}_1)_{iz} +
\left( \frac{d-2}{d} + \frac{r+1}{r-1} \right) S 
(\tilde{\mm{F}}_2)_{iz}.
\nonumber\\
\end{eqnarray}
\end{subequations}
We note again that because random stress couples {\em linearly} to
strain, the ideal nematic reference state defined by $\rv(\xv) = \xv$
is not the real ground state in the presence of this quenched random
stress.

At three dimension, a symmetric tensor field ${\boldsymbol \sigma}$,
generically contains a longitudinal part satisfying $\nabla \times \
{\boldsymbol \sigma}^L = 0$, and a transverse part satisfying $\nabla
\cdot \ {\boldsymbol \sigma}^T = 0$.  In Eq.~(\ref{H_d-sigma}),
however, because ${\boldsymbol \sigma}$ is coupled to the linearized
strain, its transverse part ${\boldsymbol \sigma}^T$ has no bulk
effect to the system, since it disappears after a simple integration
by parts.  Therefore only the {\em longitudinal} part of the random
stress are included in our model.  On the other hand, if we were to
keep higher order terms in linearized strains in our derivation, we
would have found the reduced disorder Hamiltonian to be
\begin{eqnarray}
H_{\rm d} =  - \int d^d \xv \, \sigma_{ab}(\xv) \, e_{ab}(\xv),
\end{eqnarray}
rather than Eq.~(\ref{H_d-sigma}).  That is, the random initial stress
field $\sigma_{ij}$ is coupled to the nonlinear Lagrange strain,
instead of the linearized strain $\varepsilon_{ij}$.  This result is
of course guaranteed by rotational symmetry: the linearized strain is
{\em not} rotational invariant in the embedding space, while the
nonlinear Lagrange strain is.  In the resulting model, then the
transverse part of the random initial stress, $\sigma^T_{ij}$ will
couple to the nonlinear part of the Lagrange strain as shown in
\rfs{sigma-uu}.  In this work, however, we shall only analyze the
simplified disorder Hamiltonian Eq.~(\ref{H_d-sigma}), i.e. we shall
completely ignore the transverse part of the random stress.

Because of the vanishing of quadratic term $\varepsilon_{i z}^2$ in
the elastic energy Eq.~(\ref{H_0}), it is not difficult to see (using
power-counting in previous sections) that the components $\sigma_{z
  i}$ are the most relevant part of the random stress tensor field
$\sigma_{ab}$.  We therefore focus on these components, neglecting all
others:
\begin{eqnarray}
H_{\rm d} \approx - \int d^d x \, \sum_{i}
\,\sigma_{iz}(\xv)\varepsilon_{iz} ,
	\label{H_d-sigma-2}
\end{eqnarray}
where again $i$ is restricted to the $d-1$ dimensional subspace
perpendicular to $\hat{z}$.  The random field $\sigma_{zi}(\xv)$ is
assumed to be Gaussian with a short range correlation:
\begin{equation}
\overline{\sigma_{zi}(\xv)\,\sigma_{zi}(\xv')}
 =      \Delta \, \delta_{ij}\,\delta^d(\xv-\xv'). 
        \label{disorder_variance}
\end{equation}

\section{Replica trick}
\label{App:Replica}
For completeness, in this appendix we briefly review a convenient
technology, the so-called ``replica trick'' for treating field theory
in the presence of a quenched random field $\sigma(\xv)$. Thus we
consider a general heterogeneous system described by a Hamiltonian
$H[\uv, \sigma(\xv)]$, where $\uv = \uv(\xv)$ is the physical degree
of freedom, while $\sigma(\xv)$ denotes all quenched random
parameters. We are interested in the quenched average of the free
energy
\begin{equation}
\beta\,\overline{F} = - \overline{\log Z} 
= - \overline{\log\int D\uv e^{-\beta H[\uv,\delta]}}
\label{replica-F}
\end{equation}
and of other physical quantities (correlation functions)
\begin{equation}
\overline{\langle O[\uv] \rangle} = 
\overline{\frac{1}{Z}\int D\uv O[\uv]\,e^{-\beta H[\uv,\delta]}},
\label{replica-O}
\end{equation}
as average representation of a heterogeneous sample characterized by
$\sigma(\xv)$. Above, we use $\langle O \rangle$ for the average over
thermal fluctuations (functional integral over $\uv$), and use
$\overline{O}$ to denote the average over disorder realizations
$\sigma$. The main difficulty in the calculation of
Eq.~(\ref{replica-F}) and Eq.~(\ref{replica-O}) is due to the $\log$
and $Z^{-1}$ inside the disorder average.

To overcome these obstacles, we define a $n$-replicated Hamiltonian
$$H_n[\uv^{\alpha}] = H_n[\uv^1,\uv^2,\ldots,\uv^n]$$ as well as the
associated partition function $Z_n$ and the free energy $F_n$ by:
\begin{eqnarray}
e^{-H_n/T} &=& \overline{\prod^{n}_{\alpha=1}
     e^{-H[\uv^{\alpha},\sigma]/T}},\\
Z_n &=& \int[\prod^n_{\alpha = 1}D\uv^{\alpha}] e^{-H_n/T}
     = \overline{Z^n},\\
F_n &=& -T\,\log Z_n,
\end{eqnarray} 
where $n$ is an integer.  Even though the replicated theory is only
well-defined for integer $n$, we analytically continue it to a real
number $0<n<1$ so that $F_n$ becomes a function of a continuous
variable $n$. The key advantage of this replica method is that the
resulting replicated Hamiltonian $H_n[\uv^{\alpha}]$ is disorder-free
and can therefore be studied using standard tools for treatment of a
{\em homogeneous} field theory.  Using the identity
\begin{equation}
 T\,\overline{\,\log\,Z\,} =  \lim_{n\rightarrow 0} 
        {\frac{T}{n} \log\, \overline{Z^n}},
\label{replica_identity}
\end{equation}
we easily see that the disorder averaged free energy $\overline{F}$ is
related to the replicated free energy $F_n$ by
\begin{equation}
\overline{F} = \lim_{n\rightarrow 0} 
\frac{1}{n} \,F_n. 
\end{equation}
A potential problem of the replica technique is the breakdown of
commutability of the thermodynamic limit with the replica
$n\rightarrow 0$ limit, on which identity Eq.~(\ref{replica_identity})
is based. It can be shown that for problems (like the one at hand),
where no real glass physics emerges (as in a spin glass problem),
replica trick is innocuous and is simply a convenient way to throw out
unphysical diagrams \cite{RadzihovskyNelsonPRA92}.  
  
The replica technique can also be used to calculate the
disorder-averaged correlation functions, such as
Eq.~(\ref{replica-O}).  To see this, we let $\langle O^{\alpha}\rangle
\equiv \langle O[u^{\alpha}]\rangle $ be the ``thermal'' average of a
replicated analogue of an observable $O$ in a replicated theory.  We
can show that it is related to $\overline{\,\langle O \rangle \,}$,
the same physical quantity averaged over both thermal fluctuations and
quenched disorders, by
\begin{eqnarray}
\langle O^{\alpha} \rangle &\equiv& 
\frac{1}{Z_n} \int[\prod D\uv^{\beta}]\,
O^{\alpha} e^{-H_n/T}  \nonumber\\
&=&  (\overline{Z^n})^{-1} 
\overline{\int_{\beta \neq \alpha}
e^{-\sum_{\beta \neq \alpha}H^{\beta}/T}
\int_{\alpha} O^{\alpha}\,e^{-H^{\alpha}/T} }
\nonumber\\
&=&  (\overline{Z^n})^{-1} \overline{Z^n\, \langle O[\uv] \rangle}
\rightarrow \overline{\langle O[\uv] \rangle},  
\hspace{5mm} \mbox{as $n\rightarrow 0$},    
\end{eqnarray}
where $H^{\alpha} \equiv H[\uv^{\alpha}, \sigma]$, and in the last
step we have used
\begin{equation}
\overline{Z^n} = 1 + n\, \log Z + \cdots \rightarrow 1. 
\end{equation}
Using similar technique we find
\begin{subequations}
\label{replica_identity-2}
\begin{eqnarray}
\langle A^{\alpha}\rangle &=&\overline{\langle A \rangle },\\
\langle A^{\alpha} B^{\beta}\rangle &=&
\overline{\langle A \rangle \langle B \rangle},\\
\langle A^{\alpha} B^{\beta}\cdots C^{\gamma} \rangle &=&
\overline{\langle A \rangle \langle B \rangle
\cdots\langle C \rangle },
\end{eqnarray}
\end{subequations}
where $\alpha$, $\beta$ $\cdots$, $\gamma$ are assumed to be all
different.

The correlator matrix $G^{\alpha\beta}_{ab}$ of the replicated theory,
defined by
\begin{eqnarray}
G^{\alpha\beta}_{ab} =  \langle u^{\alpha}_a u^{\beta}_b \rangle
\end{eqnarray}
therefore gives
\begin{eqnarray}
G^{\alpha\beta}_{ab} =  \overline{ \langle u_a \rangle \langle u_b \rangle}
\end{eqnarray}
for $\alpha\neq\beta$. It can be expressed as a sum of two parts:
\begin{eqnarray}
G^{\alpha\beta}_{ab} = G^{T}_{ab} \delta_{\alpha\beta} + 
G^{\Delta}_{ab}, 
\label{G-sum}
\end{eqnarray}
where 
\begin{eqnarray}
G^{T}_{ab} &=& \overline{ \langle u_a \, u_b \rangle}
-\overline{ \langle u_a \rangle \langle u_b \rangle}
\nonumber\\
&=&  \overline{(u_a - \langle u_a \rangle)
(u_b - \langle u_b \rangle)}
\end{eqnarray}
is a thermal correlator, characterizing thermal fluctuations around
the random ground state for a given quenched disorder realization,
while
\begin{eqnarray}
G^{\Delta}_{ab} = \overline{ \langle u_a \rangle \langle u_b \rangle}
\end{eqnarray}
gives the quenched correlator, characterizing the sample-to-sample
fluctuations of the ground states.  To the leading order, it can be
shown that $G^{\Delta}_{ab}$ is linear in the disorder variance
$\Delta$ but is independent of temperature $T$, while $G^T_{ab}$ is
linear in temperature.

\begin{widetext}
\section{Derivation of RG flow equations for heterogeneous elastomer}
\label{App:RG-disorder}

In this appendix we present details of the replicated RG analysis to
treat nonlinearities and random stresses of a heterogeneous
elastomer. The procedure we follow is quite similar to that in
Appendix \ref{App:RG-general} for a homogeneous system, but now
applied to a replicated heterogeneous elastomer Hamiltonian,
\rf{replicated_Hamiltonian}. To this end, we need to calculate the
cumulant expansion Eq.~(\ref{cumulants_expansion}) for the replicated
elastic Hamiltonian Eq.~(\ref{replicated_Hamiltonian}).  The harmonic
part of the Hamiltonian is given by Eq.~(\ref{har_disorder-2}), with
the temperature rescaled away, and $K_3$ set to zero.  The
nonlinearities are given by
\begin{eqnarray}
H_I[\uv^{\alpha}] = \sum_{\alpha=1}^n \left(
A_{ij}[\uv^{\alpha}]\, \ppi u_z^{\alpha} \ppj u_z^{\alpha} \right.
+ \left. B_{ijkl}\, \ppi u_z^{\alpha} \ppj u_z^{\alpha}  
 \ppk u_z^{\alpha} \ppl u_z^{\alpha} \right),
\end{eqnarray}
where $A_{ij}[\uv]$ and $B_{ijkl}$ are given in
Eqs.~(\ref{AB_vertex}).  We may represent these nonlinearities by the
Feynman diagrams shown in Fig.~\ref{vertex_replica}, with the
understanding that the replica index $\alpha$ is always summed over.

\begin{figure}[!htbp]
\begin{center}
\includegraphics[width=7cm]{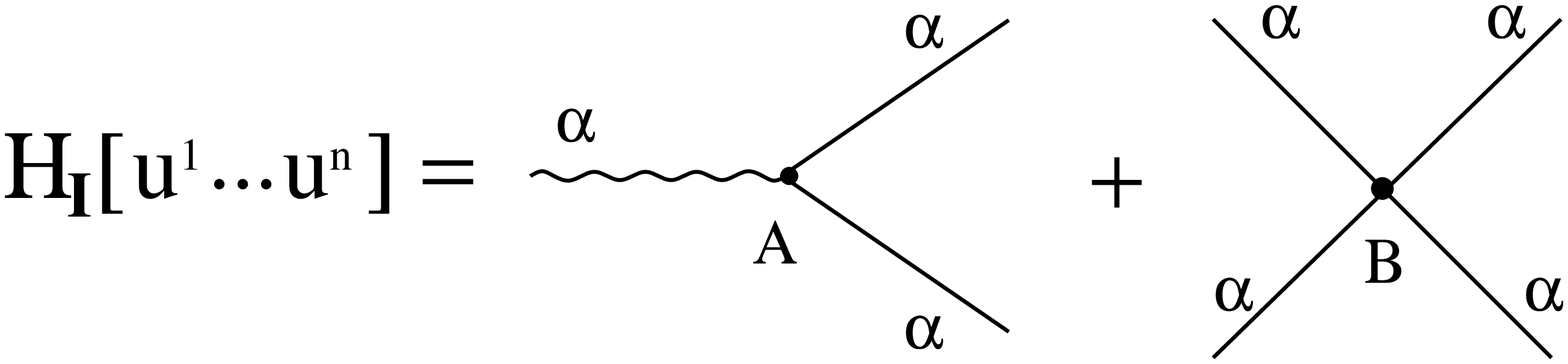}
\caption{Feynman diagrams for cubic and quartic nonlinearity in the
replicated elastic Hamiltonian Eq.~(\ref{replicated_Hamiltonian}).
The replica index $\alpha$ is summed over.  Solid lines represent
$\ppi u_z^{\alpha}$, while the wiggly line represents
$A_{ij}[\uv^{\alpha}]$.}
\label{vertex_replica}
\end{center}    
\end{figure}

As usual, we shall use an internal line (straight or wiggly) for a
high-wavevector harmonic propagator of the $u_z$ phonon field
${G}^{\alpha\beta}_{zz}$.  From Eq.~(\ref{Gzz_disorder}), we see that
this propagator is a sum of two terms.  The first term is diagonal in
replica indices describing thermal fluctuations.  The second term is
independent of replica indices and is proportional to the disorder
variance $\Delta$, describing quenched fluctuations.  We can
graphically represent this harmonic propagator (a solid line) by a sum
of two lines, corresponding to thermal ($T$) and disordered ($\Delta$)
contributions, illustrated in Fig.~\ref{propagator-sum}.

\begin{figure}[!htbp]
\begin{center}
\includegraphics[width=8cm]{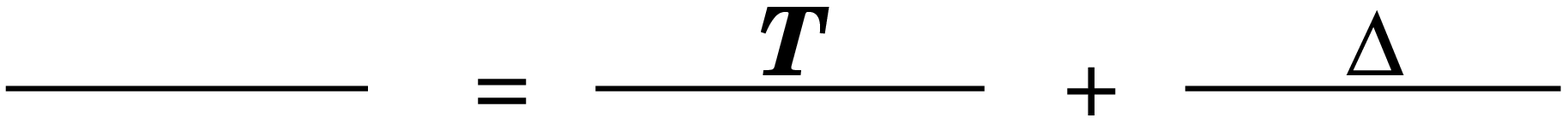}
  \vspace{0.5cm}
\caption{The replicated propagator is a sum of a thermal term (marked
with $T$) and a quenched term (marked with $\Delta$). }
\label{propagator-sum}
\end{center}    
\end{figure}

As in the thermal case, we only need to keep track of Feynman diagrams
that renormalize quadratic terms in the elastic Hamiltonian
Eq.~(\ref{har_disorder}); underlying rotational symmetry (Ward
identities) guarantee that nonlinearities have identical
renormalization, so as to preserve the form of the nonlinear strain
tensor $\mm{w}$.  We start with a Feynman diagram on the left hand
side of Fig.~\ref{quadra_replica}, which renormalize elastic moduli
$B_z$, $\lambda_{z\perp}$, $\lambda$ and $\mu$.  The corresponding
correction to the elastic Hamiltonian is therefore given by
\begin{eqnarray}
&& - \frac{1}{2}\,\cdot 2\cdot 2 \,\sum_{\alpha\beta} 
   A_{ij} [\uv^{\alpha}] A_{kl} [\uv^{\beta}] 
   \int^> \frac{d^d p}{(2\,\pi)^d} p_i\,p_j\,p_k\,p_l\,
   \left[G_{zz}(\pv) \left( \delta_{\alpha \beta}
   + \Delta p_{\perp}^2 G_{zz}(\pv) \right)\right]^2\nonumber\\
&=&  -  \sum_{\alpha\beta}
\frac{\Omega_{d-1} Q^{d+3}\delta l}{(2\,\pi)^d}
\frac{2}{(d-1)(d+1)}\left( \delta_{ij}\delta_{kl}
+ \delta_{ik}\delta_{jl} + \delta_{il}\delta_{jk} \right)
 A_{ij} [\uv^{\alpha}] A_{kl} [\uv^{\beta}] 
 \times \nonumber\\
&&\int_{-\infty}^{\infty} d p_z \left[
\delta_{\alpha\beta} (G_{zz}(\pv) + 2\,\Delta\,p_{\perp}^2
G_{zz}(\pv)^3) + \Delta^2 p_{\perp}^4 G_{zz}(\pv)^4 \right].
\label{dH01_replica}
\end{eqnarray}
\begin{figure}[!htbp]
\begin{center}
  \includegraphics[width=12cm]{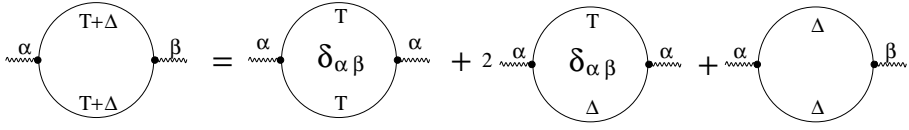}
  \vspace{0.5cm}
\caption{Feynman diagram renormalize quadratic couplings of the
elastomer model.  Each internal line represent a $u_z$ propagator,
which is itself a sum of two terms ( $T + \Delta $).  The first two
diagrams on the r.h.s. renormalize $B_z$, $\lambda_{z\perp}$,
$\lambda$ and $\mu$. The last diagram generates terms only irrelevant
terms that are therefore neglected.  The first term in r.h.s. comes
from thermal fluctuations and therefore is also irrelevant. }
\label{quadra_replica}
\end{center}    
\end{figure}

Using the decomposition shown in Fig.~\ref{propagator-sum}, the
Feynman diagram in the left side of Fig.~\ref{quadra_replica} can be
separated into three pieces, each proportional to $\Delta^0$,
$\Delta^1$ and $\Delta^2$ respectively.  This is illustrated on the
right hand side of Fig.~\ref{quadra_replica}.  It is clear that the
first two terms (first two Feynman diagrams on the right hand side of
Fig.\ref{quadra_replica}) renormalize the elastic constants $B_z$,
$\lambda_{z\perp}$, $\lambda$ and $\mu$, while the last piece
generates irrelevant contributions that will therefore be neglected.
The first piece is independent of the disorder variance $\Delta$ and
therefore describes thermal fluctuations.  According to our previous
analysis, it is less relevant than the second disorder-dependent piece
and can therefore be neglected.  We thus focus on the second piece in
Fig.~\ref{quadra_replica}, which corresponds to the second term in the
integrand in Eq.~(\ref{dH01_replica}).  Consequently
Eq.~(\ref{dH01_replica}) reduces to
\begin{eqnarray}
&-&  \sum_{\alpha} \frac{\Omega_{d-1}Q^{d-5}}
{2\,(2\,\pi)^{d-1}} \frac{ \delta l}{32 \sqrt{K^3 \hat{\mu}}} \times
 \nonumber\\
 &&
\left[ 6 (B_z-\lambda_{z\perp})^2 (\pz u_z^{\alpha})^2 
+  6 (B_z - \lambda_{z\perp})(2\lambda + \mu
 - 2\lambda_{z\perp}) (\pz u_z^{\alpha}) 
(\ppi_{\perp} \uv_i^{\alpha}) \right. \nonumber\\
&+& \left. 2 \left(3 \lambda ^2-6 \lambda_{z\perp} \lambda +3 \mu  \lambda
   +3 \lambda_{z\perp}^2+\mu ^2-3 \lambda_{z\perp} \mu
   \right) (\ppi \uv_i^{\alpha})^2
 +  \mu^2 (\ppi u_j^{\alpha})^2\right].
 \label{dH-0-1}
\end{eqnarray}
where $\hat{\mu}$ is defined in Eq.~(\ref{hatmu-def}) and we have set
$d=5$ in subsequent calculations.  Interpreting Eq.~(\ref{dH-0-1}) as
a coarse-graining correction of the harmonic part of the Hamiltonian,
i.e., equating it to
\begin{equation}
\frac{1}{2} \left[ \delta_g B_z (\pz u_z^{\alpha})^2 
+ 2 \delta_g \lambda_{z\perp} (\pz u_z^{\alpha}) (\ppi u_i^{\alpha} )
           + (\delta_g \lambda + \delta_g \mu) (\ppi u_i^{\alpha})^2 
       + \delta_g \mu (\ppi u_j^{\alpha})^2 \right].  \label{dH-0-2}
\end{equation}
we obtain the diagrammatic corrections to elastic moduli $B_z$,
$\lambda_{z\perp}$, $\lambda$ and $\mu$ to be:
\begin{eqnarray}
\delta_g B_z &=& 
	 - \psi_d \,\delta l\, \frac{3\, \Delta\, 
	(B_z - \lambda_{z\perp})^2}
        {8\, \sqrt{K^5\, \hat{\mu}}},
        \\
\delta_g \lambda_{z\perp} &=&  
	 \psi_d \,\delta l\, \frac{3\, \Delta\, 
	(B_z - \lambda_{z\perp}) 
        (2\lambda + \mu- 2\lambda_{z\perp})}
        {16\, \sqrt{K^5\, \hat{\mu}}},
	\\
\delta_g \lambda	&=&         
	-  \psi_d \,\delta l\,
	\frac{\Delta  \left(6 (\lambda -\lambda_{z\perp})^2+6 \mu 
   (\lambda -\lambda_{z\perp})+\mu ^2\right)}{16 \sqrt{K^5
   \hat{\mu }}},
   \\
\delta_g \mu		&=&
   -\frac{\Delta  \mu ^2}{16 \sqrt{K^5 \hat{\mu }}}. 
\end{eqnarray}

Feynman diagrams renormalizing $\Delta$ and $K$ are shown in
Fig.~\ref{Kdiagram_replica}.  As before, we can expand each diagram in
power of disorder variance $\Delta$.  It is not difficult to see that
the part linear in $\Delta$ renormalizes the splay constant $K$ while
the part proportional to $\Delta^2$ renormalizes $\Delta$ itself.
\begin{figure}
\begin{center}
  \includegraphics[width=9cm]{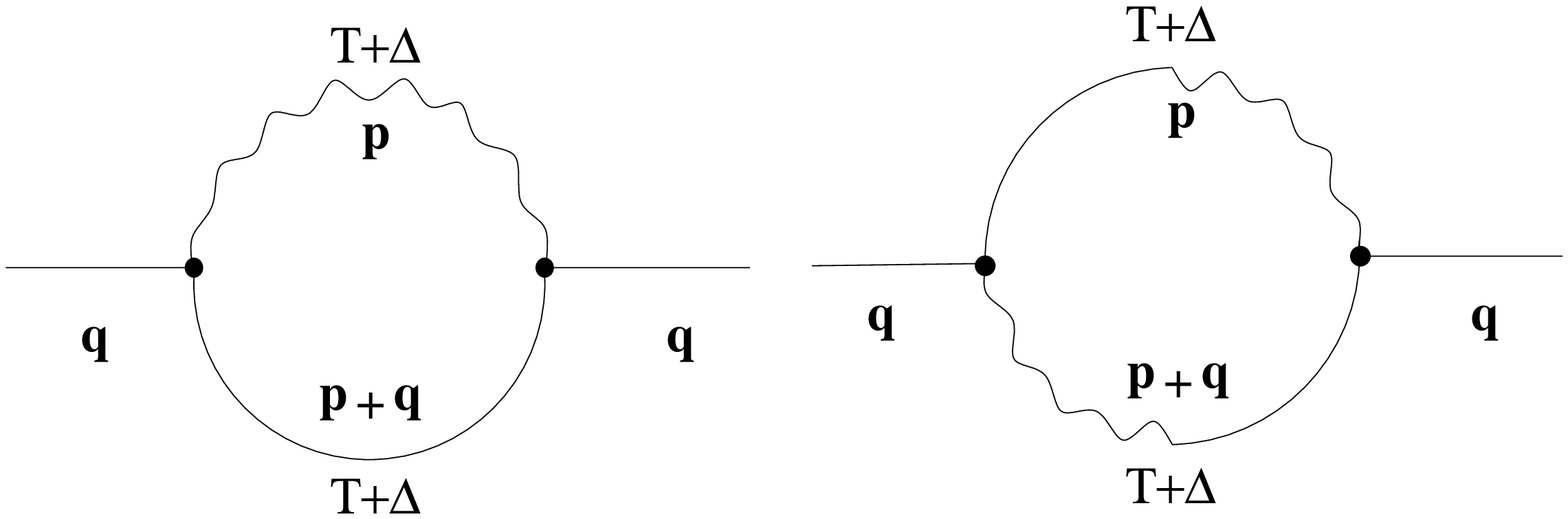}
  \vspace{0.5cm}
\caption{Feynman diagram that renormalize disorder variance $\Delta$
and splay constant $K$.}
\label{Kdiagram_replica}
\end{center}    
\end{figure}

The corresponding analytic expression (in momentum space) is given by
\begin{eqnarray}
&& - \frac{1}{2}\, \cdot 2\cdot 2\sum_{\alpha \beta}(q_i u_z^{\alpha}(\qv)) 
(q_k u_z^{\beta}(-\qv)) \int^> \frac{d^d p}{(2\,\pi)^d} \nonumber\\
&&\left( \langle A_{ij}[\uv^{\alpha}(\pv)] 
A_{kl}[\uv^{\beta}(-\pv)] \rangle_0^>
\langle (p_i+q_i) u_z^{\alpha}(\pv+\qv) 
(p_k +q_k) u_z^{\beta}(-\pv-\qv) \rangle_0^>\right. \nonumber\\
&+& \left. \langle A_{ij}[\uv^{\alpha}(\pv)]\,p_lu_z(-\pv)\rangle_0^>
  \langle (p_j+q_j) u_z^{\alpha}(\pv+\qv)
 A_{kl}[\uv^{\beta}(\pv+\qv)] \rangle_0^> \right) \nonumber\\
&=& \frac{1}{2}\sum_{\alpha \beta} \left( 
\delta_g K \,\,q_{\perp}^4 \delta_{\alpha \beta}
 + \delta_g \Delta \,\,q_{\perp}^2 \right)
u_z^{\alpha}(\qv) u_z^{\beta}(-\qv) + \mbox{irrelevant},
\end{eqnarray}
where, as illustrated in Fig.\ref{Kdiagram_replica}, we have used
$\vec{p}$ to label the large momenta that run over the momentum shell
$Q\,e^{\delta l} < p < Q$, while $\vec{q}$ is the small momentum below
the shell.

We use Mathematica to calculate the left hand side of this involved
expression and expand it in powers of the external momentum $\qv$.  As
is indicated on the right hand side of the same equation, we only need
to keep track of two types of terms: terms that are diagonal in
replica indices and proportional to $q_{\perp}^4$, thereby
renormalizing $K$, and terms that are are independent of replica
indices and proportional to $q_{\perp}^2$, thereby renormalizing the
disorder variance $\Delta$.  Furthermore, we keep only corrections to
$K$ and $\Delta$ from quenched fluctuations, as thermal fluctuations
are less relevant.  This ensures that $\delta_g K/K $ and $\delta_g
\Delta/\Delta$ are proportional to the disorder variance $\Delta$.
The final results are given by
\begin{eqnarray}
\delta_g K &=&  \frac{\Omega_{d-1} Q^{d-3}}{2\,(2 \, \pi)^{d-1}}
 \frac{\Delta \left( B_z(\lambda+2\,\mu) + 20\,\mu(\lambda+\mu) 
- \lambda_{z\perp}(\lambda_{z\perp} + 4\,\mu ) \right)}
{16(\lambda+2\,\mu)\sqrt{K^5\,\hat{\mu}}},\\
\delta_g \Delta &=& \frac{\Omega_{d-1} Q^{d-3}}{2\,(2\,\pi)^{d-1}}
\frac{\Delta^2\sqrt{\hat{\mu}}}{32\sqrt{K^5}}.
\end{eqnarray}

We follow this coarse-graining with the rescaling transformation
$R_{\parallel}(e^{\omega\,\delta l})R_{\perp}(e^{\delta l})$ to
restore the uv momentum cutoff back to $Q$. The infinitesimal
rescaling leads to the following corrections to the coupling
constants:
\begin{subequations}
\begin{eqnarray}
(\delta_r B_z, \delta_r \lambda_{z\perp}, \delta_r \lambda, \delta_r \mu)
&=& (d+3-3\,\omega) (B_z, \lambda_{z\perp},\lambda,\mu) \delta l,
\\
\delta_r K &=& (d-1-\omega) K \delta l,\\
\delta_r \Delta &=& (d+1-\omega) \Delta \delta l.
\end{eqnarray}
\end{subequations}

Putting together the graphical and rescaling corrections we find the
one-loop (lowest order in $\epsilon$) RG flow equations of elastic
constants and disorder variance $\Delta$:
\begin{subequations}
\label{RGcorrectionsdisorder}
\begin{eqnarray}
\frac{d\, B_z }{d\,l} &=& (d + 3 - 3 \,\omega) B_z  
   - \psi_d \frac{3\, \Delta\, (B_z - \lambda_{z\perp})^2}
        {8\, \sqrt{K^5\, \hat{\mu}}}
 ,\label{dBzdisorder}\\
\frac{d \lambda_{z\perp}}{d\,l} &=& 
  (d + 3 - 3 \,\omega) \lambda_{z\perp} 
   + \psi_d \frac{3\, \Delta\, 
	(B_z - \lambda_{z\perp}) 
        (2\lambda + \mu- 2\lambda_{z\perp})}
        {16\, \sqrt{K^5\, \hat{\mu}}}
         ,\label{dCdisorder}\\
\frac{d\, \lambda}{d\,l} &=& 
        (d + 3 - 3 \,\omega) \lambda 
        - \psi_d \frac{\Delta  
        \left(6 (\lambda -\lambda_{z\perp})^2
        +6 \mu (\lambda -\lambda_{z\perp})
        +\mu ^2\right)}{16 \sqrt{K^5 \hat{\mu }}}
        ,\label{dBpdisorder}\\
\frac{d\, \mu}{d\,l} &=& (d + 3 - 3 \,\omega) \mu 
        - \psi_d \frac{\Delta\, \mu^2}
        {16\, \sqrt{K^5\, \hat{\mu}}}
        ,\label{dmudisorder}\\
\frac{d\, K}{d\,l} &=&  (d-1 - \omega) K
      +  \psi_d  \frac{\Delta \left( B_z(\lambda+2\,\mu) 
        + 20\,\mu(\lambda+\mu) 
	- \lambda_{z\perp}(\lambda_{z\perp} + 4\,\mu ) \right)}
	{16(\lambda+2\,\mu)\sqrt{K^5\,\hat{\mu}}} 
        , \label{dKdisorder}\\
\frac{d\, \Delta}{d\,l} &=& (d+1 - \omega) 
        + \psi_d \frac{\Delta^2\sqrt{\hat{\mu}}}{32\, \sqrt{K^5}},
        \label{dDeltadisorder}
\end{eqnarray}
\end{subequations}
where here we defined
\begin{equation}
\psi_d = \frac{\Omega_{d-1} Q^{d-5}}{2\,(2 \, \pi)^{d-1}}. 
\end{equation}

We then define four dimensionless coupling constants $\mathsf{g}_{\rm
L}$, $\mathsf{g}_{\perp}$, $x$ and $y$ by
Eqs.~(\ref{coupling_replica}).  Their flow equations can be obtained
from Eqs.~(\ref{RGcorrectionsdisorder}), facilitated by Mathematica
to be:
\begin{subequations}
\begin{eqnarray}
\hspace{-5mm}
\frac{d \,\mathsf{g}_{\rm L}}{d \,l} &=& \epsilon \mathsf{g}_{\rm L}  
- \frac{\mathsf{g}_{\rm L} }{16}  \cdot 
	\frac{\Theta_{L}(\mathsf{g}_{\rm L} ,\mathsf{g}_{\perp},x,y )}
	{\Psi_{L}(\mathsf{g}_{\rm L} ,\mathsf{g}_{\perp},x,y )},
	\label{gL-flow}\\
\frac{d \, \mathsf{g}_{\perp}}{d \, l} &=& 
	\epsilon \mathsf{g}_{\perp} 
	- \frac{\mathsf{g}_{\perp} }{16}  \cdot
	\frac{\Theta_{\perp}(\mathsf{g}_{\rm L} ,\mathsf{g}_{\perp},x,y )}
	{\Psi_{\perp}(\mathsf{g}_{\rm L} ,\mathsf{g}_{\perp},x,y )},
	\label{gT-flow}\\
\frac{d \, x} {d \, l} &=& -\frac{3}{2} \mathsf{g}_{\rm L} \,x \,(1 - y^2),
\label{rho1flow}\\
\frac{d \, y} {d \, l} &=& -\frac{3}{4} \mathsf{g}_{\rm L}\, y\, (1-y^2)
\label{rho2flow}
\end{eqnarray}
\end{subequations}
where $\epsilon=5-d$ and
\begin{eqnarray}
\Theta_{L}(\mathsf{g}_{\rm L} ,\mathsf{g}_{\perp},x,y )    &=&
	-2500 \mathsf{g}_{\rm L}^3 y^4+6000 \mathsf{g}_{\rm L}^3 \sqrt{x} y^3
	+63500 \mathsf{g}_{\rm L}^2 \mathsf{g}_{\perp} \sqrt{x} y^3-10000 \mathsf{g}_{\rm L}^3 y^2
   \nonumber\\ 	
  &&	-115000 \mathsf{g}_{\rm L}^2 \mathsf{g}_{\perp} y^2-600 \mathsf{g}_{\rm L}^3 x y^2
  -207250 \mathsf{g}_{\rm L} \mathsf{g}_{\perp}^2 x y^2-37500 \mathsf{g}_{\rm L}^2 \mathsf{g}_{\perp} x y^2
    \nonumber\\  
&&  +150000 \mathsf{g}_{\perp}^3 x^{3/2} y+9600 \mathsf{g}_{\rm L} \mathsf{g}_{\perp}^2 x^{3/2} y
   -4320 \mathsf{g}_{\rm L}^2 \mathsf{g}_{\perp} x^{3/2} y-6000 \mathsf{g}_{\rm L}^3\sqrt{x} y
      \nonumber\\ 
&&+176250 \mathsf{g}_{\rm L} \mathsf{g}_{\perp}^2 \sqrt{x} y-36500 \mathsf{g}_{\rm L}^2 \mathsf{g}_{\perp} \sqrt{x} y
	+12500 \mathsf{g}_{\rm L}^3+159375 \mathsf{g}_{\rm L} \mathsf{g}_{\perp}^2
   \nonumber\\ 
&&	+60000 \mathsf{g}_{\perp}^3 x^2+24000 \mathsf{g}_{\rm L} \mathsf{g}_{\perp}^2 x^2
	+576 \mathsf{g}_{\rm L}^2 \mathsf{g}_{\perp} x^2+137500 \mathsf{g}_{\rm L}^2 \mathsf{g}_{\perp}
   \nonumber\\ 	
&&	+600 \mathsf{g}_{\rm L}^3 x+93750 \mathsf{g}_{\perp}^3 x
	+220750 \mathsf{g}_{\rm L} \mathsf{g}_{\perp}^2 x+38400 \mathsf{g}_{\rm L}^2 \mathsf{g}_{\perp} x,\\
\Psi_{L}(\mathsf{g}_{\rm L} ,\mathsf{g}_{\perp},x,y ) &=&   
	1000 \mathsf{g}_{\rm L}^2 \sqrt{x} y^3-2500 \mathsf{g}_{\rm L}^2 y^2
	-100 \mathsf{g}_{\rm L}^2 x y^2-6150 \mathsf{g}_{\rm L} \mathsf{g}_{\perp} x y^2
   \nonumber\\ 
&&	+9000 \mathsf{g}_{\perp}^2 x^{3/2} y-720 \mathsf{g}_{\rm L} \mathsf{g}_{\perp} x^{3/2} y
	-1000 \mathsf{g}_{\rm L}^2 \sqrt{x}y+4500 \mathsf{g}_{\rm L} \mathsf{g}_{\perp} \sqrt{x} y
   \nonumber\\ 
&&	+2500 \mathsf{g}_{\rm L}^2+3600 \mathsf{g}_{\perp}^2 x^2
	+96 \mathsf{g}_{\rm L} \mathsf{g}_{\perp} x^2+3750 \mathsf{g}_{\rm L}\mathsf{g}_{\perp}
   \nonumber\\ 
&&	+100 \mathsf{g}_{\rm L}^2 x+5625 \mathsf{g}_{\perp}^2 x+6300 \mathsf{g}_{\rm L}
   \mathsf{g}_{\perp} x, \\
\Theta_{\perp}(\mathsf{g}_{\rm L} ,\mathsf{g}_{\perp},x,y ) &=& 
	-2500 \mathsf{g}_{\rm L}^3 y^4+64500 \mathsf{g}_{\rm L}^2 \mathsf{g}_{\perp} \sqrt{x} y^3
	+5000 \mathsf{g}_{\rm L}^3 y^2-117500 \mathsf{g}_{\rm L}^2 \mathsf{g}_{\perp}y^2
	\nonumber\\
   &&-213400 \mathsf{g}_{\rm L} \mathsf{g}_{\perp}^2 x y^2-700 \mathsf{g}_{\rm L}^2 \mathsf{g}_{\perp} x y^2
   +159000 \mathsf{g}_{\perp}^3 x^{3/2} y-45120 \mathsf{g}_{\rm L}\mathsf{g}_{\perp}^2 x^{3/2} y
   \nonumber\\
 &&  +180750 \mathsf{g}_{\rm L} \mathsf{g}_{\perp}^2 \sqrt{x}y
    -64500 \mathsf{g}_{\rm L}^2 \mathsf{g}_{\perp} \sqrt{x} y
    -2500\mathsf{g}_{\rm L}^3+163125 \mathsf{g}_{\rm L} \mathsf{g}_{\perp}^2
   \nonumber\\ 
   && +63600 \mathsf{g}_{\perp}^3x^2+2496 \mathsf{g}_{\rm L} \mathsf{g}_{\perp}^2 x^2
   +117500 \mathsf{g}_{\rm L}^2\mathsf{g}_{\perp}+99375 \mathsf{g}_{\perp}^3 x
      \nonumber\\ 
  && +193300 \mathsf{g}_{\rm L} \mathsf{g}_{\perp}^2x+700 \mathsf{g}_{\rm L}^2 \mathsf{g}_{\perp} x,\\
\Psi_{\perp}(\mathsf{g}_{\rm L} ,\mathsf{g}_{\perp},x,y ) &=&
	1000 \mathsf{g}_{\rm L}^2 \sqrt{x} y^3-2500 \mathsf{g}_{\rm L}^2 y^2
	-100 \mathsf{g}_{\rm L}^2 x y^2-6150 \mathsf{g}_{\rm L} \mathsf{g}_{\perp} x y^2   
	\nonumber\\ 
 &&	+9000 \mathsf{g}_{\perp}^2 x^{3/2}y - 720 \mathsf{g}_{\rm L} \mathsf{g}_{\perp} x^{3/2} y
	-1000 \mathsf{g}_{\rm L}^2 \sqrt{x}y+4500 \mathsf{g}_{\rm L} \mathsf{g}_{\perp} \sqrt{x} y
	   \nonumber\\ 
 &&	+2500 \mathsf{g}_{\rm L}^2+3600\mathsf{g}_{\perp}^2 x^2
	+96 \mathsf{g}_{\rm L} \mathsf{g}_{\perp} x^2+3750 \mathsf{g}_{\rm L}\mathsf{g}_{\perp}
   \nonumber\\ 
 &&	+100 \mathsf{g}_{\rm L}^2 x+5625 \mathsf{g}_{\perp}^2 x+6300 \mathsf{g}_{\rm L} \mathsf{g}_{\perp} x.   
\end{eqnarray}

As noted earlier the bare values of $x$ and $y$ for typical elastomers
are much smaller than unity.  Furthermore, below five dimension, we
expect $\mathsf{g}_{\rm L}(l)$ to flow to a finite positive value
$\mathsf{g}_{\rm L}^*$.  These considerations, together with
Eq.~(\ref{rho1flow}) and Eq.~(\ref{rho2flow}) indicate that $x(l)$ and
$y(l)$ start from small values and flow to zero exponentially
according to:
\begin{subequations}
\begin{eqnarray}
x(l) &\approx& x\, e^{-3 \mathsf{g}^*_L\,l/2},\\ 
y(l) &\approx& y\, e^{-3 \mathsf{g}^*_L\,l/4}.
\end{eqnarray}\end{subequations}
Consequently we can set them to zero in the flow equations for
$\mathsf{g}_{\rm L}$ and $\mathsf{g}_{\perp}$.  This leads to a
considerable simplification of Eq.~(\ref{gL-flow}) and
Eq.~(\ref{gL-flow}):
\begin{subequations}
\begin{eqnarray}
\frac{d \,\mathsf{g}_{\rm L}}{d \,l} &=& \epsilon {\mathsf{g}_{\rm L}} 
-\frac{5 \mathsf{g}_{\rm L} \left(4 \mathsf{g}_{\rm L}^2+44 \mathsf{g}_{\perp} \mathsf{g}_{\rm L}+51
   \mathsf{g}_{\perp}^2\right)}{64 \mathsf{g}_{\rm L}+96 \mathsf{g}_{\perp}},
   \label{flow_gL3-d}\\
 \frac{d \, \mathsf{g}_{\perp}}{d \, l} &=&
  \epsilon \mathsf{g}_{\perp} 
-\frac{\mathsf{g}_{\perp} \left(-4 \mathsf{g}_{\rm L}^2+188 \mathsf{g}_{\perp} \mathsf{g}_{\rm L}+261
   \mathsf{g}_{\perp}^2\right)}{64 \mathsf{g}_{\rm L}+96 \mathsf{g}_{\perp}}.
  \qquad\label{flow_gperp3-d}
\end{eqnarray} \end{subequations}

Using Eqs.~(\ref{def_moduli_2}), Eqs.~(\ref{RGcorrectionsdisorder}),
and Eqs.~(\ref{coupling_replica}), we can derive flow equations for
elastic constants $B$, $C$, $\mu_{\rm L}$, and $\mu$. The results are
shown in Eqs.~(\ref{flow_2}).  The exponents $\eta_B$, $\eta_{\rm L}$,
$\eta_{\perp}$ are the same as in the first three equations of
Eqs.~(\ref{def_eta}), which we duplicate below:
\begin{subequations}
\label{eta_exponents-3}
\begin{eqnarray}
\eta_B  = \frac{3}{8}\, y ^2\, g_{\rm L} , \hspace{3mm}
\eta_{L}  = \frac{3}{8}\,g_{\rm L} ,  \hspace{3mm}
\eta_{\perp}  = \frac{1}{16}\, g_{\perp}.
%
\end{eqnarray}
\end{subequations}
$\eta_K$ and $\eta_{\Delta}$ are more complicated but can be
straightforwardly found with the crutch of Mathematica to help with the
algebra:
\begin{subequations}
\label{eta_exponents-4}
\begin{eqnarray}
\eta_{K} &=& \frac{5 \left(-10 \left(y^2-1\right) \mathsf{g}_{\rm L}^2+\mathsf{g}_{\perp}
   \left(24 x-80 y \sqrt{x}+175\right) \mathsf{g}_{\rm L}+100 \mathsf{g}_{\perp}^2
   x\right)}{16 K^2 \left(75 \mathsf{g}_{\perp} x+2 \mathsf{g}_{\rm L} \left(x-10 y
   \sqrt{x}+25\right)\right)},\\
\eta_{\Delta} &=&\frac{\mathsf{g}_{\rm L} \left(3 \mathsf{g}_{\perp} \left(16 x+40 y
   \sqrt{x}+25\right)-50 \mathsf{g}_{\rm L} \left(y^2-1\right)\right)}{75
   \mathsf{g}_{\perp} x+2 \mathsf{g}_{\rm L} \left(x-10 y \sqrt{x}+25\right)}. 
\end{eqnarray} 
\end{subequations}
In the limit $x$ and $y$ approaching zero, these reduce to the last
two equations in Eqs.~(\ref{def_eta}).

\end{widetext}
\bibliography{reference-NE-long}

\begin{thebibliography}{94}
\bibitem[{com({\natexlab{b}})}]{comment:glass}
\bibinfo{note}{It is fair to say that no complete description of structural
  glasses exists at the moment. One school of thought advocates glasses to be
  indistinguishable from ordinary fluids in their thermodynamic static
  properties, but with an extremely large viscosity that makes them behave
  elastically on experimentally accessible time scales.}
\bibitem[{com({\natexlab{c}})}]{comment:vulcanization}
\bibinfo{note}{It is worth to note that recently a rather sophisticated
  analytical approach for these vulcanized materials was developed
  \cite{Deam-Edwards,Vulcan_Goldbart}, which seems to be able to capture the
  network heterogeneity. Further studies within the framework of this
  vulcanization theory may yield information on details of rubber
  heterogeneity, that for the purposes of the present work we take as a working
  assumption.}
\bibitem[{com({\natexlab{d}})}]{comment:possibilities}
\bibinfo{note}{One possibility is the slowness of equilibration rates (outside
  of experimentally accessible time scales) toward a nematically-ordered
  thermodynamic ground state. Another, is the intrinsic thermodynamic
  instability of the nematic order in the presence of polymer network
  heterogeneities. Distinguishing between these two qualitatively distinct
  possibilities is in fact one of the fundamental goals of this theoretical
  research, paralleling similar efforts in random magnets
  \cite{PhysicsTodayFisher}.}
\bibitem[{com({\natexlab{f}})}]{comment:similarity}
\bibinfo{note}{The reader may notice the similarity between this soft mode of
  nematic elastomers and rotation of layers in a smectic liquid crystal.
  However, there is a qualitative difference. In a smectic liquid crystal, the
  nematic director is always locked to the layer normal, while, in contrast, in
  a nematic elastomer, the nematic director is tilted away from the principle
  axis of the elastic soft deformation.}
\bibitem[{com({\natexlab{g}})}]{comment:quantum}
\bibinfo{note}{It is worth noting that a three-dimensional system with
  power-law correlations phenomenology extending across a whole phase has been
  a holy-grail in the field of quantum strongly-correlated systems, where it is
  hoped to explain anomalous experimental observations in a variety of
  materials that do not neatly fall into conventional (e.g., Landau's Fermi
  liquid) solid state paradigm \cite{algebraicLiquidHermele}.}
\bibitem[{com({\natexlab{h}})}]{comment:first-order}
\bibinfo{note}{Such a first-order transition is not seen in real experiment,
  believed to be destroyed by gel's heterogeneity \cite{Selinger02}.}
\bibitem[{com({\natexlab{j}})}]{comment:abuse}
\bibinfo{note}{This type of abuse of notations is common for physicists.
  Rigorously speaking, we should use different notation for the functions when
  we change the variable, i.e., $\tilde{\rv}(\xv) = \rv(\Xv)$ etc.}
\bibitem[{com({\natexlab{l}})}]{comment:left-strain}
\bibinfo{note}{Which is usually called the left Cauchy-Green strain tensor in
  nonlinear elasticity theory.}
\bibitem[{com({\natexlab{m}})}]{comment:nh}
\bibinfo{note}{The reader should not confuse $\nh$ with the nematic director
  $\nh_0$ of the nematic reference state.}
\bibitem[{com({\natexlab{n}})}]{comment:volume-measure}
\bibinfo{note}{Strictly speaking we should distinguish the Lagrangian volume
  measured in referential coordinate and the Eulerian volume measure in current
  coordinate. However, since rubbery materials are nearly incompressible, the
  difference is usually negligible. We should also be careful with the terms
  elastic free energy and elastic energy. It is understood that the elastic
  energies we discuss in this section are really Landau elastic free energy
  which already included thermal fluctuations below certain length-scale, e.g.,
  that of a mesh size of the polymer network. Fluctuations on a longer
  length-scale will be studied in later sections.}
\bibitem[{com({\natexlab{o}})}]{comment:energy}
\bibinfo{note}{This elastic energy is equivalent to the one derived in
  reference \cite{LMRX}, (equation number 4.22), in spite of different
  notations used there.}
\bibitem[{com({\natexlab{p}})}]{comment:f-e}
\bibinfo{note}{Remember that we have defined $f_e$ to be the elastic energy per
  unit volume measured in the nematic referential coordinate $\xv$.}
\bibitem[{com({\natexlab{q}})}]{comment:Frank}
\bibinfo{note}{In principle, there may be higher order couplings between
  director distortion $\nabla \nh$ and the deformation gradient $\lm$. We shall
  not consider them since they are less relevant than the terms we study here.}
\bibitem[{com({\natexlab{r}})}]{comment:not-original}
\bibinfo{note}{The discussion from here to the paragraph before
  Eq.~(\ref{F_tot}) also applies to fluid and is not original. The reader can
  find similar presentations in various textbooks and monographs on micropolar
  elasticity, such as reference \cite{EL:Nowacki}.}
\bibitem[{com({\natexlab{s}})}]{comment:observation}
\bibinfo{note}{This observation is of course also true for a nematic liquid
  crystal, and actually is the starting point of nematic hydrodynamics
  \cite{LC:deGennes}.}
\bibitem[{com({\natexlab{t}})}]{comment:torque}
\bibinfo{note}{There is of course a separate contribution to the torque coming
  from the force transmitted through the surface, which acts on the {\em
  translational degrees of freedom}.}
\bibitem[{com({\natexlab{w}})}]{comment:phonon-linear}
\bibinfo{note}{In contrast, a phonon field linear in $\xv$ (is unbound and
  therefore does not have well-defined Fourier transform) corresponds to a
  macroscopic deformation, which is incompatible with the prescribed constant
  strain ensemble. Also, we note that the $\qv=0$ component of $\uv$ clearly
  corresponds to a uniform translation of the solid and thus does not
  contribute to the elastic free energy.}
\bibitem[{com({\natexlab{x}})}]{comment:S}
\bibinfo{note}{Since $\mm{S}$ is independent of spatial coordinates, we can
  always continue it to the whole region $\Omega_0$.}
\bibitem[{com({\natexlab{y}})}]{comment:analogue}
\bibinfo{note}{They are the analogues of the Helmholtz free energy and Gibbs
  free energy respectively in the case of magnetism or gas thermodynamics.}
\bibitem[{com({\natexlab{z}})}]{comment:counter-term}
\bibinfo{note}{In field theory, such terms are referred to as the ``counter
  terms''.}
\bibitem[{com({\natexlab{{}}})}]{comment:linear-terms}
\bibinfo{note}{This is quite analogous to many other ordered states, such as
  for example a ferromagnet, where to include fluctuations, one has a choice of
  either expanding about a minimum of the Hamiltonian or about a true minimum
  of the free energy.}
\bibitem[{com({\natexlab{|}})}]{comment:condition}
\bibinfo{note}{This can be easily seen from Eq.~(\ref{elast-energy-2}) and
  Eq.~(\ref{V_ni}). For a fixed geometric configuration $\rv(\xv)$, the nematic
  director fluctuation $\delta \nh$ is ``massively'' tied to the strain
  deformation through the last term in Eq.~(\ref{elast-energy-2}), as long as
  $r \neq 1$, i.e., the nematic order does not vanish. This guarantees that the
  fluctuations of the nematic director are always finite.}
\bibitem[{com({\natexlab{}})}]{comment:dimension}
\bibinfo{note}{In the case of an arbitrary $d$ dimension, it is clear that only
  the first $d$ of these invariants are independent.}
\bibitem[{com({\natexlab{~}})}]{comment:bending}
\bibinfo{note}{The bend term is still important to stabilize long wavelength
  fluctuations of the $\vec{u}_{\perp}$ field, but does not affect the RG
  analysis. We shall consider this term again in the renormalized theory.}
\bibitem[{com()}]{comment:small-region}
\bibinfo{note}{The small region in $\qv$ space where this is not true is not
  important for calculation of real space fluctuations.}
\bibitem[{com()}]{comment:natural}
\bibinfo{note}{A more natural definition of the physical phonon field is the
  displacement from the average position of mass points. See
  Sec.~\ref{Sec:reference-state} for a discussion of this issue.}
\bibitem[{com()}]{comment:rescaling}
\bibinfo{note}{We often also rescale the field $\uv^<$ appropriately.}
\end{thebibliography}
\end{document}